\documentclass[aps,prc,twocolumn,floatfix,nofootinbib,showpacs,superscriptaddress]{revtex4-1}

\usepackage{amsmath}
\usepackage{amsfonts,amssymb,bm}
\usepackage{color}
\usepackage{graphicx}
\usepackage{pzccal}

\definecolor{darkgreen}{rgb}{0,0.5,0}
\definecolor{purple}{rgb}{0.5,0,0.5}
\definecolor{nblue}{rgb}{0.0,0.0,0.50}
\definecolor{scarlet}{rgb}{1.0,0.2,0}

\newcommand{\lsim}{\mathrel{\rlap{\lower4pt\hbox{\hskip0pt$\sim$}}
\raise1pt\hbox{$<$}}}           
\newcommand{\gsim}{\mathrel{\rlap{\lower4pt\hbox{\hskip0pt$\sim$}}
\raise1pt\hbox{$>$}}}           

\newcommand{\biggL}{\mbox{$\large \left[\rule{0ex}{2ex} \right.$}}
\newcommand{\biggR}{\mbox{$\large \left.\rule{0ex}{2ex} \right]$}}

\numberwithin{equation}{section}
\numberwithin{figure}{section}
\numberwithin{table}{section}

\begin{document}

\title{Collective perspective on advances in Dyson-Schwinger Equation QCD}


\author{Adnan Bashir}
\affiliation{Instituto de F\'isica y Matem\'aticas, Universidad
Michoacana de San Nicol\'as de Hidalgo, Edificio C-3, Ciudad
Universitaria, Morelia, Michoac\'an 58040, M\'exico.}
\affiliation{Physics Division, Argonne National Laboratory, Argonne, Illinois 60439, USA}
\affiliation{Center for Nuclear Research, Department of Physics, Kent State University, Kent OH 44242, USA}

\author{Lei Chang}
\affiliation{Physics Division, Argonne National Laboratory, Argonne, Illinois 60439, USA}

\author{Ian C.~Clo\"et}
\affiliation{CSSM and CoEPP, School of Chemistry and Physics
University of Adelaide, Adelaide SA 5005, Australia}

\author{Bruno~El-Bennich}
\affiliation{Laboratorio de F\'isica Te\'orica e Computa\c{c}\~ao Cient\'ifica, Universidade Cruzeiro do Sul, 01506-000 S\~ao Paulo, SP, Brazil}
\affiliation{Instituto de F\'isica Te\'orica, Universidade Estadual Paulista, 01140-070 S\~ao Paulo, SP, Brazil}

\author{Yu-xin Liu}
\affiliation{Department of Physics, Center for High Energy Physics
and State Key Laboratory of Nuclear Physics and Technology, Peking
University, Beijing 100871, China}
\affiliation{Center of
Theoretical Nuclear Physics, National Laboratory of Heavy Ion
Accelerator, Lanzhou 730000, China}

\author{Craig D.~Roberts}
\affiliation{Physics Division, Argonne National Laboratory, Argonne, Illinois 60439, USA}
\affiliation{Department of Physics, Center for High Energy Physics and State Key Laboratory of Nuclear Physics and Technology, Peking University, Beijing 100871, China}
\affiliation{Institut f\"ur Kernphysik, Forschungszentrum J\"ulich, D-52425 J\"ulich, Germany}
\affiliation{Department of Physics, Illinois Institute of Technology, Chicago, Illinois 60616-3793, USA}

\author{Peter~C.~Tandy}
\affiliation{Center for Nuclear Research, Department of Physics, Kent State University, Kent OH 44242, USA}

\begin{abstract}
%
We survey contemporary studies of hadrons and strongly interacting quarks using QCD's Dyson-Schwinger equations, addressing:
aspects of confinement and dynamical chiral symmetry breaking; the hadron spectrum; hadron elastic and transition form factors, from small- to large-$Q^2$; parton distribution functions; the physics of hadrons containing one or more heavy quarks; and properties of the quark gluon plasma.
\medskip

\end{abstract}

\maketitle

\tableofcontents


\section{Introduction}
A hundred years and more of fundamental research in atomic and nuclear physics has shown that all matter is corpuscular, with the atoms that comprise us, themselves containing a dense nuclear core.  This core is composed of protons and neutrons, referred to collectively as nucleons, which are members of a broader class of femtometre-scale particles, called hadrons.  In working toward an understanding of hadrons, we have discovered that they are complicated bound-states of quarks and gluons.  These quarks and gluons are elementary, pointlike excitations, whose interactions are described by a Poincar\'e invariant quantum non-Abelian gauge field theory; namely, quantum chromodynamics (QCD).  The goal of hadron physics is the provision of a quantitative explanation of the properties of hadrons through a solution of QCD.

Quantum chromodynamics is the strong-interaction part of the Standard Model of Particle Physics and solving QCD presents a fundamental problem that is unique in the history of science.  Never before have we been confronted by a theory whose elementary excitations are not those degrees-of-freedom readily accessible via experiment; i.e., whose elementary excitations are \emph{confined}.  Moreover, there are numerous reasons to believe that QCD generates forces which are so strong that less-than 2\% of a nucleon's mass can be attributed to the so-called current-quark masses that appear in QCD's Lagrangian; viz., forces capable of generating mass from nothing, a phenomenon known as dynamical chiral symmetry breaking (DCSB).

Neither confinement nor DCSB is apparent in QCD's Lagrangian and yet they play the dominant role in determining the observable characteristics of real-world QCD.  The physics of hadrons is ruled by \emph{emergent phenomena} such as these, which can only be elucidated through the use of nonperturbative methods in quantum field theory.  This is both the greatest novelty and the greatest challenge within the Standard Model.  We must find essentially new ways and means to explain precisely via mathematics the observable content of QCD.

The complex of Dyson-Schwinger equations (DSEs) is a powerful tool, which has been employed with marked success to study confinement and DCSB, and their impact on hadron observables.  This will be emphasised and exemplified in this synopsis of the KITPC lecture series, which describes selected progress in the study of mesons and baryons, and strongly coupled quarks in-medium.  Our compilation complements and extends earlier and other reviews \cite{Roberts:1994dr,Roberts:2000aa,Maris:2003vk,Pennington:2005be,Holl:2006ni,%
Fischer:2006ub,Roberts:2007jh,Roberts:2007ji,Holt:2010vj,Chang:2010jq,Swanson:2010pw,%
Chang:2011vu,Boucaud:2011ug}.

\section{Hadron Physics}
\label{sect:HP}
The basic problem of hadron physics is to solve QCD.  This inspiring goal will only be achieved through a joint effort from experiment and theory because it is the feedback between them that leads most rapidly to improvements in understanding.  The hadron physics community now has a range of major facilities that are accumulating data, of unprecedented accuracy and precision, which pose important challenges for theory.  The opportunities for researchers in hadron physics promise to expand with the use of extant accelerators, and upgraded and new machines and detectors that will appear on a five-to-ten-year time-scale, in China, Germany, Japan, Switzerland and the USA.
A short list of facilities may readily be compiled: Beijing's electron-positron collider; in Germany --
COSY (J\"ulich Cooler Synchrotron),
ELSA (Bonn Electron Stretcher and Accelerator),
MAMI (Mainz Microtron), and
FAIR (Facility for Antiproton and Ion Research) under construction near Darmstadt;
in Japan -- J-PARC (Japan Proton Accelerator Research Complex) under construction in Tokai-Mura, 150km NE of Tokyo, and
KEK, Tsukuba;
in Switzerland, the ALICE and COMPASS detectors at CERN;
and in the USA, both the Thomas Jefferson National Accelerator Facility (JLab), currently being upgraded, with new generation experiments expected in 2016, and RHIC (Relativistic Heavy Ion Collider) at Brookhaven National Laboratory.  

Asymptotic coloured states have not been observed, but is it a cardinal fact that they cannot?  No solution to QCD will be complete if it does not explain confinement.  This means confinement in the real world, which contains quarks with light current-quark masses.  This is distinct from the artificial universe of pure-gauge QCD without dynamical quarks, studies of which tend merely to focus on achieving an area law for a Wilson loop and hence are irrelevant to the question of light-quark confinement.

In stepping toward an answer to the question of confinement, it will likely be necessary to map out the long-range behaviour of the interaction between light-quarks; namely, QCD's $\beta$-function at infrared momenta.  In this connection it is noteworthy that the spectrum of meson and baryon excited states, hadron elastic and transition form factors, and the phase structure of dense and hot QCD, all contribute information that is critical to elucidating the long-range interaction between light-quarks.  In addition, by studying such quantities one may expose the distribution of a hadron's characterising properties -- such as mass and momentum, linear and angular -- amongst its QCD constituents; and illuminate strong interaction properties that were critical in the universe's evolution.  The upgraded and promised future facilities will provide data that should guide the charting process.   However, to make full use of that data, it will be necessary to have Poincar\'e covariant theoretical tools that enable the reliable study of hadrons in the mass range $1$-$2\,$GeV and in the vicinity of phase boundaries.  Crucially, on these domains both confinement and DCSB are germane.

\begin{figure}[t]
\vspace*{1ex}

\centerline{
\includegraphics[clip,width=0.4\textwidth]{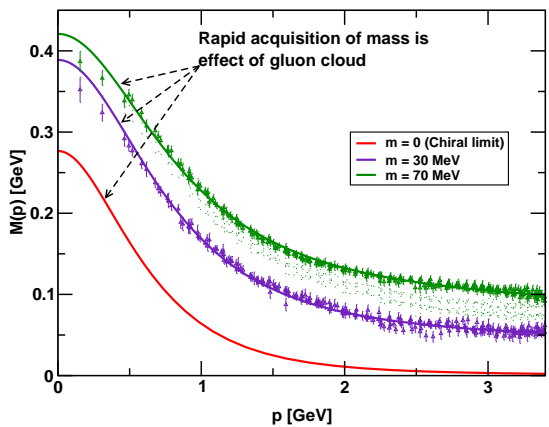}}

\caption{\label{gluoncloud}
Dressed-quark mass function, $M(p)$ in Eq.\,(\protect\ref{SgeneralN}): \emph{solid curves} -- DSE results, explained in Refs.\,\protect\cite{Bhagwat:2003vw,Bhagwat:2006tu}, ``data'' -- numerical simulations of lattice-QCD \protect\cite{Bowman:2005vx}.  (NB.\ $m=70\,$MeV is the uppermost curve and current-quark mass decreases from top to bottom.)  One observes the current-quark of perturbative QCD evolving into a constituent-quark as its momentum becomes smaller.  The constituent-quark mass arises from a cloud of low-momentum gluons attaching themselves to the current-quark.  This is dynamical chiral symmetry breaking (DCSB): an essentially nonperturbative effect that generates a quark mass \emph{from nothing}; namely, it occurs even in the chiral limit.  
}
\end{figure}

It is known that DCSB; namely, the generation of mass \emph{from nothing}, does take place in QCD.  It arises primarily because a dense cloud of gluons comes to clothe a low-momentum quark \cite{Bhagwat:2007vx}.  This is readily seen by solving the DSE for the dressed-quark propagator; i.e., the gap equation, Eq.\,\eqref{gendseN},\footnote{In our Euclidean metric:  $\{\gamma_\mu,\gamma_\nu\} = 2\delta_{\mu\nu}$; $\gamma_\mu^\dagger = \gamma_\mu$; $\gamma_5= \gamma_4\gamma_1\gamma_2\gamma_3$, tr$[\gamma_4\gamma_\mu\gamma_\nu\gamma_\rho\gamma_\sigma]=-4 \epsilon_{\mu\nu\rho\sigma}$; $\sigma_{\mu\nu}=(i/2)[\gamma_\mu,\gamma_\nu]$; $a \cdot b = \sum_{i=1}^4 a_i b_i$; and $P_\mu$ timelike $\Rightarrow$ $P^2<0$.  More information is available, e.g., in App.\,A of Ref.\,\protect\cite{Holl:2006ni}.}  which yields the result illustrated in Fig.\,\ref{gluoncloud}.
However, the origin of the interaction strength at infrared momenta, which guarantees DCSB through the gap equation, is currently unknown.  This relationship ties confinement to DCSB.  The reality of DCSB means that the Higgs mechanism is largely irrelevant to the bulk of normal matter in the universe.  Instead the single most important mass generating mechanism for light-quark hadrons is the strong interaction effect of DCSB;\footnote{In announcing hints for the Higgs boson at CERN, the following observations was made: ``The Higgs field is often said to give mass to everything.  That is wrong.  The Higgs field only gives mass to some very simple particles.  The field accounts for only one or two percent of the mass of more complex things like atoms, molecules and everyday objects, from your mobile phone to your pet llama.  The vast majority of mass comes from the energy needed to hold quarks together inside atoms.''  This recognises the fundamental role of confinement and dynamical chiral symmetry breaking, key emergent phenomena within the strongly interacting part of the Standard Model.}
e.g., one can identify it as being responsible for 98\% of a proton's mass \cite{Flambaum:2005kc,Holl:2005st}.

There is a caveat; namely, as so often, the pion is exceptional.  Its mass is given by the simple product of two terms, one of which is the ratio of two order parameters for DCSB whilst the other is determined by the current-quark mass (Sec.\,\ref{psmassformula}).  Hence the pion would be massless in the absence of a mechanism that can generate a current-mass for at least one light-quark.  The impact of a massless, strongly-interacting particle on the physics of the Universe would be dramatic.

It is natural to ask whether the connection between confinement and DCSB is accidental or causal.  There are models with DCSB but not confinement, however, an in-vacuum model with confinement but lacking DCSB has not yet been identified (see, e.g., Secs.\,2.1 and 2.2 of Ref.\,\cite{Roberts:2007jh}).  This leads to a conjecture that DCSB is a necessary consequence of confinement.  It is interesting that there are numerous models and theories which exhibit both confinement and DCSB, and possess an external control parameter such that deconfinement and chiral symmetry restoration occur simultaneously at some critical value of this parameter; e.g., quantum electrodynamics in three dimensions with $N_f$ electrons \cite{Bashir:2008fk,Bashir:2009fv,Hofmann:2010zy}, and models of QCD at nonzero temperature and chemical potential \cite{Bender:1996bm,Blaschke:1997bj,Bender:1997jf,Chen:2008zr,Fischer:2009gk,Qin:2010nq,%
Qin:2010pc,Liu:2011zz,Qin:2011zz,Ayala:2011vs}.  Whether this simultaneity is a property possessed by QCD, and/or some broader class of theories, in response to changes in: the number of light-quark flavours; temperature; or chemical potential, is a longstanding question.

The momentum-dependence of the quark mass, illustrated in Fig.\,\ref{gluoncloud}, is an essentially quantum field theoretic effect, unrealisable in quantum mechanics, and a fundamental feature of QCD.  This single curve connects the infrared and ultraviolet regimes of the theory, and establishes that the constituent-quark and current-quark masses are simply two connected points separated by a large momentum interval.  The curve shows that QCD's dressed-quark behaves as a constituent-quark, a current-quark, or something in between, depending on the momentum of the probe which explores the bound-state containing the dressed-quark.  It follows that calculations addressing momentum transfers $Q^2 \gsim M^2$, where $M$ is the mass of the hadron involved, require a Poincar\'e-covariant approach that can veraciously realise quantum field theoretical effects \cite{Cloet:2008re}.  Owing to the vector-exchange character of QCD, covariance also guarantees the existence of nonzero quark orbital angular momentum in a hadron's rest-frame \cite{Bhagwat:2006xi,Bhagwat:2006pu,Cloet:2007pi,Roberts:2007ji}.

The dressed-quark mass function has a remarkable capacity to correlate and to contribute significantly in explaining a wide range of diverse phenomena.  This brings urgency to the need to understand the relationship between parton properties in the light-front frame, whose peculiar properties simplify some theoretical analyses, and the structure of hadrons as measured in the rest frame or other smoothly related frames.  This is a problem because, e.g., DCSB, an established keystone of low-energy QCD, has not explicitly been realised in the light-front formulation.  The obstacle is the constraint $k^+:=k^0+k^3>0$ for massive quanta on the light front \cite{Brodsky:1991ir}.  It is therefore impossible to make zero momentum Fock states that contain particles and hence the vacuum is ``trivial''.
On the other hand, it is conceivable that DCSB is inextricably tied with the formation and structure of Goldstone modes and not otherwise a measurable property of the vacuum.  This conjecture is being explored \cite{Brodsky:2010xf,Brodsky:2009zd,Chang:2011mu,Glazek:2011vg} and is something about which more will be written herein (Sec.\,\ref{sec:inmeson}).
In addition, parton distribution functions, which have a probability interpretation in the infinite momentum frame, must be calculated in order to comprehend their content: parametrisation is insufficient.  It would be very interesting to know, e.g., how, if at all, the distribution functions of a Goldstone mode differ from those of other hadrons \cite{Holt:2010vj}.

\section{Confinement}
\label{Sect:Conf}
It is worth stating plainly that the potential between infinitely-heavy quarks measured in numerical simulations of quenched lattice-regularised QCD -- the so-called static potential -- is simply \emph{irrelevant} to the question of confinement in the real world, in which light quarks are ubiquitous.  In fact, it is a basic feature of QCD that light-particle creation and annihilation effects are essentially nonperturbative and therefore it is impossible in principle to compute a potential between two light quarks \cite{Bali:2005fu,Chang:2009ae}.

Drawing on a long list of sources; e.g., Refs.\,\cite{Gribov:1999ui,Munczek:1983dx,Stingl:1983pt,Cahill:1988zi}, a perspective on confinement was laid out in Ref.\,\cite{Krein:1990sf}.  Confinement can be related to the analytic properties of QCD's Schwinger functions, which are often called Euclidean-space Green functions.  For example, it can be read from the reconstruction theorem \cite{SW80,GJ81} that the only Schwinger functions which can be associated with expectation values in the Hilbert space of observables; namely, the set of measurable expectation values, are those that satisfy the axiom of reflection positivity.  This is an extremely tight constraint.  It can be shown to require as a necessary condition that the Fourier transform of the momentum-space Schwinger function is a positive-definite function of its arguments.  This condition suggests a practical confinement test, which can be used with numerical solutions of the DSEs (see, e.g., Sec.\,III.C of Ref.\,\cite{Hawes:1993ef} and Sec.\,IV of Ref.\,\cite{Maris:1995ns}).  The implications and use of reflection positivity are discussed and illustrated in Sec.~2 of Ref.\,\cite{Roberts:2007ji}.

\begin{figure}[t]

\centerline{
\includegraphics[clip,width=0.4\textwidth]{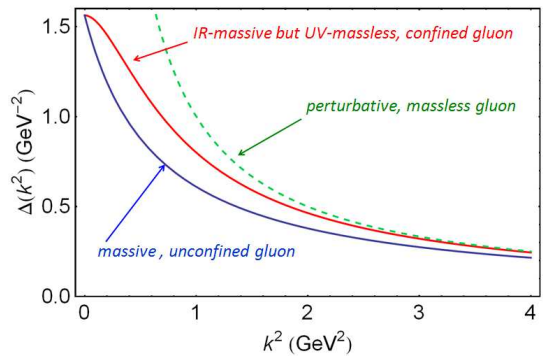}}

\caption{\label{fig:gluonrp}
$\Delta(k^2)$, the function that describes dressing of a Landau-gauge gluon propagator, plotted for three distinct cases.
A bare gluon is described by $\Delta(k^2) = 1/k^2$ (the dashed line), which is plainly convex on $k^2\in (0,\infty)$.  Such a propagator has a representation in terms of a non-negative spectral density.
In some theories, interactions generate a mass in the transverse part of the gauge-boson propagator, so that $\Delta(k^2) = 1/(k^2+m_g^2)$, which can also be represented in terms of a non-negative spectral density.
In QCD, however, self-interactions generate a momentum-dependent mass for the gluon, which is large at infrared momenta but vanishes in the ultraviolet \protect\cite{Boucaud:2011ug}.  This is illustrated by the curve labelled ``IR-massive but UV-massless.''  With the generation of a mass-\emph{function}, $\Delta(k^2)$ exhibits an inflexion point and hence cannot be expressed in terms of a non-negative spectral density.
}
\end{figure}

It is noteworthy that any 2-point Schwinger function with an inflexion point at $p^2 > 0$ must breach the axiom of reflection positivity, so that a violation of positivity can be determined by inspection of the pointwise behaviour of the Schwinger function in momentum space (Sec.\,IV.B of Ref.\,\cite{Bashir:2008fk}).
Consider then $\Delta(k^2)$, which is the single scalar function that describes the dressing of a Landau-gauge gluon propagator.  A large body of work has focused on exposing the behaviour of $\Delta(k^2)$ in the pure Yang-Mills sector of QCD.  These studies are reviewed in Ref.\,\cite{Boucaud:2011ug}.
A connection with the expression and nature of confinement in the Yang-Mills sector is indicated in Fig.\,\ref{fig:gluonrp}.  The appearance of an inflexion point in the two-point function generated by the gluon's momentum-dependent mass-function is impossible to overlook.  Hence this gluon cannot appear in the Hilbert space of observable states.  The inflexion point possessed by $M(p^2)$, visible in Fig.\,\ref{gluoncloud}, conveys the same properties on the dressed-quark propagator.

Numerical simulations of lattice-QCD confirm the appearance of an inflexion point in both the dressed-gluon and -quark propagators; e.g., see Fig.\,\ref{gluoncloud} and Ref.\,\cite{Boucaud:2011ug}.  The signal is clearest for the gluon owing to the greater simplicity of simulations in the pure Yang-Mills sector \cite{Bonnet:2000kw,Skullerud:2000un,Kamleh:2007ud}.  We emphasise that this sense of confinement is essentially quantum field theoretical in nature.  Amongst its many consequences is that light-quark confinement in QCD cannot veraciously be represented in potential models.

\begin{figure}[t]
\centerline{\includegraphics[clip,width=0.4\textwidth]{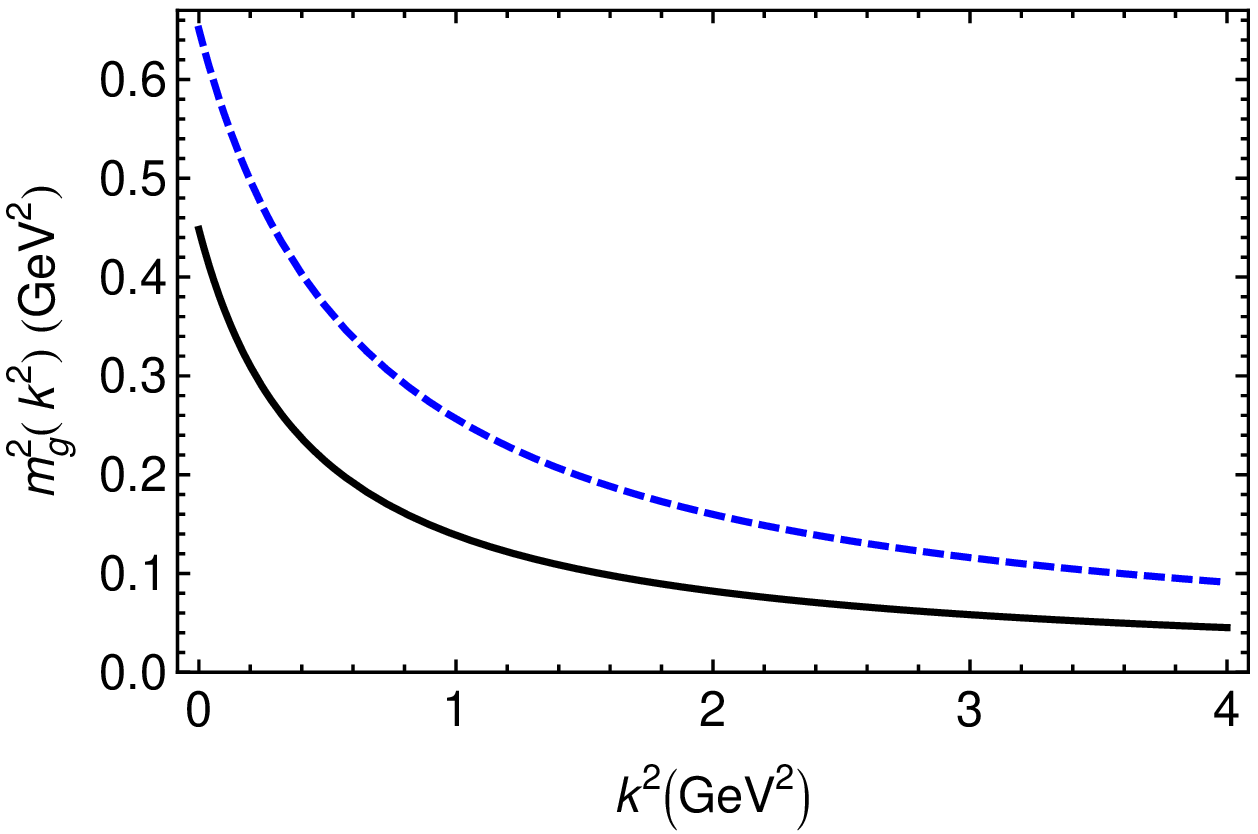}}
\centerline{\includegraphics[clip,width=0.42\textwidth]{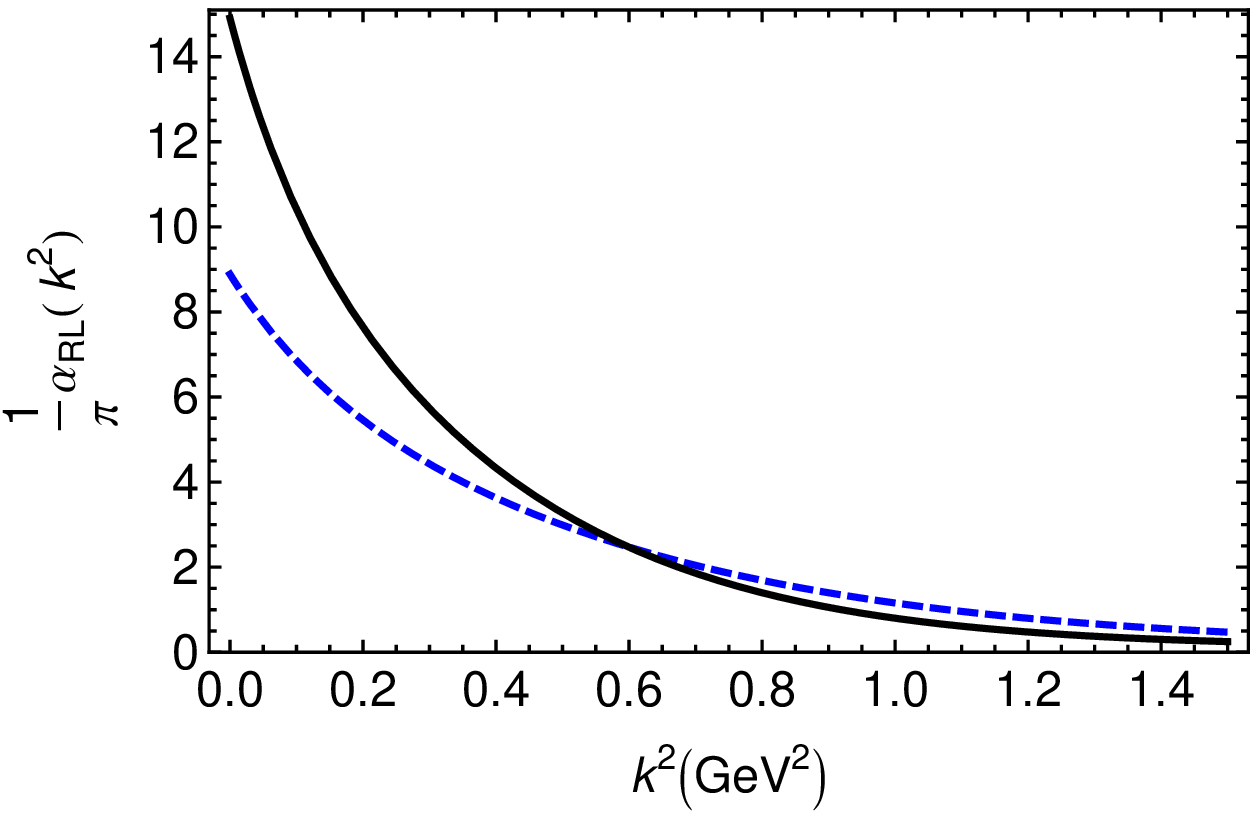}}

\caption{\label{fig:gluonrunning}
\emph{Upper panel} -- Rainbow-ladder gluon running-mass; and \emph{lower panel} -- rainbow-ladder effective running-coupling.  Matching related curves in each panel (solid to solid or dashed to dashed) gives an equivalent description of observables within the rainbow-ladder truncation.
}
\end{figure}

{F}rom the perspective that confinement can be related to the analytic properties of QCD's Schwinger functions, the question of light-quark confinement can be translated into the challenge of charting the infrared behavior of QCD's \emph{universal} $\beta$-function.  (Although this function may depend on the scheme chosen to renormalise the theory, it is unique within a given scheme \protect\cite{Celmaster:1979km}.  Of course, the behaviour of the $\beta$-function on the perturbative domain is well known.)  This is a well-posed problem whose solution is an elemental goal of modern hadron physics (e.g., Refs.\,\cite{Qin:2011dd,Brodsky:2010ur,Aguilar:2010gm}) and which can be addressed in any framework enabling the nonperturbative evaluation of renormalisation constants.  It is the $\beta$-function that is responsible for the behaviour evident in Fig.\,\ref{gluoncloud} and Fig.\,\ref{fig:gluonrunning}, below; and one of the more interesting of contemporary questions is whether it is possible to reconstruct the $\beta$-function, or at least constrain it tightly, given empirical information on the gluon and quark mass functions.

In this connection we note that the DSEs connect the $\beta$-function to experimental observables.  Hence, comparison between DSE computations and observations of the hadron mass spectrum, and elastic and transition form factors, can be used to chart $\beta$-function's long-range behaviour.  Extant studies show that the properties of hadron excited states are a great deal more sensitive to the long-range behaviour of the $\beta$-function than those of ground states.  This is illustrated in Refs.\,\cite{Qin:2011dd,Qin:2011xq}, which through a study of ground-state, radially-excited and exotic scalar-, vector- and flavoured-pseudoscalar-mesons in rainbow-ladder truncation, which is leading order in the most widely used nonperturbative scheme \cite{Munczek:1994zz,Bender:1996bb}, produced the effective coupling and running gluon mass depicted in Fig.\,\ref{fig:gluonrunning}.

\section{Gap and Bethe-Salpeter Equations}
\label{sec:gap}
In order to proceed it is necessary to describe explicitly the best known and simplest DSE.  The Dyson or gap equation determines how quark propagation is influenced by interactions; viz., for a quark of flavour $f$,
\begin{eqnarray}
\lefteqn{
\nonumber S_f(p)^{-1} = Z_2 \,(i\gamma\cdot p + m_f^{\rm bm})}\\
&& + Z_1 \int^\Lambda_q\!\! g^2 D_{\mu\nu}(p-q)\frac{\lambda^a}{2}\gamma_\mu S_f(q) \frac{\lambda^a}{2}\Gamma^f_\nu(q,p) , \rule{1em}{0ex}
\label{gendseN}
\end{eqnarray}
where: $D_{\mu\nu}$ is the gluon propagator; $\Gamma^f_\nu$, the quark-gluon vertex; $\int^\Lambda_q$, a symbol that represents a Poincar\'e invariant regularization of the four-dimensional Euclidean integral, with $\Lambda$ the regularization mass-scale; $m_f^{\rm bm}(\Lambda)$, the current-quark bare mass; and $Z_{1,2}(\zeta^2,\Lambda^2)$, respectively, the vertex and quark wave-function renormalisation constants, with $\zeta$ the renormalisation point -- dependence upon which is not usually made explicit.

The gap equation's solution is the dressed-quark propagator,
\begin{subequations}
\label{SgeneralN}
\begin{eqnarray}
 S(p) & =&
- i \gamma\cdot p \,\sigma_V(p^2,\zeta^2) + \sigma_S(p^2,\zeta^2) \,,\\
& = &  \frac{1}{i \gamma\cdot p \, A(p^2,\zeta^2) + B(p^2,\zeta^2)}\\
&=& \frac{Z(p^2,\zeta^2)}{i\gamma\cdot p + M(p^2)}\,,
\end{eqnarray}
\end{subequations}
which is obtained from Eq.\,(\ref{gendseN}) augmented by a renormalisation condition.  A mass-independent scheme is a useful choice and can be implemented by fixing all renormalisation constants in the chiral limit.\footnote{See, e.g., Ref.\,\cite{Chang:2008ec} and references therein; or Ref.\,\protect\cite{tarrach} for a detailed discussion of renormalisation.}

The mass function, $M(p^2)=B(p^2,\zeta^2)/A(p^2,\zeta^2)$, is independent of the renormalisation point, $\zeta$; and the renormalised current-quark mass,
\begin{equation}
\label{mzeta}
m_f^\zeta = Z_m(\zeta,\Lambda) \, m^{\rm bm}(\Lambda) = Z_4^{-1} Z_2\, m_f^{\rm bm},
\end{equation}
wherein $Z_4$ is the renormalisation constant associated with the Lagrangian's mass-term. Like the running coupling constant, this ``running mass'' is familiar from textbooks.  However, it is not commonly appreciated that $m^\zeta$ is simply the dressed-quark mass function evaluated at one particular deep spacelike point; viz,
\begin{equation}
m_f^\zeta = M_f(\zeta^2)\,.
\end{equation}
The renormalisation-group invariant current-quark mass may be inferred via
\begin{equation}
\label{mfhat}
\hat m_f = \lim_{p^2\to\infty} \left[\frac{1}{2}\ln \frac{p^2}{\Lambda^2_{\rm QCD}}\right]^{\gamma_m} M_f(p^2)\,,
\end{equation}
where $\gamma_m = 12/(33-2 N_f)$: $N_f$ is the number of quark flavours employed in computing the running coupling; and $\Lambda_{\rm QCD}$ is QCD's dynamically-generated renormalisation-group-invariant mass-scale.  The chiral limit is expressed by
\begin{equation}
\hat m_f = 0\,.
\end{equation}
Moreover,
\begin{equation}
\forall \zeta \gg \Lambda_{\rm QCD}, \;
\frac{m_{f_1}^\zeta}{m^\zeta_{f_2}}=\frac{\hat m_{f_1}}{\hat m_{f_2}}\,.
\end{equation}
However, we would like to emphasise that in the presence of DCSB the ratio
\begin{equation}
\frac{m_{f_1}^{\zeta=p^2}}{m^{\zeta=p^2}_{f_2}}=\frac{M_{f_1}(p^2)}{M_{f_2}(p^2)}
\end{equation}
is not independent of $p^2$: in the infrared; i.e., $\forall p^2 \lesssim \Lambda_{\rm QCD}^2$, it then expresses a ratio of constituent-like quark masses, which, for light quarks, are two orders-of-magnitude larger than their current-masses and nonlinearly related to them \cite{Flambaum:2005kc,Holl:2005st}.

The gap equation illustrates the features and flaws of each DSE.  It is a nonlinear
integral equation for the dressed-quark propagator and hence can yield much-needed nonperturbative information.  However, the kernel involves the two-point function $D_{\mu\nu}$ and the three-point function $\Gamma^f_\nu$.  The gap equation is therefore coupled to the DSEs satisfied by these functions, which in turn involve higher $n$-point functions.  Hence the DSEs are a tower of coupled integral equations, with a tractable problem obtained only once a truncation scheme is specified.  It is unsurprising that the best known truncation scheme is the weak coupling expansion, which reproduces every
diagram in perturbation theory.  This scheme is systematic and valuable in the analysis
of large momentum transfer phenomena because QCD is asymptotically free but it precludes any possibility of obtaining nonperturbative information.

Given the importance of DCSB in QCD, it is significant that the dressed-quark propagator features in the axial-vector Ward-Takahashi identity, which expresses chiral symmetry and its breaking pattern:\footnote{Section~\protect\ref{flavourless} discusses the important differences encountered in treating flavourless pseudoscalar mesons.}
\begin{eqnarray}
\nonumber
&& P_\mu \Gamma_{5\mu}^{fg}(k;P) + \, i\,[m_f(\zeta)+m_g(\zeta)] \,\Gamma_5^{fg}(k;P)\\
&=&S_f^{-1}(k_+) i \gamma_5 +  i \gamma_5 S_g^{-1}(k_-) \,,
\label{avwtimN}
\end{eqnarray}
where $P=p_1+p_2$ is the total-momentum entering the vertex and $k$ is the relative-momentum between the amputated quark legs.\footnote{To be explicit, $k=(1-\eta) p_1 + \eta p_2$, with $\eta \in [0,1]$, and hence $k_+ = p_1 = k + \eta P$, $k_- = p_2 = k - (1-\eta) P$.  In a Poincar\'e covariant approach, such as presented by a proper use of DSEs, no observable can depend on $\eta$; i.e., the definition of the relative momentum.
}
In this equation, $\Gamma_{5\mu}^{fg}$ and $\Gamma_5^{fg}$ are, respectively, the amputated axial-vector and pseudoscalar vertices.  They are both obtained from an inhomogeneous Bethe-Salpeter equation (BSE), which is exemplified here using a textbook expression \cite{Salpeter:1951sz}:
\begin{eqnarray}
\nonumber
\lefteqn{[\Gamma_{5\mu}(k;P)]_{tu} = Z_2 [\gamma_5 \gamma_\mu]_{tu}}  \\
&&
+ \int_q^\Lambda [ S(q_+) \Gamma_{5\mu}(q;P) S(q_-) ]_{sr} K_{tu}^{rs}(q,k;P), \rule{1.5em}{0ex}
\label{bsetextbook}
\end{eqnarray}
in which $K$ is the fully-amputated quark-antiquark scattering kernel, and
the colour-, Dirac- and flavour-matrix structure of the elements in the equation is  denoted by the indices $r,s,t,u$.  N.B.\ By definition, $K$ does not contain quark-antiquark to single gauge-boson annihilation diagrams, nor diagrams that become disconnected by cutting one quark and one antiquark line.

The Ward-Takahashi identity, Eq.\,(\ref{avwtimN}), entails that an intimate relation exists between the kernel in the gap equation and that in the BSE.  (This is another example of the coupling between DSEs.)  Therefore an understanding of chiral symmetry and its dynamical breaking can only be obtained with a truncation scheme that preserves this relation, and hence guarantees Eq.\,(\ref{avwtimN}) without a fine-tuning of model-dependent parameters.

\section{Nonperturbative Truncation}
\label{spectrum1}
Through the gap and Bethe-Salpeter equations the pointwise behaviour of the $\beta$-function determines the pattern of chiral symmetry breaking; e.g., the behaviour in Fig.\,\ref{gluoncloud}.  Moreover, the fact that these and other DSEs connect the $\beta$-function to experimental observables entails that comparison between computations and observations of the hadron mass spectrum, and hadron elastic and transition form factors, can be used to constrain the $\beta$-function's long-range behaviour.

In order to realise this goal, a nonperturbative symmetry-preserving DSE truncation is necessary.  Steady quantitative progress can be made with a scheme that is systematically improvable \cite{Munczek:1994zz,Bender:1996bb}.  In fact, the mere existence of such a scheme has enabled the proof of exact nonperturbative results in QCD.

Before describing a number of these in some detail, it is worth explicating the range of applications.  For example, there are:
veracious statements about the pion $\sigma$-term \cite{Flambaum:2005kc},
radially-excited and hybrid pseudoscalar mesons \cite{Holl:2004fr,Holl:2005vu},
heavy-light \cite{Ivanov:1998ms} and heavy-heavy mesons \cite{Bhagwat:2006xi}
novel results for the pion susceptibility obtained via analysis of the isovector-pseudoscalar vacuum polarisation \cite{Chang:2009at}, which bear upon the essential content of the so-called ``Mexican hat'' potential, which is used in building models for QCD;
and a derivation \cite{Chang:2008sp} of the Weinberg sum rule \cite{Weinberg:1967kj}.

\subsection{Pseudoscalar meson mass formula}
\label{psmassformula}
Turning now to a fuller illustration, the first of the results was introduced in Ref.\,\cite{Maris:1997hd}; namely, a mass formula that is exact for flavour non-diagonal pseudoscalar mesons:
\begin{equation}
\label{mrtrelation}
f_{H_{0^-}} m_{H_{0^-}}^2 = (m_{f_1}^\zeta + m_{f_2}^\zeta) \rho_{H_{0^-}}^\zeta,
\end{equation}
where: $m_{f_i}^\zeta$ are the current-masses of the quarks constituting the meson; and
\begin{eqnarray}
\lefteqn{
f_{H_{0^-}} P_\mu = \langle 0 | \bar q_{f_2} \gamma_5 \gamma_\mu q_{f_1} |H_{0^-}\rangle} \\
& = & Z_2\; {\rm tr}_{\rm CD}
\int_q^\Lambda i\gamma_5\gamma_\mu S_{f_1}(q_+) \Gamma_{H_{0^-}}(q;P) S_{f_2}(q_-)\,, \rule{2em}{0ex} \label{fpigen}
\\
\lefteqn{i\rho_{H_{0^-}} = -\langle 0 | \bar q_{f_2} i\gamma_5 q_{f_1} |H_{0^-} \rangle} \\
& = & Z_4\; {\rm tr}_{\rm CD}
\int_q^\Lambda \gamma_5 S_{f_1}(q_+) \Gamma_{H_{0^-}}(q;P) S_{f_2}(q_-) \,,
\label{rhogen}
\end{eqnarray}
where $\Gamma_{H_{0^-}}$ is the pseudoscalar meson's bound-state Bethe-Salpeter amplitude:
\begin{eqnarray}
\nonumber
\lefteqn{\Gamma_{H_{0^-}}(k;P) = \gamma_5 \left[ i E_{H_{0^-}}(k;P) + \gamma\cdot P F_{H_{0^-}}(k;P) \right.}\\
&& \left. + \gamma\cdot k \, G_{H_{0^-}}(k;P) - \sigma_{\mu\nu} k_\mu P_\nu H_{H_{0^-}}(k;P) \right],\rule{1em}{0ex}
\label{genGpi}
\end{eqnarray}
which is determined from the associated homogeneous BSE.  N.B.\ An analogous formula for scalar mesons is presented in Ref.\,\cite{Chang:2011mu}.

It is worth emphasising that the quark wave-function and Lagrangian mass renormalisation constants, $Z_{2,4}(\zeta,\Lambda)$, respectively, depend on the gauge parameter in precisely the manner needed to ensure that the right-hand sides of Eqs.\,(\ref{fpigen}), (\ref{rhogen}) are gauge-invariant.  Moreover, $Z_2(\zeta,\Lambda)$ ensures that the right-hand side of Eq.\,(\ref{fpigen}) is independent of both $\zeta$ and $\Lambda$, so that $f_{H_{0^-}}$ is truly an observable; and $Z_4(\zeta,\Lambda)$ ensures that $\rho_{H_{0^-}}^\zeta$ is independent of $\Lambda$ and evolves with $\zeta$ in just the way necessary to guarantee that the product $m^\zeta \rho_{H_{0^-}}^\zeta$ is
renormalisation-point-independent.  In addition, it should be noted that Eq.\,(\ref{mrtrelation}) is valid for every pseudoscalar meson and for any value of the current-quark masses; viz., $\hat m_{f_i} \in [ 0,\infty)$, $i=1,2$.  This includes arbitrarily large values and also the chiral limit, in whose neighbourhood Eq.\,(\ref{mrtrelation}) can be shown \cite{Maris:1997hd} to reproduce the familiar Gell-Mann--Oakes--Renner relation.

The axial-vector Ward-Takahashi identity, Eq.\,(\ref{avwtimN}), is a crucial bridge to Eqs.\,(\ref{mrtrelation}) -- (\ref{rhogen}); and on the way one can also prove the following Goldberger-Treiman-like relations \cite{Maris:1997hd}:
\begin{eqnarray}
\label{gtlrelE}
f_{H_{0^-}}^0 E_{H_{0^-}}(k;0) &=& B^0(k^2)\,,\\
\label{gtlrelF}
F_R(k;0) + 2 f_{H_{0^-}}^0 F_{H_{0^-}}(k;0) &=& A^0(k^2)\,,\\
\label{gtlrelG}
G_R(k;0) + 2 f_{H_{0^-}}^0 G_{H_{0^-}}(k;0) &=& \frac{d}{dk^2}A^0(k^2)\,,\\
\label{gtlrelH}
H_R(k;0) + 2 f_{H_{0^-}}^0 H_{H_{0^-}}(k;0) &=& 0\,,
\end{eqnarray}
wherein the superscript indicates that the associated quantity is evaluated in the chiral limit, and $F_R$, $G_R$, $H_R$ are analogues in the inhomogeneous axial-vector vertex of the scalar functions in the $H_{0^-}$-meson's Bethe-Salpeter amplitude.

These identities are of critical importance in QCD.
The first, Eq.\,(\ref{gtlrelE}), can be used to prove that a massless pseudoscalar meson appears in the chiral-limit spectrum if, and only if, chiral symmetry is dynamically broken.  Moreover, it exposes the fascinating consequence that the solution of the two-body pseudoscalar bound-state problem is almost completely known once the one-body problem is solved for the dressed-quark propagator, with the relative momentum within the bound-state identified unambiguously with the momentum of the dressed-quark.  This latter emphasises that Goldstone's theorem has a pointwise expression in QCD.
The remaining three identities are also important because they show that a pseudoscalar meson \emph{must} contain components of pseudovector origin.  This result overturned a misapprehension of twenty-years standing; namely, that only $E_{H_{0^-}}(k;0)$ is nonzero \cite{delbourgo:1979me}.  These pseudovector components materially influence the observable properties of pseudoscalar mesons \cite{Maris:1998hc,GutierrezGuerrero:2010md,Roberts:2010rn,Roberts:2011wy,Nguyen:2011jy}, Sec.\,\ref{FF1}, as do their analogues in other mesons \cite{Maris:1999nt,Maris:1999bh,Maris:2000sk}.

It is natural to reiterate here a prediction for the properties of non-ground-state pseudoscalar mesons, which follows from the exact results described above; namely, in the chiral limit \cite{Holl:2004fr,Holl:2005vu}
\begin{equation}
\label{fpin}
f_{\pi_n} \equiv 0 \,, \forall n\geq 1\,,
\end{equation}
where $n$ is a state label and $n=0$ denotes the ground state.  This is the statement that Goldstone modes are the only pseudoscalar mesons to possess a nonzero leptonic decay constant in the chiral limit when chiral symmetry is dynamically broken.  The decay constants of all other pseudoscalar mesons on this trajectory, e.g., radial excitations, vanish.  On the other hand, in the absence of DCSB the leptonic decay constant of each such pseudoscalar meson vanishes in the chiral limit; i.e, Eq.\,(\ref{fpin}) is true $\forall n \geq 0$.

\begin{figure}[t]

\centerline{
\includegraphics[clip,width=0.45\textwidth]{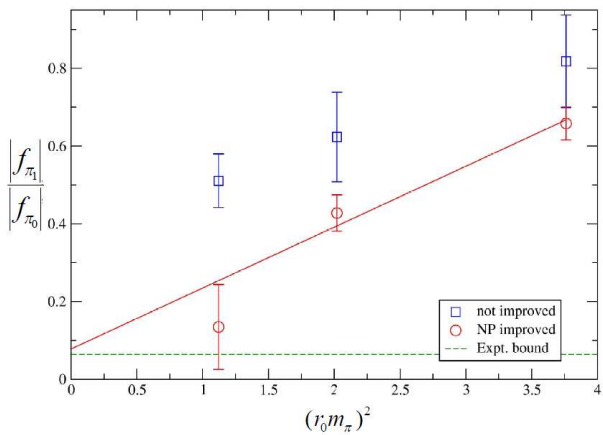}}

\caption{\label{fig:fRadial}
Lattice-QCD results for the ratio of the decay constants for the first-excited- and ground-state pseudoscalar mesons as a function of the pion mass squared.  (Lattice parameters: volume\,$=16^3\times 32$; $\beta = 5.2$, spacing $a\simeq 0.1\,$fm, two flavours of degenerate sea quarks; Wilson gauge action and clover fermions.)
The ``not improved'' results were obtained from a fermion action with poor chiral symmetry properties.  In this case $|f_{\pi_1}/f_{\pi_0}|\approx 0.4$, consistent with expectations based on quantum mechanics.
The ``improved'' results were obtained through implementation of the full ALPHA method for the nonperturbative improvement of the fermion action, which greatly improves the simulation's chiral symmetry properties.  In this case, $|f_{\pi_1}/f_{\pi_0}|\approx 0.01$.  (NB. The sign of the ratio was not determined in the simulation but must be negative on general grounds \protect\cite{Holl:2004fr}.  Figure adapted from Ref.\,\protect\cite{McNeile:2006qy}.)}
\end{figure}

From the perspective of quantum mechanics, Eq.\,(\ref{fpin}) is a surprising fact.  The leptonic decay constant for $S$-wave states is typically proportional to the wave-function at the origin.  Compared with the ground state, this is smaller for
an excited state because the wave-function is broader in configuration space and wave-functions are normalised.  However, it is a modest effect.  For example, consider the $e^+e^-$ decay of vector mesons.  A calculation in relativistic quantum mechanics based on light-front dynamics \cite{deMelo:2005cy} yields $|f_{\rho_1}/f_{\rho_0}| = 0.5$, consistent with the value inferred from experiment and DSEs in rainbow-ladder truncation \cite{Qin:2011dd}: $|f_{\rho_1}/f_{\rho_0}| = 0.45$.  Thus, it is not uncommon for Eq.\,(\ref{fpin}) to be perceived as ``remarkable'' or ``unbelievable.''  Notwithstanding this, in connection with the pion's first radial excitation, the value of $f_{\pi_1}= -2\,$MeV predicted in Ref.\,\cite{Holl:2004fr} is consistent with experiment \cite{Diehl:2001xe} and simulations of lattice-QCD \cite{McNeile:2006qy}, as illustrated in Fig.\,\ref{fig:fRadial}.  It is now recognised that the suppression of $f_{\pi_1}$ is a useful benchmark, which can be used to tune and validate lattice QCD techniques that try to determine the properties of excited states mesons.

\subsection{Condensates are confined within hadrons}
\label{sec:inmeson}
Dynamical chiral symmetry breaking and its connection with the generation of hadron masses was first considered in Ref.\,\cite{Nambu:1961tp}.  The effect was represented as a vacuum phenomenon.  Two essentially inequivalent classes of ground-state were identified in the mean-field treatment of a meson-nucleon field theory: symmetry preserving (Wigner phase); and symmetry breaking (Nambu phase).  Notably, within the symmetry breaking class, each of an uncountable infinity of distinct configurations is related to every other by a chiral rotation.  This is arguably the origin of the concept that strongly-interacting quantum field theories possess a nontrivial vacuum.

\begin{figure}[t]
\vspace*{-1ex}

\includegraphics[clip,width=0.40\textwidth]{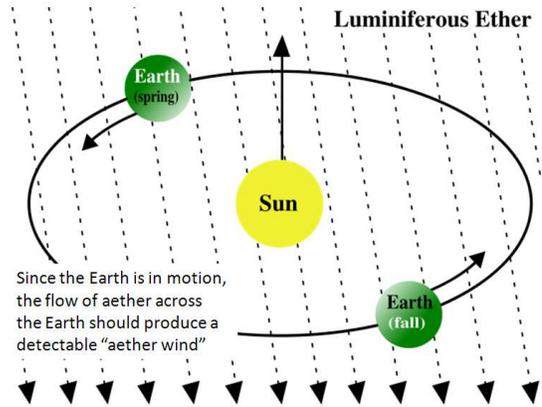}

\caption{\label{fig:aether} Physics theories of the late 19th century postulated that, just as water waves must have a medium to move across (water), and audible sound waves require a medium to move through (such as air or water), so also light waves require a medium, the ``luminiferous aether.''  This was apparently unassailable logic until, of course, one of the most famous failed experiments in the history of science to date \protect\cite{MichelsonMorley}.
}

\end{figure}

With the introduction of the parton model for the description of deep inelastic scattering (DIS), this notion was challenged via an argument \cite{Casher:1974xd} that DCSB can be realised as an intrinsic property of hadrons, instead of through a nontrivial vacuum exterior to the observable degrees of freedom.  Such a perspective is tenable because the essential ingredient required for dynamical symmetry breaking in a composite system is the existence of a divergent number of constituents and DIS provided evidence for the existence within every hadron of a sea of low-momentum partons.  This view has, however, received scant attention.  On the contrary, the introduction of QCD sum rules as a theoretical artifice to estimate nonperturbative strong-interaction matrix elements entrenched the belief that the QCD vacuum is characterised by numerous distinct, spacetime-independent condensates.  Faith in empirical vacuum condensates might be compared with an earlier misguided conviction that the universe was filled with a luminiferous aether, Fig.\,\ref{fig:aether}.

Notwithstanding the prevalence of the belief in empirical vacuum condensates, it does lead to problems; e.g., entailing, as explained below, a cosmological constant that is $10^{46}$-times greater than that which is observed \cite{Brodsky:2009zd,Turner:2001yu}.  This unwelcome consequence is partly responsible for reconsideration of the possibility that the so-called vacuum condensates are in fact an intrinsic property of hadrons.  Namely, in a confining theory -- and confinement is essential to this view -- condensates are not constant, physical mass-scales that fill all spacetime; instead, they are merely mass-dimensioned parameters that serve a practical purpose in some theoretical truncation schemes but otherwise do not have an existence independent of hadrons \cite{Brodsky:2009zd,Brodsky:2010xf,Chang:2011mu,%
Burkardt:1998dd,Brodsky:2008be,Glazek:2011vg}.

To account for the gross overestimate, recall that the expansion of the universe is accelerating \cite{Turner:2001yu}.  One piece of evidence is provided by the observations of type Ia supernovae reported in Refs.\,\cite{1538-3881-116-3-1009,0004-637X-517-2-565}; and another by measurements of the composition of the universe, which point to a missing energy component with negative pressure.  To explain the latter, the curvature of the universe may be characterised by a mass-energy density, $\rho_U$.  There is a critical value of this density for which the universe is flat: $\rho_F$.  Observations of the cosmic microwave background's power spectrum, e.g., Ref.\,\cite{Netterfield:2001yq}, indicate that $\Omega_0=\rho_U/\rho_F=1 \pm 0.04$.  In a flat universe, the matter density, $\Omega_M$, and the energy density must sum to the critical density.  However, matter only contributes about one-third of the critical density, $\Omega_M=0.33\pm0.04$.  Thus two-thirds of the critical density is missing.

In order to have escaped detection, the missing energy must be smoothly distributed.  In order not to interfere with the formation of structure (by inhibiting the growth of density perturbations) the energy density in this component must change more slowly than matter (so that it was subdominant in the past).  The universe's accelerated expansion can be accommodated in general relativity through the cosmological constant, $\Lambda$, and observations determine an associated density, $\rho_{\Lambda}^{\rm obs}$.  (Recall that Einstein introduced the repulsive effect of the cosmological constant in order to balance the attractive gravity of matter so that a static universe was possible.  However, he promptly discarded it after the discovery of the expansion of the Universe.)

It has been suggested that the advent of quantum field theory makes consideration of the cosmological constant obligatory not optional \cite{Turner:2001yu}.  Indeed, the only possible covariant form for the energy of the (quantum) vacuum; viz.,
\begin{equation}
T_{\mu\nu}^{\rm VAC} = \rho_{\rm VAC}\, \delta_{\mu\nu}
\end{equation}
is mathematically equivalent to the cosmological constant.  The vacuum is \cite{Turner:2001yu} ``\ldots a perfect fluid and precisely spatially uniform \ldots'' so that ``Vacuum energy is almost the perfect candidate for dark energy.''  Now, if the ground state of QCD is truly expressed in a nonzero spacetime-independent expectation value $\langle\bar q q\rangle$, then the energy difference between the symmetric and broken phases is of order $M_{\rm QCD} \sim 0.3\,$GeV, as indicated by Fig.\,\ref{gluoncloud}.  One obtains therefrom:
\begin{equation}
\rho_\Lambda^{\rm QCD} = 10^{46} \rho_\Lambda^{\rm obs}.
\end{equation}
In fact, the discrepancy is far greater if the Higgs vacuum expectation value is treated in a similar manner.

This mismatch has been called ``the greatest embarrassment in theoretical physics.''  However, it vanishes if one discards the notion that condensates have a physical existence, which is independent of the hadrons that express QCD's asymptotically realisable degrees of freedom \cite{Brodsky:2009zd}; namely, if one accepts that such condensates are merely mass-dimensioned parameters in one or another theoretical truncation scheme.  This appears mandatory in a confining theory \cite{Brodsky:2010xf,Chang:2011mu}, a perspective one may embed in a broader context by considering just what is observable in quantum field theory \cite{Weinberg:1978kz}: ``\ldots although individual quantum field theories have of course a good deal of content, quantum field theory itself has no content beyond analyticity, unitarity, cluster decomposition and symmetry.''  If QCD is a confining theory, then the principle of cluster decomposition is only realised for colour singlet states \cite{Krein:1990sf} and all observable consequences of the theory, including its ground state, can be expressed via an hadronic basis.  This is quark-hadron duality.

It is worthwhile to recapitulate upon the arguments in Refs.\,\cite{Brodsky:2010xf,Chang:2011mu}.  To begin, note that Eq.\,(\ref{fpigen}) is the exact expression in QCD for the leptonic decay constant of a pseudoscalar meson.  It is a property of the pion and, as consideration of the integral expression reveals, it can be described as the pseudovector projection of the pion's Bethe-Salpeter wave-function onto the origin in configuration space.  Note that the product $\psi = S \Gamma S$ is called the Bethe-Salpeter wave-function because, when a nonrelativistic limit can validly be performed, the quantity $\psi$ at fixed time becomes the quantum mechanical wave-function for the system under consideration.  (N.B.\ In the neighborhood of the chiral limit, a value for $f_{H_{0^-}}$ can be estimated via either of two approximation formulae \protect\cite{Cahill:1985mh,Pagels:1979hd,Chang:2009zb}.  These formulae both illustrate and emphasize the role of $f_{H_{0^-}}$ as an order parameter for DCSB.)

If chiral symmetry were not dynamically broken, then in the neighborhood of the chiral limit $f_{H_{0^-}} \propto \hat m$ \cite{Holl:2004fr}.  Of course, chiral symmetry is dynamically broken in QCD \cite{Bhagwat:2003vw,Bhagwat:2006tu,Bhagwat:2007vx,Bowman:2005vx} and so for the ground-state pseudoscalar
\begin{equation}
\lim_{\hat m\to 0} f_{H_{0^-}}(\hat m) = f^0_{H_{0^-}} \neq 0\,.
\end{equation}
Taken together, these last two observations express the fact that $f_{H_{0^-}}$, which is an intrinsic property of the pseudoscalar meson, is a \emph{bona fide} order parameter for DCSB.  An analysis within chiral perturbation theory \cite{Bijnens:2006zp} suggests that the chiral limit value, $f^0_{H_{0^-}}$, is $\sim 5$\% below the measured value of 92.4\,MeV; and efficacious DSE studies give a 3\% chiral-limit reduction~\cite{Maris:1997tm}.

Now, Eq.\,(\ref{rhogen}) is kindred to Eq.\,(\ref{fpigen}); it is the expression in quantum field theory which describes the \emph{pseudoscalar} projection of the pseudoscalar meson's Bethe-Salpeter wave-function onto the origin in configuration space.  It is thus truly just another type of pseudoscalar meson decay constant.  

In this connection it is therefore notable that one may rigorously define an ``in-meson'' condensate; viz.\,\cite{Maris:1997hd,Maris:1997tm}:
\begin{eqnarray}
\label{inpiqbq}
-\langle \bar q_{f_2} q_{f_1} \rangle^\zeta_{H_{0^-}} &\equiv& -
f_{H_{0^-}} \langle 0 | \bar q_{f_2} \gamma_5 q_{f_1} |H_{0^-} \rangle\\
&=& f_{H_{0^-}} \rho_{H_{0^-}}^\zeta =: \kappa_{H_{0^-}}^\zeta(\hat m)\,.
\end{eqnarray}
Now, using Eq.\,(\ref{gtlrelE}), one finds \cite{Maris:1997hd}
\begin{equation}
\lim_{\hat m\to 0} \kappa^\zeta_{H_{0^-}}(\hat m)
=
Z_4 \, {\rm tr}_{\rm CD}\int^\Lambda \!\!\!\! \mbox{\footnotesize $\displaystyle\frac{d^4 q}{(2\pi)^4}$} S^0(q;\zeta) =  -\langle \bar q q \rangle_\zeta^0\,.
\label{qbqpiqbq0}
\end{equation}
Hence the so-called vacuum quark condensate is, in fact, the chiral-limit value of the in-meson condensate; i.e., it describes a property of the chiral-limit pseudoscalar meson.  One can therefore argue that this condensate is no more a property of the ``vacuum'' than the pseudoscalar meson's chiral-limit leptonic decay constant.  Moreover, Ref.\,\cite{Langfeld:2003ye} establishes the equivalence of all three definitions of the so-called vacuum quark condensate: a constant in the operator product expansion \cite{Lane:1974he,Politzer:1976tv}; via the Banks-Casher formula \cite{Banks:1979yr}; and the trace of the chiral-limit dressed-quark propagator.  Hence, they are all related to the in-meson condensate via Eq.\,\eqref{qbqpiqbq0} and none is defined essentially in connection with the vacuum.


It is worth remarking that in the presence of confinement it is impossible to write a valid nonperturbative definition of a single quark or gluon annihilation operator; and therefore impossible to rigorously define a second quantised vacuum (ground state) for QCD upon a foundation of gluon and quark (quasiparticle) operators.   To do so would be to answer the question: What is the state that is annihilated by an operator which is - as appears at present - unknowable?  However, with the assumptions that confinement is absolute and that it entails quark-hadron duality, the question changes completely.  In this case, the nonperturbative Hamiltonian of observable phenomena in QCD is diagonalised by colour-singlet states alone.  The ground state of this nonperturbative strong-interaction Hamiltonian is the state with zero hadrons.  One may picture the creation and annihilation operators for such states as rigorously defined via smeared sources on a spacetime lattice.  The ground-state is defined with reference to such operators, employing, e.g., the Gell-Mann - Low theorem \cite{GellMann:1951rw}, which is applicable in this case because there are well-defined asymptotic states and associated annihilation and creation operators.

In learning that the so-called vacuum quark condensate is actually the chiral-limit value of an in-pion property, some respond as follows.
The electromagnetic radius of any hadron which couples to pseudoscalar mesons must diverge in the chiral limit.  This long-known effect arises because the propagation of \emph{massless} on-shell colour-singlet pseudoscalar mesons is undamped \cite{Beg:1973sc,Pervushin:1974nm,Gasser:1983yg,Alkofer:1993gu}.
Therefore, does not each pion grow to fill the universe; so that, in this limit, the in-pion condensate reproduces the conventional paradigm?

Confinement, again, vitiates this objection.  Both DSE- and lattice-QCD studies indicate that confinement entails dynamical mass generation for both gluons and quarks, see Sec.\,\ref{Sect:Conf}.  The dynamical gluon and quark masses remain large in the limit of vanishing current-quark mass.  In fact, the dynamical masses are almost independent of the current-quark mass in the neighbourhood of the chiral limit.  It follows that for any hadron the quark-gluon containment-radius does not diverge in the chiral limit.  Instead, it is almost insensitive to the magnitude of the current-quark mass because the dynamical masses of the hadron's constituents are frozen at large values; viz., $2 - 3\,\Lambda_{\rm QCD}$.  These considerations show that the divergence of the electromagnetic radius does not correspond to expansion of a condensate from within the pion but rather to the copious production and subsequent propagation of composite pions, each of which contains a condensate whose value is essentially unchanged from its nonzero current-quark mass value within a containment-domain whose size is similarly unaffected.

There is more to be said in connection with the definition and consequences of a chiral limit.  Plainly, the existence of strongly-interacting massless composites would have an enormous impact on the evolution of the universe; and it is naive to imagine that one can simply set $\hat m_{u,d}=0$ and consider a circumscribed range of manageable consequences whilst ignoring the wider implications for hadrons, the Standard Model and beyond.  For example, with all else held constant, Big Bang Nucleosynthesis is very sensitive to the value of the pion-mass \cite{Flambaum:2007mj}.  We are fortunate that the absence of quarks with zero current-quark mass has produced a universe in which we exist so that we may carefully ponder the alternative.

The discussion of Ref.\,\cite{Brodsky:2010xf} was restricted to pseudoscalar mesons.  It is expanded in Ref.\,\cite{Chang:2011mu} via a demonstration that the in-pseudoscalar-meson condensate can be represented through the pseudoscalar-meson's scalar form factor at zero momentum transfer.  With the aid of a mass formula for scalar mesons, revealed therein, the analogue was shown to be true for in-scalar-meson condensates.  As argued, the concept is readily extended to all hadrons so that, via the zero momentum transfer value of any hadron's scalar form factor, one can readily extract the value for a quark condensate in that hadron which is a measure of dynamical chiral symmetry breaking.

Given that quark condensates are an intrinsic property of hadrons, one arrives at a new paradigm, as observed in the popular science press \cite{Courtland:2010zz}: ``EMPTY space may really be empty.  Though quantum theory suggests that a vacuum should be fizzing with particle activity, it turns out that this paradoxical picture of nothingness may not be needed.  A calmer view of the vacuum would also help resolve a nagging inconsistency with dark energy, the elusive force thought to be speeding up the expansion of the universe.''  In connection with the cosmological constant, putting QCD condensates back into hadrons reduces the mismatch between experiment and theory by a factor of $10^{46}$.  If technicolour-like theories are the correct scheme for extending the Standard Model \cite{Andersen:2011yj}, then the impact of the notion of in-hadron condensates is far greater still. 

\subsection{Flavourless pseudoscalar mesons}
\label{flavourless}
\begin{figure}[t]

\centerline{
\includegraphics[clip,width=0.35\textwidth]{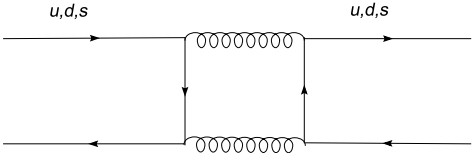}}

\caption{\label{fig:nonanomaly}
This simple flavour-mixing diagram is immaterial to the resolution of the $\eta$-$\eta^\prime$ conundrum, as is any collection of processes for which the figure may serve as a skeleton diagram.  (Straight lines denote quarks and springs denote gluons.)
}
\end{figure}

In connection with electric-charge-neutral pseudoscalar mesons, Eq.\,(\ref{mrtrelation}) is strongly modified owing to the non-Abelian anomaly.  This entails that whilst the classical action associated with QCD is invariant under $U_A(N_f)$ (non-Abelian axial transformations generated by  $\lambda_0 \gamma_5$, where $\lambda_0 \propto{\rm diag}[1,\ldots ,1_{N_f}]$), the quantum field theory is not.  The modification is particularly important to properties of $\eta$ and $\eta^\prime$ mesons.  The latter is obviously a peculiar pseudoscalar meson because its mass is far greater than that of any other light-quark pseudoscalar meson; e.g., $m_{\eta^\prime} = 1.75\, m_{\eta}$.  We note that the diagram depicted in Fig.\,\ref{fig:nonanomaly} is often cited as central to a solution of the $\eta$-$\eta^\prime$ puzzle.  However, as will become clear below, whilst it does contribute to flavour-mixing, the process is immaterial in resolving the $\eta$-$\eta^\prime$ conundrum, as is any collection of processes for which the figure may serve as a skeleton diagram.

The correct mass formula for flavourless pseudoscalars follows from consideration of the complete $U_A(N_f)$ Ward-Takahashi identity:
\begin{eqnarray}
%
%
%
\nonumber P_\mu \Gamma_{5\mu}^a(k;P)& =& {\cal S}^{-1}(k_+) i \gamma_5 {\cal F}^a
+ i \gamma_5 {\cal F}^a {\cal S}^{-1}(k_-)
\\
 & &
- 2 i {\cal M}^{ab}\Gamma_5^b(k;P)  - {\cal A}^a(k;P)\,,
\label{avwtiG}
\end{eqnarray}
which generalises Eq.\,(\ref{avwtimN}).  In Eq.\,(\ref{avwtiG}),
$\{{\cal F}^a | \, a=0,\ldots,N_f^2-1\}$ are the generators of $U(N_f)$ in the fundamental representation, orthonormalised according to tr${\cal F}^a {\cal F}^b= \frac{1}{2}\delta^{ab}$;
the dressed-quark propagator ${\cal S}=\,$diag$[S_u,S_d,S_s,S_c,S_b,\ldots]$ is matrix-valued;
and
\begin{equation}
{\cal M}^{ab} = {\rm tr}_F \left[ \{ {\cal F}^a , {\cal M}^\zeta \} {\cal F}^b \right],
\end{equation}
where ${\cal M}^\zeta$ is a matrix of renormalised current-quark masses and the trace is over flavour indices.

The final term in the last line of Eq.\,(\ref{avwtiG}) expresses the non-Abelian axial anomaly.  It can be written
\begin{equation}
\label{amputate}
{\cal A}^a(k;P) =  {\cal S}^{-1}(k_+) \,\delta^{a0}\, {\cal A}_U(k;P) {\cal S}^{-1}(k_-)\,,
\end{equation}
with
\begin{equation}
{\cal A}_U(k;P) = \!\!  \int\!\! d^4xd^4y\, e^{i(k_+\cdot x - k_- \cdot y)} N_f \left\langle  {\cal F}^0\,q(x)  \, {\cal Q}(0) \,   \bar q(y)
\right\rangle, \label{AU}
\end{equation}
and since ${\cal A}^{a=0}(k;P)$ is a pseudoscalar, it has the general form
\begin{eqnarray}
\nonumber \lefteqn{{\cal A}^0(k;P) = {\cal F}^0\gamma_5 \left[ i {\cal E}_{\cal A}(k;P) + \gamma\cdot P {\cal F}_{\cal A}(k;P) \right.} \\
&& \left. +\, \gamma\cdot k \, k\cdot P {\cal G}_{\cal A}(k;P) + \sigma_{\mu\nu} k_\mu P_\nu {\cal H}_{\cal A}(k;P)\right].\rule{1em}{0ex}
\end{eqnarray}
The matrix element in Eq.\,(\ref{AU}) represents an operator expectation value in full QCD; the operation in Eq.\,(\ref{amputate}) amputates the external quark lines; and
\begin{equation}
{\cal Q}(x) = i \frac{\alpha_s }{4 \pi} {\rm tr}_{C}\left[ \epsilon_{\mu\nu\rho\sigma} F_{\mu\nu} F_{\rho\sigma}(x)\right]  \label{topQ}\\
= \partial_\mu K_\mu(x)
\end{equation}
is the topological charge density operator, where the trace is over colour indices and $F_{\mu\nu}=\frac{1}{2}\lambda^a F_{\mu\nu}^a$ is the matrix-valued gluon field strength tensor.  It is plain and important that only ${\cal A}^{a=0}$ is nonzero.  NB.\ While ${\cal Q}(x)$ is gauge invariant, the associated Chern-Simons current, $K_\mu$, is not.  Thus in QCD no physical state can couple to $K_\mu$ and hence no state which appears in the observable spectrum can contribute to a resolution of the so-called $U_A(1)$-problem; i.e., physical states cannot play any role in ensuring that the $\eta^\prime$ is not a Goldstone mode.

As described in Sec.\,\ref{psmassformula}, if one imagines there are $N_f$ massless quarks, then DCSB is a necessary and sufficient condition for the $a\neq 0$ components of Eq.\,(\ref{avwtiG}) to guarantee the existence of $N_f^2-1$ massless bound-states of a dressed-quark and -antiquark.  However, owing to Eq.\,(\ref{amputate}), $a=0$ in Eq.\,(\ref{avwtiG}) requires special consideration.  One case is easily covered; viz., it is clear that if ${\cal A}^{0} \equiv 0$, then the $a=0$ component of Eq.\,(\ref{avwtiG}) is no different to the others and there is an additional massless bound-state in the chiral limit.

On the other hand, the large disparity between the mass of the $\eta^\prime$-meson and the octet pseudoscalars suggests that ${\cal A}^{0} \neq 0$ in real-world QCD.  If one carefully considers that possibility, then the Goldberger-Treiman relations in Eqs.\,(\ref{gtlrelE}) -- (\ref{gtlrelH}) become \cite{Bhagwat:2007ha}
\begin{eqnarray}
\nonumber
2 f_{H_{0^-}}^0 E_{BS}(k;0) &= & 2 B^{0}(k^2) - {\cal E}_{\cal A}(k;0),\\
&& \label{ewti}\\
\nonumber
F_R^0(k;0) + 2 f_{H_{0^-}}^0 F_{BS}(k;0) & = & A^{0}(k^2) - {\cal F}_{\cal A}(k;0),\rule{1em}{0ex}\\
&& \label{fwti}\\
\nonumber
G_R^0(k;0) + 2 f_{H_{0^-}}^0 G_{BS}(k;0) & = & 2 \frac{d}{dk^2}A^{0}(k^2) - {\cal G}_{\cal A}(k;0),\\
&&\\
\label{hwti}
H_R^0(k;0) + 2 f_{H_{0^-}}^0 H_{BS}(k;0) & = & - {\cal H}_{\cal A}(k;0),
\end{eqnarray}
It follows that the relationship
\begin{equation}
\label{calEB}
{\cal E}_{\cal A}(k;0) = 2 B^{0}(k^2) \,,
\end{equation}
is necessary and sufficient to guarantee that $\Gamma_{5\mu}^0(k;P)$, the flavourless pseudoscalar vertex, does not possess a massless pole in the chiral limit; i.e., that there are only $N_f^2-1$ massless Goldstone bosons.  Now, in the chiral limit, $B^{0}(k^2) \neq 0 $ if, and only if, chiral symmetry is dynamically broken.   Hence, the absence of an additional massless bound-state is only assured through the existence of an intimate connection between DCSB and an expectation value involving the topological charge density.

This critical connection is further highlighted by the following result, obtained through a few straightforward manipulations of Eqs.\,(\ref{avwtiG}), (\ref{amputate}) and (\ref{AU}):
\begin{eqnarray}
\lefteqn{
\langle \bar q q \rangle_\zeta^0 = - \lim_{\hat m \to 0} \kappa_{H_{0^-}}^\zeta(\hat m)}\\
& = & -\lim_{\Lambda\to \infty}Z_4(\zeta^2,\Lambda^2)\, {\rm tr}_{\rm CD}\int^\Lambda_q\!
S^{0}(q,\zeta)  \\
& = &
\mbox{\footnotesize $\displaystyle \frac{N_f}{2}$} \int d^4 x\, \langle \bar q(x) i\gamma_5  q(x) {\cal Q}(0)\rangle^0.
\end{eqnarray}
The absence of a Goldstone boson in the $a=0$ channel is only guaranteed if this explicit identity between the chiral-limit in-meson condensate and a mixed vacuum polarisation involving the topological charge density is satisfied.

Mass formulae valid for all pseudoscalar mesons have also been obtained \cite{Bhagwat:2007ha}
\begin{equation}
\label{newmass}
%
f_{H_{0^-}}^a m_{H_{0^-}}^2 = 2\,{\cal M}^{ab} \rho_{H_{0^-}}^b + \delta^{a0} \, n_{H_{0^-}}\,,
\end{equation}
where
\begin{eqnarray}
\label{fpia} f_{H_{0^-}}^a \,  P_\mu &=& Z_2\,{\rm tr} \int^\Lambda_q
{\cal F}^a \gamma_5\gamma_\mu\, \chi_{H_{0^-}}(q;P) \,, \\
\label{cpres} i  \rho_{H_{0^-}}^a\!(\zeta)  &=& Z_4\,{\rm tr}
\int^\Lambda_q {\cal F}^a \gamma_5 \, \chi_{H_{0^-}}(q;P)\,,\\
n_{H_{0^-}} &=& \mbox{\footnotesize $\displaystyle \sqrt{\frac{N_f}{2}}$} \, \nu_{H_{0^-}} \,, \; \nu_{H_{0^-}}= \langle 0 | {\cal Q} | H_{0^-}\rangle \,.\rule{1em}{0ex}
\end{eqnarray}
For charged pseudoscalar mesons, Eq.\,(\ref{newmass}) is equivalent to Eq.\,(\ref{mrtrelation}), but the novelty of Eq.\,(\ref{newmass}) is what it expresses for flavourless pseudoscalars.  To illustrate, consider the case of a $U(N_f=3)$-symmetric mass matrix, in which all $N_f=3$ current-quark masses assume the single value $m^\zeta$, then this formula yields:
\begin{equation}
\label{etapchiral}
m_{\eta^\prime}^2 f_{\eta^\prime}^0 = n_{\eta^\prime} + 2 m^\zeta\rho_{\eta^\prime}^{0 \zeta} \,.
\end{equation}
Plainly, the $\eta^\prime$ is split from the Goldstone modes so long as $n_{\eta^\prime} \neq 0$.  Numerical simulations of lattice-QCD have confirmed this identity \protect\cite{Bardeen:2000cz,Ahmad:2005dr}.

It is important to elucidate the physical content of $n_{\eta^\prime}$.  Returning to the definition; viz.,
\begin{equation}
\nu_{\eta^\prime}= \mbox{\footnotesize $\displaystyle \sqrt{\frac{3}{2}}$} \, \langle 0 | {\cal Q} | \eta^\prime \rangle \,,
\end{equation}
it is readily seen to be another type of in-meson condensate.  It is analogous to those discussed in Sec.\,\ref{sec:inmeson} but in this case the hadron-to-vacuum transition amplitude measures the topological content of the $\eta^\prime$.  One may therefore state that the $\eta^\prime$ is split from the Goldstone modes so long as its wavefunction possesses nonzero topological content.  This is plainly very different to requiring the QCD vacuum be topologically nontrivial.

\begin{figure}[t]
\vspace*{-30ex}

\centerline{
\includegraphics[clip,width=0.5\textwidth]{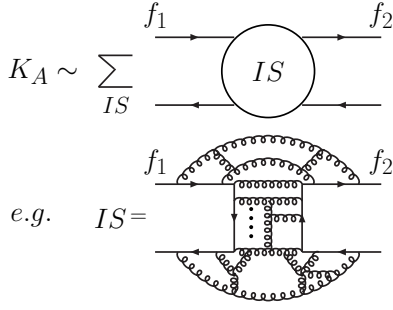}}
\vspace*{-31ex}

\caption{\label{etaglue} An illustration of the nature of the contribution to the Bethe-Salpeter kernel associated with the non-Abelian anomaly.  All terms have the ``hairpin'' structure illustrated in the lower panel.  No finite sum of such intermediate states is sufficient.  (Straight lines denote quarks, with $f_1$ and $f_2$ independent, and springs denote gluons.)}
\end{figure}

Within QCD the properties of the $\eta^\prime$ can be computed via the BSE, just like other mesons.  A nonzero value of $n_{\eta^\prime}$ can be achieved with a Bethe-Salpeter kernel that contains the contribution depicted in Fig.\,\ref{etaglue} because one may argue from Eqs.\,(\ref{AU}) and (\ref{topQ}) that an anomaly-related contribution to a meson's Bethe-Salpeter kernel cannot contain external quark or antiquark lines that are connected to the incoming lines: purely gluonic configurations must mediate, as illustrated in Fig.\,\ref{etaglue}.  Furthermore, it is straightforward to see that no finite sum of gluon exchanges can serve this purpose.  Indeed, consider any one such single contribution in the chiral limit.  It will be proportional to the total momentum and hence vanish for $P=0$, in conflict with Eq.\,(\ref{etapchiral}).  This lies behind the need for something like the Kogut-Susskind \emph{ghost}; i.e., the coupling of a massless axial-vector gauge-like field to the Chern-Simons current, which does not appear in the particle spectrum of QCD because the current is not gauge invariant.  (See Ref.\,\protect\cite{Kogut:1974kt} and Sec.\,5.1 of Ref.\,\cite{Christos:1984tu}.)

It is argued \cite{Witten:1979vv,Veneziano:1979ec} that in QCD with $N_c$ colours,
\begin{equation}
n_{\eta^\prime} \sim \frac{1}{\sqrt{N_c}}\,,
\end{equation}
and it can be seen to follow from the gap equation, the homogeneous BSE and Eqs.\,(\ref{fpia}), (\ref{cpres}) that
\begin{equation}
f_{\eta^\prime}^0 \sim \sqrt{N_c} \sim \rho_{\eta^\prime}^0(\zeta)\,.
\end{equation}
One thus obtains
\begin{equation}
m_{\eta^\prime}^2 =  \frac{n_{\eta^\prime}}{f_{\eta^\prime}^0} + 2 m(\zeta) \frac{\rho_{\eta^\prime}^0(\zeta)}{f_{\eta^\prime}^0} \,.
\end{equation}
The first term vanishes in the limit $N_c\to \infty$ while the second remains finite.  Subsequently taking the chiral limit, the $\eta^\prime$ mass approaches zero in the manner characteristic of all Goldstone modes.  (N.B.\ One must take the limit $N_c\to \infty$ before the chiral limit because the procedures do not commute \cite{Narayanan:2004cp}.)  These results are realised in the effective Lagrangian of Ref.\,\cite{DiVecchia:1979bf} in a fashion that is consistent with all the constraints of the anomalous Ward identity.  N.B.\ This is not true of the so-called 't\,Hooft determinant \protect\cite{Crewther:1977ce,Crewther:1978zz,Christos:1984tu}.

Implications of the mass formula in Eq.\,(\ref{newmass}) were exemplified in Ref.\,\cite{Bhagwat:2007ha} using an elementary dynamical model that includes a one-parameter \emph{Ansatz} for that part of the Bethe-Salpeter kernel related to the non-Abelian anomaly, an illustration of which is provided in Fig.\,\ref{etaglue}.  The study compares ground-state pseudoscalar- and vector-mesons constituted from all known quarks, excluding the $t$-quark.  Amongst the notable results is a prediction for the mixing angles between neutral mesons; e.g.,
\begin{equation}
\label{valmixing}
\theta_\eta = -15.4^\circ\,,\;
\theta_{\eta^\prime} = -15.7^\circ\,.
\end{equation}
N.B.\ There are necessarily two mixing angles, with each determined at the appropriate pole position in the inhomogeneous vertex.  It is interesting that the angles are approximately equal and compare well with the value inferred from a single mixing angle analysis \cite{Bini:2007zza} $\theta = -13.3^\circ \pm 1.0^\circ$.

It is worth explicating the nature of the flavour-induced difference between the $\pi^0$ and $\pi^\pm$ masses.  If one ignores mixing with mesons containing other than $u,d$-quarks; viz., works solely within $SU(N_f=2)$, then $m_{\pi^0}-m_{\pi^+}=-0.04\,$MeV.  On the other hand, the full calculation yields $m_{\pi^0}-m_{\pi^+}=-0.4\,$MeV, a factor of ten greater, and one obtains a $\pi^0$-$\eta$ mixing angle, whose value at the neutral pion mass shell is
\begin{equation}
\theta_{\pi \eta}(m_{\pi^0}^2)=1.2^\circ.
\end{equation}
For comparison, Ref.\,\cite{Green:2003qw} infers a mixing angle of $0.6^\circ \pm 0.3^\circ$ from a $K$-matrix analysis of the process $p\, d \rightarrow\, ^3$He$\,\pi^0$.  Plainly, mixing with the $\eta$-meson is the dominant non-electromagnetic contribution to the $\pi^\pm$-$\pi^0$ mass splitting.  The analogous angle at the $\eta$ mass-shell is
\begin{equation}
\theta_{\pi \eta}(m_{\eta}^2)=1.3^\circ.
\end{equation}

The angles in Eq.\,(\ref{valmixing}) correspond to
\begin{eqnarray}
\label{pi0f}
|\pi^0\rangle & \sim & 0.72 \, \bar u u - 0.69 \, \bar d d - 0.013 \, \bar s s\,, \\
\label{pi8f}
|\eta\rangle & \sim & 0.53\, \bar u u + 0.57 \, \bar d d - 0.63 \, \bar s s\,, \\
\label{pi9f}
|\eta^\prime\rangle & \sim & 0.44\, \bar u u + 0.45 \, \bar d d + 0.78 \, \bar s s \,.
\end{eqnarray}
Evidently, in the presence of a sensible amount of isospin breaking, the $\pi^0$ is still predominantly characterised by ${\cal F}^3$ but there is a small admixture of $\bar ss$.  It is found in Ref.\,\cite{Bhagwat:2007ha} that mixing with the $\pi^0$ has a similarly modest impact on the flavour content of the $\eta$ and $\eta^\prime$.  It's effect on their masses is far less.

\section{Many Facets of DCSB}
\label{sec:Facets}
%
The importance and interconnection of confinement and DCSB are summarised in Secs.\,\ref{Sect:Conf}, \ref{sec:gap}; and some of the profound implications of DCSB for pseudoscalar mesons are detailed in Sec.\,\ref{spectrum1}.  The latter could be proved owing to the existence of at least one systematic nonperturbative symmetry-preserving DSE truncation scheme \cite{Munczek:1994zz,Bender:1996bb}.  On the other hand, the practical application of this particular scheme has numerous shortcomings.
For example, at leading-order (rainbow-ladder) the truncation is accurate for ground-state vector- and electrically-charged pseudoscalar-mesons because corrections in these channels largely cancel, owing to parameter-free preservation of the Ward-Takahashi identities.  However, they do not cancel in other channels \cite{Roberts:1996jx,Roberts:1997vs,Bender:2002as,Bhagwat:2004hn}.  Hence studies based on the rainbow-ladder truncation, or low-order improvements thereof, have usually provided poor results for scalar- and axial-vector-mesons \cite{Burden:1996nh,Watson:2004kd,Maris:2006ea,Cloet:2007pi,Fischer:2009jm,%
Krassnigg:2009zh}, produced masses for exotic states that are too low in comparison with other estimates \cite{Qin:2011dd,Qin:2011xq,Burden:1996nh,Cloet:2007pi,Krassnigg:2009zh}, and exhibit gross sensitivity to model parameters for tensor-mesons \cite{Krassnigg:2010mh} and excited states \cite{Qin:2011dd,Qin:2011xq,Holl:2004fr,Holl:2004un}.  In these circumstances one must conclude that physics important to these states is omitted.
One anticipates therefore that significant qualitative advances in understanding the essence of QCD could be made with symmetry-preserving kernel \emph{Ans\"atze} that express important additional nonperturbative effects, which are impossible to capture in any finite sum of contributions.  Such an approach has recently become available \cite{Chang:2009zb} and will be summarised below.  

\subsection{DCSB in the Bethe-Salpeter kernel}
\label{sec:building}
\begin{figure}[t]
\centerline{%
\includegraphics[width=0.30\textwidth]{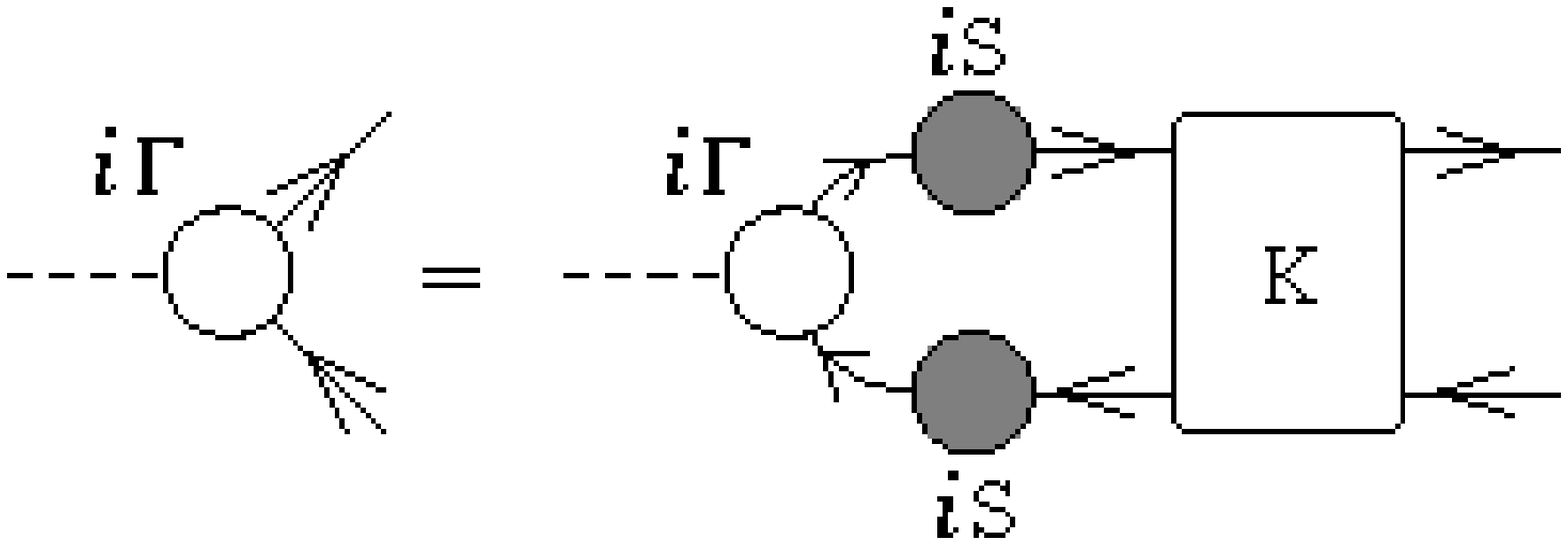}}
\centerline{%
\includegraphics[width=0.48\textwidth]{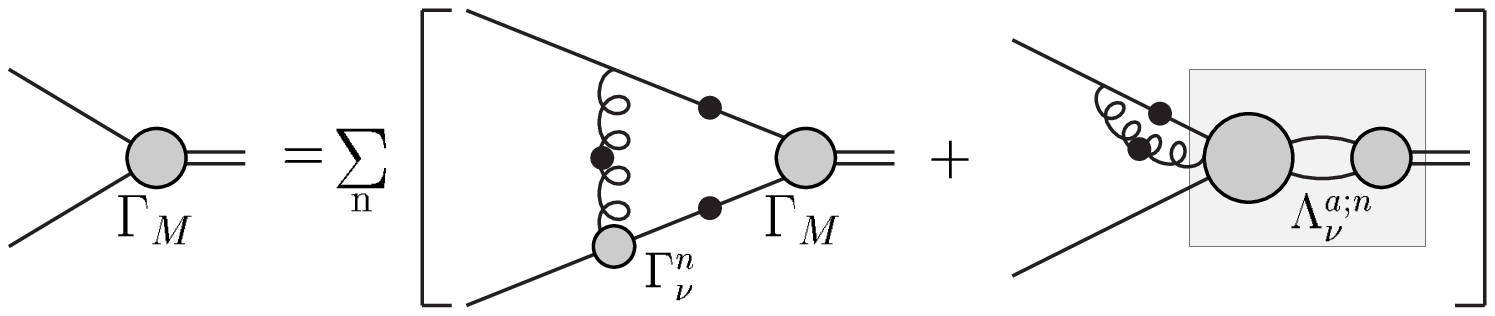}}
\caption{\label{detmoldkernel}
Omitting the inhomogeneity, the \emph{upper panel} illustrates the textbook form of the Bethe-Salpeter equation, Eq.\,(\protect\ref{bsetextbook}), whereas the \emph{lower panel} depicts the form expressed in Eq.\,(\protect\ref{genbse}).  The reversal of the total-momentum's flow is immaterial here.
N.B.\ In any symmetry-preserving truncation, beyond the leading-order identified in Ref.\,\protect\cite{Bender:1996bb}; i.e., rainbow-ladder, the Bethe-Salpeter kernel is nonplanar even if the vertex in the gap equation is planar \protect\cite{Bender:2002as}. }
\end{figure}

In order to illustrate the decisive importance of DCSB in the Bethe-Salpeter kernel, consider, e.g., flavoured pseudoscalar and axial-vector mesons, which appear as poles in the inhomogeneous BSE for the axial-vector vertex, $\Gamma_{5\mu}^{fg}$, where $f,g$ are flavour labels.  An exact form of that equation is ($k$, $q$ are relative momenta, $P$ is the total momentum flowing into the vertex, and $q_\pm = q\pm P/2$, etc.)
\begin{eqnarray}
\nonumber
\lefteqn{\Gamma_{5\mu}^{fg}(k;P) = Z_2 \gamma_5\gamma_\mu - \int_q^\Lambda g^2 D_{\alpha\beta}(k-q)\,}\\
\nonumber
&&
\times \frac{\lambda^a}{2}\,\gamma_\alpha S_f(q_+) \Gamma_{5\mu}^{fg}(q;P) S_g(q_-) \frac{\lambda^a}{2}\,\Gamma_\beta^g(q_-,k_-) \\
\nonumber
&  + &\int^\Lambda_q g^2D_{\alpha\beta}(k-q)\, \frac{\lambda^a}{2}\,\gamma_\alpha S_f(q_+) \frac{\lambda^a}{2} \Lambda_{5\mu\beta}^{fg}(k,q;P),\\
&& \rule{2em}{0ex} \label{genbse}
\end{eqnarray}
where $\Lambda_{5\mu\beta}^{fg}$ is a 4-point Schwinger function.  [The pseudoscalar vertex satisfies an analogue of Eq.\,(\ref{genbse}).]  This form of the BSE was first written in Ref.\,\cite{Bender:2002as} and is illustrated in the lower-panel of Fig.\,\ref{detmoldkernel}.  The diagrammatic content of the right-hand-side is completely equivalent to that of Eq.\,(\ref{bsetextbook}), which is depicted in the upper-panel of the figure.  However, in striking qualitative opposition to that textbook equation, Eq.\,(\ref{genbse}) partly embeds the solution vertex in the four-point function, $\Lambda$, whilst simultaneously explicating a part of the effect of the dressed-quark-gluon vertex.  This has the invaluable consequence of enabling the derivation of both an integral equation for the new Bethe-Salpeter kernel, $\Lambda$, in which the driving term is the dressed-quark-gluon vertex \cite{Bender:2002as}, and a Ward-Takahashi identity relating $\Lambda$ to that vertex \cite{Chang:2009zb}.  No similar equations have yet been found for $K$ and hence the textbook form of the BSE, whilst tidy, is very limited in its capacity to expose the effects of DCSB in bound-state physics.

As emphasised above, no study of light-quark hadrons is dependable if it fails to comply with the axial-vector Ward-Takahashi identity, Eq.\,(\ref{avwtimN}).  The condition
\begin{eqnarray}
\nonumber && P_\mu \Lambda_{5\mu\beta}^{fg}(k,q;P) + i [m_f(\zeta)+m_g(\zeta)] \Lambda_{5\beta}^{fg}(k,q;P)\\
&=& \Gamma_\beta^f(q_+,k_+) \, i\gamma_5+ i\gamma_5 \, \Gamma_\beta^g(q_-,k_-) \,, \label{LavwtiGamma}
\end{eqnarray}
where $\Lambda_{5\beta}^{fg}$ is the analogue of $\Lambda_{5\mu\beta}^{fg}$ in the pseudoscalar equation, is necessary and sufficient to ensure the Ward-Takahashi identity is satisfied by the solution of Eqs.\,(\ref{gendseN}) and (\ref{genbse}) \cite{Chang:2009zb}.

Consider Eq.\,(\ref{LavwtiGamma}).  Rainbow-ladder is the lead\-ing-or\-der term in the systematic DSE truncation scheme of Refs.\,\cite{Munczek:1994zz,Bender:1996bb}.  It corresponds to $\Gamma_\nu^f=\gamma_\nu$, in which case Eq.\,(\ref{LavwtiGamma}) is solved by $\Lambda_{5\mu\beta}^{fg}\equiv 0 \equiv \Lambda_{5\beta}^{fg}$.  This is the solution that indeed provides the rainbow-ladder forms of Eq.\,(\ref{genbse}).  Such consistency will be apparent in any valid systematic term-by-term improvement of the rainbow-ladder truncation.

However, since the two-point functions of elementary excitations are strongly modified in the infrared, one must accept that the same is generally true for three-point functions; i.e., the vertices.  Hence the bare vertex will be a poor approximation to the complete result unless there are extenuating circumstances.  This is readily made apparent, for with a dressed-quark propagator of the form in Eq.\,\eqref{SgeneralN}, one finds immediately that the Ward-Takahashi identity is breached; viz.,
\begin{equation}
P_\mu i \gamma_\mu \neq S^{-1}(k+P/2) - S^{-1}(k-P/2)\,,
\end{equation}
and the violation is significant whenever and wherever the mass function in Fig.\,\ref{gluoncloud} is large.  This was actually realised early on, with studies of the fermion--gauge-boson vertex in Abelian gauge theories \cite{Ball:1980ay} that have inspired numerous ensuing analyses.  The importance of this dressing to the reliable computation of hadron physics observables was exposed in Refs.\,\cite{Frank:1994mf,Roberts:1994hh}, insights from which have subsequently been exploited effectively; e.g., Refs.\,\cite{Chang:2011vu,Maris:1997hd,Maris:2000sk,Eichmann:2008ef,Cloet:2008re,Chang:2010hb,%
Eichmann:2011vu,Chang:2011ei,Wilson:2011rj}.

The most important feature of the perturbative or bare vertex is that it cannot cause spin-flip transitions; namely, it is an helicity conserving interaction.  However, one must expect that DCSB introduces nonperturbatively generated structures that very strongly break helicity conservation.  These contributions will be large when the dressed-quark mass-function is large.  Conversely, they will vanish in the ultraviolet; i.e., on the perturbative domain.  The exact form of the vertex contributions is still the subject of study but their existence is model-independent.

Critical now is a realisation that Eq.\,(\ref{LavwtiGamma}) is far more than just a device for checking a truncation's consistency.  For, just as the vector Ward-Takahashi identity has long been used to build \emph{Ans\"atze} for the dressed-quark-photon vertex \cite{Roberts:1994dr,Ball:1980ay,Kizilersu:2009kg,Bashir:2011dp}, Eq.\,(\ref{LavwtiGamma}) provides a tool for constructing a symmetry-preserving kernel of the BSE that is matched to any reasonable form for the dressed-quark-gluon vertex which appears in the gap equation.  With this powerful capacity, Eq.\,(\ref{LavwtiGamma}) achieves a goal that has been sought ever since the Bethe-Salpeter equation was introduced \cite{Salpeter:1951sz}.  As will become apparent, it produces a symmetry-preserving kernel that promises to enable the first reliable Poincar\'e invariant calculation of the spectrum of mesons with masses larger than 1\,GeV.

The utility of Eq.\,(\ref{LavwtiGamma}) was illustrated in Ref.\,\cite{Chang:2009zb} through an application to ground state pseudoscalar and scalar mesons composed of equal-mass $u$- and $d$-quarks.  To this end, it was supposed that in Eq.\,(\ref{gendseN}) one employs an \emph{Ansatz} for the quark-gluon vertex which satisfies
\begin{equation}
P_\mu i \Gamma_\mu^f(k_+,k_-) = {\cal B}(P^2)\left[ S_f^{-1}(k_+) - S_f^{-1}(k_-)\right] , \label{wtiAnsatz}
\end{equation}
with ${\cal B}$ flavour-independent.  (N.B.\ While the true quark-gluon vertex does not satisfy this identity, owing to the form of the Slavnov-Taylor identity which it does satisfy, it is plausible that a solution of Eq.\,(\protect\ref{wtiAnsatz}) can provide a reasonable pointwise approximation to the true vertex \cite{Bhagwat:2004kj}.)  Given Eq.\,(\ref{wtiAnsatz}), then Eq.\,(\ref{LavwtiGamma}) entails ($l=q-k$)
\begin{equation}
i l_\beta \Lambda_{5\beta}^{fg}(k,q;P) =
{\cal B}(l^2)\left[ \Gamma_{5}^{fg}(q;P) - \Gamma_{5}^{fg}(k;P)\right], \label{L5beta}
\end{equation}
with an analogous equation for $P_\mu l_\beta i\Lambda_{5\mu\beta}^{fg}(k,q;P)$.  This identity can be solved to obtain
\begin{equation}
\Lambda_{5\beta}^{fg}(k,q;P)  :=  {\cal B}((k-q)^2)\, \gamma_5\,\overline{ \Lambda}_{\beta}^{fg}(k,q;P) \,, \label{AnsatzL5a}
\end{equation}
with, using an obvious analogue of Eq.\,(\ref{genGpi}),
\begin{eqnarray}
\nonumber
\lefteqn{\overline{ \Lambda}_{\beta}^{fg}(k,q;P) =
2 \ell_\beta \, [ i \Delta_{E_5}(q,k;P)+ \gamma\cdot P \Delta_{F_5}(q,k;P) ]}\\
\nonumber
&+&  \gamma_\beta \, \Sigma_{G_5}(q,k;P) +  2 \ell_\beta \,  \gamma\cdot\ell\, \Delta_{G_5}(q,k;P)+[ \gamma_\beta,\gamma\cdot P] \\
& & \times
\Sigma_{H_5}(q,k;P) + 2 \ell_\beta  [ \gamma\cdot\ell ,\gamma\cdot P]  \Delta_{H_5}(q,k;P) \,,\label{AnsatzL5b}
\end{eqnarray}
where $\ell=(q+k)/2$, $\Sigma_{\Phi}(q,k;P) = [\Phi(q;P)+\Phi(k;P)]/2$ and $\Delta_{\Phi}(q,k;P) = [\Phi(q;P)-\Phi(k;P)]/[q^2-k^2]$.

Now, given any \emph{Ansatz} for the quark-gluon vertex that satisfies Eq.\,(\ref{wtiAnsatz}), then the pseudoscalar analogue of Eq.\,(\ref{genbse}), and Eqs.\,(\ref{gendseN}), (\ref{AnsatzL5a}), (\ref{AnsatzL5b}) provide a symmetry-preserving closed system whose solution predicts the properties of pseudoscalar mesons.
The relevant scalar meson equations are readily derived.  With these systems one can anticipate, elucidate and understand the influence on hadron properties of the rich nonperturbative structure expected of the fully-dressed quark-gluon vertex in QCD: in particular, that of the dynamically generated dressed-quark mass function, whose impact is quashed at any finite order in the truncation scheme of Ref.\,\cite{Bender:1996bb}, or any kindred scheme.

To proceed one need only specify the gap equation's kernel because, as noted above, the BSEs are completely defined therefrom.  To complete the illustration \cite{Chang:2009zb} a simplified form of the effective interaction in Ref.\,\cite{Maris:1997tm} was employed and two vertex \emph{Ans\"atze} were compared; viz., the bare vertex $\Gamma_\mu^g = \gamma_\mu$, which defines the rainbow-ladder truncation of the DSEs and omits vertex dressing; and the Ball-Chiu (BC) vertex \cite{Ball:1980ay}, which nonperturbatively incorporates some of the vertex dressing associated with DCSB:
\begin{eqnarray}
\nonumber
\lefteqn{i\Gamma^g_\mu(q,k)  =i\Sigma_{A^g}(q^2,k^2)\,\gamma_\mu}\\
&+& 2 \ell_\mu \biggL i\gamma\cdot \ell\,
\Delta_{A^g}(q^2,k^2) + \Delta_{B^g}(q^2,k^2)\biggR \!. 
\label{bcvtx}
\end{eqnarray}

\begin{figure}[t]
\vspace*{-3ex}

\includegraphics[clip,width=0.35\textheight]{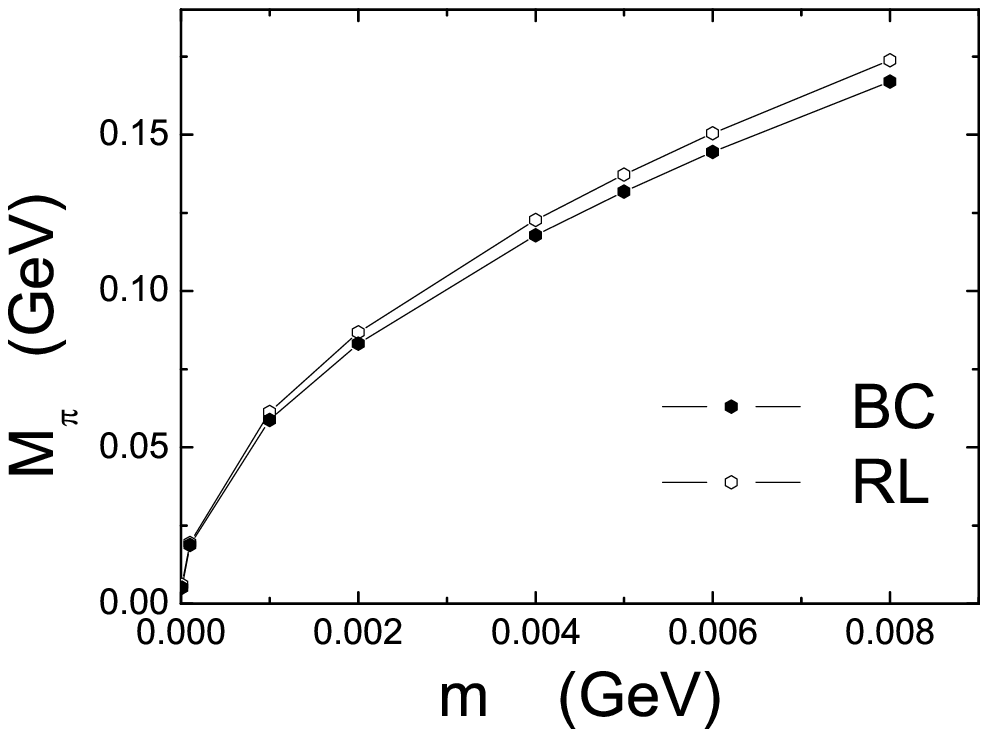}
\vspace*{-7ex}

\includegraphics[clip,width=0.35\textheight]{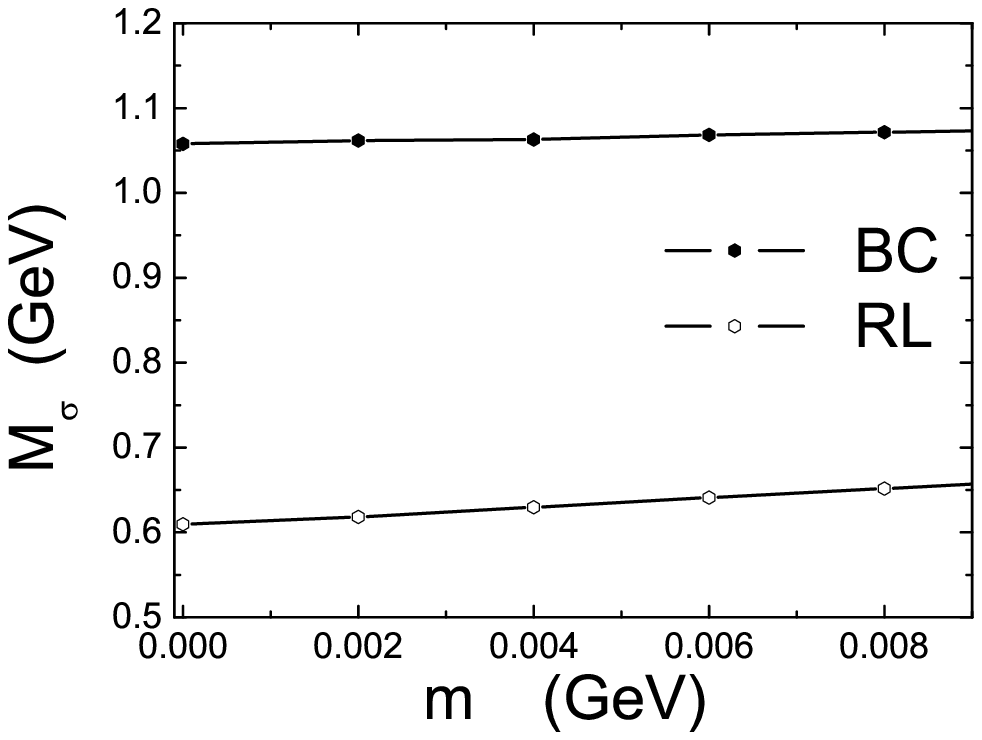}
\vspace*{-5ex}



\caption{\label{massDlarge} Dependence of pseudoscalar (upper panel) and scalar (lower) meson masses on the current-quark mass, $m$.  The Ball-Chiu vertex (BC) result is compared with the rainbow-ladder (RL) result.  (Figure adapted from Ref.\,\protect\cite{Chang:2009zb}.)}
\end{figure}

A particular novelty of the study is that one can calculate the current-quark-mass-dependence of meson masses using a symmetry-preserving DSE truncation whose diagrammatic content is unknown. That dependence is depicted in Fig.\,\ref{massDlarge} and compared with the rainbow-ladder result.  The $m$-dependence of the pseudoscalar meson's mass provides numerical confirmation of the algebraic fact that the axial-vector Ward-Takahashi identity is preserved by both the rainbow-ladder truncation and the BC-consistent \emph{Ansatz} for the Bethe-Salpeter kernel.  The figure also shows that the axial-vector Ward-Takahashi identity and DCSB conspire to shield the pion's mass from material variation in response to dressing the quark-gluon vertex \cite{Roberts:2007jh,Bender:2002as,Bhagwat:2004hn}.

As noted in Ref.\,\cite{Chang:2009zb}, a rainbow-ladder kernel with realistic interaction strength yields
\begin{equation}
\label{epsilonRL}
\varepsilon_\sigma^{\rm RL} := \frac{2 M(0) - m_\sigma }{2 M(0)}
\rule[-2.5ex]{0.1ex}{6ex}_{\rm RL}
= (0.3 \pm 0.1)\,,
\end{equation}
which can be contrasted with the value obtained using the BC-consistent Bethe-Salpeter kernel; viz.,
\begin{equation}
\label{epsilonBC}
\varepsilon_\sigma^{\rm BC} \lesssim 0.1\,.
\end{equation}
Plainly, significant additional repulsion is present in the BC-consistent truncation of the scalar BSE.

\begin{figure}[t]

\includegraphics[clip,width=0.45\textwidth]{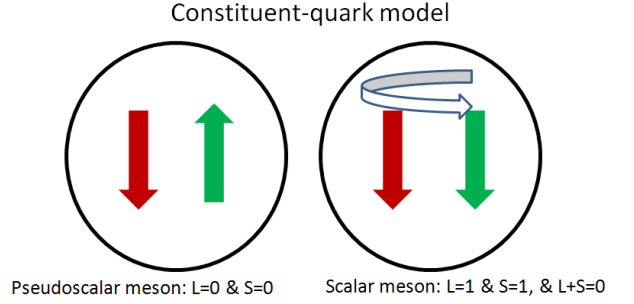}

\caption{\label{fig:ScalarCQM}
In constituent-quark-like models, pseudoscalar mesons are $^1\!S_0$ states -- constituent spins antiparallel and zero orbital angular momentum; and scalar mesons are $^3\!P_0$ states -- constituent spins parallel and one unit of orbital angular momentum.  Hence a scalar is a spin and orbital excitation of a pseudoscalar meson.}

\end{figure}

Scalar mesons are commonly identified as $^3\!P_0$ states, see Fig.\,\ref{fig:ScalarCQM}.  This assignment expresses a constituent-quark-model perspective, from which a $J^{PC}=0^{++}$ fermion-antifermion bound-state must have the constituents' spins aligned and one unit of constituent orbital angular momentum.  Hence a scalar is a spin and orbital excitation of a pseudoscalar meson.  Of course, no constituent-quark-model can be connected systematically with QCD.  Nevertheless, the presence of orbital angular momentum in a hadron's rest frame is a necessary consequence of Poincar\'e covariance and the momentum-dependent vector-boson-exchange character of QCD \cite{Roberts:2007ji,Bhagwat:2006xi,Bhagwat:2006pu}, so there is a realisation in QCD of the quark-model anticipation.

Extant studies of realistic corrections to the rainbow-ladder truncation show that they reduce hyperfine splitting \cite{Bhagwat:2004hn}.  Hence, with the comparison between Eqs.\,(\ref{epsilonRL}) and (\ref{epsilonBC}) one has a clear indication that in a Poincar\'e covariant treatment the BC-consistent truncation magnifies spin-orbit splitting, an effect which can be attributed to the influence of the quark's dynamically-enhanced scalar self-energy \cite{Roberts:2007ji} in the Bethe-Salpeter kernel.

\subsection{Quark anomalous magnetic moments}
\label{spectrum2}
It was conjectured in Ref.\,\cite{Chang:2009zb} that a full realisation of DCSB in the Bethe-Salpeter kernel will have a big impact on mesons with mass greater than 1\,GeV.  Moreover, that it can overcome a longstanding failure of theoretical hadron physics.  Namely, no extant continuum hadron spectrum calculation is believable because all symmetry preserving studies produce a splitting between vector and axial-vector mesons that is far too small: just one-quarter of the experimental value (see, e.g., Refs.\,\cite{Watson:2004kd,Maris:2006ea,Cloet:2007pi,Fischer:2009jm}).  Significant developments have followed that conjecture \cite{Chang:2010hb,Chang:2011ei} and will now be related.

In Dirac's relativistic quantum mechanics, a fermion with charge $q$ and mass $m$, interacting with an electromagnetic field, has a magnetic moment $\mu= q/[2 m]$ \cite{Dirac:1928hu}.  For the electron, this prediction held true for twenty years, until improvements in experimental techniques enabled the discovery of a small deviation \cite{Foley:1948zz}, with the moment increased by a multiplicative factor: $1.00119\pm 0.00005$.  This correction was explained by the first systematic computation using renormalised quantum electrodynamics (QED) \cite{Schwinger:1948iu}:
\begin{equation}
\label{anommme}
\frac{q}{2m} \to \left(1 + \frac{\alpha}{2\pi}\right) \frac{q}{2m}\,,
\end{equation}
where $\alpha$ is QED's fine structure constant.  The agreement with experiment established quantum electrodynamics as a valid tool.  The correction defines the electron's \emph{anomalous magnetic moment}, which is now known with extraordinary precision and agrees with theory at O$(\alpha^5)$ \cite{Mohr:2008fa}.

The fermion-photon coupling in QED is described by:
\begin{equation}
\label{QEDinteraction}
\int d^4\! x\, i q \,\bar\psi(x) \gamma_\mu \psi(x)\,A_\mu(x)\,,
\end{equation}
where $\psi(x)$, $\bar\psi(x)$ describe the fermion field and $A_\mu(x)$ describes the photon.  This interaction generates the following electromagnetic current for an on-shell Dirac fermion ($k=p_f -p_i$),
\begin{equation}
\label{ecurrent2}
i q \, \bar u(p_f) \left[ \gamma_\mu F_1(k^2)+ \frac{1}{2 m} \sigma_{\mu\nu} k_\nu F_2(k^2)\right] u(p_i)\,,
\end{equation}
where: $F_1(k^2)$, $F_2(k^2)$ are form factors; and $u(p)$, $\bar u(p)$ are electron spinors.  Using their Euclidean space definition, one can derive a Gordon-identity; viz., with $2 \ell=p_f + p_i$,
\begin{equation}
\label{Gordon}
2 m \, \bar u(p_f) i \gamma_\mu u(p_i) = \bar u(p_f)\left[ 2 \ell_\mu + i \sigma_{\mu\nu} k_\nu \right]u(p_i)\,.
\end{equation}
With this rearrangement one sees that for massive fermions the interaction can be decomposed into two terms: the first describes the spin-independent part of the fermion-photon interaction, and is common to spin-zero and spin-half particles, whilst the second expresses the spin-dependent, helicity flipping part.
Moreover, one reads from Eqs.\,(\ref{ecurrent2}) and (\ref{Gordon}) that a point-particle in the absence of radiative corrections has $F_1 \equiv 1$ and $F_2 \equiv 0$, and hence Dirac's value for the magnetic moment.  The anomalous magnetic moment in Eq.\,(\ref{anommme}) corresponds to $F_2(0) = \alpha/2\pi$.

One infers from Eq.\,(\ref{Gordon}) that an anomalous contribution to the magnetic moment can be associated with an additional interaction term:
\begin{equation}
\label{anominteraction}
\int d^4\! x\, \mbox{\small $\frac{1}{2}$} q \, \bar \psi(x) \sigma_{\mu\nu} \psi(x) F_{\mu\nu}(x)\,,
\end{equation}
where $F_{\mu\nu}(x)$ is the gauge-boson field strength tensor.  This term is invariant under local $U(1)$ gauge transformations but is not generated by minimal substitution in the action for a free Dirac field.

Consider the effect of the global chiral transformation $\psi(x) \to \exp(i \theta\gamma_5) \psi(x)$.  The term in Eq.\,(\ref{QEDinteraction}) is invariant.  However, the interaction of Eq.\,(\ref{anominteraction}) is not.  These observations facilitate the understanding of a general result: $F_2\equiv 0$ for a massless fermion in a quantum field theory with chiral symmetry realized in the Wigner mode; i.e., when the symmetry is not dynamically broken.  A firmer conclusion can be drawn.  For $m=0$ it follows from Eq.\,(\ref{Gordon}) that Eq.\,(\ref{QEDinteraction}) does not mix with the helicity-flipping interaction of Eq.\,(\ref{anominteraction}) and hence a massless fermion does not possess a measurable magnetic moment.

A reconsideration of Ref.\,\cite{Schwinger:1948iu} reveals no manifest conflict with these facts.  The perturbative expression for $F_2(0)$ contains a multiplicative numerator factor of $m$ and the usual analysis of the denominator involves steps that are only valid for $m\neq 0$.  Fundamentally, there is no conundrum because QED is not an asymptotically free theory and hence alone does not have a well-defined nonperturbative chiral limit. (N.B.\ Four-fermion operators become relevant in strong-coupling QED and must be included in order to obtain a well-defined continuum limit \cite{Rakow:1990jv,Reenders:1999bg}.)

On the other hand, in QCD the chiral limit is rigorously defined nonperturbatively \cite{Maris:1997tm}.  (It remains to be seen whether the theory thus obtained is meaningful, as indicated in the antepenultimate paragraph of Sec.\,\ref{sec:inmeson}.)  The analogue of Schwinger's one-loop calculation can then be carried out to find an anomalous \emph{chromo}-magnetic moment for the quark.  There are two diagrams in this case: one similar in form to that in QED; and another owing to the gluon self-interaction.  One reads from Ref.\,\cite{Davydychev:2000rt} that the perturbative result vanishes in the chiral limit.  However, Fig.\,\ref{gluoncloud} demonstrates that chiral symmetry is dynamically broken in QCD and one must therefore ask whether this affects the chromomagnetic moment.

Of course, it does; and it is now known that this is signalled by the appearance of $\Delta_{B^g}$ in Eq.\,(\ref{bcvtx}).  If one writes the quark-gluon vertex as
\begin{eqnarray}
\lefteqn{i \Gamma_\mu(p_f,p_i;k) = \lambda_1(p_f,p_i;k)\, i \gamma_\mu} \\
\nonumber &+& 2 \ell_\mu \biggL i\gamma\cdot \ell\,
\lambda_2(p_f,p_i;k) + \lambda_3(p_f,p_i;k) \biggR  + \ldots \,, \rule{2em}{0ex}
\end{eqnarray}
then contemporary simulations of lattice-regularised QCD \cite{Skullerud:2003qu} and DSE studies \cite{Bhagwat:2004kj} agree that
\begin{equation}
\label{alpha3B}
\lambda_3(p,p;0) \approx \frac{d}{dp^2} B(p^2,\zeta)
\end{equation}
and also on the form of $\lambda_1(p,p;0)$, which is functionally similar to $A(p^2,\zeta)$.  However, owing to non-orthogonality of the tensors accompanying $\lambda_1$ and $\lambda_2$, it is difficult to obtain a lattice signal for $\lambda_2$.  One must therefore consider the DSE prediction for $\lambda_2$ in Ref.\,\cite{Bhagwat:2004kj} more reliable.

As pointed out above, perturbative massless-QCD conserves helicity so the quark-gluon vertex cannot perturbatively have a term with the helicity-flipping characteristics of $\lambda_3$.  Equation~(\ref{alpha3B}) is thus remarkable, showing that the dressed-quark-gluon vertex contains at least one chirally-asymmetric component whose origin and size owe solely to DCSB; and Sec.\,\ref{sec:building} illustrates that $\lambda_3$ has a material impact on the hadron spectrum.

This reasoning is extended in Ref.\,\cite{Chang:2010hb}: massless fermions in gauge field theories cannot possess an anomalous chromo/electro-magnetic moment because the term that describes it couples left- and right-handed fermions; however, if chiral symmetry is strongly broken dynamically, then the fermions should also posses large anomalous magnetic moments.  Such an effect is expressed in the dressed-quark-gluon vertex via a term
\begin{equation}
\label{qcdanom1}
\Gamma_\mu^{\rm acm_5} (p_f,p_i;k) = \sigma_{\mu\nu} k_\nu \, \tau_5(p_f,p_i,k)\,.
\end{equation}

\begin{figure}[t]
\includegraphics[width=0.46\textwidth]{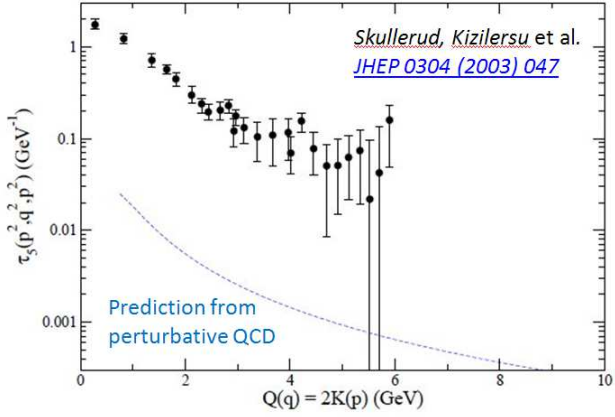}
\caption{\label{latticetau5}
Direct anomalous chromomagnetic moment contribution to the dressed-quark-gluon vertex computed in quenched-QCD with current-quark mass $m\sim 100\,$MeV \protect\cite{Skullerud:2003qu}.  The one-loop perturbative result is shown for comparison.
Plainly, the nonperturbative result is two orders-of-magnitude larger than the perturbative computation.  This level of magnification is typical of DCSB; e.g., with a current-quark mass of 4\,MeV, one obtains $M(p^2=0)\sim 400\,$MeV.
(Figure adapted from Ref.\,\protect\cite{Skullerud:2003qu}.)}
\end{figure}

That QCD generates a strongly momentum-dependent chromomagnetic form factor in the quark-gluon vertex, $\tau_5$, with a large DCSB-component, is confirmed in Ref.\,\cite{Skullerud:2003qu}.  Only a particular kinematic arrangement was readily accessible in that lattice simulation but this is enough to learn that, at the current-quark mass considered: $\tau_5$ is roughly two orders-of-magnitude larger than the perturbative form (see Fig.\,\ref{latticetau5}); and
\begin{equation}
\label{boundtau5}
\forall p^2>0: \; |\tau_5(p,-p;2 p)| \gtrsim |\lambda_3(p,p;0)|\,.
\end{equation}
The magnitude of the lattice result is consistent with instanton-liquid model estimates \cite{Kochelev:1996pv,Diakonov:2002fq}.

This large chromomagnetic moment is likely to have a broad impact on the properties of light-quark systems \cite{Diakonov:2002fq,Ebert:2005es}.  In particular, as will be illustrated in Sec.\,\ref{sec:a1rho}, it can explain the longstanding puzzle of the mass splitting between the $a_1$- and $\rho$-mesons in the hadron spectrum \cite{Chang:2011ei}.  Here a different novelty will be elucidated; viz., the manner in which the quark's chromomagnetic moment generates a quark anomalous \emph{electro}magnetic moment.  This demonstration is only possible now that the method of Ref.\,\cite{Chang:2009zb} is available.  It was accomplished \cite{Chang:2010hb} using the same simplification of the effective interaction in Ref.\,\cite{Maris:1997tm} that produced Figs.\,\ref{massDlarge}.

In order to understand the vertex \emph{Ansatz} used in Ref.\,\cite{Chang:2010hb}, it is necessary to return to perturbation theory.  As mentioned above Eq.\,\eqref{alpha3B}, one can determine from Ref.\,\cite{Davydychev:2000rt} that at leading-order in the coupling, $\alpha_s$, the three-gluon vertex does not contribute to the QCD analogue of Eq.\,(\ref{anommme}) and the Abelian-like diagram produces the finite and negative correction $(-\alpha_s/[12 \pi])$.
The complete cancellation of ultraviolet divergences occurs only because of the dynamical generation of another term in the transverse part of the quark-gluon vertex; namely,
\begin{equation}
\Gamma_\mu^{\rm acm_4}(p_f,p_i) = [ \ell_\mu^{\rm T} \gamma\cdot  k + i \gamma_\mu^{\rm T} \sigma_{\nu\rho}\ell_\nu k_\rho] \tau_4(p_f,p_i)\,,
\end{equation}
with $T_{\mu\nu} = \delta_{\mu\nu} - k_\mu k_\nu/k^2$, $a_\mu^{\rm T} := T_{\mu\nu}a_\nu$. (N.B.\ The tensor denominated $\Gamma_\mu^4$ here is labelled $T^8$ in Refs.\,\cite{Kizilersu:2009kg,Bashir:2011dp}.)
Cognisant of this, one may use a simple \emph{Ansatz} to express the dynamical generation of an anomalous chromomagnetic moment via the dressed-quark gluon vertex; viz.,
\begin{eqnarray}
\label{ourvtx}
\tilde\Gamma_\mu(p_f,p_i)  & = & \Gamma_\mu^{\rm BC}(p_f,p_i) +
\Gamma_\mu^{\rm acm}(p_f,p_i)\,,\\
\Gamma_\mu^{\rm acm}(p_f,p_i) &=& \Gamma_\mu^{\rm acm_4}(p_f,p_i) + \Gamma_\mu^{\rm acm_5}(p_f,p_i)\,,\rule{1.5em}{0ex}
\label{ourvtxacm}
\end{eqnarray}
with $\tau_5(p_f,p_i) =  (-7/4)\Delta_B(p_f^2,p_i^2)$ and
\begin{equation}
\tau_4(p_f,p_i) = {\cal F}(z) \bigg[  \frac{1-2\eta}{M_E}\Delta_B(p_f^2,p_i^2) - \Delta_A(p_f^2,p_i^2) \bigg]. \label{tau4}
\end{equation}
The damping factor ${\cal F}(z)=(1- \exp(-z))/z$, $z=(p_i^2 + p_f^2- 2 M_E^2)/\Lambda_{\cal F}^2$, $\Lambda_{\cal F}=1\,$GeV, simplifies numerical analysis but is otherwise irrelevant; and $M_E=\{ p| p>0, p^2 = M^2(p^2)\}$ is the Euclidean constituent-quark mass.
A confined quark does not possess a mass-shell (Sec.\,\ref{Sect:Conf}).  Hence, one cannot unambiguously assign a single value to its anomalous magnetic moment.  One can nonetheless compute a magnetic moment distribution.  At each value of $p^2$, spinors can be defined to satisfy the free-particle Euclidean Dirac equation with mass $m\to M(p^2)=:\varsigma$, so that
\begin{eqnarray}
\nonumber
\lefteqn{\bar u(p_f;\varsigma) \, \Gamma_\mu( p_f,p_i;k)\,  u(p_i;\varsigma)}\\
& = &
\bar u(p_f) [ F_1(k^2) \gamma_\mu + \frac{1}{2 \varsigma} \,\sigma_{\mu \nu} k_\nu F_2(k^2)] u(p_i) \rule{1.5em}{0ex}
\label{GenSpinors}
\end{eqnarray}
and then, from Eqs.\,(\ref{ourvtx}) -- (\ref{tau4}),
\begin{equation}
\label{kappaacm}
\kappa^{\rm acm}(\varsigma) = \frac{ - 2 \varsigma \, \eta \delta_B^{\varsigma}}
    {\sigma_A^{\varsigma} - 2 \varsigma^2 \delta_A^{\varsigma}+ 2 \varsigma \delta_B^{\varsigma} }\,,
\end{equation}
where $\sigma_A^{\varsigma} = \Sigma_A(\varsigma,\varsigma)$, $\delta_A^{\varsigma} = \Delta_A(\varsigma,\varsigma)$, etc.  The numerator's simplicity owes to a premeditated cancellation between $\tau_4$ and $\tau_5$, which replicates the one at leading-order in perturbation theory.
Where a comparison of terms is possible, this vertex \emph{Ansatz} is semi-quantitatively in agreement with Refs.\,\cite{Skullerud:2003qu,Bhagwat:2004kj}.  However, the presence and understanding of the role of $\Gamma_\mu^{\rm acm_4}$ is a novel contribution by Ref.\,\cite{Chang:2010hb}.  N.B.\ It is apparent from Eq.\,(\ref{kappaacm}) that $\kappa^{\rm acm} \propto m^2$ in the absence of DCSB, so that $\kappa^{\rm acm}/[2m]\to 0$ in the chiral limit.

The BSE for the quark-photon vertex can be written following the method of Ref.\,\cite{Chang:2009zb}.  Since the method guarantees preservation of the Ward-Takahashi identities, the general form of the solution is
\begin{eqnarray}
\Gamma_\mu^\gamma(p_f,p_i) & = & \Gamma_\mu^{\rm BC}(p_f,p_i) + \Gamma_\mu^{\rm T}(p_f,p_i)\,,\\
\nonumber
\Gamma_\mu^{\rm T}(p_f,p_i) & = &
\gamma_\mu^{\rm T} \hat F_1
+ \sigma_{\mu\nu} k_\nu \hat F_2
+ T_{\mu\rho} \sigma_{\rho\nu} \ell_\nu \,\ell\cdot k\, \hat F_3\\
\nonumber & &
+ [ \ell_\mu^{\rm T} \gamma\cdot  k + i \gamma_\mu^{\rm T} \sigma_{\nu\rho}\ell_\nu k_\rho] \hat F_4 - i \ell_\mu^{\rm T} \hat F_5\\
\nonumber & & +\, \ell_\mu^{\rm T} \gamma\cdot k \, \ell \cdot  k\, \hat F_6 - \ell_\mu^{\rm T} \gamma\cdot \ell \, \hat F_7  \\
%
& & + \ell_\mu^T \sigma_{\nu\rho} \ell_\nu k_\rho \hat F_8\,,
\end{eqnarray}
where $\{\hat F_i|i=1,\ldots,8\}$ are scalar functions of Lorentz-invariants constructed from $p_f$, $p_i$, $k$.  The Ward-Takahashi identity is plainly satisfied; viz., 
\begin{equation}
k_\mu i \Gamma_\mu(p_f,p_i) = k_\mu i \Gamma_\mu^{\rm BC} (p_f,p_i)
= S^{-1}(p_f) - S^{-1}(p_i)\,.
\end{equation}

\begin{figure}[t]
\vspace*{-3ex}

\centerline{\includegraphics[clip,width=0.43\textwidth]{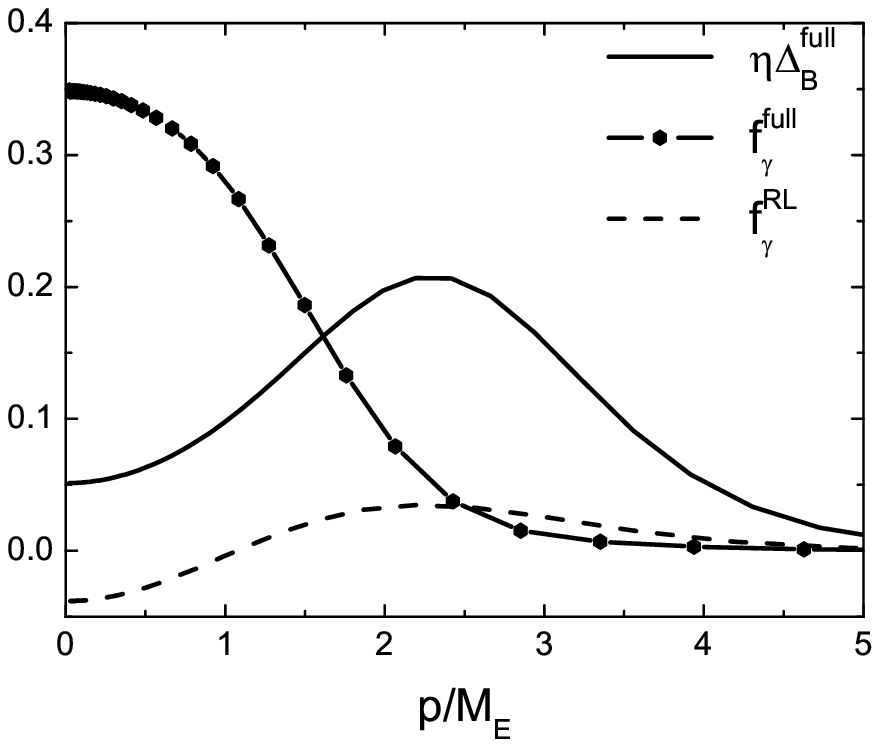}}
\vspace*{-6ex}

\centerline{\includegraphics[clip,width=0.45\textwidth]{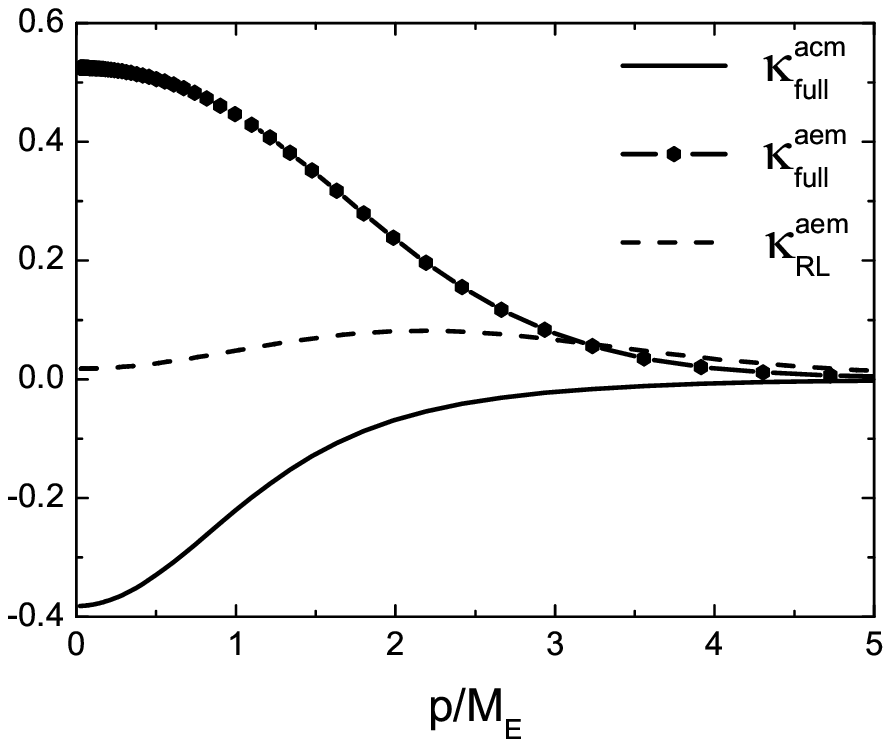}\rule{1em}{0em}}
\vspace*{-5ex}



\caption{\label{figACM}
\emph{Left panel} -- $f_\gamma$ (GeV$^{-1}$) in Eq.\,(\protect\ref{fgamma})
cf.\ $(-7/4)\Delta_B(p^2,p^2)$, both computed using Eq.\,(\protect\ref{ourvtx}) and the same simplification of the interaction in Ref.\,\protect\cite{Maris:1997tm}.
\emph{Right panel} -- Anomalous chromo- and electro-magnetic moment distributions for a dressed-quark, computed using Eq.\,(\protect\ref{kappavalue}).
The dashed-curve in both panels is the rainbow-ladder (RL) truncation result.}
\end{figure}

Figure~\ref{figACM} depicts the results obtained for the quark's anomalous electromagnetic moment form factor
\begin{equation}
f_\gamma(p) :=  \lim_{p_f\to p}\frac{-1}{12\,k^2}{\rm tr}\, \sigma_{\mu\nu} k_\mu \Gamma_\nu^\gamma(p_f,p)
=  \hat F_2+ \frac{1}{3}p^2 \hat F_8\,. \label{fgamma}
\end{equation}
The result is evidently sizable.
It is worth reiterating that $f_\gamma$ is completely nonperturbative: in the chiral limit, at any finite order in perturbation theory, $f_\gamma\equiv 0$.  For contrast the figure also displays the result computed in the rainbow-ladder truncation of QCD's DSEs. As the leading-order in a systematic but stepwise symmetry-preserving scheme \cite{Bender:1996bb}, this truncation only partially expresses DCSB: it is exhibited by the dressed-quark propagator but not present in the quark-gluon vertex.  In this case $f_\gamma$ is nonzero but small.  These are artefacts of the truncation that cannot be remedied at any finite order of the procedure in Ref.\,\cite{Bender:1996bb} or a kindred scheme.

Employing Eq.\,(\ref{GenSpinors}), in connection with the dressed-quark-photon vertex,
one can write an expression for the quark's anomalous electromagnetic moment distribution
\begin{equation}
\label{kappavalue}
\kappa(\varsigma)=\frac{2 \varsigma \hat F_{2} + 2 \varsigma^2 \hat F_4  +\Lambda_{\kappa}(\varsigma)}
{\sigma_{A}^{\varsigma} + \hat F_{1}-\Lambda_{\kappa}(\varsigma)}\,,
\end{equation}
where: $\Lambda_{\kappa}(\varsigma)= 2\varsigma^{2}\delta_{A}^\varsigma-2 \varsigma \delta_{B}^\varsigma -\varsigma \hat F_5 - \varsigma^2 \hat F_7$; and the $\hat F_i$ are evaluated at $p_f^2=p_i^2=M(p_f^2)^2=:\varsigma^2$, $k^2=0$.  Plainly, $\kappa(\varsigma)\equiv 0$ in the chiral limit when chiral symmetry is not dynamically broken.  Moreover, as a consequence of asymptotic freedom, $\kappa(\varsigma) \to 0$ rapidly with increasing momentum.
The computed distribution is depicted in Fig.\,\ref{figACM}.  It yields Euclidean mass-shell values:
\begin{equation}
\begin{array}{llll}
& M_{\rm full}^E = 0.44\,{\rm GeV},& \kappa_{\rm full}^{\rm acm}= -0.22\,, &
\kappa_{\rm full}^{\rm aem}= 0.45\,\\[1ex]
{\rm cf}. & M_{\rm RL}^E = 0.35\,{\rm GeV}, & \kappa_{\rm RL}^{\rm acm}= 0\,, & \kappa_{\rm RL}^{\rm aem}= 0.048 .
\end{array}
\end{equation}

It is thus apparent that DCSB produces a dressed light-quark with a momentum-dependent anomalous chromomagnetic moment, which is large at infrared momenta.  Significant amongst the consequences is the generation of an anomalous electromagnetic moment for the dressed light-quark with commensurate size but opposite sign.  (N.B.\ This result was anticipated in Ref.\,\protect\cite{Bicudo:1998qb}, which argued that DCSB usually triggers the generation of a measurable anomalous magnetic moment for light-quarks.)
The infrared dimension of both moments is determined by the Euclidean constituent-quark mass.  This is two orders-of-magnitude greater than the physical light-quark current-mass, which sets the scale of the perturbative result for both these quantities.

There are two more notable features; namely, the rainbow-ladder truncation, and low-order stepwise improvements thereof, underestimate these effects by an order of magnitude; and both the $\tau_4$ and $\tau_5$ terms in the dressed-quark-gluon vertex are indispensable for a realistic description of hadron phenomena.  Whilst a simple interaction was used to illustrate these outcomes, they are robust.

These results are stimulating a reanalysis of hadron elastic and transition electromagnetic form factors \cite{Wilson:2011rj,Chang:2011tx}, and the hadron spectrum, results of which will be described below.
Furthermore, given the magnitude of the muon ``$g_\mu-2$ anomaly'' and its assumed importance as an harbinger of physics beyond the Standard Model \cite{Jegerlehner:2009ry}, it might also be worthwhile to make a quantitative estimate of the contribution to $g_\mu-2$ from the quark's DCSB-induced anomalous moments following, e.g., the computational pattern for the hadronic light-by-light scattering component of the photon polarization tensor indicated in Ref.\,\cite{Goecke:2011pe}.

\subsection{\mbox{\boldmath $a_1$}-\mbox{\boldmath $\rho$} mass splitting}
\label{sec:a1rho}
The analysis in Ref.\,\cite{Chang:2009zb} enables one to construct a symmetry-preserving kernel for the BSE given any form for $\Gamma_\mu$.  Owing to the importance of symmetries in forming the spectrum of a quantum field theory, this is a pivotal advance.  One may now use all information available, from any reliable source, to construct the best possible vertex \emph{Ansatz}.  The last two subsections illustrated that this enables one to incorporate crucial nonperturbative effects, which any finite sum of contributions is incapable of capturing, and thereby prove that DCSB generates material, momentum-dependent anomalous chromo- and electro-magnetic moments for dressed light-quarks.

The vertex described in Sec.\,\ref{spectrum2} contains a great deal of information about DCSB.  It is the best motivated \emph{Ansatz} to date, has stimulated a detailed reanalysis of the quark-photon coupling \cite{Bashir:2011dp}, and may be used in the calculation of the masses of ground-state spin-zero and -one light-quark mesons in order to illuminate the impact of DCSB on the hadron spectrum.  This analysis expands significantly on the discussion of scalar and pseudoscalar mesons in Sec.\,\ref{sec:building}.

Given a vertex, a prediction for the spectrum follows once the gap equation's kernel is specified and the Ward-Identity solved for $\Lambda_{5\mu\beta}^{fg}$.  In the pseudoscalar and axial-vector channels the Ward-Takahashi identity for the Bethe-Salpeter kernel is solved by
\begin{eqnarray}
\nonumber
2 \Lambda_{5\beta(\mu)} &= & [\tilde \Gamma_{\beta}(q_{+},k_{+})+\gamma_{5}\tilde \Gamma_{\beta}(q_{-},k_{-})
\gamma_{5}] \\
\nonumber
&& \times \frac{1}{S^{-1}(k_{+})+S^{-1}(-k_{-})}\Gamma_{5(\mu)}(k;P)\nonumber\\
\nonumber
&+&\Gamma_{5(\mu)}(q;P)\frac{1}{S^{-1}(-q_{+})+S^{-1}(q_{-})}\\
&& \times
[\gamma_{5}\tilde\Gamma_{\beta}(q_{+},k_{+})\gamma_{5}
+\tilde\Gamma_{\beta}(q_{-},k_{-})], \rule{2em}{0ex}
\end{eqnarray}
where $\tilde \Gamma$ is the chosen \emph{Ansatz} for the quark-gluon vertex.  Kernels for other channels are readily constructed.

\begin{figure}[t]

\includegraphics[clip,width=0.46\textwidth]{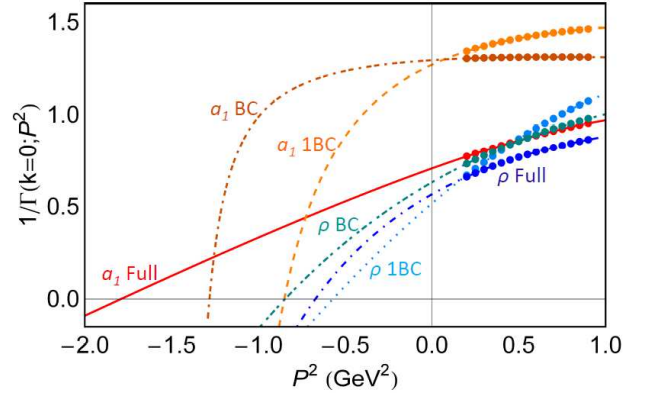}

\caption{\label{F1a1}
Illustration of the procedure used to determine meson masses,
which is fully described in Ref.\,\cite{Bhagwat:2007rj} and analogous to the method used in lattice-QCD.
\emph{Solid curve} -- $a_1$-meson, nonperturbative kernel; \emph{dot-dash-dash} -- $a_1$, kernel derived from Eq.\,(\protect\ref{bcvtx}) only (Ball-Chiu, BC); and \emph{dash} -- $a_1$, kernel derived from just the first term in Eq.\,(\ref{bcvtx}) (1BC, a minimal renormalization improvement \protect\cite{Bloch:2002eq} of the leading-order -- RL, rainbow-ladder -- kernel \protect\cite{Bender:1996bb}).
\emph{Dot-dash curve} -- $\rho$-meson, nonperturbative kernel; \emph{Dot-dash-dot} -- $\rho$, BC-kernel; and \emph{dotted} -- $\rho$, 1BC-kernel.
\emph{Points} -- values of $1/\Gamma(k=0;P^2)$ in the given channel computed with the kernel described.  Pad\'e approximants are constructed in each case; and the location of the zero is identified with $(-m_{\rm meson}^2)$.}
\end{figure}

Reference~\protect\cite{Chang:2011ei} computes ground-state masses using the interaction described in Ref.\,\cite{Qin:2011dd}, which produced Fig.\,\ref{fig:gluonrunning}; the vertex model explicated in Sec.\,\ref{spectrum2}; and the method for solving the inhomogeneous Bethe-Salpeter equation that is detailed in Ref.\,\cite{Bhagwat:2007rj}, which ensures one need only solve the gap and Bethe-Salpeter equations at spacelike momenta.  This simplifies the numerical problem.  To explain, the inhomogeneous BSE is solved for the complete Bethe-Salpeter amplitude in a particular channel on a domain of spacelike total-momenta, $P^2>0$.  Any bound-state in that channel appears as a pole in the solution at $P^2=-m_{\rm meson}^2$.  Denoting the leading Chebyshev moment of the amplitude's dominant Dirac structure by $\Gamma(k;P)$, where $k$ is the relative momentum, then $1/\Gamma(k=0;P^2)$ exhibits a zero at $(-m_{\rm meson}^2)$.  The location of that zero is determined via extrapolation of a Pad\'e approximant to the spacelike-behavior of $1/\Gamma(k=0;P^2)$.  This is illustrated for the $\rho$- and $a_1$-channels in Fig.\,\ref{F1a1}.

A full set of results is listed in Table~\ref{tablemasses}, wherein
the level of agreement between Cols.~3 and 4 illustrates the efficacy of the method used to compute masses: no difference is greater than 1\%.
Next consider $m_\sigma$ and compare Cols.~1--3.  It is an algebraic result that in the RL-truncation of QCD's DSEs, $m_\sigma \approx 2 M$, where $M$ is a constituent-like quark mass \cite{Roberts:2011cf}.  On the other hand, incorporating the quark mass function into the Bethe-Salpeter kernel via $\Gamma_\mu^{\rm BC}$ generates a strong spin-orbit interaction, which significantly boosts $m_\sigma$ \cite{Chang:2009zb}.  This feature is evidently unaffected by the inclusion of $\Gamma_\mu^{\rm acm}$; i.e., those terms associated with a dressed-quark anomalous chromomagnetic moment.
Since terms associated with pion final-state interactions were deliberately omitted from the nonperturbative kernel derived in Ref.\,\cite{Chang:2011ei}, it is noteworthy that $m_\sigma$ in Col.~1 matches estimates for the mass of the dressed-quark-core component of the $\sigma$-meson obtained using unitarised chiral perturbation theory \cite{Pelaez:2006nj,RuizdeElvira:2010cs}.  N.B.\ In addition to providing a width, such final-state interactions necessarily reduce the real part of the mass \cite{Holl:2005st,Cloet:2008fw,Chang:2009ae}.

\begin{table}[t]
\begin{center}
\caption{\label{tablemasses}
%
Col.~1: Spectrum obtained with the full nonperturbative Bethe-Salpeter kernels described herein, which express effects of DCSB.
The method of Ref.\,\protect\cite{Bhagwat:2007rj} was used:  the error reveals the sensitivity to varying the order of Pad\'e approximant.
Col.~2 -- 
Experimental values; computed, except $m_\sigma$, from isospin mass-squared averages \protect\cite{Nakamura:2010zzi}.
%
Col.~3 -- Masses determined from the inhomogeneous BSE at leading-order in the DSE truncation scheme of Ref.\,\protect\cite{Bender:1996bb} using the interaction in Ref.\,\cite{Chang:2009zb} (with this simple kernel, the Pad\'e error is negligible);
and Col.~4 -- results in Ref.\,\protect\cite{Alkofer:2002bp}, obtained directly from the homogeneous BSE at the same order of truncation.
%
}
\begin{tabular*}
{\hsize}
{|l@{\extracolsep{0ptplus1fil}}
|l@{\extracolsep{0ptplus1fil}}
|l@{\extracolsep{0ptplus1fil}}
|l@{\extracolsep{0ptplus1fil}}
|l@{\extracolsep{0ptplus1fil}}|}\hline
\rule{0em}{3ex}
    & Ref.\,\protect\cite{Chang:2011ei} & Expt.~ & \emph{RL-Pad\'e}~ & \emph{RL-direct}~ \\\hline
$m_\pi$   & $0.138 $~ & 0.138 & 0.138~ & 0.137~ \\
$m_\rho$  & $0.817 \pm 0.016$~ & 0.777 & 0.754~ & 0.758~ \\
$m_\sigma$& $0.90 \pm 0.05$  & $0.4$ -- $1.2$~ & 0.645~ & 0.645~ \\
$m_{a_1}$ & $1.30 \pm 0.11$~ & $1.24 \pm 0.04$~ & 0.938~ & 0.927~  \\
$m_{b_1}$ & $1.15 \pm 0.07$~ & $1.21 \pm 0.02$~ & $0.904 $~ & 0.912~ \\
$m_{a_1}-m_\rho$~ & $0.48 \pm 0.12$ & $0.46 \pm 0.04$ & 0.18 & 0.17 \\
$m_{b_1}-m_\rho$~ & $0.33 \pm 0.09$ & $0.43 \pm 0.02$ & 0.15 & 0.15 \\\hline
\end{tabular*}
\end{center}
\end{table}

Now compare the entries in Rows~2, 4--6.  The $\rho$- and $a_1$-mesons have been known for more than thirty years and are typically judged to be parity-partners; i.e., they would be degenerate if chiral symmetry were manifest in QCD.  Plainly, they are not, being split by roughly $450\,$MeV (i.e., $> m_\rho/2$).  It is suspected that this large splitting owes to DCSB.  Hitherto, however, no symmetry-preserving bound-state treatment could explain it.  This is illustrated by Cols.~3, 4, which show that whilst a good estimate of $m_\rho$ is readily obtained at leading-order in the systematic DSE truncation scheme of Ref.\,\cite{Bender:1996bb}, the axial-vector masses are much underestimated.  The flaw persists at next-to-leading-order \cite{Watson:2004kd,Fischer:2009jm}.

The analysis in Ref.\,\cite{Chang:2011ei} points to a remedy for this longstanding failure.  Using the Poincar\'e-covariant, symmetry preserving formulation of the meson bound-state problem enabled by Ref.\,\cite{Chang:2009zb}, with nonperturbative kernels for the gap and Bethe-Salpeter equations, which incorporate effects of DCSB that are impossible to capture in any step-by-step procedure for improving upon the rainbow-ladder truncation, it provides realistic estimates of axial-vector meson masses.
In obtaining these results, Ref.\,\cite{Chang:2011ei} showed that the vertex \emph{Ansatz} used most widely in studies of DCSB, $\Gamma_\mu^{BC}$, is inadequate as a tool in hadron physics.  Used alone, it increases both $m_\rho$ and $m_{a_1}$ but yields $m_{a_1}-m_\rho=0.21\,$GeV, qualitatively unchanged from the rainbow-ladder-like result (see Fig.\,\ref{F1a1}).
A good description of axial-vector mesons is only achieved by including interactions derived from $\Gamma_\mu^{\rm acm}$; i.e., connected with the dressed-quark anomalous chromomagnetic moment \cite{Chang:2010hb}.  Moreover, used alone, neither term in $\Gamma_\mu^{\rm acm}$, Eq.\,\eqref{ourvtxacm}, can produce a satisfactory result.  The full vertex \emph{Ansatz} and the associated gap and Bethe-Salpeter kernels described in Sec.\,\ref{spectrum2} are the minimum required.

Row~5 contains additional information.  The leading-covariant in the $b_1$-meson channel is $\gamma_5 k_\mu$.  The appearance of $k_\mu$ suggests that dressed-quark orbital angular momentum will play a significant role in this meson's structure, even more so than in the $a_1$-channel for which the dominant covariant is $\gamma_5\gamma_\mu$.
(N.B.\ In a simple quark-model, constituent spins are parallel within the $a_1$ but antiparallel within the $b_1$.  Constituents of the $b_1$ may therefore become closer, so that spin-orbit repulsion can exert a greater influence.)
This expectation is borne out by the following: with the full kernel, $m_{b_1}$ is far more sensitive to the interaction's momentum-space range parameter than any other state, decreasing rapidly as the interaction's spatial-variation is increasingly suppressed.
%

The results reviewed in this subsection rest on an \emph{Ansatz} for the quark-gluon vertex and whilst the best available information was used in its construction, improvement is nonetheless possible.  That will involve elucidating the role of Dirac covariants in the quark-gluon vertex which have not yet been considered, as in Ref.\,\cite{Bashir:2011dp}, and of resonant contributions; viz., meson loop effects that give widths to some of the states considered.  In cases for which empirical width-to-mass ratios are $\lesssim 25$\%, one might judge that such contributions can reliably be obtained via bound-state perturbation theory \cite{Pichowsky:1999mu}.  Contemporary studies indicate that these effects reduce bound-state masses but the reduction can uniformly be compensated by a modest inflation of the interaction's mass-scale \cite{Roberts:2011cf,Eichmann:2008ae}, so that the masses in Table~\ref{tablemasses} are semiquantitatively unchanged.  The case of the $\sigma$-meson is more complicated.  However, the prediction of a large mass for this meson's dressed-quark core leaves sufficient room for a strong reduction by resonant contributions \cite{Pelaez:2006nj,RuizdeElvira:2010cs}.

This section reviewed a continuum framework for computing and explaining the meson spectrum, which combines a veracious description of pion properties with estimates for masses of light-quark mesons heavier than $m_\rho$.
(A contemporary lattice-QCD perspective on this problem may be drawn from Refs.\,\protect\cite{Dudek:2011bn,Engel:2011aa}.)
The method therefore offers the promise of a first reliable Poincar\'e-invariant, symmetry-preserving computation of the spectrum of light-quark hybrids and exotics; i.e., those putative states which are impossible to construct in a quantum mechanics based upon constituent-quark degrees-of-freedom.  So long as the promise is promptly fulfilled, the approach will provide predictions to guide the forthcoming generation of facilities.

\section{Pion Electromagnetic Form Factors}
\label{FF1}
In charting the long-range interaction between light-quarks via the feedback between experiment and theory, hadron elastic and transition form factors can provide unique information, beyond that obtained through studies of the hadron spectrum.

\subsection{Charged pion}
This is demonstrated very clearly by an analysis of the electromagnetic pion form factor, $F^{\rm em}_{\pi}(Q^2)$, because the pion has a unique place in the Standard Model.  It is a bound-state of a dressed-quark and -antiquark, and also that almost-massless collective excitation which is the Goldstone mode arising from the dynamical breaking of chiral symmetry.  This dichotomy can only be understood through the symmetry-preserving analysis of two-body bound-states \cite{Maris:1997hd}.  Furthermore, the possibility that this dichotomous nature could have wide-ranging effects on pion properties has made the empirical investigation of these properties highly desirable, despite the difficulty in preparing a system that can act as a pion target and the concomitant complexities in the interpretation of the experiments; e.g., \cite{Volmer:2000ek,Horn:2006tm,Tadevosyan:2007yd,Wijesooriya:2005ir}.

The merit of using $F^{\rm em}_{\pi}(Q^2)$ to elucidate the potential of an interplay between experiment and nonperturbative theory as a means of constraining the long-range behaviour of QCD's $\beta$-function is amplified by the existence of a prediction that $Q^2 F_{\pi}(Q^2)\approx\,$constant for $Q^2\gg m_\pi^2$ in a theory whose interaction is mediated by massless vector-bosons.  To be explicit \cite{Farrar:1979aw,Efremov:1979qk,Lepage:1980fj}:
\begin{equation}
Q^2 F_{\pi}(Q^2) \stackrel{Q^2\gg m_\pi^2}{\simeq} 16 \pi f_\pi^2 \alpha(Q^2),
\end{equation}
which takes the value $0.13\,$GeV$^2$ at $Q^2=10\,$GeV$^2$ if one uses the one-loop result $\alpha(Q^2=10\,{\rm GeV}^2)\approx 0.3$.  The verification of this prediction is a strong motivation for modern experiment \cite{Volmer:2000ek,Horn:2006tm,Tadevosyan:2007yd}, which can also be viewed as an attempt to constrain and map experimentally the pointwise behaviour of the exchange interaction that binds the pion.

Section~\ref{psmassformula} details some extraordinary consequences of DCSB, amongst them the Goldberger-Treiman relations of Eqs.\,(\ref{gtlrelE}) -- (\ref{gtlrelH}).  Of these, Eqs.\,(\ref{gtlrelF}) and (\ref{gtlrelG}) entail that the pion possesses components of pseudovector origin which alter the asymptotic form of $F_{\pi}^{\rm em}(Q^2)$ by a multiplicative factor of $Q^2$ cf.\ the result obtained in their absence \cite{Maris:1998hc}.

QCD-based DSE calculations of $F^{\rm em}_\pi(Q^2)$ exist \cite{Maris:1998hc,Maris:2000sk}, the most systematic of which \cite{Maris:2000sk} predicted the measured form factor \cite{Volmer:2000ek}.  Germane to this discourse, however, is an elucidation of the sensitivity of $F^{\rm em}_\pi(Q^2)$ to the pointwise behaviour of the interaction between quarks.  We therefore recapitulate on Refs.\,\cite{GutierrezGuerrero:2010md,Roberts:2011wy}, which explored how predictions for pion properties change if quarks interact not via massless vector-boson exchange but instead through a contact interaction; viz.,
\begin{equation}
\label{njlgluon}
g^2 D_{\mu \nu}(p-q)
= \delta_{\mu \nu} \frac{1}{\mathpzc{m}_G^2}
= \delta_{\mu \nu} \frac{4 \pi \alpha_{\rm IR}}{m_G^2} \,,
\end{equation}
wherein $m_G=0.8\,$GeV is a gluon mass-scale (such a scale is generated dynamically in QCD \cite{Boucaud:2011ug,Bowman:2004jm}) and $\alpha_{\rm IR}$ is a parameter that specifies the interaction strength, and proceeded by embedding this interaction in a rainbow-ladder truncation of the DSEs.

In this case, using a confining regularisation scheme \cite{Ebert:1996vx}, the gap equation, which determines this interaction's momentum-independent dressed-quark mass, can be written
\begin{equation}
M = m +  M  \frac{4 \alpha_{\rm IR}}{3\pi m_G^2} \,{\cal C}(M^2;\tau_{\rm ir},\tau_{\rm uv})\,, \label{gap-1}
\end{equation}
where $m$ is the current-quark mass, $\tau_{\rm ir}=1/\Lambda_{\rm ir}$, $\tau_{\rm uv}=1/\Lambda_{\rm uv}$, and
\begin{equation}
\label{calC}
{\cal C}(M^2;\tau_{\rm ir},\tau_{\rm uv}) = M^2[ \Gamma(-1,M^2 \tau_{\rm uv}^2) - \Gamma(-1,M^2 \tau_{\rm ir}^2)],
\end{equation}
with $\Gamma(\alpha,y)$ being the incomplete gamma-function.  Results are presented in Table\,\ref{Table:static}.

\begin{table}[t]
\caption{Results obtained with \protect\cite{Roberts:2011wy} $\alpha_{\rm IR}/\pi=0.93$ in Eq.\,\eqref{njlgluon}, and (in GeV): $m=0.007$, $\Lambda_{\rm ir} = 0.24\,$, $\Lambda_{\rm uv}=0.905$ in Eq.\,\eqref{gap-1}.
(Dimensioned quantities are listed in GeV.)
\label{Table:static}
}
\begin{center}
\begin{tabular*}
{\hsize}
{
l@{\extracolsep{0ptplus1fil}}
|c@{\extracolsep{0ptplus1fil}}
c@{\extracolsep{0ptplus1fil}}
c@{\extracolsep{0ptplus1fil}}
|c@{\extracolsep{0ptplus1fil}}
c@{\extracolsep{0ptplus1fil}}
c@{\extracolsep{0ptplus1fil}}
c@{\extracolsep{0ptplus1fil}}
c@{\extracolsep{0ptplus1fil}}
c@{\extracolsep{0ptplus1fil}}}\hline
$m$ & $E_\pi$ & $F_\pi$ & $E_\rho$ & $M$ & $\kappa_\pi^{1/3}$ & $m_\pi$ & $m_\rho$ & $f_\pi$ & $f_\rho$ \\\hline
0 & 3.568 & 0.459 & 1.520 & 0.358 & 0.241 & 0\,~~~~~ & 0.919 & 0.100 & 0.130\rule{0ex}{2.5ex}\\
0.007 & 3.639 & 0.481 & 1.531 & 0.368 & 0.243 & 0.140 & 0.928 & 0.101 & 0.129\\\hline
\end{tabular*}
\end{center}
\end{table}

With a symmetry-preserving regularisation of the interaction in Eq.\,(\ref{njlgluon}), the Bethe-Salpeter amplitude cannot depend on relative momentum.  Hence Eq.\,(\ref{genGpi}) becomes
\begin{equation}
\label{pointpionBSA}
\Gamma_\pi(P) = \gamma_5 \left[ i E_\pi(P) + \frac{1}{M} \gamma\cdot P F_\pi(P) \right]
\end{equation}
and the explicit form of the model's ladder BSE is
\begin{equation}
\label{bsefinal0}
\left[
\begin{array}{c}
E_\pi(P)\\
F_\pi(P)
\end{array}
\right]
= \frac{4 \alpha_{\rm IR}}{3\pi m_G^2}
\left[
\begin{array}{cc}
{\cal K}_{EE} & {\cal K}_{EF} \\
{\cal K}_{FE} & {\cal K}_{FF}
\end{array}\right]
\left[\begin{array}{c}
E_\pi(P)\\
F_\pi(P)
\end{array}
\right],
\end{equation}
where, with $m=0=P^2$, anticipating the Goldstone character of the pion,
\begin{equation}
\begin{array}{cl}
{\cal K}_{EE}  =  {\cal C}(M^2;\tau_{\rm ir}^2,\tau_{\rm uv}^2)\,, &  {\cal K}_{EF}  =  0\,,\\
2 {\cal K}_{FE} = {\cal C}_1(M^2;\tau_{\rm ir}^2,\tau_{\rm uv}^2) \,,&
{\cal K}_{FF} = - 2 {\cal K}_{FE}\,,
\end{array}
\end{equation}
and ${\cal C}_1(z) = - z {\cal C}^\prime(z)$, where we have suppressed the dependence on $\tau_{\rm ir,uv}$.  The solution of Eq.\,(\ref{bsefinal0}) gives the pion's chiral-limit Bethe-Salpeter amplitude, which, for the computation of observables, should be normalised canonically; viz.,
\begin{equation}
%
 P_\mu = N_c\, {\rm tr} \int\! \frac{d^4q}{(2\pi)^4}\Gamma_\pi(-P)
 \frac{\partial}{\partial P_\mu} S(q+P) \, \Gamma_\pi(P)\, S(q)\,. \label{Ndef}
\end{equation}
Hence, in the chiral limit,
\begin{equation}
1 = \frac{N_c}{4\pi^2} \frac{1}{M^2} \, {\cal C}_1(M^2;\tau_{\rm ir}^2,\tau_{\rm uv}^2)
E_\pi [ E_\pi - 2 F_\pi],
\label{Norm0}
\end{equation}
and the pion's leptonic decay constant is
\begin{equation}
\label{fpi0}
f_\pi^0 = \frac{N_c}{4\pi^2} \frac{1}{M} {\cal C}_1(M^2;\tau_{\rm ir}^2,\tau_{\rm uv}^2)  [ E_\pi - 2 F_\pi ]\,.
\end{equation}

If one has preserved Eq.\,(\ref{avwtimN}), then, for $m=0$ in the neighbourhood of $P^2=0$, the solution of the axial-vector BSE has the form:
\begin{equation}
\Gamma_{5\mu}(k_+,k) = \frac{P_\mu}{P^2} \, 2 f_\pi^0 \, \Gamma_\pi(P) + \gamma_5\gamma_\mu F_R(P)
\end{equation}
and the following subset of Eqs.\,(\ref{gtlrelE}) -- (\ref{gtlrelH}) will hold:
\begin{equation}
\label{GTI}
f_\pi^0 E_\pi = M \,,\; 2\frac{F_\pi}{E_\pi}+ F_R = 1\,.
\end{equation}
Hence $F_\pi(P)$ is necessarily nonzero in a vector exchange theory, irrespective of the pointwise behaviour of the interaction.  It has a measurable impact on the value of $f_\pi$ 
and on the form factor, as we shall see.

\begin{figure}[t] 
\centerline{\includegraphics[clip,width=0.45\textwidth]{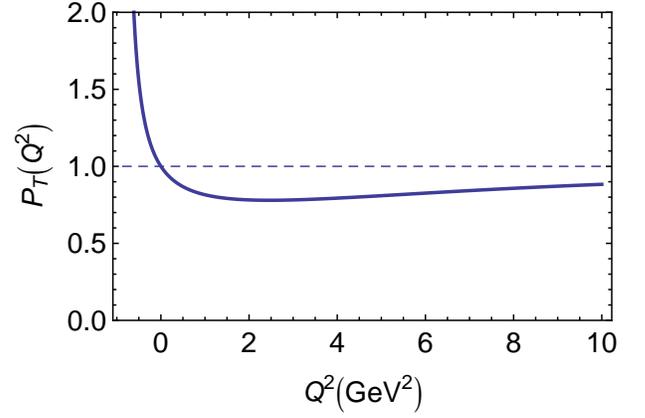}}
\caption{\label{qgammavertex}  Dressing function for the transverse piece of the quark-photon vertex, computed using the parameter values described in connection with Table~\protect\ref{Table:static}.  The pole associated with the ground-state vector meson is clear.}
\end{figure}

Based upon these results, one can proceed to compute the electromagnetic pion form factor in the generalised impulse approximation \cite{GutierrezGuerrero:2010md,Roberts:2011wy}; i.e., at leading-order in a symmetry-preserving DSE truncation scheme \cite{Maris:1998hc,Maris:2000sk,Roberts:1994hh}.  Namely, for an incoming pion with momentum $p_1=K-Q/2$, which absorbs a photon with space-like momentum $Q$, so that the outgoing pion has momentum $p_2=K+Q/2$,
\begin{eqnarray}
\nonumber
\lefteqn{
2 K_{\mu} F_{\pi}^{\rm em}(Q^2) = 2 N_c \int\frac{d^4t}{(2\pi)^4}
{\rm tr_D} \Bigg[ i \Gamma_{\pi}(-p_2) S(t+p_2)} \rule{2em}{0ex} \\
& \times&  i \gamma_{\mu} P_{\rm T}(Q^2) S(t+p_1) \; i \Gamma_{\pi}(p_1) \; S(t) \Bigg], \rule{2em}{0ex}
\label{KF}
\end{eqnarray}
where $P_{\rm T}(Q^2)$, depicted in Fig.\,\ref{qgammavertex}, describes the full extent of dressing on the quark-photon vertex produced by a contact interaction in the rainbow-ladder truncation \cite{Roberts:2011wy}.  The form factor is expressible as follows:
\begin{eqnarray}
F_{\pi}^{\rm em}(Q^2) &=&P_{\rm T}(Q^2)\, F_{\pi,\not \rho}^{\rm em}(Q^2),\\
F_{\pi,\not \rho}^{\rm em}(Q^2)&=& F_{\pi,EE}^{{\rm em}} + F_{\pi,EF}^{{\rm em}} + F_{\pi,FF}^{{\rm em}},\rule{1em}{0ex}\\
\nonumber
& =&  E_\pi^{\rm c}E_\pi^{\rm c}  T^{\pi}_{EE}(Q^2) + E_\pi^{\rm c} F_\pi^{\rm c} T^{\pi}_{EF}(Q^2)\\
&& + F_\pi^{\rm c}F_\pi^{\rm c} T^{\pi}_{FF}(Q^2) ,
\label{F123}
\end{eqnarray}
where $F_{\pi, \not \rho}^{\rm em}(Q^2)$ is that part of the form factor produced by the undressed quark-photon vertex and the functions $T^{\pi}$ have simple algebraic forms in this calculation \cite{Roberts:2011wy}.

\begin{figure}[t] 
\includegraphics[clip,height=0.32\textheight,angle=-90]{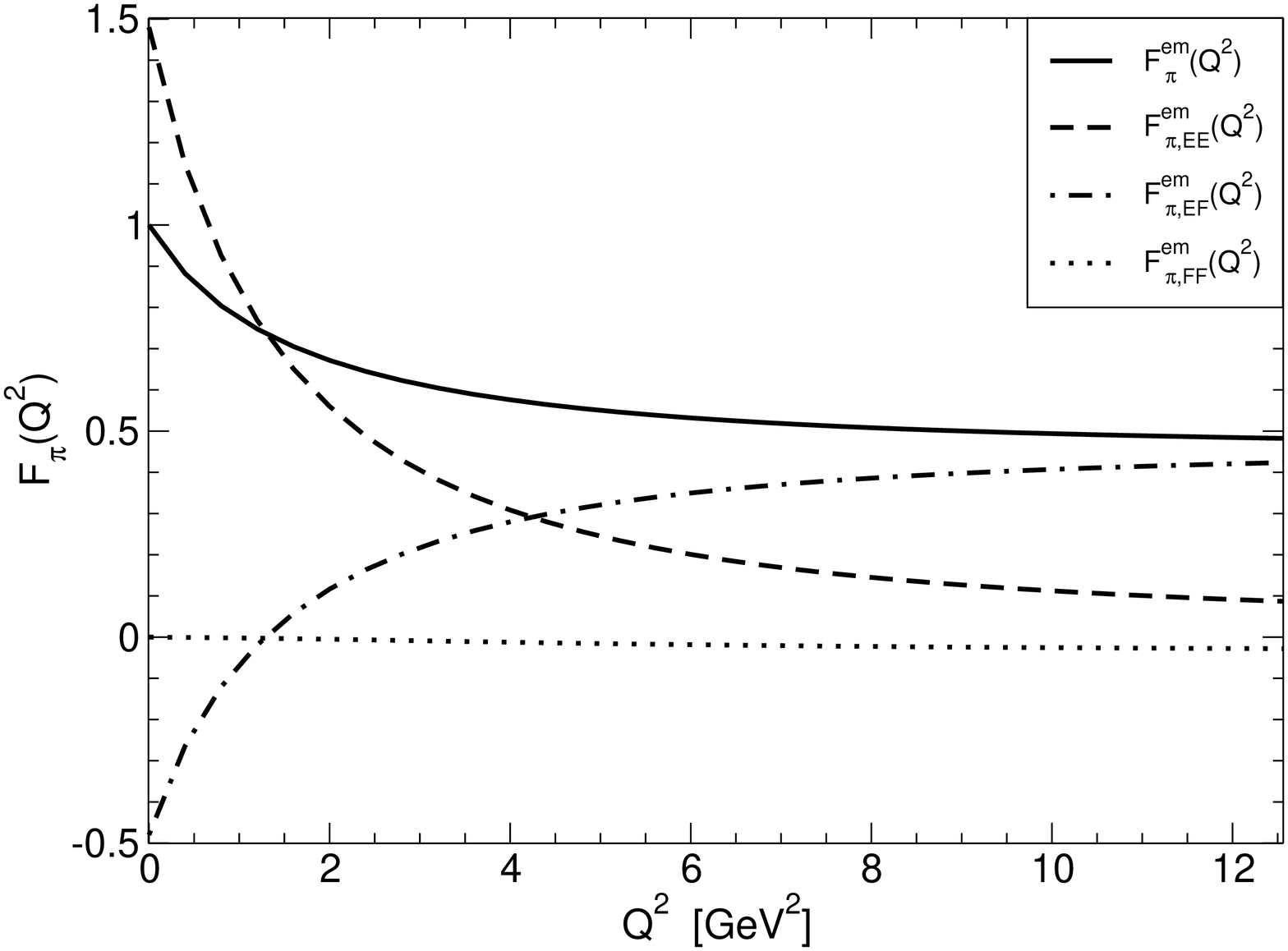}\\
\includegraphics[clip,height=0.22\textheight]{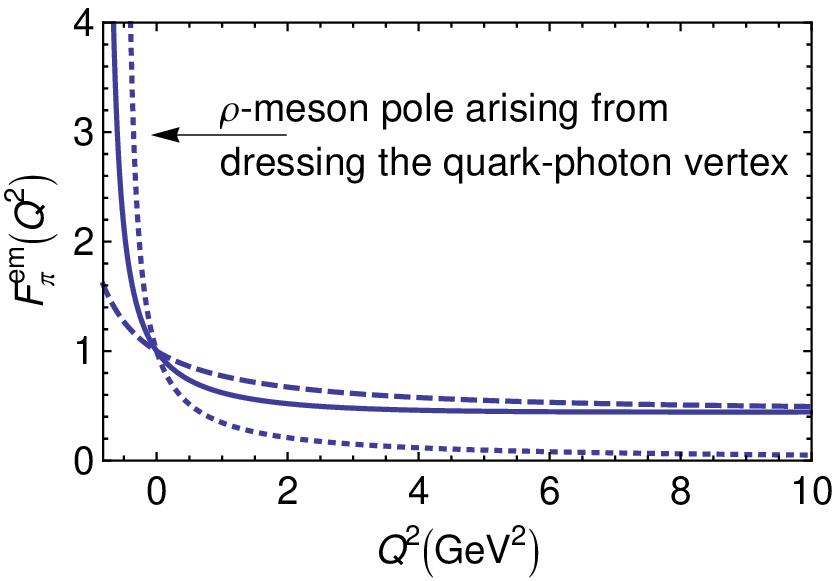}
\caption{\label{fig4}
\emph{Upper panel}.
$F^{\rm em}_{\pi,\not\rho}(Q^2)$ and the separate contributions introduced in Eq.\,(\protect\ref{F123}).
$F^{\rm em}_{\pi}(Q^2=0)=1$, without fine-tuning, because a symmetry-preserving regularisation of the interaction in Eq.\,(\protect\ref{njlgluon}) was implemented.
\emph{Lower panel}.
$F^{\rm em}_{\pi}(Q^2)$ computed in rainbow-ladder truncation from the interaction in Eq.\,(\protect\ref{njlgluon}): \emph{solid curve} -- fully consistent, i.e., with a dressed-quark-photon vertex so that the $\rho$-pole appears; and \emph{dashed curve} -- computed using a bare quark-photon vertex, namely, $F_{\pi,\not\rho}^{\rm em}(Q^2)$.  \emph{Dotted curve} -- fit to the result in Ref.\,\protect\cite{Maris:2000sk}, which also included a consistently-dressed quark-photon vertex and serves to illustrate the trend of contemporary data.
%
}
\end{figure}

In the upper panel of Fig.\,\ref{fig4} we present $F_{\pi,\not\rho}^{\rm em}(Q^2)$ and the three separate contributions defined in Eq.\,(\ref{F123}).  It is evident that, in magnitude, $F_{\pi,EF}^{\rm em}$ contributes roughly one-third of the pion's unit charge.  This could have been anticipated from Eq.\,(\ref{Norm0}).
More dramatically, perhaps: the interaction in Eq.\,(\ref{njlgluon}) generates \begin{equation}
\label{Fpiconstant}
F_\pi^{\rm em}(Q^2 \to\infty) =\,{\rm constant.}
\end{equation}
Both results originate in the nonzero value of $F_\pi(P)$, which is a straightforward consequence of the symmetry-preserving treatment of a vector exchange theory \cite{Maris:1997hd}.  Equation~(\ref{Fpiconstant}) should not come as a surprise: with a symmetry-preserving regularisation of the interaction in Eq.\,(\ref{njlgluon}), the pion's Bethe-Salpeter amplitude cannot depend on the constituent's relative momentum.  This is characteristic of a pointlike particle, which must have a hard form factor.
The lower panel of the figure illustrates that the necessary inclusion of $P_{\rm T}(Q^2)$ is critical in the timelike region and has a measurable quantitative impact for a significant range of spacelike momenta.  It does not, however, affect the truly ultraviolet behaviour.

\begin{figure}[t] 
\includegraphics[clip,height=0.32\textheight,angle=-90]{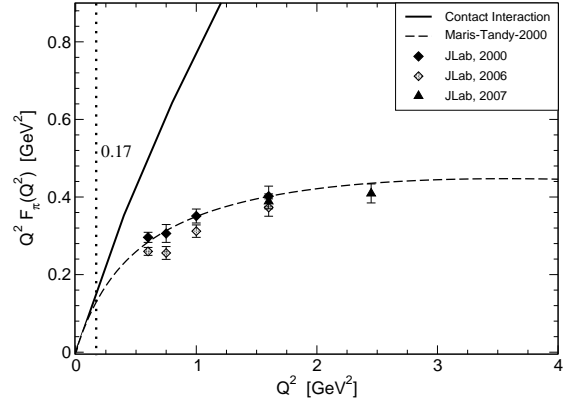}
\caption{\label{FpiUV}
\underline{Solid curve}: $Q^2 F_{\pi,\not\rho}(Q^2)$ obtained with Eq.\,(\protect\ref{njlgluon}).
%
\underline{Dashed curve}: DSE prediction \protect\cite{Maris:2000sk}, which employed a momentum-dependent renormalisation-group-improved gluon exchange interaction.
For $Q^2>0.17\,$GeV$^2\approx M^2$, marked by the vertical \emph{dotted line}, the contact interaction result for $F^{\rm em}_{\pi,\not\rho}(Q^2)$ differs from that in  Ref.\,\protect\cite{Maris:2000sk} by more than 20\%.
The data are from Refs.\,\protect\cite{Volmer:2000ek,Horn:2006tm,Tadevosyan:2007yd}.
%
}
\end{figure}

In Fig.\,\ref{FpiUV} we compare the form factor computed from Eq.\,(\ref{njlgluon}) with contemporary experimental data \cite{Volmer:2000ek,Horn:2006tm,Tadevosyan:2007yd} and a QCD-based DSE prediction \cite{Maris:2000sk}.  Both the QCD-based result and that obtained from the momentum-independent interaction yield the same values for the pion's static properties \cite{GutierrezGuerrero:2010md,Roberts:2010rn,Roberts:2011wy}.  However, for $Q^2>0$ the form factor computed using $\sim 1/k^2$-vector-boson exchange is immediately distinguishable empirically from that produced by a momentum-independent interaction.  Indeed, the figure shows that for $F_\pi^{\rm em}$, existing experiments can already distinguish between different possibilities for the quark-quark interaction.

Combining Figs.\,\ref{fig4} and \ref{FpiUV} it becomes apparent that $F_{\pi,EE}^{\rm em}$ is only a good approximation to the net pion form factor for $Q^2 \lsim M^2$.  $F_{\pi,EE}^{\rm em}$ and $F_{\pi,EF}^{\rm em}$ evolve with equal rapidity -- there is no reason for this to be otherwise, as they are determined by the same mass-scales -- but a nonzero constant comes quickly to dominate over a form factor that falls swiftly to zero.

It is plain now that when a momentum-independent vector-exchange interaction is regularised in a symmetry-preserving manner, the results are directly comparable with experiment, computations based on well-defined and systematically-improvable truncations of QCD's DSEs \cite{Maris:2000sk}, and perturbative QCD.  In this context it will be apparent that a contact interaction, whilst capable of describing pion static properties well, Table\,\ref{Table:static}, generates a form factor whose evolution with $Q^2$ deviates markedly from experiment for $Q^2>0.17\,$GeV$^2\approx M^2$ and produces asymptotic power-law behaviour, Eq.\,(\ref{Fpiconstant}), in serious conflict with QCD \cite{Farrar:1979aw,Efremov:1979qk,Lepage:1980fj}.

In that connection Fig.\,\ref{gluoncloud} and Eqs.\,(\ref{gtlrelE}) -- (\ref{gtlrelH}) are relevant again.
In the electromagnetic elastic scattering process, the momentum transfer, $Q$, is primarily shared equally between the pion's constituents because the bound-state Bethe-Salpeter amplitude is peaked at zero relative momentum.  Thus, one can consider $k\sim Q/2$.
The Goldberger-Treiman-like relations express a mapping between the relative momentum of the pion's constituents and the one-body momentum of dressed-quark; and the momentum dependence of the dressed-quark mass function is well-described by perturbation theory when $k^2>2\,$GeV$^2$.
Hence, one should expect a perturbative-QCD analysis of the pion form factor to be valid for $k^2=Q^2/4 \gtrsim 2\,$GeV$^2$; i.e.,
\begin{equation}
F_\pi^{\rm em}(Q^2) \approx F_\pi^{\rm em\, pQCD}(Q^2) \; \mbox{for} \; Q^2\gtrsim 8\,{\rm GeV}^2.
\end{equation}
This explains the result in Ref.\,\cite{Maris:1998hc}.
A similar argument for baryons suggests that the nucleon form factors should be perturbative for $Q^2\gtrsim 18\,$GeV$^2$.

It is worth reiterating that the contact interaction produces a momentum-independent dressed-quark mass function, in contrast to QCD-based DSE studies \cite{Roberts:2007ji,Bhagwat:2006tu} and lattice-QCD \cite{Bowman:2005vx}.  This is fundamentally the origin of the marked discrepancy between the form factor it produces and extant experiment.  Hence Refs.\,\cite{GutierrezGuerrero:2010md,Roberts:2010rn,Roberts:2011wy} highlight that form factor observables, measured at an upgraded JLab, e.g., are capable of mapping the running of the dressed-quark mass function.  Efforts are underway to establish the signals of the running mass in baryon elastic and transition form factors.  They are reviewed in Sec.\,\ref{sec:Baryons}.

\subsection{Neutral pion}
The process $\gamma^\ast \gamma \to \pi^0$ is also of great interest because in order to explain the associated transition form factor within the Standard Model on the full domain of momentum transfer, one must combine, using a single internally-consistent framework, an explanation of the essentially nonperturbative Abelian anomaly with the features of perturbative QCD.  The case for attempting this received a significant boost with the publication of data from the BaBar Collaboration \cite{Aubert:2009mc} because, while they agree with earlier experiments on their common domain of squared-momentum-transfer \cite{Behrend:1990sr,Gronberg:1997fj}, the BaBar data are unexpectedly far \emph{above} the prediction of perturbative QCD at larger values of $Q^2$.

This so-called ``Babar anomaly'' was considered in Ref.\,\cite{Roberts:2010rn}, wherein it is argued that in fully-self-consistent treatments of pion: static properties; and elastic and transition form factors, the asymptotic limit of the product $Q^2 G_{\gamma^\ast\gamma \pi^0}(Q^2)$, which is determined \emph{a priori} by the interaction employed, is not exceeded at any finite value of spacelike momentum transfer: the product is a monotonically-increasing concave function.  A consistent approach is one in which: a given quark-quark scattering kernel is specified and solved in a well-defined, symmetry-preserving truncation scheme; the interaction's parameter(s) are fixed by requiring a uniformly good description of the pion's static properties; and relationships between computed quantities are faithfully maintained.

Within such an approach it is nevertheless possible for $Q^2 F^{\rm em}_\pi(Q^2)$ to exceed its perturbative-QCD asymptotic limit because the leading-order matrix-element involves two Bethe-Salpeter amplitudes.  This permits an interference between dynamically-generated infrared mass-scales in the computation.  Moreover, for $F^{\rm em}_\pi(Q^2)$ the perturbative QCD limit is more than an order-of-magnitude smaller than $m_\rho^2$.  Owing to the proximity of the $\rho$-meson pole to $Q^2=0$, the latter mass-scale must provide a fair first-estimate for the small-$Q^2$ evolution of $F^{\rm em}_\pi(Q^2)$.  A monopole based on this mass-scale exceeds the pQCD limit $\forall Q^2>0$.  For the transition form factor, however, the opposite is true because $m_\rho^2$ is less-than the pQCD limit; viz. \cite{Lepage:1980fj},
\begin{equation}
\label{BLuv}
\lim_{Q^2\to\infty} Q^2 2 G(Q^2) = 8\pi^2 f_\pi^2.
\end{equation}

\begin{figure}[t] 
\centerline{\includegraphics[clip,width=0.45\textwidth]{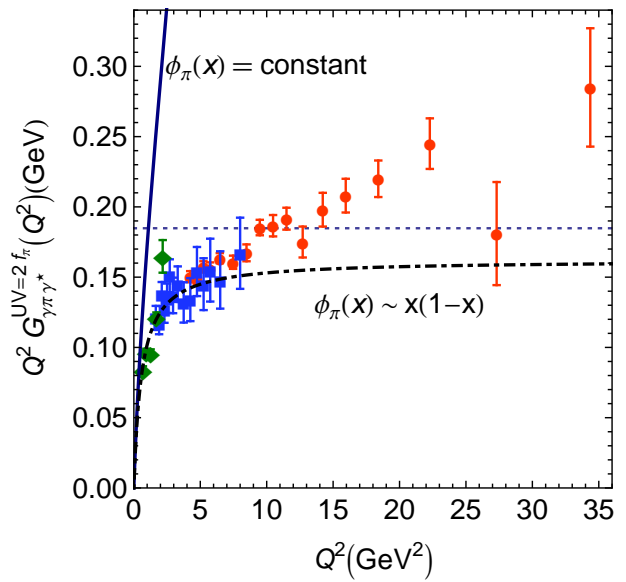}}
\caption{\label{transitionFO}
$Q^2$-weighted $\gamma^\ast \gamma \to \pi^0$ transition form factor.
Data: red circles, Ref.\,\protect\cite{Aubert:2009mc}; green diamonds, Ref.\,\protect\cite{Behrend:1990sr}; and blue squares, Ref.\protect\cite{Gronberg:1997fj}.
\emph{Solid curve} -- $Q^2 G(Q^2)$ computed using the symmetry-preserving, fully-self-consistent rainbow-ladder treatment of the contact interaction in Eq.\,(\protect\ref{njlgluon}), which produces a pion distribution amplitude $\phi_\pi(x)=\,$constant;
and \emph{dot-dashed curve} -- fit to the $\gamma^\ast\gamma \to \pi^0$ transition form factor computed in a QCD-based rainbow-ladder-truncation DSE study \cite{Maris:2002mz}.
Both curves have been divided by $(2\pi^2 f_\pi)$ in order to match the data's normalisation.
%
%
%
}
\end{figure}

The vector current-current contact-interaction canvassed in this Section may be described as a vector-boson exchange theory with vector-field propagator $(1/k^2)^\kappa$, $\kappa=0$.  It was shown \cite{Roberts:2010rn} that the consistent treatment of such an interaction produces a $\gamma^\ast\gamma \to \pi^0$ transition form factor that disagrees with \emph{all} available data.  On the other hand, precisely the same treatment of an interaction which preserves the one-loop renormalisation group behaviour of QCD, produces a form factor in good agreement with all but the large-$Q^2$ data from the BaBar Collaboration \cite{Aubert:2009mc}.  These points are illustrated in Fig.\,\ref{transitionFO}.

Studies exist which interpret the BaBar data as an indication that the pion's distribution amplitude, $\phi_\pi(x)$, deviates dramatically from its QCD asymptotic form, indeed, that $\phi_\pi(x)=\,$constant, or is at least flat and nonvanishing at $x=0,1$ \cite{Radyushkin:2009zg,Polyakov:2009je}.  However, it has often been explained \cite{Hecht:2000xa,GutierrezGuerrero:2010md,Holt:2010vj,Roberts:2010rn,Roberts:2011wy} that such a distribution amplitude characterises an essentially-pointlike pion; and, as we have seen, when used in a fully-consistent treatment, it produces results for pion elastic and transition form factors that are in striking disagreement with experiment.  Reiterating, a bound-state pion with a pointlike component will produce the hardest possible form factors; i.e., form factors which become constant at large-$Q^2$.

On the other hand, QCD-based studies produce soft pions, a valence-quark distribution amplitude for the pion that vanishes as $\sim (1-x)^2$ for $x\sim 1$, and results that agree well with the bulk of existing data.  We will return to this in the next section.

The analysis in Ref.\,\cite{Roberts:2010rn} shows that the large-$Q^2$ BaBar data is inconsistent with QCD and also inconsistent with a vector current-current contact interaction.  It supports a conclusion that the large-$Q^2$ data reported by BaBar is not a true representation of the $\gamma^\ast\gamma \to \pi^0$ transition form factor, a perspective also developed elsewhere \cite{Mikhailov:2009sa,Brodsky:2011yv,Bakulev:2011rp,Brodsky:2011xx}.  There is experimental evidence in support of this view; namely, the $\gamma^\ast \to \eta \gamma$ and $\gamma^\ast \to \eta^\prime \gamma$ transition form factors have also been measured by the BaBar Collaboration \cite{Aubert:2006cy}, at $Q^2=112\,$GeV$^2$, and in these cases the results from CLEO \cite{Gronberg:1997fj} and BaBar are fully consistent with perturbative-QCD expectations.

\section{Pion and kaon valence-quark distributions}
\label{FF2}
Quarks were discovered in deep inelastic scattering (DIS) experiments at SLAC, performed during the period 1966-1978.  DIS is completely different to elastic scattering.  In this process, one disintegrates the target instead of keeping only those events in which it remains intact.  On a well-defined kinematic domain; namely, the Bjorken limit:
\begin{equation}
q^2 \to \infty\,,\; P\cdot q \to -\infty\,,\; x_{\rm Bjorken}:=-\frac{q^2}{2 P\cdot q} = {\rm fixed},
\end{equation}
where $P$ is the target's four-momentum and $q$ is the momentum transfer, the cross-section can rigorously be interpreted as a measurement of the momentum-fraction probability distribution for quarks and gluons within the target hadron: $q(x)$, $g(x)$.  These quantities describe the probability that a quark/gluon within the target will carry a fraction $x$ of the bound-state's momentum, as defined in the infinite-momentum or light-front frame.  (The light-front formulation of quantum field theory is built upon Dirac's front form of relativistic quantum dynamics \cite{Dirac:1949cp}.)

The past forty years have seen a tremendous effort to deduce the parton distribution
functions (PDFs) of the most accessible hadrons -- the proton, neutron and pion.  There are many reasons for this long sustained and thriving interest \cite{Holt:2010vj} but in large part it is motivated by the suspected process-independence of the usual parton distribution functions and hence an ability to unify many hadronic processes through their computation.  In connection with uncovering the essence of the strong interaction, the behaviour of the valence-quark distribution functions at large Bjorken-$x$ is most relevant.
Furthermore, an accurate determination of the behavior of distribution functions in the valence region is also important to high-energy physics.  Particle discovery experiments and Standard Model tests with colliders are only possible if the QCD background is completely understood.  QCD evolution, apparent in the so-called scaling violations by parton distribution functions,\footnote{DGLAP evolution is described in Sec.IID of Ref.\,\cite{Holt:2010vj}.  The evolution equations are derived in perturbative QCD and determine the rate of change of parton densities when the energy-scale chosen for their definition is varied.}
entails that with increasing center-of-mass energy, the support at large-$x$ in the distributions evolves to small-$x$ and thereby contributes materially to the collider background.
N.B.\ The nucleon PDFs are now fairly well determined for $x\lesssim 0.8$ but the pion and kaon PDFs remain poorly known on the entire domain of $x$.

Owing to the dichotomous nature of Goldstone bosons, understanding the valence-quark distribution functions in the pion and kaon is of great importance.  Moreover, given the large value of the ratio of $s$-to-$u$ current-quark masses, a comparison between the pion and kaon structure functions offers the chance to chart effects of explicit chiral symmetry breaking on the structure of would-be Goldstone modes.  There is also the prediction \cite{Ezawa:1974wm,Farrar:1975yb} that a theory in which the quarks interact via $1/k^2$-vector-boson exchange will produce valence-quark distribution functions for which
\begin{equation}
\label{pQCDuvx}
q_{\rm v}(x) \propto (1-x)^{2+\gamma} \,,\; x\gtrsim 0.85\,,
\end{equation}
where $\gamma\gtrsim 0$ is an anomalous dimension that grows with increasing momentum transfer.  (See Sec.VI.B.3 of Ref.\,\cite{Holt:2010vj} for a detailed discussion.)

\begin{figure}[t] 
\includegraphics[clip,height=0.12\textheight]{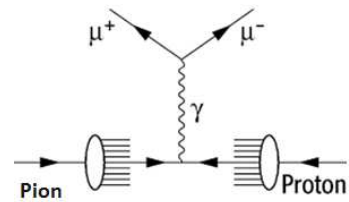}
\caption{\label{piNDY} $\pi N$ Drell-Yan process, in which, e.g., a valence antiquark from the pion annihilates with a valence quark in the nucleon to produce a $\mu^+ \mu^-$ pair.
}
\end{figure}

Owing to the absence of pseudoscalar meson targets, experimental knowledge of the parton structure of the pion and kaon arises primarily from pionic or kaonic Drell-Yan processes, illustrated in Fig.\,\ref{piNDY}, involving nucleons in heavy nuclei \cite{Badier:1980jq,Badier:1983mj,Betev:1985pg,Conway:1989fs,Wijesooriya:2005ir}.  Theoretically, given that DCSB plays a crucial role in connection with pseudoscalar mesons, one must employ an approach that realistically expresses this phenomenon.  The DSEs therefore provide a natural framework: studies of the pion and kaon exist and will be reviewed here.  The first \cite{Hecht:2000xa} computed pion PDFs, using efficacious parametrisations of both the Bethe-Salpeter amplitude and dressed-quark propagators \cite{Roberts:1994hh,Burden:1995ve,ElBennich:2011py}.  The second \cite{Nguyen:2011jy} employed direct, numerical DSE solutions in the computation of the pion and kaon PDFs, adapting the approach employed in successful predictions of electromagnetic form factors \cite{Holl:2005vu,Maris:1999bh,Maris:2000sk,Maris:2002mz}; and also studied the ratio $u_K(x)/u_\pi(x)$ in order to elucidate aspects of the influence of an hadronic environment.

In rainbow-ladder truncation, one obtains the pion's valence-quark distribution from
\begin{eqnarray}
\nonumber
\lefteqn{u_\pi(x) = -\frac{1}{2} \int \frac{d^4 \ell}{(2\pi)^4}  {\rm tr}_{\rm cd}\, \left[ \Gamma_\pi(\ell,-P)\right.}\\
&& \left. \times \,S_u(\ell)\, \Gamma^n(\ell;x) \, S_u(\ell)\, \Gamma_\pi(\ell,P)\, S_d(\ell-P) \right] , \rule{2em}{0ex}
\label{Eucl_pdf_LR_Ward}
\end{eqnarray}
wherein the Bethe-Salpeter amplitudes and dressed-quark propagators are discussed above and $\Gamma^n(\ell;x)$ is a generalization of the dressed-quark-photon vertex, describing a dressed-quark scattering from a zero momentum photon.  It satisfies a BSE (here with a rainbow-ladder kernel) with the inhomogeneous term  $i\gamma\cdot n \, \delta(\ell \cdot n - x P\cdot n)$.  In Eq.\,\eqref{Eucl_pdf_LR_Ward}, $n_\mu$ is a light-like vector satisfying \mbox{$n^2 = 0$}.
In choosing rainbow-ladder truncation one implements a precise parallel to the symmetry-preserving treatment of the pion charge form factor at \mbox{$Q^2 = 0 $}, wherein the vector current is conserved by use of ladder dynamics at all three vertices and rainbow dynamics for all three quark propagators \cite{Maris:1998hc,Maris:1999bh,Maris:2000sk,Roberts:1994hh}.   Equation~(\ref{Eucl_pdf_LR_Ward}) ensures automatically that
\begin{equation}
 \langle x_f^0 \rangle := \int_0^1 dx \,  q^v_f(x) = 1\; \mbox{for}\; f = u, \bar d\,,
\end{equation}
since $ \int dx \, \Gamma^n(\ell;x)$ gives the Ward-identity vertex and the Bethe-Salpeter amplitudes are canonically normalised.

\begin{figure}[t]
\includegraphics[clip,width=0.4\textwidth]{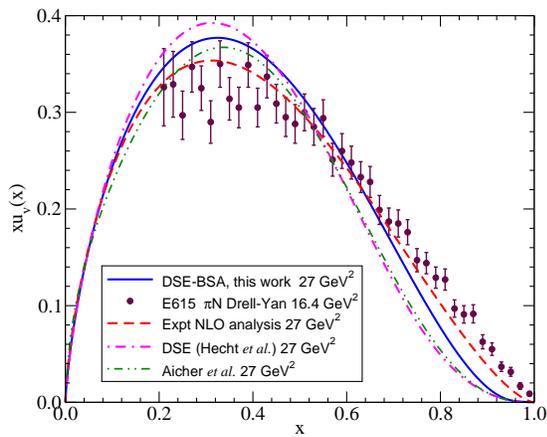}

\caption{ Pion valence quark distribution function evolved to (5.2~GeV)$^2$.  \emph{Solid curve} -- full DSE calculation \protect\cite{Nguyen:2011jy}; \emph{dot-dashed curve} -- semi-phenomenological DSE-based calculation in Ref.\,\protect\cite{Hecht:2000xa}; \emph{filled circles} -- experimental data from Ref.\,\protect\cite{Conway:1989fs}, at scale (4.05\,{\rm GeV})$^2$;
\emph{dashed curve} -- NLO reanalysis of the experimental data \protect\cite{Wijesooriya:2005ir};
and \emph{dot-dot-dashed curve} -- NLO reanalysis of experimental data with inclusion of soft-gluon resummation \protect\cite{Aicher:2010cb}.
\label{fig:pi_DSE}}
\end{figure}

Figure~\ref{fig:pi_DSE} displays the DSE results for the pion's valence $u$-quark distribution, evolved from a resolving scale $Q_0^2=(0.6\,$GeV)$^2$ to $Q^2 = (5.2~{\rm GeV})^2$ using leading-order DGLAP evolution (Sec.IID of Ref.\,\cite{Holt:2010vj}), and a comparison with $\pi N$ Drell-Yan data \cite{Conway:1989fs} at a scale  $Q^2 \sim (4.05~{\rm GeV})^2$, inferred via a leading-order analysis.  The computation's resolving scale, $Q_0$, was fixed by matching the $\langle x^n\rangle^\pi$ moments for $n=1,2,3$ to an experimental analysis at (2\,{\rm GeV})$^2$ \cite{Sutton:1991ay}.

It is notable that at $Q_0$ the DSE results yield
\begin{equation}
\label{momcons}
2 \, \langle x \rangle^\pi_{Q_0} = 0.7\,,\;
2 \, \langle x \rangle^K_{Q_0} = 0.8\,.
\end{equation}
(For comparison, the parametrised valence-like pion parton distributions of Ref.\,\protect\cite{Gluck:1998xa} yield a gluon momentum fraction of $\langle x_g\rangle^\pi_{Q_0=0.51} = 0.3$.)
In each case the remainder of the hadron's momentum is carried by gluons, which effect binding of the meson bound state and are invisible to the electromagnetic probe.  Some fraction of the hadron's momentum is carried by gluons at all resolving scales unless the hadron is a point particle \cite{Holt:2010vj}.  Indeed, it is a simple algebraic exercise to demonstrate that the only non-increasing, convex function which can produce $\langle x^0\rangle =1$ and $\langle x \rangle = \frac{1}{2}$, is the distribution $u(x)=1$, which is uniquely connected with a pointlike meson; viz., a meson whose Bethe-Salpeter amplitude is momentum-independent.  Thus Eqs.\,(\ref{momcons}) are an essential consequence of momentum conservation.

Whilst the DSE results in Fig.\,\ref{fig:pi_DSE} are both consistent with Eq.\,(\ref{pQCDuvx}); i.e., they produce algebraically the precise behaviour predicted by perturbative QCD, on the valence-quark domain -- which is uniquely sensitive to the behaviour of the dressed-quark mass-function, $M(p^2)$ -- it is evident that they disagree markedly with the Drell-Yan data reported in Ref.\,\cite{Conway:1989fs}.  This tension was long seen as a crucial mystery for a QCD description of the lightest and subtlest hadron \cite{Holt:2010vj}.  Its re-emergence with Ref.\,\cite{Hecht:2000xa} motivated a NLO reanalysis of the Drell-Yan data \cite{Wijesooriya:2005ir}, the result of which is also displayed in Fig.\,\ref{fig:pi_DSE}.  At NLO the extracted PDF is softer at high-$x$ but the discrepancy nevertheless remains.
To be precise, Ref.\,\cite{Wijesooriya:2005ir} determined a high-$x$ exponent of $\beta \simeq 1.5$ whereas the exponents produced by the DSE studies \cite{Hecht:2000xa,Nguyen:2011jy} are, respectively, $2.1$ and $2.4$ at the common model scale.  They do not allow much room for a harder PDF at high-$x$.

Following the highlighting of this discrepancy in Ref.\,\cite{Holt:2010vj}, a resolution of the conflict between data and well-constrained theory was proposed.  In Ref.\,\cite{Aicher:2010cb} a long-overlooked effect was incorporated; namely, ``soft-gluon resummation.'' With the inclusion of this next-to-leading-logarithmic threshold resummation effect in the calculation of the Drell-Yan cross section, a considerably softer valence-quark distribution was obtained at high-$x$.
This is readily understood.  The Drell-Yan cross-section factorises into two pieces: one hard and the other soft.  The soft piece involves the PDF and the hard piece is calculable in perturbation theory.  Adding additional interactions to the latter, which are important at large-$x$; viz., soft gluons, provides greater strength in the hard piece on the valence-quark domain.  Hence a description of the data is obtained with a softer PDF.
Indeed, the distribution obtained thereby matches precisely the expectations based on perturbative-QCD and obtained using DSEs.  This is evident in a comparison between the \emph{dash-dot} and \emph{dash-dot-dot} curves in Fig.\,\ref{fig:pi_DSE}.
This outcome again emphasises the predictive power and strength of using a single internally-consistent, well-constrained framework to correlate and unify the description of hadron observables.

\begin{figure}[t]
\includegraphics[clip,width=0.43\textwidth]{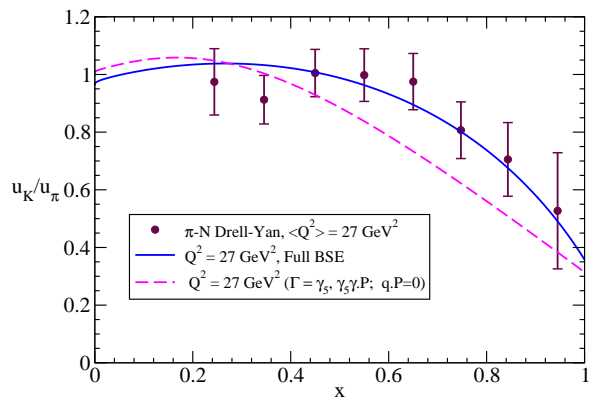}

\caption{\label{fig:pi_DSE_ratio}
DSE prediction for the ratio of $u$-quark distributions in the kaon and pion \protect\cite{Holt:2010vj,Nguyen:2011jy}.  The full Bethe-Salpeter amplitude produces the \emph{solid} curve; a reduced BSE vertex produces the \emph{dashed} curve.  The reduced amplitude retains only the invariants and amplitudes  involving pseudoscalar and axial vector Dirac matrices, and ignores dependence on the variable $q\cdot P$.  These are part of the reductions that occur in a pointlike treatment of pseudoscalar mesons.  The experimental data is from \protect\cite{Badier:1980jq,Badier:1983mj}.}
\end{figure}

The ratio $u_K/u_\pi$ measures the effect of the local hadronic environment.  In the kaon, the $u$-quark is bound with a heavier partner than in the pion ($m_s \approx 25 m_d$) and this should cause $u_K(x)$ to peak at lower-$x$ than $u_\pi(x)$.
In fact, one finds that the $s$-quark distribution peaks at a value of $x$ which is just 15\% larger than that of the $u$-quark.  Hence, even though DIS is a high-$Q^2$ process, constituent-quark-like mass-scales explain this shift: $M_s/M_u \approx 1.25$.
The DSE prediction of $u_K/u_\pi$ \cite{Holt:2010vj,Nguyen:2011jy} is shown in Fig.\,\ref{fig:pi_DSE_ratio} along with available Drell-Yan data \cite{Badier:1980jq,Badier:1983mj}.  The parameter-free DSE result agrees well with the data.
We note that
\begin{equation}
\frac{u_K(0)}{u_\pi(0)} \stackrel{\mbox{\footnotesize\rm DGLAP:}Q^2\to \infty}{\to} 1\,;
\end{equation}
viz, the ratio approaches one under evolution to larger resolving scales owing to the increasingly large population of sea-quarks produced thereby \cite{Chang:2010xs}.  On the other hand, the value at $x=1$ is a fixed-point under evolution: $\forall Q_1^2>Q_0^2$,
\begin{eqnarray}
\frac{u_K(1)}{u_\pi(1)}\rule[-2.5ex]{0.1ex}{6ex}_{Q_1^2}
&=& \mbox{\footnotesize\rm DGLAP}_{Q_1^2\leftarrow  Q_0^2}\left[\frac{u_K(1)}{u_\pi(1)}\right]_{Q_0^2}\\
&=& \frac{u_K(1)}{u_\pi(1)} \rule[-2.5ex]{0.1ex}{6ex}_{Q_0^2}
\end{eqnarray}
i.e., it is the same at every value of the resolving scale $Q^2$, and is therefore a persistent probe of nonperturbative dynamics \cite{Holt:2010vj}.

With Ref.\,\cite{Nguyen:2011jy} a significant milestone was achieved; viz., unification of the computation of distribution functions that arise in analyses of deep inelastic scattering with that of numerous other properties of pseudoscalar mesons, including meson-meson scattering \cite{Bicudo:2001aw,Bicudo:2001jq} and the successful prediction of electromagnetic elastic and transition form factors.  The results confirm the large-$x$ behavior of distribution functions predicted by the QCD parton model; provide a good account of the $\pi$-$N$ Drell-Yan data for $u_\pi(x)$; and a parameter-free prediction for the ratio $u_K(x)/u_\pi(x)$ that agrees with extant data, showing a strong environment-dependence of the $u$-quark distribution.  The new Drell-Yan experiment running at FNAL is capable of validating this comparison, as is the COMPASS~II experiment at CERN.  Such an experiment should be done so that complete understanding of QCD's Goldstone modes can be claimed.

\section{Charm and Beauty mesons}
\label{charmbeauty}

\subsection{CP violation and strong phases}

The past two decades have brought important advances in flavour physics and in particular the understanding of weak decays of heavy-mesons.  From the first observation of a $B$ meson by the CLEO  Collaboration in 1981 at the Cornell Electron Storage Ring \cite{Bebek:1980zd}, to the dedicated $B$-physics facilities at SLAC in California and KEK in Japan, much progress has been made.  Although $B$ physics was the main focus of the Belle and BaBar Collaborations at KEK and SLAC, respectively, and of the CDF experiment at Fermilab, considerable efforts have also been devoted to studies of $D$-meson decays, charmonium and $\tau$ physics.

The driving force is the confirmation of the electroweak sector of the Standard Model, which has established itself as the foremost paradigm to describe $CP$ violation in terms of the Cabibbo-Kobayashi-Maskawa (CKM) mechanism.  The aim of experimental data analyses of branching fractions, polarisations or $CP$-violating amplitudes in, today, a large variety of decay channels, is determination of the precise area of the ``CKM triangle'' and the weak $CP$ violating phase encoded within its angles.

From a theoretical perspective, heavy mesons can be used to test simultaneously all manifestations of the Standard Model, namely the interplay between electroweak and strong interactions.  It is noteworthy that no $CP$-violating amplitude can be generated without {\em relative\/} strong phases, i.e. a single strong phase is insufficient to produce a $CP$ asymmetry.  This can be understood as follows: suppose a heavy-meson decays, $H\to M$, where $M=M_1,M_2 ...$ denotes the final-state mesons, and that the Lagrangian of the Standard Model contributes two terms (e.g., two Feynman diagrams) to this process.  The decay amplitude and its corresponding $CP$ conjugate are, most generally,
\begin{eqnarray}
  \mathcal{A}(H\to M) & = & \lambda_1 A_1 e^{i\varphi_1} + \lambda_2 A_2 e^{i\varphi_2}
  \label{AH2M}\\
   \bar{\mathcal{A}}(\bar H \to \bar M) & = & \lambda_1^* A_1 e^{i\varphi_1} + \lambda_2^* A_2 e^{i\varphi_2} \ .
  \label{AH2Mbar}
\end{eqnarray}
The weak coupling $\lambda_i$ is a combination of possibly complex CKM matrix elements and $A_j e^{i\varphi}$ denote the strong component of the transition amplitude.  We emphasise that they, too, can have both a real part, or magnitude, and a phase, or absorptive part, owing to final rescattering of quarks and mesons.  These $CP$-related intermediate states must contribute the same absorptive part to the two decays,
therefore the strong phases $\varphi_i$ are the same in Eqs.~\eqref{AH2M} and \eqref{AH2Mbar}.  The difference of the absolute squares, known as direct $CP$ violation, is given by
\begin{equation}
 |\mathcal{A}|^2  -  |\bar{\mathcal{A}}|^2   =   2 A_1 A_2\, \mathrm{Im} (\lambda_1 \lambda_2^*)  \sin (\varphi_1-\varphi_2 ) \, .
 \label{CPasym}
\end{equation}
It is plain that $CP$ violation, $ |\bar{\mathcal{A}}/\mathcal{A} | \neq 1$, cannot occur if all weak couplings are real or if the decay amplitude
contains no relative strong phases. Thus, it is crucial to evaluate {\em reliably\/} any strong contributions to the decay in order to extract weak CKM phases with precision.

\subsection{Flavourful hadrons \label{flavourful}}

Heavy mesons also provide an excellent opportunity to reassess the nonperturbative features of QCD discussed above for light hadrons.  A daunting challenge, however, is presented by the widely disparate energy scales involved in heavy-meson decays.  Not surprisingly, factorisation theorems, which allow for a disentanglement of short-distance or {\em hard\/} physics from long-distance or {\em soft\/} physics, are a central aspect of heavy-meson phenomenology.

Soft matrix elements, commonly expressed as hadronic form factors, couplings or decay constants, usually consist of multiple Green functions between two or more physical mass states, with one Green function describing the propagation of the heavy quark.  These matrix elements are formally obtained via the same expedients introduced for the flavoured light mesons: Eqs.\,\eqref{fpigen} to \eqref{rhogen} yield the heavy-pseudoscalar's mass formula and weak decay constant and, likewise, the related Bethe-Salpeter amplitude of a heavy pseudoscalar meson is given by Eq.~\eqref{genGpi}.

\begin{figure}[t]
\begin{center}
\includegraphics*[clip,width=0.42\textwidth]{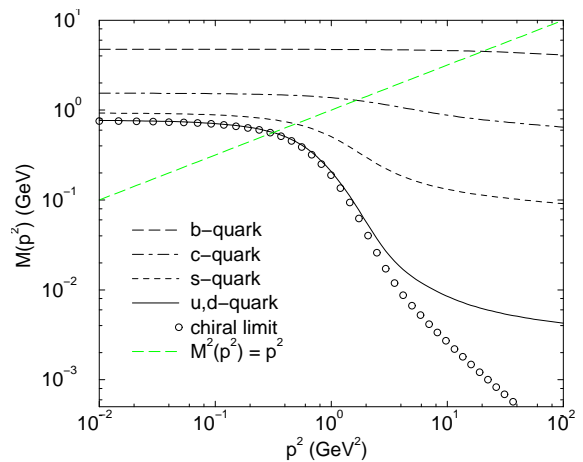}
\caption{Dynamical quark mass function obtained in rainbow-ladder truncation for the flavours $q_f=u,d,s,c,b$ and in the chiral limit.  See, for example, Refs.~\cite{Ivanov:1998ms,Roberts:2007jh} for details.}
\label{flavourmassfunc}
\end{center}
\end{figure}

To this end, the extension of the dynamical dressed-quark mass-function, $M(p^2)$, to heavy flavours via solutions of Eq.\,\eqref{SgeneralN} is necessary \cite{Ivanov:1998ms,Krassnigg:2004if,Maris:2005tt}.  This is depicted in Fig.~\ref{flavourmassfunc} for different current quark masses and the chiral limit.  It is clear from this figure that whilst DCSB is at the origin of a rapid increase of a light quarks's mass-function to the order of several hundred MeV in the infrared, the effect of dressing the $c$ quark is modest and barely noticeable for the $b$ quark, whose large current-quark mass almost entirely suppresses momentum-dependent dressing.  Thus, $M_b(p^2)$ is nearly constant on a large domain.  This is true to a lesser extent for the charm quark.

This can also be appreciated from the definition of a single quantitative measure, namely the renormalisation-point invariant ratio $\varsigma_f := \sigma_f/M_f^E$, where $ \sigma_f$ is a constituent-quark $\sigma$-term~\cite{Holl:2005st},
\begin{equation}
 \sigma_f := m_f(\zeta ) \frac{\partial M_f^E}{\partial m_f(\zeta ) } \ ,
\end{equation}
which probes the impact of explicit chiral symmetry breaking on the mass function.  The Euclidean constituent-quark mass is defined via
\begin{equation}
  (M^E)^2 :=  \{ s  | s>0, s = M^2(s) \} .
\end{equation}
For the solutions depicted in Fig.~\ref{flavourmassfunc}, one finds:
\begin{equation}
\begin{array}{c|ccccc}
  f & \mathrm{chiral} & u, d  & s  &  c &  b \\ \hline
  M_f^E/GeV  &  0.42 & 0.42 & 0.56 & 1.57 & 4.68\\
  \varsigma_f & 0~~~ & 0.02 & 0.23 & 0.65 & 0.8~ \\
\end{array}
\end{equation}
The ratio $\varsigma_f$ quantifies the effect of explicit chiral symmetry breaking on the dressed-quark mass-function compared with the sum of the effects of explicit and dynamical chiral symmetry breaking.  Its values are readily understood.  In the neighbourhood of the chiral limit, $\varsigma_f$ is small because the magnitude of the light-quark constituent-mass owes primarily to DCSB, whilst $\varsigma_f$ approaches unity for heavy quarks because explicit chiral symmetry breaking becomes the dominant source of their mass.

\subsection{Quarkonia}
We first review some aspects of heavy equal-mass, $\bar QQ$, pairs before discussing heavy-light flavoured mesons, in the following section, for which DCSB is decisively more relevant.  First explorations for mass-symmetric ${\bar Q Q}$ states combining a consistent use of the rainbow-ladder truncation in the kernels of the gap- and Bethe-Salpeter-equations were undertaken for $\bar cc$ bound states in Refs.~\cite{Krassnigg:2004if,Bhagwat:2006xi}, with extrapolations to the $\bar bb$ systems in Ref.\,\cite{Maris:2005tt}.  A full numerical solution for flavour-singlet pseudoscalar mesons yielded charmonium and bottonium masses and decay constants in very good agreement with experimental data; predictions for states with exotic quantum numbers were also made \cite{Maris:2006ea,Krassnigg:2009zh}.  The effect of the quark-gluon interaction in the gap equation and the vertex-consistent Bethe-Salpeter kernel was investigated in Ref.~\cite{Bhagwat:2004hn}.  (This is discussed further in Sec.\,\ref{hlmesons}.)

For a quark flavour $Q$, a constituent-quark spectrum-mass may be defined through
\begin{equation}
  M^S_Q = M_Q(p^2= \zeta^2_{H_0}), \ \zeta^2_{H_0} = - \tfrac{1}{4} M^2_{H_0}  \, ,
\end{equation}
where $M_Q (p^2)$ is the renormalisation-point-invariant dressed-quark mass-function in Eq.\,\eqref{SgeneralN}, obtained as the solution of Eq.\,\eqref{gendseN} when a heavy-quark current-mass is used via Eq.\,\eqref{mzeta}.  As the  renormalisation-group invariant current-quark mass, $\hat m_Q$ in Eq.\,\eqref{mfhat}, is increased, $M^S_Q$ becomes equivalent to the so-called pole-mass in non-relativistic QCD (NRQCD).  For $0^-$ quarkonia one can prove \cite{Bhagwat:2006xi}
\begin{equation}
   M^{\bar Q Q}_{H_n}  = 2 M^S_Q\, \left [ 1 + \epsilon^{\bar Q Q }_{H_n} / M^S_Q \right ] ,
\end{equation}
where $ \epsilon^{\bar Q Q }_{\pi^n} $ is a binding-energy that does not grow with $M_Q^S$.  Here $H_0$ denotes the lowest-mass pseudoscalar and increasing $n$ labels bound-states of increasing mass.

Using the renormalisation-group-invariance of the product $m(\zeta^2)\rho_{H_n}(\zeta^2)$, Eq.\,\eqref{fpigen} predicts
\begin{equation}
\label{rhoHn}
 \rho_{H_n}^{\bar Q Q} \stackrel{\hat m_Q\to \infty}{=} f_{H_n}^{\bar Q Q} M_{H_n}^{\bar Q Q} \ ,
\end{equation}
which establishes an identity between the pseudoscalar and pseudovector projections of the meson's Bethe-Salpeter wave function at the origin in configuration space.  The elements in Eq.~\eqref{rhoHn} are each gauge invariant and renormalisation point independent.  In case of heavy-light systems it can be shown algebraically \cite{Ivanov:1997iu,Ivanov:1997yg,Ivanov:1998ms} that Eq.\,\eqref{rhoHn} is realised via
\begin{equation}
   \rho_{H_n}^{\bar Q q}  \propto \surd M_{H_n}^{\bar Q q}   ; \quad
   f_{H_n}^{\bar Q q}  \propto 1/ \surd  M_{H_n}^{\bar Q q}  ,
 \label{frhorel}
\end{equation}
which follows from the expansion in $\epsilon^{\bar Q q }_{H_n} / M^S_Q$ and $w^{\bar Q q}_{H_n}/ M^S_Q$ of the integrands in Eqs.\,\eqref{fpigen} to \eqref{rhogen}.  Here, $w^{\bar Q q }_{H_n}$ is the width of the meson's Bethe-Salpeter wave function defined as the value of the relative momentum, $k$, whereat the first Chebyshev moment of the amplitude $\mathcal{E}^{\bar Q q}_{H_n}$ falls to one-half of its maximum value.  In heavy-light mesons, $k \sim w^{\bar Q q }_{H_n}$ is the typical momentum of the light-quark.  A remarkable feature is that $w^{\bar Q q }_{H_n}$ reaches a finite nonzero value in the limit $M^S_Q \to \infty$ which implies that a heavy-light meson always has a nonzero spatial extent.

\begin{figure}[t!]
\centering
\includegraphics*[width=0.42\textwidth]{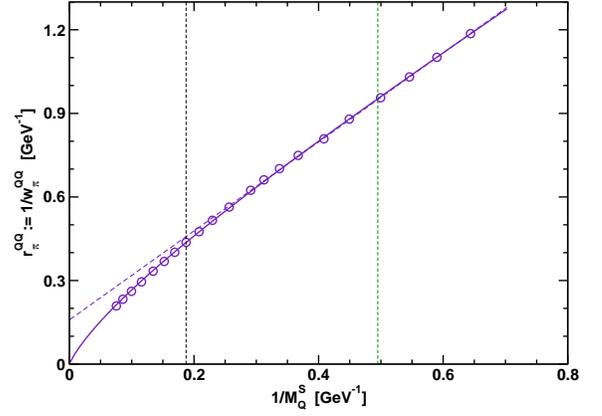}
\caption{Evolution of the spatial size, $r_{H_n}^{\bar Q Q}$, with $1/M_Q^S$.  \emph{Circles} -- $r_{H_0}^{\bar Q Q}$ calculated using the interaction model of Ref.~\cite{Maris:2002mt}; \emph{solid curve}, as described by Eq.\,(\protect\ref{rQQ}); \emph{dashed curve} -- linear fit to calculated result in neighbourhood of $M_c^S$.
Dashed vertical lines mark, from left, $1/M_b^S$ and $1/M_c^S$.  Note that as $M^S_Q \to\infty$, $r_{H_n}^{\bar Q Q}\to 0$.
\label{qqbarsize}}
\end{figure}

As shown in Ref.\,\cite{Maris:2002mt}, this is not the case for $\bar QQ$ systems. Figure~\ref{qqbarsize} depicts the evolution of the heavy-heavy meson's spatial size as a function of the inverse constituent-quark spectrum-mass, $M_Q^S$.\footnote{The evolution was computed using the renormalisation-group-improved rainbow-ladder DSE truncation and the interaction model in Ref.\,\protect\cite{Maris:2002mt}.  As shown, e.g., in Ref.\,\protect\cite{Bhagwat:2004hn}, all corrections to this truncation vanish in
the heavy-heavy limit.}
The curve in Fig.\,\ref{qqbarsize} is the function
\begin{equation}
  r_{H_n}^{\bar Q Q} = \frac{\gamma_M}{M_Q^S} \ln \left [ \tau_M +\frac{M^E}{\Lambda_\mathrm{QCD}} \right ]; \ \gamma_M = 0.68,\, \tau_M = 8.56 ,
\label{rQQ}
\end{equation}
with $\Lambda_\mathrm{QCD}= 0.234\,$GeV. One deduces that with increasing constituent-quark mass, the heavy-heavy system becomes ``point-like'' in configuration space and hence delocalised in momentum space.

Consequently, the evolution with $M^S_Q$ of an observable like the decay constant, $f_{H_n}^{\bar Q Q}$, is sensitive to the $\bar QQ$ interaction over a wide range of momentum scales and therefore a useful probe of that interaction.
In this connection we note that NRQCD predicts the matrix elements for various spin states of a given quarkonium system are equal up to corrections of order $v_Q^2\approx (w_{\pi_n}^{\bar QQ}/M_Q^S)^2$, where $k\sim w_{\pi_n}^{\bar QQ}$ is the typical
magnitude of the heavy-constituent's three-momentum in the meson's rest frame \cite{Bodwin:1994jh}.  In this picture, $0^{-+}$ and $1^{--}$ mesons, which differ because spins are anti-aligned in the pseudoscalar and aligned in the vector, become degenerate in the limit $M^S_Q\to\infty$ and their leptonic decay constants
become identical; i.e., $f_{\rho_n}^{\bar Q Q}=f_{\pi_n}^{\bar Q Q}$.  It is noteworthy, however, that Eq.\,(\ref{rQQ}) gives $v^2_c \approx 0.27$ and $v^2_b\approx 0.18$.  Moreover, $v^2_Q$ falls only as $\alpha_s^2(M_Q^S)$.  Hence a quantitative discrepancy between $f_{\rho_n}^{\bar Q Q}$ and $f_{\pi_n}^{\bar Q Q}$ can conceivably exist until
rather large quark masses.

\subsection{Heavy-light mesons}
\label{hlmesons}
The mass asymmetry in flavoured $\bar Qq$ mesons leads to a diverse array of energy scales, a feature that leads to difficulties in DSE studies not encountered in the calculation of either light or heavy equal-mass systems within the rainbow-ladder
truncation.  For illustration, Table~\ref{massdecayheavy} unmistakably states that while masses of flavoured pseudoscalar mesons, such as $D_{(s)}$ and $B_{(s)}$ mesons, are in  good agreement with experimental data, this is not so for their respective weak decay
constants \cite{Maris:2005tt,Nguyen:2010yh}.

\begin{table}[t]
\caption{\label{massdecayheavy}
Calculated masses and electroweak decay constants for ground state pseudoscalar and vector heavy-light mesons compared with respective experimental data when available \protect\cite{Nakamura:2010zzi}, except for $f_{B_s}$ for which the most recent numerical value from lattice-QCD is indicated by a dagger \protect\cite{Gamiz:2009ku} (all values are GeV).  In the rows labelled ``R.L.'', the heavy quark in the rainbow-ladder
truncation is described by a constituent-quark mass, which is fit to the lightest pseudoscalar (the $D$ and $B$ mesons).  In the rows labelled ``$k_\mathrm{min}$'',
the heavy quark is dressed through the DSE with an infrared suppression of the gluon momentum as described in the text.  No such infrared suppression is applied to the dressing of the light quark or the binding kernel.
(Adapted from Ref.\,\protect\cite{Nguyen:2010yh}.)
}\medskip
\hspace*{-1em}\begin{minipage}[t]{0.45\textwidth}
\begin{tabular}{|c|cc|cc|cc|cc|cc|} \hline
 $M_{0^-(1^-)}^{\bar Qq}$ & $D$  & $D^*$ & $D_s$ & $D^*_s$ & $B$ & $B^*$ & $B_s$ & $B_s^*$ & $B_c$ & $B^*_c$   \\  \hline \hline
Exp. & 1.86 & 2.01 &1.97 &  2.11 & 5.28  & 5.33 & 5.37  & 5.41 & 6.29 & --  \\ \hline
R.L.  &  1.85 & 2.04 & 1.97 & 2.17 & 5.27 & 5.32 & 5.38 & 5.42 & 6.36 & 6.44  \\ \hline
 $k_\mathrm{min}$ & 1.88 & -- & 1.90 & -- & 5.15 & -- & 4.75  & -- & 5.83 & -- \\ \hline
\end{tabular}
\end{minipage}\\
\hspace*{-1em}\begin{minipage}[t]{0.45\textwidth}
\begin{tabular}{|c|cc|cc|cc|cc|cc|} \hline
 $f_{0^-(1^-)}^{\bar Qq}$ & $D$  & $D^*$ & $D_s$ & $D^*_s$ & $B$ & $B^*$ & $B_s$ & $B_s^*$ & $B_c$ & $B^*_c$   \\  \hline \hline
Exp. & 0.22 & -- & 0.29 &  -- & 0.18  & -- & 0.23$^\dagger$ & -- & -- & --  \\ \hline
R.L.  &  0.15 & 0.16 &  0.20 & 0.18 & 0.11 & 0.18 & 0.14 & 0.20 & 0.21 & 0.18  \\ \hline
 $k_\mathrm{min}$ & 0.26 & -- & 0.28 & -- & 0.27 & -- & 0.16  & -- & 0.45 & -- \\ \hline
\end{tabular}
\end{minipage}
\end{table}

More precisely, if the masses and decay constants of heavy-light ground state pseudoscalars and vectors involving a $c$- or $b$-quark are calculated using DSE solutions for the dressed light-quarks and a constituent-quark propagator for the heavy quark, the various meson masses are readily reproduced (see values in the row labelled R.L.).  This is expected since, as we saw in Sec.\,\ref{flavourful}, heavy quarks,
and in particular the $b$ quark, have mass functions whose momentum dependence is nearly constant on a large domain.  However, the weak decay constants obtained from the constituent-like mass approximation are 30-50\% below the experimental values.

Using fully-dressed quark propagators, both heavy and light, in the rainbow-ladder model does not much improve the situation.  Indeed, within the context of fully-dressed propagators, it was numerically shown \cite{Maris:2005tt} that for small current-quark masses, weak decay constants of pseudoscalar and vector heavy-heavy and heavy-light mesons increase with the quark mass, yet tend to level off between the $s$- and $c$-quark mass.  This behaviour is consistent with the heavy-quark limit; namely, a decrease of the decay constant with increasing meson mass like $ f_{H_n}^{\bar Q q}  \propto 1/\surd{m_{H_n}^{\bar Q q}}$, Eq.\,\eqref{frhorel}.  This asymptotic behaviour might occur as low as $Q=c$ for the $\bar Qu$ mesons.  However, as in the case of constituent-quark propagators, $f_D$ and $f_{D_s}$ are about 20\% below their experimental values~\cite{Maris:2005tt}.  Since a $\bar Qq$ decay constant depends on the norm of the Bethe-Salpeter amplitude, which in turn depends on the derivative of the quark propagators, it might be anticipated that the decay constants are more sensitive to details of the model than meson masses and are thus better indicators of deficiencies in the modelling.

These observations suggest strongly that the rainbow-ladder truncation is not a reliable tool for the study of heavy-light mesons the charm-quark region (or larger), despite the fact that experimental meson masses are well reproduced.  In fact, in such systems cancellations, which largely mask the effect of dressing the quark-gluon vertices in light-light mesons, are blocked by the dressed-propagator asymmetry.

\begin{figure}[t!]
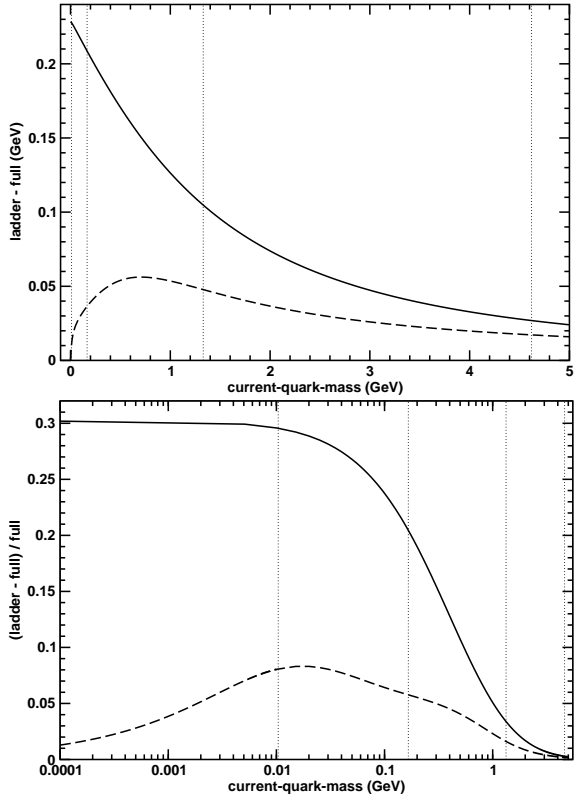

\centering
\includegraphics*[clip,width=0.42\textwidth]{ladderfulldiffA.eps}
\includegraphics*[clip,width=0.42\textwidth]{ladderfulldiffL}
\caption{Evolution with current-quark mass of the difference between the meson mass calculated in the rainbow-ladder truncation and the exact value; namely, that obtained using the completely resummed dressed-quark-gluon vertex in the gap equation and the vertex-consistent Bethe-Salpeter kernel.  The upper panel depicts the absolute error and the lower panel, the relative error.  \emph{Solid lines}: vector meson trajectories;
and \emph{dashed-lines}; pseudoscalar meson trajectories. The dotted vertical lines mark the $m_{u,d}$, $m_s$, $m_c$ and $m_b$ current-quark masses.}
\label{ladderfulldiff}
\vspace*{-3mm}
\end{figure}

In this context, as discussed in Ref.\,\cite{Bhagwat:2004hn}, Fig.\,\ref{ladderfulldiff} is instructive: it shows the evolution with current-quark mass of the difference between the $\bar qq$ meson mass calculated in the rainbow-ladder truncation and the
value obtained using a completely resummed dressed-quark-gluon vertex in the gap equation and vertex-consistent Bethe-Salpeter kernel.
With growing current-quark mass, the rainbow-ladder truncation provides an increasingly reliable estimate of the meson masses.  The accuracy is best for pseudoscalar $\bar qq$ mesons; e.g., the absolute difference reaches its peak of $\approx 60\,$MeV at $m \sim 4 m_s$ whereat the relative error is only 3\%.
This behaviour in the pseudoscalar sector is fundamentally important because it expresses Goldstone's theorem: no truncation is valid unless it guarantees that the axial-vector Ward-Takahashi identity be preserved, something which guarantees a massless pseudoscalar meson in the chiral limit.  Of signficance here is the observation that for large current-quark masses, say $m\sim m_b$, the contributions from vertex corrections, which produce nonplanar diagrams in the Bethe-Salpeter kernel, are suppressed.  Nonetheless,
Fig.~\ref{ladderfulldiff} emphasises that the rainbow-ladder truncation is not appropriate for charmed mesons; moreover, even for heavier mesons, corrections to that truncation ought to be included in precision spectroscopic calculations.

An {\em ad hoc\/} approach to alter the infrared sector of the model kernel was recently devised~\cite{Nguyen:2010yh}.  In essence it suppresses infrared contributions from a quark-gluon-vertex model originally introduced for light quarks.  This introduces a sharp cut-off, $k_\mathrm{min}$, for the gluon momentum in the DSE kernel of the heavy quark, while the original kernel is kept for the dressing of the light quark and the binding.  The results are displayed in the rows of Table~\ref{massdecayheavy} labelled by ``$k_{\rm min}$''.  The decay constant values for pseudoscalar mesons (vector mesons were not investigated in this scheme) increase dramatically for the $D$, $D_s$ and $B$ meson, and are somewhat closer to experiment than the values obtained with the constituent-mass approximation.  The expedient has a somewhat detrimental effect on the computed masses.

To summarise, it has become clear that realistic calculations of heavy-light Bethe-Salpeter amplitudes must involve a truncation that goes beyond rainbow-ladder.  The exact form of the Bethe-Salpeter equation developed in Ref.\,\cite{Chang:2009zb} might provide the advance needed.

\subsection{Flavoured form factors and couplings}
As a corollary of Eq.\,\eqref{CPasym}, complex weak phases are a necessary but not sufficient condition for $CP$-violating amplitudes.  Any decay must involve non-vanishing relative strong phases.  Their contributions may originate at any stage of the heavy-meson decay and can arise through: hard radiative corrections to dimension-six and higher operators in heavy quark effective theory (HQET); hadronic transition matrix elements between initial- and final-state mesons; as well as final-state interactions between daughter hadrons \cite{ElBennich:2006yi,Boito:2008zk,ElBennich:2009da,Magalhaes:2011sh}.
The development of perturbative QCD factorization \cite{Beneke:1999br,Beneke:2000ry,Bauer:2000yr,Beneke:2002ph,Bauer:2004tj} provides  a means of simplifying this problem by enabling a systematic approximation for a given process in terms of products of soft and hard matrix elements.

Transition form factors that characterise the decays of $B$-mesons into light pseudoscalar and vector mesons are basic to an understanding of $B$-meson exclusive non-leptonic, semi-leptonic and rare radiative decays.  An understanding of the entirety of decay processes is essential to the reliable determination of CKM matrix elements, which is also meant to provide a means of searching for non Standard Model effects and CP violation.  Considering all these factors, it is not surprising that heavy-light form factors are the subject of much experimental and theoretical scrutiny.

An immediate consequence of QCD-factorisation theorems for $B\to M_1 M_2$ decays, $m_b \gg \Lambda_\mathrm{QCD}$, is the emergence of two hadronic matrix elements,
\begin{eqnarray}
  \lefteqn{\langle M_1 M_2 | O_i | B \rangle\  = \   \langle M_1| j_1 | B \rangle \langle M_2 | j_2 | 0 \rangle }  \nonumber \\
    & &   \hspace*{5mm}  \times \ \left [ 1 + \sum_n r_n \alpha_s^n + \mathcal{O} ( \Lambda_\mathrm{QCD}/m_b ) \right ] ,
\label{QCDfac}
\end{eqnarray}
where $j_1$ and $j_2$ are bilinear currents.  This factorisation has been verified to leading order in $\alpha_s$ \cite{Beneke:1999br,Bauer:2004tj} and including the one-loop correction, $\alpha_s^2$, to tree-diagram scattering between the emitted meson and the one containing the spectator- (light-) quark in the $B$-meson \cite{Beneke:2005vv}.  It is evident from Eq.\,\eqref{QCDfac} that higher orders in $\alpha_s$ violate the factorisation but the corrections can systematically be included.  The analogy with perturbative factorisation for exclusive processes in QCD at large-momentum
transfer is not accidental \cite{Lepage:1980fj}.  Of course, semi-leptonic decays are much cleaner and, in the absence of interactions with a second meson, the transition form factor $\langle M_1 | O_i | B \rangle$ completely describes the decay amplitude.

References~\cite{Ivanov:1997iu,Ivanov:1998ms,Ivanov:2007cw} compute heavy-to-light transition amplitudes, $\langle M(p_2)|\bar q\,\Gamma_I h |H(p_1)\rangle$, in generalised impulse approximation, which is the leading-order in the systematic, nonperturbative
and symmetry-preserving DSE truncation scheme of Refs.\,\cite{Munczek:1994zz,Bender:1996bb}.  The transition amplitude is given by
\begin{eqnarray}
  \hspace*{-7mm}
  \lefteqn{
  \mathcal{A}(p_1,p_2) = \mathrm{tr}_\mathrm{CD}\!\  \!\int\! \frac{d^4k}{(2\pi)^4} \, \bar \Gamma_{M}^{(\mu)}(k;-p_2)  S_q(k+p_2)}
  \nonumber   \\
  &   \times & \Gamma_I(p_1,p_2)  S_Q (k+p_1)  \Gamma_H(k; p_1) S_{q'} (k) , \rule{1.5em}{0ex}
 \label{heavylightamp}
\end{eqnarray}
where: $S(p)$ are dressed quark propagators, $Q=c,b$, $q=q'=u,d,s$; and $M=S,P,V,A$,
with the index $\mu$ indicating a possible vector structure in the final-state BSA. $\Gamma_I$ is the interaction vertex whose Lorentz structure depends on the operator $O_i$ in the HQET and $\Gamma_H$ is the heavy meson BSA.  The trace is over Dirac and colour indices.  For $M=P,V$ and $\Gamma_I=\gamma_\mu(1-\gamma_5), \sigma_{\mu\alpha} q^\alpha$, the amplitude $\mathcal{A}(p_1,p_2)$ can be decomposed into Lorentz vectors and tensors as follows ($q_l=u,d,s,c$),
\begin{eqnarray}
\langle  P(p_2) | \bar q_l \gamma_\mu b | B \rangle
&  =  & F_+(q^2)  P_\mu   +   F_-(q^2) q_\mu ,  \label{HPSformfac}  \\
\nonumber
\langle  P(p_2) | \bar q_l \sigma_{\mu\alpha} q_\alpha b | B \rangle   & = & \frac{i \, F_T(q^2) }{m_1+m_2}   \{ q^2 P_\mu  -  q \cdot P\,  q_\mu \},\\
&& \label{Ftrans}
\end{eqnarray}
$\,$\\[-8ex]
\begin{eqnarray}
\nonumber
\lefteqn{
\langle  V(p_2,\epsilon_2) | \bar q_l \gamma_\mu (1-\gamma_5) b | B \rangle
= \frac{i}{m_1+m_2} \epsilon_2^{\dagger\nu}
} \rule{3em}{0ex}\\
&\times& 
\{ -\delta_{\mu\nu}\, P\!\cdot\! q A_0 (q^2)
  + P_\mu P_\nu  A_+(q^2) \rule{2em}{0ex} \label{HLformfactors} \\
\nonumber
&&  +\, q_\mu P_\nu A_-(q^2)  +   \varepsilon_{\mu\nu\alpha\beta}  P^\alpha q^\beta V(q^2) \big  \}  ,
\end{eqnarray}
which defines the heavy-to-light transition form factors $F_\pm(q^2)$, $F_T(q^2)$, $A_0(q^2)$, $A_\pm(q^2)$ and $V(q^2)$, and where $P=p_1+p_2$, $q=p_1-p_2$, $\epsilon_2^\nu$ is the polarisation vector of the vector meson, and $m_1, m_2$ are the respective initial- and final-state meson masses.  Similar decompositions hold for $M=S,A$ and other four-quark interaction operators $\Gamma_I$; see, e.g. Ref.\,\cite{Ivanov:2007cw}.

\begin{figure}[t]
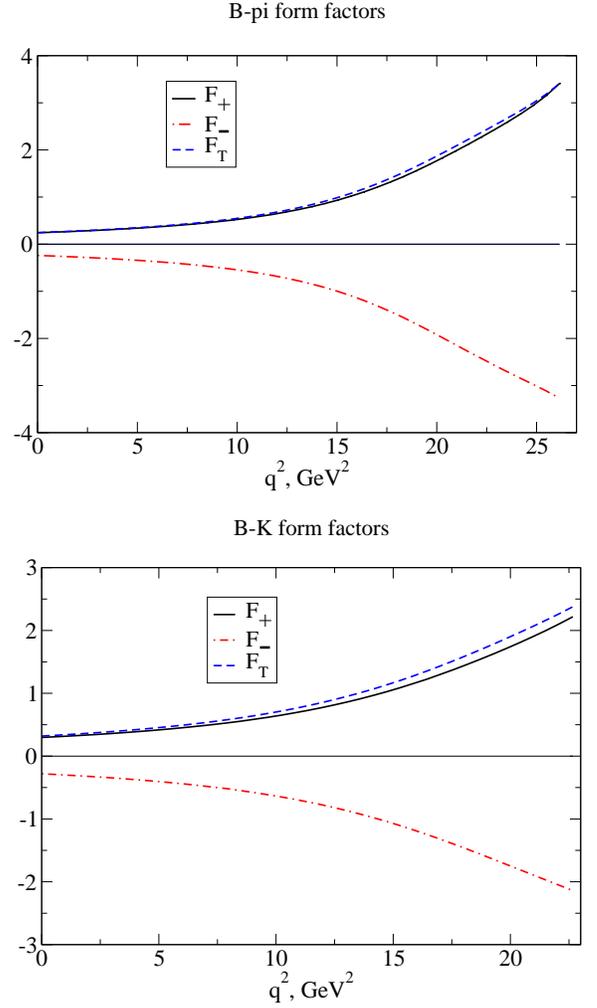

\centering
\includegraphics*[clip,width=0.42\textwidth]{BpiKU}
\vspace*{1em}

\includegraphics*[clip,width=0.42\textwidth]{BpiKL}
\caption{Results for the form factors $F_\pm (q^2)$ in Eq.\,\eqref{HPSformfac} and $F_T(q^2)$ in Eq.\,\eqref{Ftrans}.  \emph{Top panel}, $B\to \pi$; and \emph{bottom panel}, $B\to K$: solid $F_+$, dot-dashed $F_-$, dashed $F_T$.
}
\label{btopik}
\end{figure}

\begin{figure}[t]
\centering
\includegraphics*[clip,width=0.45\textwidth]{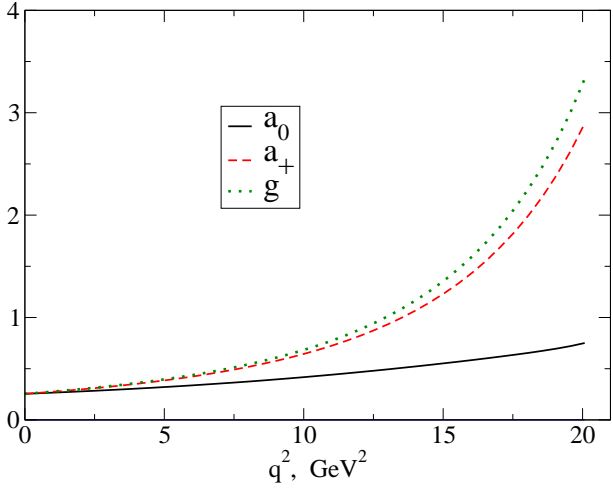}
\caption{Results for the $B\to \rho$ form factors $A_0 (q^2)$, $A_\pm (q^2)$ and $V(q^2)$ in Eq.\,\eqref{HLformfactors}: \emph{solid curve}, $A_0$; \emph{dashed}, $A_+$; \emph{dot-dashed}, $A_-$; and \emph{dotted} $V$.
}
\label{btorho}
\end{figure}

As noted in Sec.\,\ref{flavourful}, the ratio $\varsigma_b$ indicates that the mass function of the $b$ quark varies little on a large momentum-squared domain, whilst the alteration for the $c$ quark is modest.  It is therefore sensible to replace solutions for the dressed quark propagators from DSEs by constituent-mass propagators whose dressed mass is constant:
\begin{equation}
  S_{Q =c,b} \approx \frac{1}{i\gamma\cdot p + \hat M_Q}\ .
\label{heavypropag}
\end{equation}
Furthermore, in Sec.\,\ref{hlmesons} we discussed obstacles to defining a realistic symmetry-preserving truncation scheme for the interaction kernel of heavy-light mesons, which has so far impeded efforts to build adequate Bethe-Salpeter amplitudes for heavy-light $0^{-+}$ and $1^{--}$ mesons.  Calculations are for now limited to Bethe-Salpeter amplitude \emph{Ans\"atze} with a single width parameter that is fixed by a fit to extant hadronic data.  Nonetheless, the form factors calculated with such a DSE-based model are obtained for the entire physical momentum domain without any extrapolations and the chiral limit is directly accessible \cite{Ivanov:2007cw,ElBennich:2009vx}.  Predictions for the $B\to \pi$, $B\to K$ and $B\to \rho$ transition form factors using Eqs.\,\eqref{heavylightamp}, \eqref{HPSformfac}, \eqref{Ftrans} and \eqref{HLformfactors} are shown in Figs.\,\ref{btopik} and \ref{btorho}.

To make contact with HQET, the heavy-quark limit has also been investigated.  In this limit, the heavy-quark carries all the momentum of the heavy meson, which constrains the momentum of the light-quark in the Bethe-Salpeter amplitude.  The heavy-meson momentum is $P_\mu = (\hat M_Q +E )v_\mu$, where $E\equiv M_H - \hat M_Q$, $M_H$ is the mass of the heavy meson and $v_\mu$ is a time-like unit vector, $v^2=-1$.  A heavy quark has four-momentum $P_\mu+k_\mu$, where the residual momentum $k \approx \Lambda_\mathrm{QCD}$ is associated with motion of the light-quark within the bound-state.  In the limit $\hat M_Q\to \infty$, the expansion of the propagator in Eq.\,\eqref{heavypropag} in powers of $1/\hat M_Q$ yields
\begin{equation}
    S_Q  (k + P) =  \frac{1}{2}\, \frac{1- i\gamma\cdot v }{k\cdot v - E} + \mathcal{O}\! \left ( \frac{|k|}{\hat M_Q},  \frac{E}{\hat M_Q} \right ) ,
\end{equation}
given that $|k|/\hat M_Q \ll 1$.  If we employ a two-covariant {\em Ansatz} for the heavy pseudoscalar meson's Bethe-Salpeter amplitude; viz.,
\begin{equation}
  \Gamma_{H_{0^-}} (k;P) = \gamma_5 \left ( 1 - \tfrac{1}{2} i\, \gamma\cdot v \right )  \frac{1}{\mathcal{N}_H}\, \varphi (k^2) \  ,
\end{equation}
where $\mathcal{N}_H$ is the canonical normalisation constant, then the weak decay constant, Eq.\,\eqref{fpigen}, becomes, with $z=u-2E\sqrt{u}$:
\begin{eqnarray}
   f_{H_{0^-}}  & = & \frac{\kappa_f}{\sqrt{M_H}} \frac{N_c}{8\pi^2} \int_0^\infty \!\!du \left (\sqrt{u} -E \right ) \varphi^2(z) \nonumber \\
&&\times \left [ \sigma^f_S(z) + \tfrac{1}{2}\sqrt{u}\, \sigma_V^f(z) \right ] .
\end{eqnarray}
Here we have introduced the $M_H$-independent canonical normalisation, $\kappa_f$, in order to make explicit the previously mentioned relation, Eq.\,\eqref{frhorel}: $f_{H_{0^-}}\sqrt{M_H} =$ constant.

Similarly, at leading order in $1/\hat M_Q$, the $B\to P$ transition form factors in Eq.\,\eqref{HPSformfac} are
\begin{equation}
   F_{\pm} (q^2) = \frac{1}{2}\frac{M_P \pm M_B}{\sqrt{M_P M_B}}\,  \xi (w) ;
\end{equation}
i.e. the form factors depend on a single universal function (the so-called Isgur-Wise function):
\begin{align}
   \xi (w) &  = \kappa_f^2 \frac{N_c}{32\pi^2} \int_0^1\!\! d\tau \frac{1}{W} \int_0^\infty \!\!\! du\, \varphi^2(z_w) \nonumber \\
   & \times \left [ \sigma^f_S(z) + \frac{\sqrt{u}}{W} \, \sigma_V^f(z) \right ] \ ,
\end{align}
with $W = 1+2\tau(1-\tau )(w-1)$,  $z_W = u-2E \sqrt{u/W}$ and $w = - v_B\cdot v_P = (M_B^2+M_P^2 - q^2)/(2M_B M_P)$.  Owing to the canonical normalisation of the Bethe-Salpeter amplitudes, $\xi(w=1) =1$.

The DSEs plainly reproduce the results of heavy-quark symmetry and, in addition, provide a means by which to examine the fidelity of the formulae obtained in the heavy-quark limit.  In other words: one may pose and answer the question: under which conditions is heavy-quark symmetry applicable?  Based on the analysis in Ref.\,\cite{Ivanov:1998ms}, which presents a unified and uniformly accurate description of a broad range of light- and heavy-meson observables, one concludes that corrections to the heavy-quark symmetry limit of $\lesssim 30$\% are encountered in $b\to c$ transitions but that these corrections can be as large as a factor of 2 in $c\to d$ transitions.

Radiative and strong decay amplitudes have also been calculated in the impulse approximation, and their heavy-quark limits exposed \cite{Ivanov:1998ms,ElBennich:2009vx}.  The amplitude for the decay $H^*(p_1) \to H(p_2) \gamma(k)$, with $p_1^2 = -M_{H^*}^2$,  $p_1^2 = -M_H^2$ and $k^2=0$, is given by
\begin{align}
   \mathcal{A} (H^*\to H\gamma)  = \epsilon_\mu^{\lambda_{H^*}}(p_1)  & \epsilon_\nu^{\lambda_\gamma}(k)   \nonumber \\
    \times \big [ e_Q M_{\mu\nu}^Q (p_1,p_2)   & + e_{q_f} M_{\mu\nu}^{q_f}(p_1,p_2) \big ]  \ ,
 \label{hstargamma}
\end{align}
where $e_{Q,q_f}$ is the fractional charge of the active quark in units of the positron charge.  The sum indicates that the decay occurs via a spin-flip transition by either the heavy or light quark.  The hadronic tensors in Eq.\,\eqref{hstargamma} are:
\begin{align}
M_{\mu\nu}^Q (p_1,p_2)  =  & \ \mathrm{tr}_\mathrm{CD}\!\  \!\int\! \frac{d^4\ell}{(2\pi)^4} \, \bar \Gamma^H (\ell;-p_2)  S_Q(\ell_2)
  \nonumber   \\
 \times & \, i \Gamma^Q_\nu (\ell_2,\ell_1)  S_Q (\ell_1)  \Gamma_\mu^{H^*} (\ell; p_1) S_{q_f} (\ell)   \\
M_{\mu\nu}^{q_f} (p_1,p_2)  =  & \ \mathrm{tr}_\mathrm{CD}\!\  \!\int\! \frac{d^4\ell}{(2\pi)^4} \, \bar \Gamma^H (\ell;-p_2)  S_Q(\ell_1)
  \nonumber   \\
 \times  \,\Gamma_\mu^{H^*}\!  (\ell; &\,  p_1)  S_{q_f} (\ell)\,  i \Gamma^q_\nu (\ell,\ell+k)  S_{q_f} (\ell + k)\,  ,
\end{align}
where $Q=c, b$, $q_f = u, s$  and $\ell_{1,2} = \ell+ p_{1,2}$. $\Gamma^f_\nu (\ell_1,\ell_2)$ is the dressed-quark-photon vertex, for which
the BC {\em Ansatz\/}, Eq.\,\eqref{bcvtx}, was used because it satisfies the vector Ward-Takahashi identity and thus ensures current conservation.  In numerical calculations it is helpful that, in case of a heavy constituent, $\Gamma^f_\nu (\ell_1,\ell_2) = \gamma_\nu$ owing to Eq.~\eqref{heavypropag}: $A_Q =1$ and $B_Q = \hat M_Q$.

Likewise, in the impulse approximation, the strong decay $H^*(p_1) \to H(p_2) \pi(q)$ is described by the invariant amplitude,
\begin{align}
   \mathcal{A} (H^*\to H\pi)  &   = \epsilon_\mu^{\lambda_{H^*}}\!(p_1) M_{\mu\nu}^{H^*H\pi}            \nonumber \\
                                                 &  :=\epsilon_\mu^{\lambda_{H^*}}\!(p_1) \,  p_{2\mu} \  g_{H^*H\pi} .
\label{Hstarpicoupl}
\end{align}
The dimensionless coupling $g_{H^*H\pi}$, which can in principle be determined via the decay width $\Gamma_{H^*H\pi}$, is related to the putative universal strong coupling, $\hat g$, between heavy-light-vector and -pseudoscalar mesons in a heavy-meson chiral
Lagrangian \cite{Casalbuoni:1996pg}.  At tree level, the couplings $g_{H^*H\pi}$ and $\hat g$ are related as,
\begin{equation}
    g_{H^*H\pi}  = \frac{\sqrt{m_H m_{H^*}}}{f_\pi} \hat g \, .
\end{equation}
Practically, the matrix element in Eq.\,\eqref{Hstarpicoupl} describes the physical process $D^*\to D\pi$, with both the final pseudoscalar mesons on-shell.  It also serves to compute the unphysical soft-pion emission amplitude $B^*\to B\pi$ in the chiral limit ($m_{B^*} - m_B < m_\pi$), which defines  $g_{B^*\!B\pi}$.  A comparison between these two couplings is an indication of the degree to which notions of heavy-quark symmetry
can be applied in the charm sector.  This was done in Ref.\,\cite{ElBennich:2010ha}, wherein it was demonstrated that the difference in either extracting $\hat g$ from $g_{D^*\! D\pi}$ or $g_{B^*\!B\pi}$ is material:
\begin{align}
 g_{D^*\!D\pi}  & =15.8\ \Rightarrow \ \hat g = 0.53  \,,     \\
 g_{B^*\!B\pi} & = 30.0 \ \Rightarrow \ \hat g = 0.37 \, .
\end{align}
Here again, the results emphasise that when the $c$-quark is a system's heaviest constituent, $\Lambda_\mathrm{QCD}/m_c$-corrections are not well controlled.  Since the DSE approach lends itself to the calculation of other amplitudes involving light and heavy mesons; e.g., couplings of phenomenological interest for charmonium production and D-mesic nuclei \cite{ElBennich:2011py}, it should help to overcome the limitations of an expansion in $1/\hat M_c$. 

\section{Describing Baryons and Mesons Simultaneously}
\label{sec:Baryons}
%
While a symmetry-preserving description of mesons is essential, it is only part of the physics that nonperturbative QCD must describe because Nature also presents us with baryons: light-quarks in three-particle composites.  An explanation of the spectrum of baryons and the nature of interactions between them is basic to understanding the Standard Model.  The present and planned experimental programmes at JLab, and other facilities worldwide, are critical elements in this effort.

No approach to QCD is comprehensive if it cannot provide a unified explanation of both mesons and baryons.  We have explained that DCSB is a keystone of the Standard Model, which is evident in the momentum-dependence of the dressed-quark mass function -- Fig.\,\ref{gluoncloud}: it is just as important to baryons as it is to mesons.  Since constituent-quark-like models cannot incorporate the momentum-dependent dressed-quark mass-function, they are not a viable tool for use in this programme.  The DSEs furnish the only extant continuum framework that can simultaneously connect both meson and baryon observables with this basic feature of QCD, having provided, e.g., a direct correlation of meson and baryon properties via a single interaction kernel, which preserves QCD's one-loop renormalisation group behaviour and can systematically be improved.  This is evident in the preceding sections and their combination with Refs.\,\cite{Eichmann:2008ef,Eichmann:2011vu,Eichmann:2008ae,Eichmann:2011ej}.

In order to illustrate this programme, we will review the computation of baryon masses, and nucleon elastic and transition electromagnetic form factors.  In this connection, we recall that, for a structureless or simply-structured fermion $F_1(Q^2) \equiv 1$ and $F_2(Q^2) \equiv 0$ [see after Eq.\,\eqref{Gordon}], so that $G_E(Q^2) \equiv G_M(Q^2)$ and the distribution of charge and magnetisation is identical within that composite.  This was believed to be the case for the proton until 1999.  In that year, enabled by the high luminosity of the accelerator at JLab, a new method was employed to measure the ratio $G_E^p(Q^2)/G_M^p(Q^2)$ \cite{Jones:1999rz}.  The result astonished the community: whilst $G_E^p(Q^2)/G_M^p(Q^2)\approx 1$ for $Q^2< 1\,$GeV$^2$, $G_E^p(Q^2)/G_M^p(Q^2)$ is a rapidly decreasing function for $Q^2>1\,$GeV$^2$ (see Fig.\,\ref{fig:GEGMp}).  As one of the highlights herein, we will indicate how this may be understood.

\subsection{Faddeev equation}
\label{sec:FE}
In quantum field theory a baryon appears as a pole in a six-point quark Green function.  The residue is proportional to the baryon's Faddeev amplitude, which is obtained from a Poincar\'e covariant Faddeev equation that sums all possible exchanges and interactions that can take place between three dressed-quarks.  A tractable Faddeev equation for baryons \cite{Cahill:1988dx} is founded on the observation that an interaction which describes colour-singlet mesons also generates nonpointlike quark-quark (diquark) correlations in the colour-$\bar 3$ (antitriplet) channel \cite{Cahill:1987qr}.  The dominant correlations for ground state octet and decuplet baryons are scalar ($0^+$) and axial-vector ($1^+$) diquarks because, for example, the associated mass-scales are smaller than the baryons' masses \cite{Burden:1996nh,Maris:2002yu} and their parity matches that of these baryons.  It follows that only they need be retained in approximating the quark-quark scattering matrix which appears as part of the Faddeev equation \cite{Eichmann:2008ef,Cloet:2008re,Roberts:2011cf}.  On the other hand, pseudoscalar ($0^-$) and vector ($1^-$) diquarks dominate in the parity-partners of ground state octet and decuplet baryons \cite{Roberts:2011cf}.  The DSE approach treats mesons and baryons on the same footing and, in particular, enables the impact of DCSB to be expressed in the prediction of baryon properties.

In quantum field theory a baryon appears as a pole in a six-point quark Green function.  The residue is proportional to the baryon's Faddeev amplitude, which is obtained from a Poincar\'e covariant Faddeev equation that sums all possible exchanges and interactions that can take place between three dressed-quarks.  A tractable Faddeev equation for baryons \cite{Cahill:1988dx} is founded on the observation that an interaction which describes colour-singlet mesons also generates nonpointlike quark-quark (diquark) correlations in the colour-$\bar 3$ (antitriplet) channel \cite{Cahill:1987qr}.  The dominant correlations for ground state octet and decuplet baryons are scalar ($0^+$) and axial-vector ($1^+$) diquarks because, for example, the associated mass-scales are smaller than the baryons' masses \cite{Burden:1996nh,Maris:2002yu} and their parity matches that of these baryons.  It follows that only they need be retained in approximating the quark-quark scattering matrix which appears as part of the Faddeev equation \cite{Eichmann:2008ef,Cloet:2008re,Roberts:2011cf}.  On the other hand, pseudoscalar ($0^-$) and vector ($1^-$) diquarks dominate in the parity-partners of ground state octet and decuplet baryons \cite{Roberts:2011cf}.  The DSE approach treats mesons and baryons on the same footing and, in particular, enables the impact of DCSB to be expressed in the prediction of baryon properties.

It is important to appreciate that diquarks do not appear in the strong interaction spectrum \cite{Bender:1996bb,Bhagwat:2004hn}.  However, the attraction between quarks in this channel justifies a picture of baryons in which two quarks within a baryon are always correlated as a colour-$\bar 3$ diquark pseudoparticle, and binding is effected by the iterated exchange of roles between the bystander and diquark-participant quarks.   Here it is important to emphasise strongly that QCD supports \emph{nonpointlike} diquark correlations \cite{Roberts:2011wy,Maris:2004bp}, as we shall see in Sec.\,\ref{sec:diquarkFFs}.  Hence models that employ pointlike diquark degrees of freedom have little connection with QCD.

\begin{figure}[t]
\includegraphics[clip,width=0.45\textwidth]{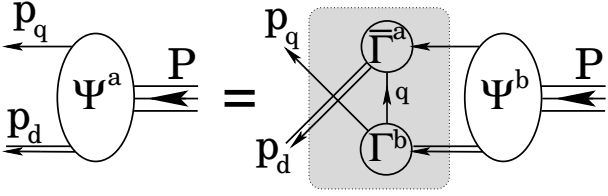}
\caption{\label{fig:Faddeev} Diagrammatic representation of a Poincar\'e covariant Faddeev equation for a baryon.  $\Psi$ is the Faddeev amplitude for a baryon of total momentum $P= p_q + p_d$.  It expresses the relative momentum correlation between the dressed-quark and -diquarks within the baryon.
The shaded region demarcates the kernel of the Faddeev equation, in which: the \emph{single line} denotes the dressed-quark propagator, $\Gamma$ is the diquark Bethe-Salpeter amplitude; and the \emph{double line} is the diquark propagator.}
\end{figure}

The Faddeev equation, illustrated in Fig.\,\ref{fig:Faddeev}, is a linear homogeneous matrix equation, in many respects similar to a Bethe-Salpeter equation.  Its solution is the nucleon's Poincar\'e-covariant Faddeev amplitude, which describes quark-diquark relative motion within the nucleon.
The composite nature of the diquark correlations, and their dynamical breakup and reformation through dressed-quark exchange, pictured in Fig.\,\ref{fig:Faddeev}, is crucial to maintaining fermion statistics for the nucleon bound-state.
Furthermore, owing to the critical importance of both scalar- and axial-vector-diquark correlations, the nucleon's rest-frame wave-function possesses $S$-, $P$- and  $D$-wave correlations; i.e., a nucleon should \emph{a priori} be expected to contain significant dressed-quark orbital angular momentum.
This is verified in Ref.\,\cite{Cloet:2007pi}, which shows that in the nucleon's
rest frame just 37\% of the total spin of the nucleon is contained within components of
the Faddeev amplitude which possess zero quark orbital angular momentum.

It is worthwhile expressing one Faddeev equation concretely and we choose that for the $\Delta$-resonance, a $J=3/2$ bound-state of three valence light-quarks.  It is possible to obtain a simple yet extremely informative equation if one employs the interaction presented in Eq.\,\eqref{njlgluon} and an additional simplification; i.e., representing the quark exchanged between the diquarks as
\begin{equation}
S^{\rm T}(k) \to \frac{g_\Delta^2}{M}\,,
\label{staticexchange}
\end{equation}
where $g_\Delta=1.56$ \cite{Roberts:2011cf}.  This is a variant of the so-called ``static approximation,'' which itself was introduced in Ref.\,\cite{Buck:1992wz} and has subsequently been used in studying a range of nucleon properties \cite{Bentz:2007zs}.  In combination with diquark correlations generated by Eq.\,(\ref{njlgluon}), whose Bethe-Salpeter amplitudes are momentum-independent, Eq.\,(\ref{staticexchange}) generates Faddeev equation kernels which themselves are independent of the external relative momentum variable.  The dramatic simplifications which this produces are the merit of Eq.\,(\ref{staticexchange}).

Following this through, one derives \cite{Roberts:2011cf}
\begin{eqnarray}
\nonumber 1 &=& 8 \frac{g_\Delta^2}{M } \frac{E_{qq_{1^+}}^2}{m_{qq_{1^+}}^2}
\int\frac{d^4\ell^\prime}{(2\pi)^4} \int_0^1 d\alpha\,\\
&& \times \frac{(m_{qq_{1^+}}^2 + (1-\alpha)^2 m_\Delta^2)(\alpha m_\Delta + M)}
{[\ell^{^\prime 2} + \sigma_\Delta(\alpha,M,m_{qq_{1^+}},m_\Delta)]^2}\\
\nonumber
&=& \frac{g_\Delta^2}{M}\frac{E_{qq_{1^+}}^2}{m_{qq_{1^+}}^2}\frac{1}{2\pi^2}
\int_0^1 d\alpha\, (m_{qq_{1^+}}^2 + (1-\alpha)^2 m_\Delta^2)\\
&& \times (\alpha m_\Delta + M)\overline{\cal C}^{\rm iu}_1(\sigma_\Delta(\alpha,M,m_{qq_{1^+}},m_\Delta)),
\label{DeltaFaddeev}
\end{eqnarray}
where
$E_{qq_{1^+}}$ is the canonically-normalised axial-vector diquark Bethe-Salpeter amplitude,
$m_{qq_{1^+}}$ is the computed mass for this correlation,
\begin{eqnarray}
\nonumber
\lefteqn{\sigma_\Delta(\alpha,M,m_{qq_{1^+}},m_\Delta)}\\
&= & (1-\alpha)\, M^2 + \alpha \, m_{qq_{1^+}} - \alpha (1-\alpha)\, m_\Delta^2,
\end{eqnarray}
$\overline{\cal C}^{\rm iu}_1(z) = {\cal C}_1(z)/z$, with
${\cal C}^{\rm iu}_1(z) = - z (d/dz){\cal C}^{\rm iu}(z)$
and ${\cal C}^{\rm iu}(z)$ defined in Eq.\,\eqref{calC}.
Equation~\eqref{DeltaFaddeev} is an eigenvalue problem whose solution yields the mass for the dressed-quark-core of the $\Delta$-resonance, $m_\Delta$.  It is one dimensional because only the axial-vector diquark correlation contributes to the structure of the $\Delta$.  (The nucleon, on the other hand, is constituted from scalar- and axial-vector-diquarks and presents a five-dimensional eigenvalue problem.)

\begin{figure}[t]
\centerline{\includegraphics[clip,width=0.50\textwidth]{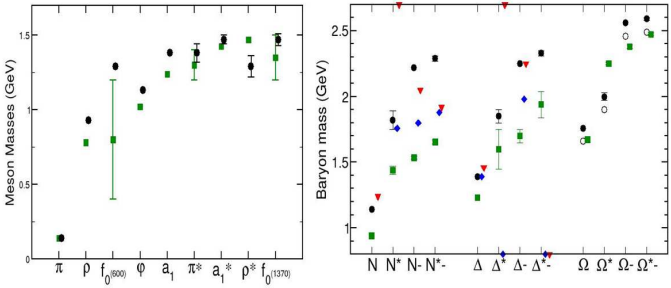}}
\caption{\label{fig:Fig2}
Comparison between DSE-computed hadron masses (\emph{filled circles}) and: bare baryon masses from Ref.\,\protect\cite{Suzuki:2009nj} (\emph{filled diamonds}) and Ref.\,\protect\cite{Gasparyan:2003fp} (\emph{filled triangles}); and experiment \protect\cite{Nakamura:2010zzi}, \emph{filled-squares}.
For the coupled-channels models a symbol at the lower extremity indicates that no associated state is found in the analysis, whilst a symbol at the upper extremity indicates that the analysis reports a dynamically-generated resonance with no corresponding bare-baryon state.
In connection with $\Omega$-baryons the \emph{open-circles} represent a shift downward in the computed results by $100\,$MeV.  This is an estimate of the effect produced by pseudoscalar-meson loop corrections in $\Delta$-like systems at a $s$-quark current-mass.
}
\end{figure}

With experience, one can look at Eq.\,\eqref{DeltaFaddeev} and see that increasing the current-quark mass will boost the mass of the bound-state.  This is just one of the Faddeev equations in Ref.\,\cite{Roberts:2011cf}.  In fact, building on lessons from meson studies \cite{Chang:2011vu}, a unified spectrum of $u,d$-quark hadrons was obtained therein using the symmetry-preserving regularization of a vector$\,\times\,$vector contact interaction that we have briefly described herein.  Reference~\cite{Roberts:2011cf} reports a study that simultaneously correlates the masses of meson and baryon ground- and excited-states within a single framework.  In comparison with relevant quantities, the computation produces $\overline{\mbox{rms}}$=13\%, where $\overline{\mbox{rms}}$ is the root-mean-square-relative-error$/$degree-of freedom.  As evident in Fig.\,\ref{fig:Fig2}, the prediction uniformly overestimates the PDG values of meson and baryon masses \cite{Nakamura:2010zzi}.  Given that the employed truncation deliberately omitted meson-cloud effects in the Faddeev kernel, this is a good outcome, since inclusion of such contributions acts to reduce the computed masses.

\begin{figure}[t]
\includegraphics[clip,width=0.45\textwidth]{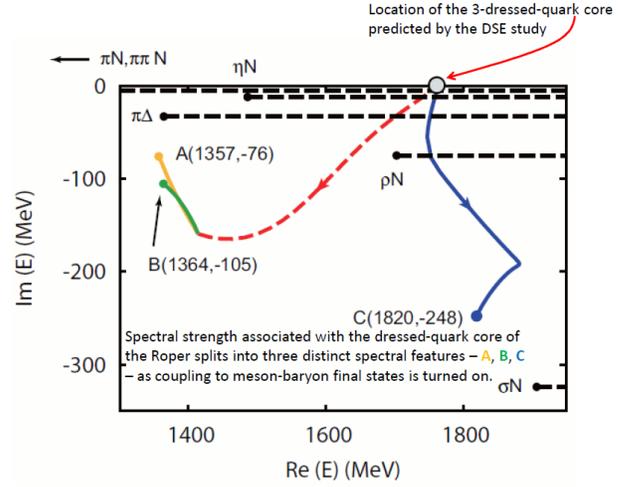}

\caption{\label{ebac}
The Excited Baryon Analysis Center (EBAC) examined the $P_{11}$-channel and found that the two poles associated with the Roper resonance and the next higher resonance were all associated with the same seed dressed-quark state.  Coupling to the continuum of meson-baryon final states induces multiple observed resonances from the same bare state.  In EBAC's analysis, all PDG-identified resonances were found to consist of a core state plus meson-baryon components.  (Adapted from Ref.\,\protect\cite{Suzuki:2009nj}.)}

\end{figure}

Following this line of reasoning, a striking result is agreement between the DSE-computed baryon masses \cite{Roberts:2011cf} and the bare masses employed in modern coupled-channels models of pion-nucleon reactions \cite{Suzuki:2009nj,Gasparyan:2003fp}, see Fig.\,\ref{fig:Fig2} and also Ref.\,\cite{Wilson:2011rj}.  The Roper resonance is very interesting.  The DSE study \cite{Roberts:2011cf} produces an excitation of the nucleon at $1.82\pm0.07\,$GeV.  This state is predominantly a radial excitation of the quark-diquark system, with both the scalar- and axial-vector diquark correlations in their ground state.  Its predicted mass lies precisely at the value determined in the analysis of Ref.\,\cite{Suzuki:2009nj}.  This is significant because for almost 50 years the ``Roper resonance'' has defied understanding.  Discovered in 1963, it appears to be an exact copy of the proton except that its mass is 50\% greater.  The mass was the problem: hitherto it could not be explained by any symmetry-preserving QCD-based tool.  That has now changed.  Combined, see Fig.\,\ref{ebac}, Refs.\,\cite{Roberts:2011cf,Suzuki:2009nj} demonstrate that the Roper resonance is indeed the proton's first radial excitation, and that its mass is far lighter than normal for such an excitation because the Roper obscures its dressed-quark-core with a dense cloud of pions and other mesons.  Such feedback between QCD-based theory and reaction models is critical now and for the foreseeable future, especially since analyses of experimental data on nucleon-resonance electrocouplings suggest strongly that this structure is typical; i.e., most low-lying $N^\ast$-states can best be understood as an internal quark-core dressed additionally by a meson cloud \cite{Aznauryan:2011td,Gothe:2011up}.

Additional analysis \cite{Wilson:2011rj} suggests a fascinating new feature of the Roper.  To elucidate, we focus first on the nucleon, whose Faddeev amplitude describes a ground-state that is dominated by its scalar diquark component (78\%).  The axial-vector component is significantly smaller but nevertheless important.  This heavy weighting of the scalar diquark component persists in solutions obtained with more sophisticated Faddeev equation kernels (see, e.g., Table~2 in Ref.\,\cite{Cloet:2008re}).  From a perspective provided by the nucleon's parity partner and the radial excitation of that state, in which the scalar and axial-vector diquark probabilities are \cite{Roberts:2011ym} 51\%-49\% and 43\%-57\%, respectively, the scalar diquark component of the ground-state nucleon actually appears to be unnaturally large.

One can nevertheless understand the structure of the nucleon.  As with so much else, the composition of the nucleon is intimately connected with dynamical chiral symmetry breaking.  In a two-color version of QCD, the scalar diquark is a Goldstone mode, just like the pion \cite{Roberts:1996jx}.  (This is a long-known result of Pauli-G\"ursey symmetry \cite{Pauli:1957,Gursey:1958}.)  A memory of this persists in the three-color theory and is evident in many ways.  Amongst them, through a large value of the canonically normalized Bethe-Salpeter amplitude and hence a strong quark$+$quark$-$diquark coupling within the nucleon.  (A qualitatively identical effect explains the large value of the $\pi N$ coupling constant.) There is no such enhancement mechanism associated with the axial-vector diquark.  Therefore the scalar diquark dominates the nucleon.

With the Faddeev equation treatment described herein, the effect on the Roper is dramatic: orthogonality of the ground- and excited-states forces the Roper to be constituted almost entirely (81\%) from the axial-vector diquark correlation.  It is important to check whether this outcome survives with a Faddeev equation kernel built from a momentum-dependent interaction.

With masses and Faddeev amplitudes in hand, it is possible to compute baryon electromagnetic form factors.  For the nucleon, studies of the Faddeev equation exist that are based on the one-loop renormalisation-group-improved interaction that was used efficaciously in the study of mesons \cite{Eichmann:2008ef,Cloet:2008re}.  These studies retain the scalar and axial-vector diquark correlations, for the reasons explained in Sec.\,\ref{sec:FE}.

\begin{figure}[t]
\begin{minipage}[t]{0.45\textwidth}
\begin{minipage}[t]{0.45\textwidth}
\leftline{\includegraphics[width=0.90\textwidth]{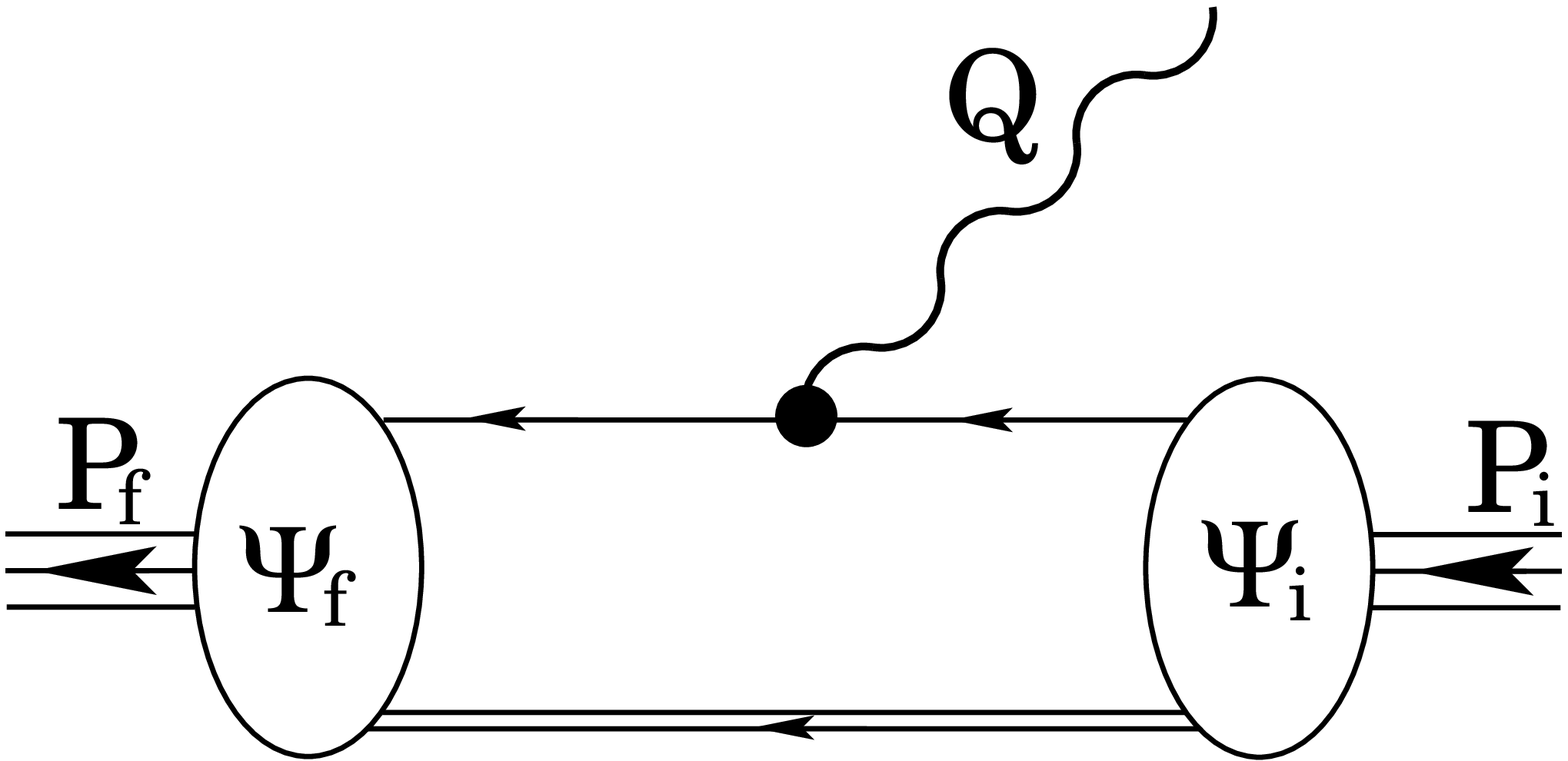}}
\end{minipage}
\begin{minipage}[t]{0.45\textwidth}
\rightline{\includegraphics[width=0.90\textwidth]{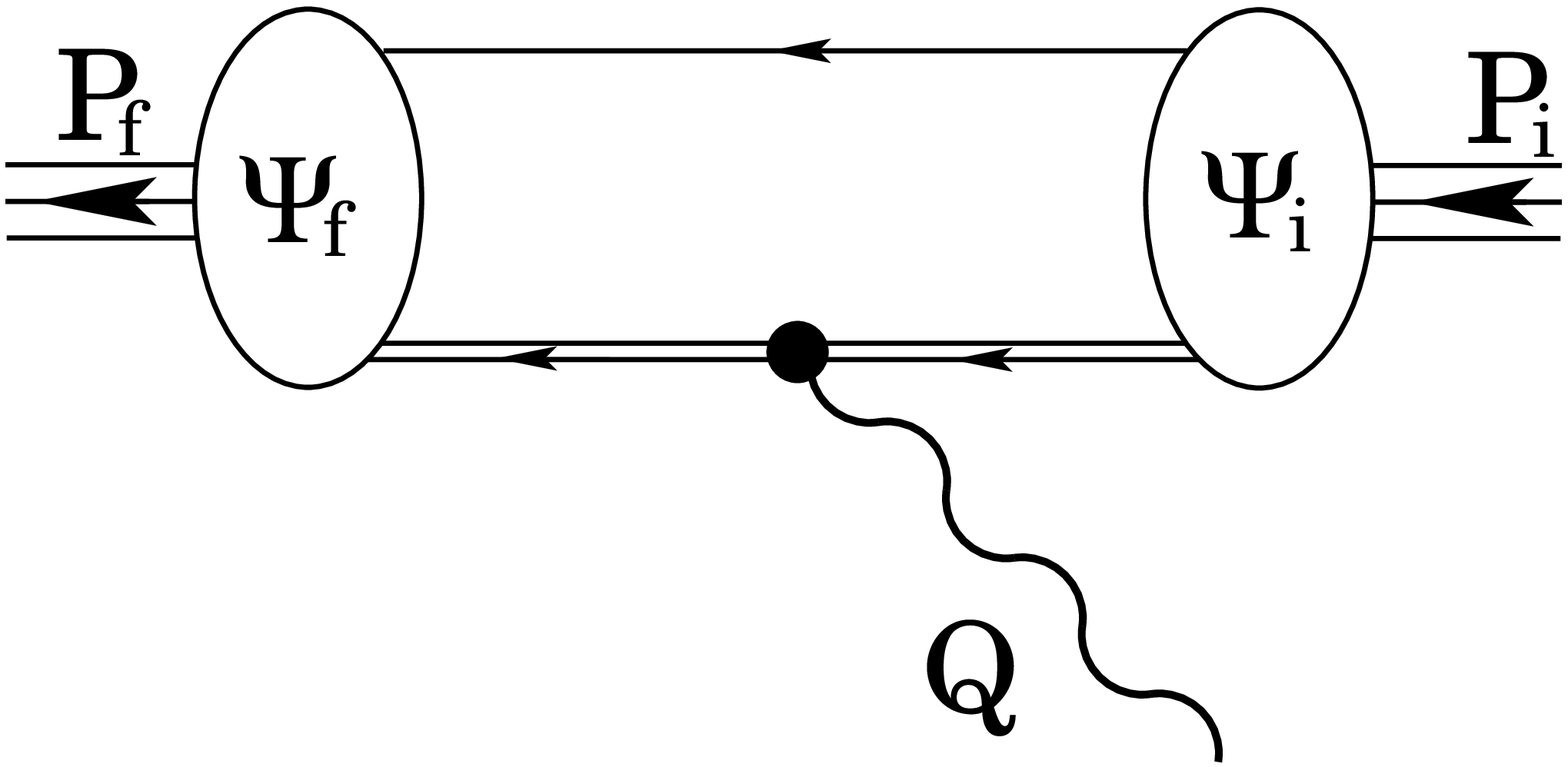}}
\end{minipage}\vspace*{3ex}

\begin{minipage}[t]{0.45\textwidth}
\leftline{\includegraphics[width=0.90\textwidth]{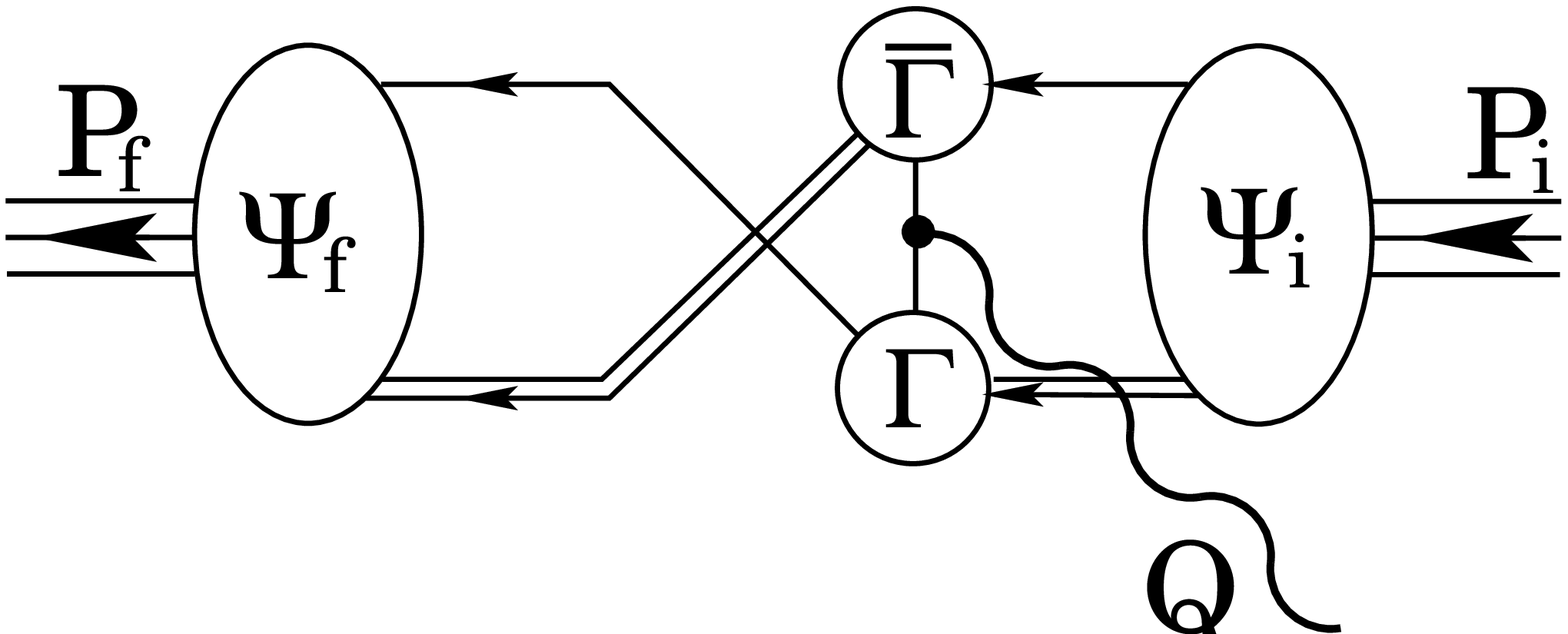}}
\end{minipage}
\begin{minipage}[t]{0.45\textwidth}
\rightline{\includegraphics[width=0.90\textwidth]{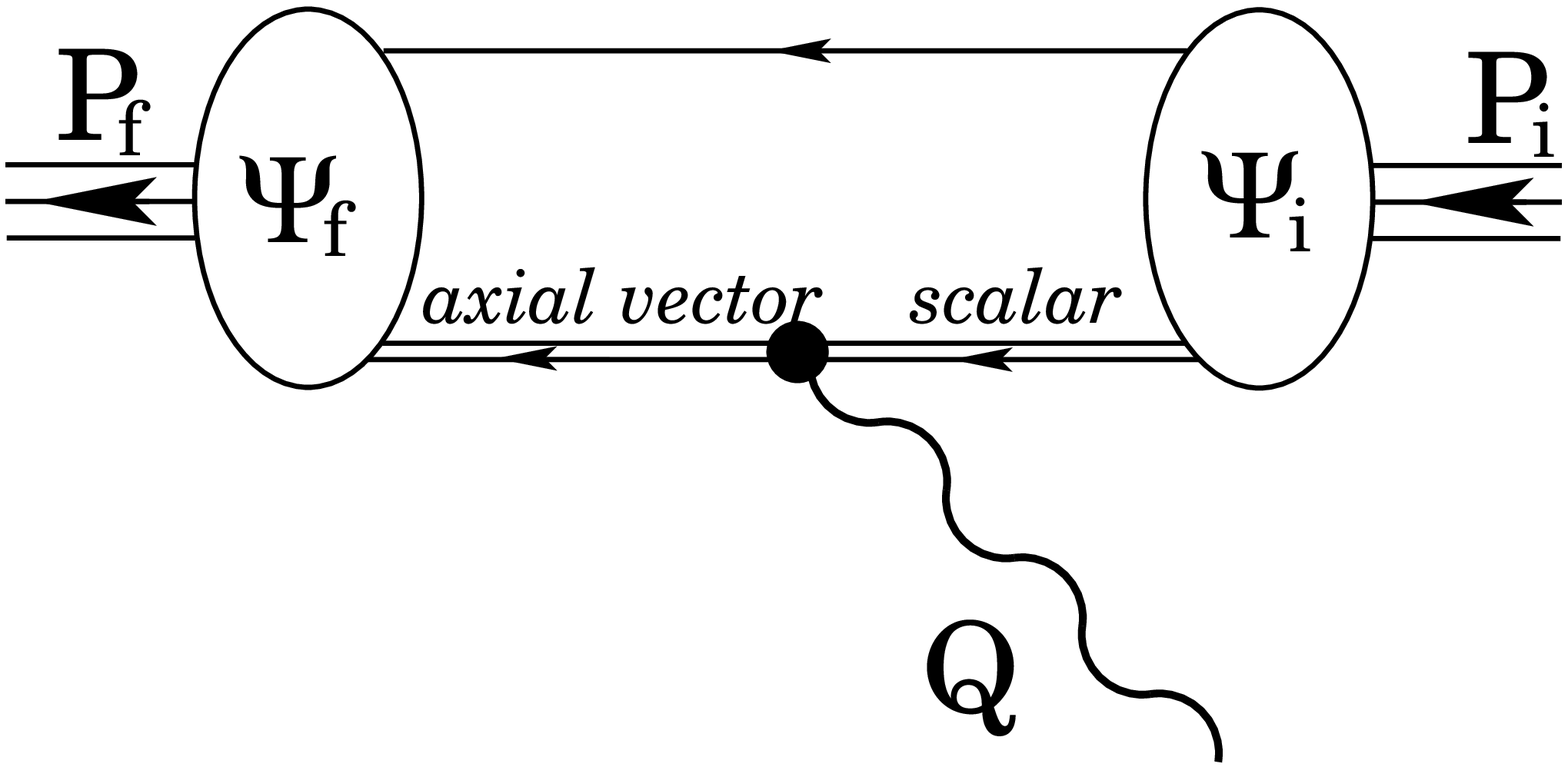}}
\end{minipage}\vspace*{3ex}

\begin{minipage}[t]{0.45\textwidth}
\leftline{\includegraphics[width=0.90\textwidth]{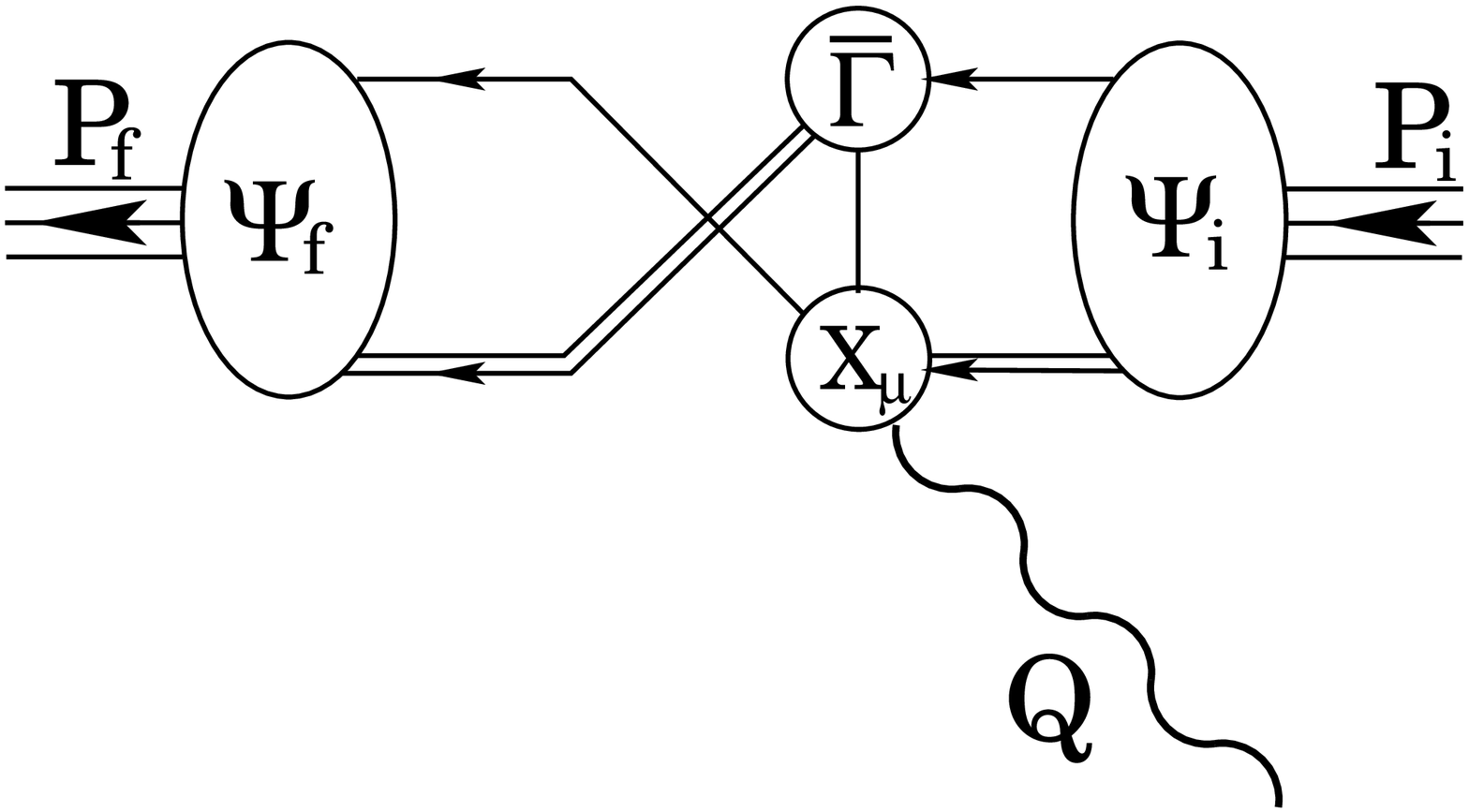}}
\end{minipage}
\begin{minipage}[t]{0.45\textwidth}
\rightline{\includegraphics[width=0.90\textwidth]{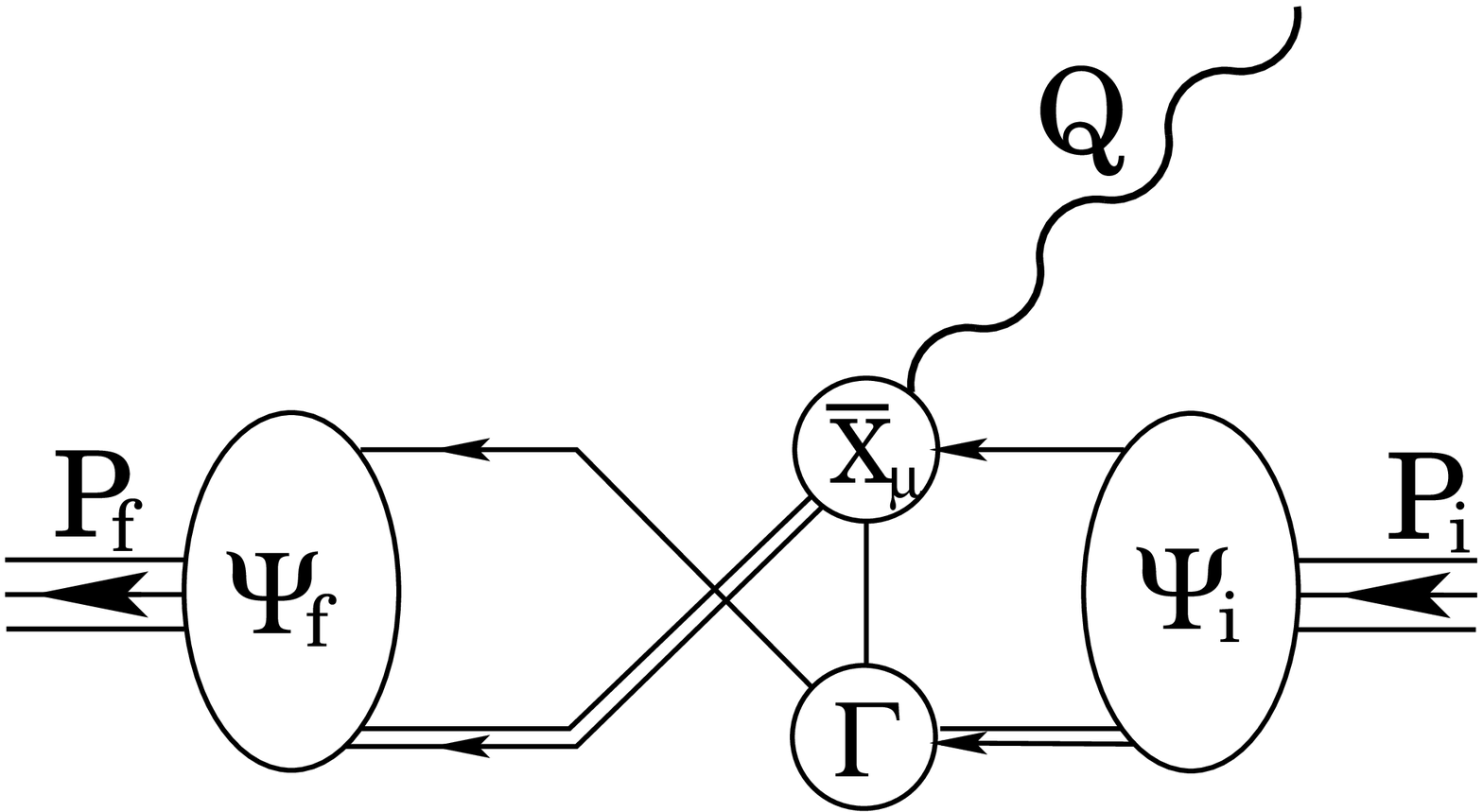}}
\end{minipage}
\end{minipage}
\caption{\label{fig:vertex} Vertex which ensures a conserved current for on-shell nucleons described by the Faddeev amplitudes, $\Psi_{i,f}$, obtained from the equation depicted in Fig.\,\protect\ref{fig:Faddeev}.  The photon probing the nucleon introduces momentum $Q$ and is represented by the wiggly line; the single line represents $S(p)$, the dressed-quark propagator, and the double line, the diquark propagator; and $\Gamma$ is the diquark Bethe-Salpeter amplitude.  The vertex in Diagram~4 (second row, right column) represents an electromagnetically induced transition between the scalar and axial-vector diquarks; and the vertices in row~3 are ``seagull terms,'' which appear as partners to Diagram~3 and arise because binding in the nucleons' Faddeev equations is, in general, effected by the exchange of a momentum-carrying dressed-quark between \textit{nonpointlike} diquark correlations \cite{Oettel:1999gc}.}
\end{figure}

A nonpointlike composite nucleon must interact with the photon via a sophisticated current, whose form is constrained by vector Ward-Takahashi identities.  This continues a thread that pervades these notes.  For the bound-state described by the Faddeev equation in Fig.\,\ref{fig:Faddeev} that current is described in Ref.\,\cite{Oettel:1999gc} and depicted in Fig.\,\ref{fig:vertex}: Diagrams~4--6 represent eight-dimensional integrals, which can be evaluated using Monte-Carlo techniques.

The pattern of the computation should now be clear.  In principle, one specifies an interaction kernel for the gap equation and solves for the dressed-quark propagator.  This completes the specification of the Bethe-Salpeter kernel, so one may compute the masses and amplitudes for the diquark correlations.  With all these quantities determined, the Faddeev equation is defined and can be solved for a baryon's mass and amplitude.  In combination with the current, Fig.\,\ref{fig:vertex}, it is a straightforward numerical task to compute the form factors.  That is, at least, once one has information about the diquark electromagnetic couplings which appear in the current.

\subsection{Diquark and meson elastic and transition Form Factors}
\label{sec:diquarkFFs}
\centerline{\emph{\mbox{\boldmath $\rho$}-meson elastic form factors}}
This is an ideal place to detail the properties of nonpointlike diquarks,
At leading-order in a symmetry preserving truncation of the DSEs \cite{Munczek:1994zz,Bender:1996bb}, simple changes in the equations describing $\pi$- and $\rho$ mesons yield expressions that provide detailed information about the scalar and axial-vector diquarks; e.g., their masses \cite{Cahill:1987qr,Praschifka:1989fd,Burden:1996nh,Maris:2002yu,Roberts:2011cf,%
Roberts:2010hu}, and electromagnetic elastic \cite{Maris:2004bp} and transition form factors, which are critical elements in the computation of a baryon's kindred properties via the current in Fig.\,\ref{fig:vertex}.  It is therefore natural to elucidate concurrently the properties of $\pi$- and $\rho$-mesons and those of the scalar and axial-vector diquark correlations because it opens the way to a unified, symmetry-preserving explanation of meson and baryon properties as they are predicted by a single interaction.  The potential of this approach is apparent in Refs.\,\cite{Eichmann:2008ae,Eichmann:2008ef} but it has yet to be fully realised.  For the present the best connection is provided by the less rigorous approach of Ref.\,\cite{Cloet:2008re}, which uses more parameters to express features of QCD but also predicts and describes simultaneously a larger array of phenomena
\cite{deJager:2009xs,Puckett:2010ac,Riordan:2010id}.

A wide-ranging study of $u/d$-quark meson and diquark form factors, using the contact interaction explained in Sec.\,\ref{FF1}, is described in Ref.\,\cite{Roberts:2011wy}.  Here we recapitulate on some interesting points that are relevant to the study of baryons.   To begin, we note that in analogy with Eq.\,(\ref{pointpionBSA}), one has the following general form of the Bethe-Salpeter amplitude for the $\rho$-meson obtained with a symmetry-preserving regularisation of a contact-interaction:
\begin{eqnarray}
\Gamma_\mu^\rho(P) & = & \gamma^T_\mu E_\rho(P)+ \frac{1}{M}
\sigma_{\mu\nu} P_\nu F_\rho(P)\,. \label{rhobsa}
\end{eqnarray}
However, as a peculiar consequence of the rainbow-ladder truncation:
\begin{equation}
\label{Frhozero} F_\rho(P) \stackrel{\mbox{\footnotesize
ladder}}{\equiv} 0\,.
\end{equation}
One has $F_\rho(P)\neq 0$ in any symmetry-preserving truncation that goes beyond this leading-order \cite{Bender:1996bb}.  The accident expressed in Eq.\,(\ref{Frhozero}) has material consequences.

Owing to the connection between properties of vector mesons and axial-vector diquarks, a computation of the $\rho$-meson elastic electromagnetic form factors is an important step along the way to diquark form factors.  The $J^{PC}=1^{--}$ $\rho$-meson has three elastic form factors and we follow Ref.\,\cite{Bhagwat:2006pu} in defining them.  Denoting the incoming photon momentum by $Q$, and the incoming and outgoing $\rho$-meson momenta by $p^i= K-Q/2$ and $p^f= K+Q/2$, then $K\cdot Q=0$, $K^2+Q^2/4= -m_\rho^2$ and the $\rho$-$\gamma$-$\rho$ vertex can be expressed:
\begin{eqnarray}
\label{Lambdarho}
\Lambda_{\lambda,\mu\nu}(K,Q) & = & \sum_{j=1}^3 T_{\lambda,\mu\nu}^j(K,Q) \, F_j(Q^2)\,,\\
T_{\lambda,\mu\nu}^1(K,Q) & = & 2 K_\lambda\, {\cal P}^T_{\mu\alpha}(p^i) \, {\cal P}^T_{\alpha\nu}(p^f)\,,\\
\nonumber
T_{\lambda,\mu\nu}^2(K,Q) & = & \left[Q_\mu - p^i_\mu \frac{Q^2}{2 m_\rho^2}\right] {\cal P}^T_{\lambda\nu}(p^f) \\
&& - \left[Q_\nu + p^f_\nu \frac{Q^2}{2 m_\rho^2}\right] {\cal P}^T_{\lambda\mu}(p^i)\,, \\
\nonumber
T_{\lambda,\mu\nu}^3(K,Q) & = & \frac{K_\lambda}{m_\rho^2}\, \left[Q_\mu - p^i_\mu \frac{Q^2}{2 m_\rho^2}\right] \left[Q_\nu + p^f_\nu \frac{Q^2}{2 m_\rho^2}\right] \,,\\
&&
\end{eqnarray}
where ${\cal P}^T_{\mu\nu}(p) = \delta_{\mu\nu} - p_\mu p_\nu/p^2$.  A symmetry-preserving regularisation scheme is essential so that the following Ward-Takahashi identities are preserved throughout the analysis:
\begin{eqnarray}
Q_\lambda \Lambda_{\lambda,\mu\nu}(K,Q) &=& 0,\\
p^i_\mu \Lambda_{\lambda,\mu\nu}(K,Q) &=& 0 = p^f_\nu
\Lambda_{\lambda,\mu\nu}(K,Q)\,.
\end{eqnarray}

The electric, magnetic and quadrupole form factors are constructed
as follows:
\begin{eqnarray}
G_E(Q^2) & = & F_1(Q^2)+\frac{2}{3} \eta G_Q(Q^2)\,,\\
G_M(Q^2) & = & - F_2(Q^2)\,,\\
%
G_Q(Q^2) & = & F_1(Q^2) + F_2(Q^2) + \left[1+\eta\right]
F_3(Q^2) \,,\; \rule{2em}{0ex}
\end{eqnarray}
where $\eta=Q^2/[4 m_\rho^2]$.  In the limit $Q^2\to 0$, these form factors define the charge, and magnetic and quadrupole moments of the $\rho$-meson; viz.,
\begin{eqnarray}
\label{chargenorm}
G_E^\rho(Q^2=0) & = & 1\,, \\
G_M^\rho(Q^2=0) & = & \mu_\rho\,,\; G_Q^\rho(Q^2=0) = Q_\rho\,.
\end{eqnarray}
It is readily seen that Eq.\,(\ref{chargenorm}) is a symmetry constraint.  One has $G_E(Q^2=0)=F_1(Q^2=0)$ and
\begin{equation}
\Lambda(K,Q) \stackrel{Q^2\to 0}{=} 2 K_\lambda\, {\cal
P}^T_{\mu\alpha}(K) \, {\cal P}^T_{\alpha\nu}(K)\, F_1(0)\,.
\end{equation}
Using the vector WTI for the quark-photon vertex and the explicit form for $P_T(Q^2)$, Fig.\,\ref{qgammavertex},
this becomes
\begin{eqnarray}
\nonumber
\lefteqn{ K_\lambda\, {\cal P}^T_{\mu\nu}(K) F_1(0) }\\
&=& N_c E_{\rho}^2 {\rm tr}_{\rm D}\int\frac{d^4 q}{(2\pi^4)}
i\gamma_\nu\, \frac{\partial}{\partial K_\lambda} S(\ell+K)
i\gamma_\mu\, S(\ell)\,.\rule{3em}{0ex}
\end{eqnarray}
The right-hand-side (rhs) is simply the analogue of Eq.\,(\ref{Ndef}) for the rainbow-ladder vector meson.  Hence, when the Bethe-Salpeter amplitude is canonically-normalised
and so long as one employs a symmetry-preserving regularisation procedure, the rhs is equal to $K_\lambda\, {\cal P}^T_{\mu\nu}(K)$ and thus $F_1(0)=1$.

\begin{table}[bt]
\caption{\emph{Row 1}: Form factor radii (in fm), and magnetic and quadrupole moments for the $\rho$-meson, $G_M^\rho(Q^2=0)$ and $G_Q^\rho(Q^2=0)$ respectively, computed with the parameter values in Table~\ref{Table:static}
%
For a structureless vector meson, $\mu=2$ and ${\cal Q}=-1$.
Experimentally, $r_\pi=0.672\pm 0.008\,$fm \cite{Nakamura:2010zzi}. \emph{Row 2}: Results for the scalar and axial-vector diquark correlations.  Here the magnetic and
quadrupole moments should be multiplied by the relevant charge factor; viz., $e_{\{uu\}}=\frac{4}{3}$, $e_{\{ud\}}=\frac{1}{3}$ and $e_{\{dd\}}=-\frac{2}{3}$.
%
%
\label{Table:radii} }
\begin{center}
\begin{tabular*}
{\hsize} { l@{\extracolsep{0ptplus1fil}}
c@{\extracolsep{0ptplus1fil}} c@{\extracolsep{0ptplus1fil}}
c@{\extracolsep{0ptplus1fil}} c@{\extracolsep{0ptplus1fil}}
c@{\extracolsep{0ptplus1fil}} c@{\extracolsep{0ptplus1fil}}}\hline
$r_\pi$ & $r_\rho^E $ & $r_\rho^M$ & $r_\rho^E$ & $\mu_\rho$ & ${\cal Q}_\rho$ \\
0.45 & 0.56 & 0.51 & 0.51 & 2.11 & -0.85 \\
$r_{0^+}$ & $r_{1^+}^E $ & $r_{1^+}^M$ & $r_{1^+}^E$ & $\mu_{1^+}$ & ${\cal Q}_{1^+}$ \\
0.49 &  0.55 &  0.51 & 0.51 & 2.13 & -0.81
\\\hline
\end{tabular*}
\end{center}
\end{table}

Table~\ref{Table:radii} reports computed form factor radii, and the magnetic and quadrupole moments.  An interpretation of the ratio $r_\pi/r_\rho=0.80$ determined from the Table is complicated by the fact that rainbow-ladder truncation was used consistently; but in this case alone $F_\rho(P)=0$, whereas $F_\pi(P)\neq 0$ always and $F_\rho(P)\neq 0$ in all other truncations.  It is useful, therefore, to observe that $r_\pi = 0.51\,$fm if one artificially sets $F_\pi(P)=0$, in which case $r_\pi/r_\rho=0.92$.  Moreover, the DSE computation in Ref.\,\cite{Bhagwat:2006pu},
which employs a QCD-based interaction, produces $r_\pi/r_\rho=0.90$; and in combination, the more phenomenological DSE studies of Refs.\,\cite{Burden:1995ve,Hawes:1998bz} yield
$r_\pi/r_\rho=0.92$.

\begin{figure}[t] 
\centerline{\includegraphics[clip,width=0.45\textwidth]{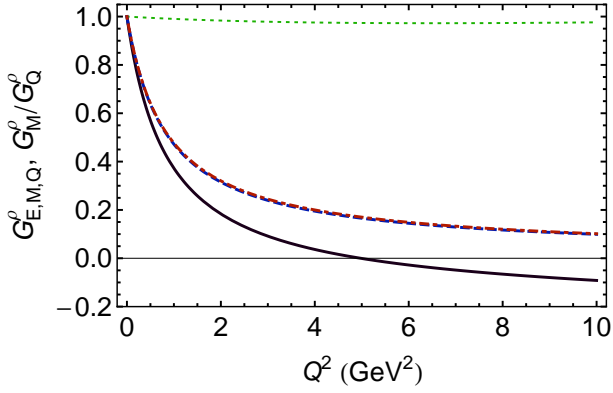}}
\caption{\label{FGAllrho} \emph{Solid curve} -- $\rho$-meson electric form factor, $G_E^\rho(Q^2)$, which exhibits a zero at $Q^2=5.0\,$GeV$^2$.  (It is notable that
$1-\frac{2}{3}\eta = 0 $ for $Q^2=6 m_\rho^2= 5.2\,$GeV$^2$.)  The \emph{dashed curve}, $G_M^\rho(Q^2)/\mu_\rho$, and \emph{dot-dashed curve}, $G_Q^\rho(Q^2)/{\cal Q}_\rho$, are almost indistinguishable, as emphasised by the \emph{dotted curve}, $[G_M^\rho(Q^2)/\mu_\rho]/[G_Q^\rho(Q^2)/{\cal Q}_\rho]$.
The charge radii, and magnetic and quadrupole moments are given in Table~\protect\ref{Table:radii}.
NB.\ All form factors exhibit a pole at $Q^2=-m_\rho^2$ because the quark-photon vertex is dressed as described in Sec.\,\protect\ref{FF1}.
%
}
\end{figure}

The computed $\rho$-meson electric form factor is plotted in Fig.\,\ref{FGAllrho}.  It displays a zero at $Q^2=5.0\,$GeV$^2$ and remains negative thereafter.  Given that the deuteron is a weakly-bound $J=1$ system, constituted from two fermions, and its electric form factor possesses a zero \cite{Kohl:2008zz}, it is unsurprising that $G_E^\rho(Q^2)$ exhibits a zero.  It is notable in addition that the deuteron's zero is located at $z_Q^{\rm D}:=\surd Q^2 = 0.8\,$GeV, so that
\begin{equation}
z_Q^{\rm D}r_{\rm D} \approx z_Q^\rho r_\rho^E\,,
\end{equation}
where $r_{\rm D}$ is the deuteron's radius.  An interpolation valid on $Q^2\in[-m_\rho^2,10\,{\rm GeV}^2]$ is
\begin{equation}
G_E^\rho(Q^2) \stackrel{\rm interpolation}{=}
\frac{1-0.20\,Q^2}{1+1.15 \, Q^2-0.013\,Q^4}\,.
\end{equation}

Figure~\ref{FGAllrho} also depicts the magnetic and quadrupole form factors of the $\rho$-meson,  both normalised by their values $Q^2=0$.  Notably, neither of these two form factors change sign: for $Q^2>-m_\rho^2$, $G_M^\rho(Q^2)$ is positive definite and
$G_E^\rho(Q^2)$ is negative definite.  Furthermore, over this entire domain of $Q^2$, these form factors exhibit a very similar $Q^2$-dependence, which is made especially apparent via the dotted-curve in Fig.\,\ref{FGAllrho}.  Interpolations valid on
$Q^2\in[-m_\rho^2,10\,{\rm GeV}^2]$ are
%
\begin{eqnarray}
G_M^\rho(Q^2) &\stackrel{\rm interp.}{=} &
\frac{2.11+0.021\,Q^2}{1+1.15 \, Q^2-0.015\,Q^4}\,,\\
G_Q^\rho(Q^2) &\stackrel{\rm interp.}{=} &
-\frac{0.85+0.038\,Q^2}{1+1.17 \, Q^2+0.014\,Q^4}\,.
\end{eqnarray}

The similar momentum-dependence of $G_M^\rho$ and $G_Q^\rho$ recalls a prediction in Ref.\,\cite{Brodsky:1992px}; namely,
\begin{equation}
\label{hillerratios} G_E(Q^2):G_M(Q^2): G_Q(Q^2) \stackrel{Q^2\to
\infty}{=} 1- \frac{2}{3}\eta  : 2 : -1
\end{equation}
in theories with a vector-vector interaction mediated via bosons propagating as $1/k^2$  at large-$k^2$.  The computed ratio $r_{M/Q}:=G_M^\rho(Q^2)/G_Q^\rho(Q^2)$ conforms approximately with this prediction on a large domain of $Q^2$; e.g.,
\begin{equation}
\begin{array}{lcccc}
Q^2 & 0 & 10 & 10^2 & 10^3 \\
r_{M/Q}& -2.48 & -2.54 & -2.38 & -2.17
\end{array}\,.
\end{equation}
However, at $Q^2=10^4\,{\rm GeV}^2$, $r_{M/Q}=-1.28$.  Moreover, the remaining two ratios are always in conflict with the prediction; and closer inspection reveals that even the apparent agreement for $G_M^\rho(Q^2)/G_Q^\rho(Q^2)$ is accidental, since Eqs.\,(\ref{hillerratios}) are true if, and only if,
\begin{equation}
\label{hillerratiosF} F_1(Q^2):F_2(Q^2): Q^2 F_3(Q^2)
\stackrel{Q^2\to \infty}{=} 1  : -2 : 0\,;
\end{equation}
and none of these predictions are satisfied in this computation.

The mismatch originates, of course, with Eq.\,(\ref{njlgluon}) and the concomitant need for a regularisation procedure in which the ultraviolet cutoff plays a dynamical role.  If one carefully removes $\Lambda_{\rm uv} \to \infty$, Eqs.\,(\ref{hillerratiosF}) are recovered but at the cost of a logarithmic divergence in the individual form factors.
One concludes therefore that a vector-vector contact interaction cannot reasonably be regularised in a manner consistent with Eq.\,(\ref{hillerratios}).

In closing this subsection it is worth reiterating that it is only in the rainbow-ladder truncation that $F_\rho(P)\equiv 0$.  Therefore in connection with the $\rho$-meson's form factors, material changes should be anticipated when proceeding beyond this leading-order truncation.

\begin{figure}[t] 
\centerline{\includegraphics[clip,width=0.45\textwidth]{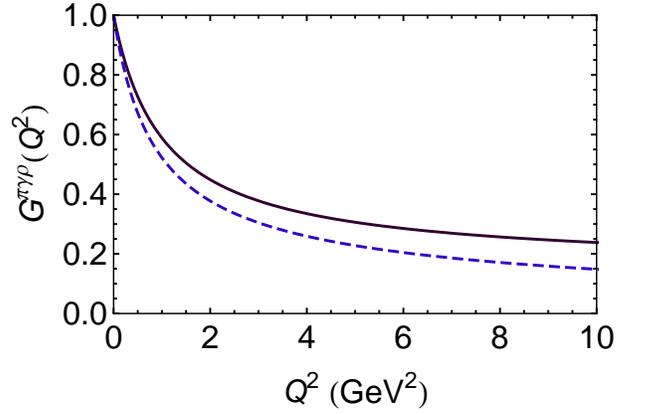}}
\caption{\label{FGpgr} \emph{Solid curve} -- the full result for $G^{\pi\gamma\rho}(Q^2)$; and \emph{dashed curve} -- $G^{\pi\gamma\rho}(Q^2)$ obtained with $F_\pi(P)\equiv 0$.  Experimentally \protect\cite{Nakamura:2010zzi}, the partial width for $\rho^+ \to \pi^+ \gamma$ is $68\pm 7\,$keV, which corresponds to \protect\cite{Maris:2002mz} $g_{\pi\gamma\rho} = (0.74\pm0.05)\,m_\rho $.  This is in fair agreement with the computed result; viz., $g_{\pi\gamma\rho} = 0.63\,m_\rho$.
%
}
\end{figure}

\medskip

\centerline{\emph{\mbox{\boldmath $\rho$}-\mbox{\boldmath $\pi$} transition form factor}}
This transition is closely related to the $\gamma^\ast \pi \gamma$ transition form factor, whose behaviour in connection with Eq.\,(\ref{njlgluon}) was analysed in Ref.\,\cite{Roberts:2010rn}.  The corresponding interaction vertex defines a single form factor; viz.,
\begin{equation}
T_{\mu\nu}^{\pi\gamma\rho}(k_1,k_2) =
\frac{g_{\pi\gamma\rho}}{m_\rho} \,
\epsilon_{\mu\nu\alpha\beta}k_{1\alpha} k_{2\beta}\,
G^{\pi\gamma\rho}(Q^2)\,,
\end{equation}
where $k_1^2=Q^2$, $k_2^2=-m_\rho^2$.  The coupling constant, $g_{\pi\gamma\rho}$, is defined such that $G^{\pi\gamma\rho}(Q^2=0)=1$.

The contact-interaction result for the form factor \cite{Roberts:2011wy} is depicted in Fig.\,\ref{FGpgr}. Naturally, because the quark-photon vertex is dressed (see Fig.\,\ref{qgammavertex}), the transition form factor exhibits a pole at $Q^2=-m_\rho^2$, which is not displayed.  An interpolation valid on $Q^2\in[-m_\rho^2,10\,{\rm GeV}^2]$ is
\begin{equation}
G^{\pi\gamma\rho}(Q^2) \stackrel{\rm interp.}{=}
\frac{1+0.37\,Q^2+0.024 Q^4}{1+1.29 \, Q^2+0.015\,Q^4}\,.\\
\end{equation}

In the neighbourhood of $Q^2=0$, the form factor is characterised by a radius-like length-scale; viz.,
\begin{equation}
\label{rpgr} r_{\pi\gamma\rho}^2 := -6 \left. \frac{d}{dQ^2}
G^{\pi\gamma\rho}(Q^2)\right|_{Q^2=0} = (0.46\,{\rm fm})^2,
\end{equation}
which is almost indistinguishable from both $r_\pi=0.45\,$fm in Table~\ref{Table:radii} and the anomaly interaction radius defined in Ref.\,\cite{Roberts:2010rn}; viz., $r_{\pi^0}^\ast=0.48\,$fm.  On the other hand
\begin{equation}
\label{rpgUV}
\lim_{Q^2\to \infty} G^{\pi\gamma\rho}(Q^2)= 0.11 \,, 
\end{equation}
owing to the presence of the pion's pseudovector component, a result in keeping with the pointlike nature of bound-states generated by a contact-interaction
\cite{GutierrezGuerrero:2010md,Roberts:2010rn}.

\begin{figure}[t] 
\centerline{\includegraphics[clip,width=0.45\textwidth]{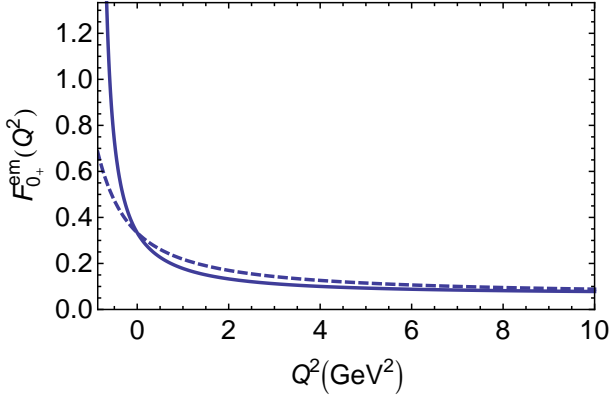}}
\caption{\label{Fem0} \emph{Solid curve} -- full result for scalar-diquark elastic electromagnetic form factor; and \emph{dashed curve} -- result obtained without dressing the quark-photon vertex.  The computed mass of the diquark is $m_{qq_{0^+}}=0.776\,$GeV and the charge radius is given in Table~\protect\ref{Table:radii}.
}
\end{figure}

\medskip

%
\centerline{\emph{Scalar-diquark elastic form factor}}
%
In the context of the interaction in Eq.\,(\ref{njlgluon}), a detailed discussion of the relationship between pseudoscalar- and vector-mesons and scalar- and axial-vector-diquark correlations may be found in Ref.\,\cite{Roberts:2011cf}.  Using the information provided therein, it is straightforward to show that in rainbow-ladder truncation the electromagnetic form factor of a scalar diquark is readily obtained from the expression for $F^{\rm em}_\pi(Q^2)$.  Namely,
\begin{equation}
F^{\rm em}_{0^+}(Q^2) = \frac{1}{3} \left. F^{\rm em}_{\pi}(Q^2)
\right|^{(E_\pi,F_\pi)\to
\sqrt{\frac{2}{3}}(E_{qq_{0^+}},F_{qq_{0^+}})}_{m_\pi\to
m_{qq_{0^+}}},
\end{equation}
where the scalar-diquark Bethe-Salpeter amplitude is expressed via ($C=\gamma_2\gamma_4$ is the charge-conjugation matrix)
\begin{equation}
\Gamma_{qq_{0^+}}(P) C^\dagger = \gamma_5 \left[ i E_{qq_{0^+}}(P)
+ \frac{1}{M} \gamma\cdot P F_{qq_{0^+}}(P) \right].
\end{equation}

The result for the scalar diquark elastic electromagnetic form factor is presented in Fig.\,\ref{Fem0}.  An interpolation valid on $Q^2\in [-m_\rho^2,10\,{\rm GeV}^2]$ is
\begin{equation}
F^{\rm em}_{0^+}(Q^2) \stackrel{\rm interp.}{=} \frac{1}{3}
\frac{1+0.25\,Q^2+0.027\,Q^4}{1+1.27 Q^2 + 0.13 \,Q^4}\,.
\end{equation}
The normalisation is different but the momentum-dependence is similar to that of $F_\pi^{\rm em}$.  This is indicated, too, by the ratio of charge radii; viz., $r_{0^+}/r_\pi=1.08$, which may be compared to the value of $1.09$ obtained in Ref.\,\cite{Maris:2004bp} and contrasted with the value of $0.8$ in \cite{Bloch:1999ke}.  In the absence of the scalar-diquark Bethe-Salpeter amplitude's pseudovector component,
$F_{qq_{0^+}}\equiv 0$, we find $r_{0^+}=0.51\,$fm; i.e., an increase of 6\%.

\begin{figure}[t] 
\centerline{\includegraphics[clip,width=0.45\textwidth]{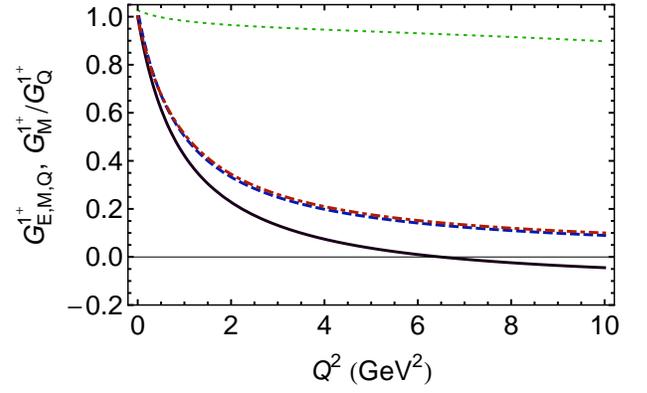}}
\caption{\label{Fem1} 
\emph{Solid curve} -- Pseudovector-diquark electric form factor, $G_E^{1^+}(Q^2)$, which exhibits a zero at $Q^2=6.5\,$GeV$^2$.  (In this case $1-\frac{2}{3}\eta = 0 $ for $Q^2=6 m_{1^+}^2= 6.7\,$GeV$^2$, given the computed mass of $1.06\,$GeV.)  The \emph{dashed curve}, $G_M^{1^+}(Q^2)/\mu_{1^+}$, and \emph{dot-dashed curve}, $G_Q^{1^+}(Q^2)/{\cal Q}_{1^+}$, are almost indistinguishable, as emphasised by the \emph{dotted curve}, $[G_M^{1^+}(Q^2)/\mu_\rho]/[G_Q^{1^+}(Q^2)/{\cal Q}_{1^+}]$.
The charge radii, and magnetic and quadrupole moments are given in Table~\protect\ref{Table:radii}.
NB.\ All form factors exhibit a pole at $Q^2=-m_\rho^2$ because the quark-photon vertex is dressed as described in Sec.\,\protect\ref{FF1}.
%
}
\end{figure}

\medskip

\centerline{\emph{Pseudovector-diquark elastic form factors}}
From the above observations it will be apparent that the rainbow-ladder results for the $\{u d\}$ axial-vector diquark elastic form factors may be obtained directly from those of the $\rho$-meson through the substitutions
\begin{equation}
F^{\rm em}_{1^+_{\{ud\}},j}(Q^2) = \frac{1}{3} \left. F_{j}(Q^2)
\right|^{E_\rho\to \sqrt{\frac{2}{3}} E_{qq_{1^+}}}_{m_\pi\to
m_{qq_{1^+}}}\,.
\end{equation}
The momentum-dependence of the form factors for the $\{uu\}$ and $\{dd\}$ correlations is identical but in these cases the normalisations are, respectively, $\frac{4}{3}$ and
$-\frac{2}{3}$.

The axial-vector diquark form factors are depicted in Fig.\,\ref{Fem1}.  They are similar to but distinguishable from those of the $\rho$-meson, falling-off a little less rapidly owing to the larger mass of the axial-vector diquark.  Interpolations valid on $Q^2\in[-m_\rho^2,10\,{\rm GeV}^2]$ are
%
%
\begin{eqnarray}
G_E^{1^+}(Q^2) &\stackrel{\rm interp.}{=} &
\frac{1-0.16\,Q^2}{1+1.17 \, Q^2 + 0.012 \,Q^4}\,,\rule{1.5em}{0ex}\\
G_M^{1^+}(Q^2) &\stackrel{\rm interp.}{=}&
\frac{2.13-0.19\,Q^2}{1+1.07 \, Q^2 - 0.10 \,Q^4}\,,\rule{1.5em}{0ex}\\
G_Q^{1^+}(Q^2) &\stackrel{\rm interp.}{=}&
-\frac{0.81-0.029\,Q^2}{1+1.11 \, Q^2 - 0.054 \,Q^4}\,,\rule{1.5em}{0ex}
\end{eqnarray}
from which the particular pseudovector diquark form factors are obtained after multiplication by the appropriate charge factors, listed in Table~\ref{Table:radii}.

\pagebreak

\centerline{\emph{\mbox{\boldmath $1^+$}-\mbox{\boldmath $0^+$} diquark transition form factor}}
Owing to the flavour structure of the scalar diquark, this transition can only involve the $\{ud\}$ axial-vector diquark.  It is described by a single form factor, which can be introduced through
\begin{equation}
T_{\mu\nu}^{0^+ \gamma 1^+}(k_1,k_2) = \frac{1}{3} \frac{g_{0^+
\gamma 1^+}}{m_{qq_{1^+}}} \,
\epsilon_{\mu\nu\alpha\beta}k_{1\alpha} k_{2\beta}\, G^{0^+ \gamma
1^+}(Q^2)\,,
\end{equation}
and one may readily determine that in rainbow-ladder truncation
\begin{eqnarray}
\nonumber
\lefteqn{G^{0^+ \gamma 1^+}(Q^2)}\\
& =& \left. G^{\pi\gamma\rho}(Q^2)
\right|^{(E_\pi,F_\pi,E_\rho)\to
\sqrt{\frac{2}{3}}(E_{qq_{0^+}},F_{qq_{0^+}},E_{qq_{1^+}})}_{m_\pi\to
m_{qq_{0^+}},m_\rho\to m_{qq_{1^+}}}\!\!\!\!. \rule{2em}{0ex}
\end{eqnarray}

\begin{figure}[t] 
\centerline{\includegraphics[clip,width=0.45\textwidth]{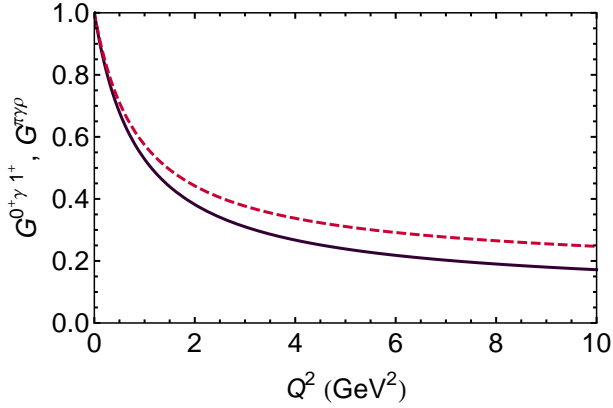}}
\caption{\label{F10g}
\emph{Solid curve} -- momentum-dependence of full result for axial-vector--scalar-diquark transition form factor, $G^{0^+\! \gamma 1^+}(Q^2)$; and \emph{dashed curve} -- result for $G^{\pi\gamma\rho}(Q^2)$ in Fig.\,\protect\ref{FGpgr}.  The different rates of evolution are typical of meson cf.\ diquark form factors.
Note that $e_{\{ud\}}g_{0^+\gamma 1^+} m_{qq_{1^+}} = e_{\{ud\}} 0.74 = 0.25$.
%
}
\end{figure}

The computed result for the transition form factor \cite{Roberts:2011wy} is depicted in Fig.\,\ref{F10g}.  An interpolation valid on $Q^2
\in [-m_\rho^2,10\,{\rm GeV}^2]$ is
\begin{equation}
G^{0^+ \gamma 1^+}(Q^2) \stackrel{\rm interp.}{=}
\frac{1+0.10\,Q^2}{1+1.073\,Q^2}\,.
\end{equation}
The associated transition radius is
\begin{equation}
r_{0^+\gamma 1^+} = 0.48\,{\rm fm},
\end{equation}
which is 5\% larger than $r_{\pi\gamma \rho}$ in Eq.\,(\ref{rpgr}), and
\begin{equation}
\lim_{Q^2\to \infty} G^{0^+\gamma 1^+}(Q^2)= 0.049 \,, 
\end{equation}
just under one-half of the value in Eq.\,(\ref{rpgUV}).

\begin{figure}[t]
\begin{centering}
\includegraphics[clip,width=0.45\textwidth]{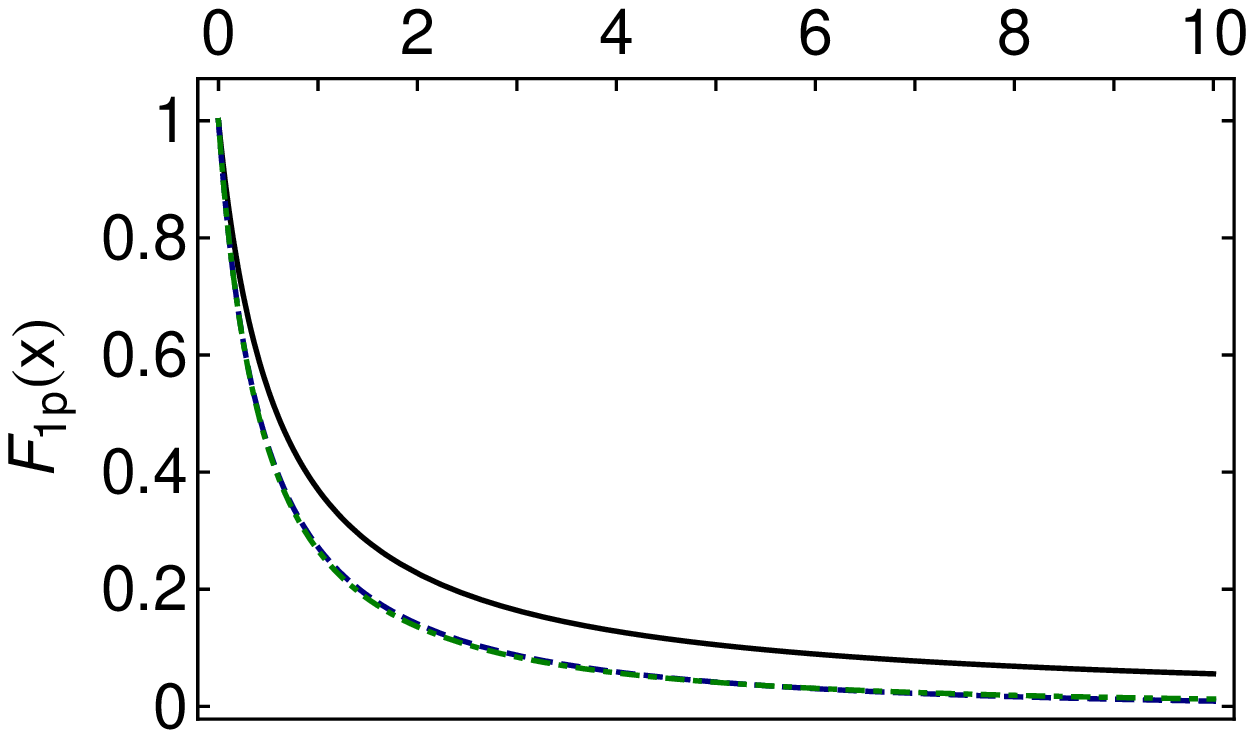}
\vspace*{-4.5ex}

\includegraphics[clip,width=0.45\textwidth]{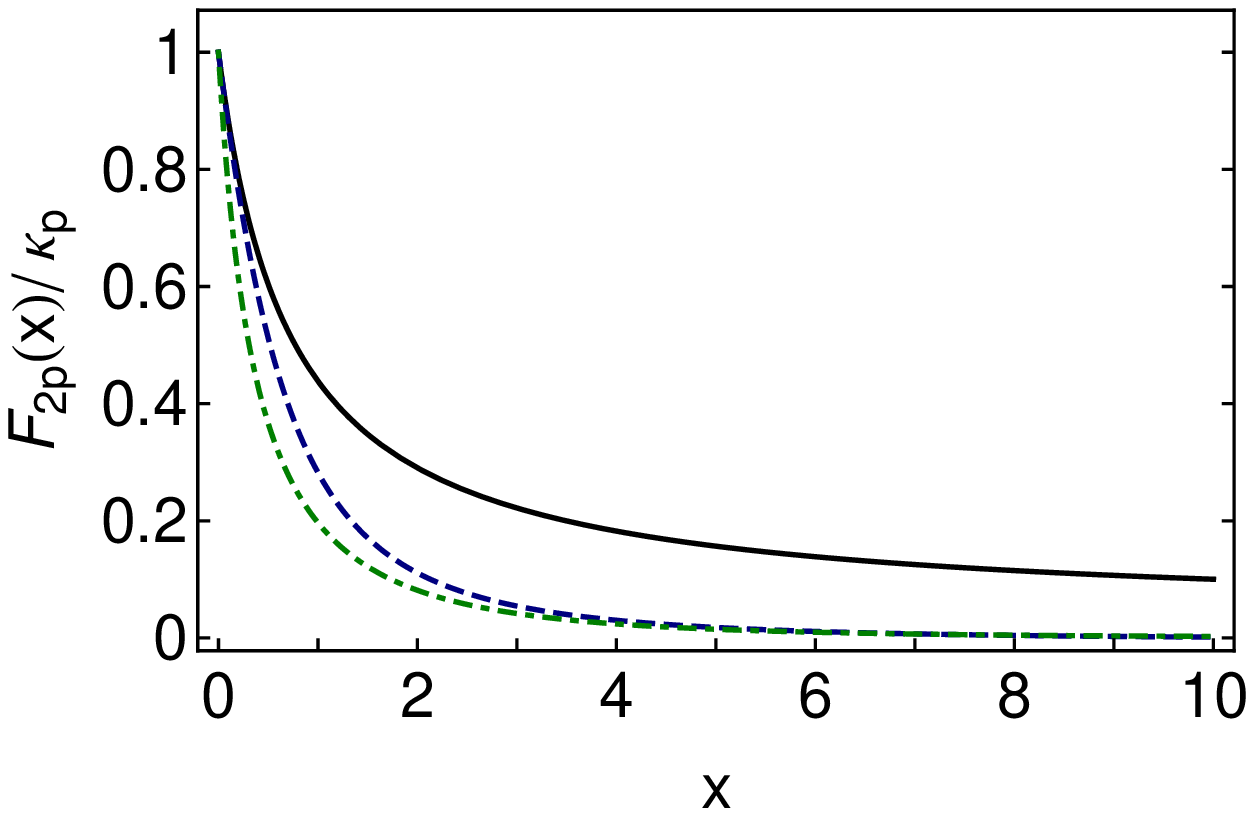}
\end{centering}
\caption{\label{fig:ProtonF1F2} Proton Dirac (upper panel) and Pauli (lower panel) form factors, as a function of $x=Q^2/m_N^2$.  \emph{Solid curve} -- result obtained \protect\cite{Wilson:2011rj} using the contact-interaction, Eq.\,\protect\eqref{njlgluon}, and hence a dressed-quark mass-function and diquark Bethe-Salpeter amplitudes that are momentum-independent; \emph{dashed curve} -- result obtained in Ref.\,\protect\cite{Cloet:2008re}, which employed QCD-like momentum-dependence for the dressed-quark propagators and diquark Bethe-Salpeter amplitudes in solving the Faddeev equation; \emph{dot-dashed curve} -- a parametrisation of experimental data \protect\cite{Kelly:2004hm}.}
\end{figure}

\begin{figure}[t]
\begin{centering}
\includegraphics[clip,width=0.45\textwidth]{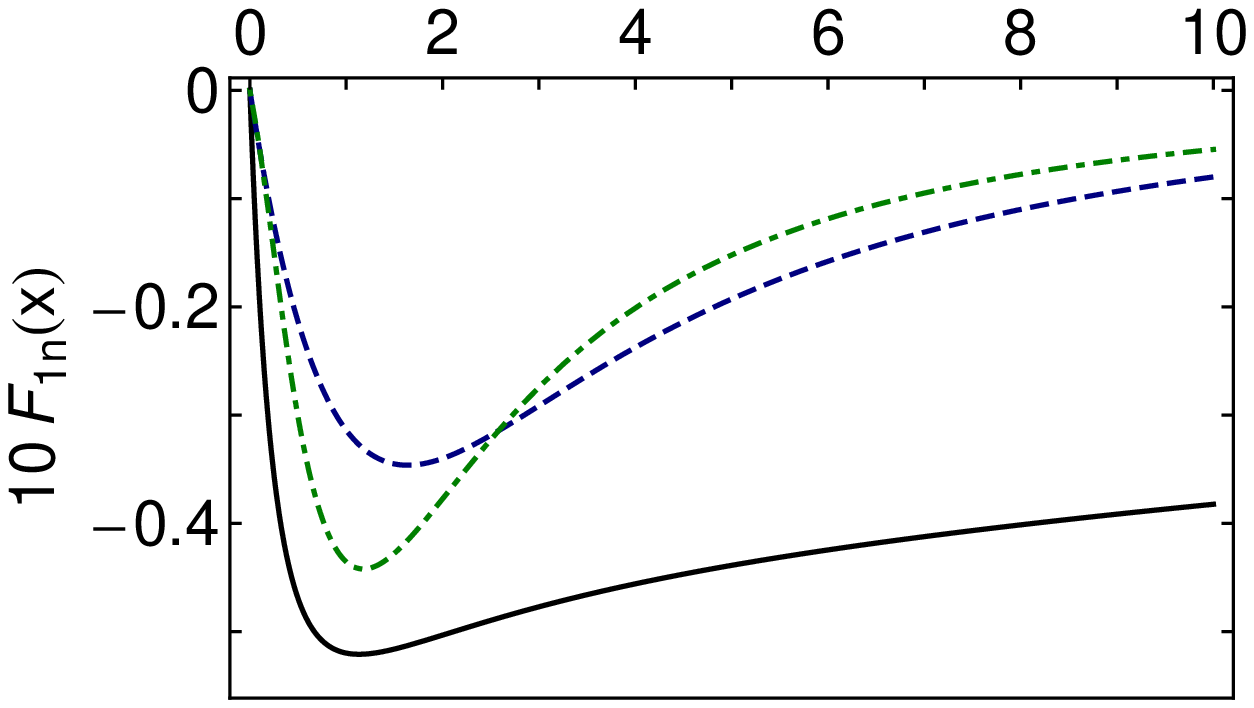}
\vspace*{-4.5ex}

\includegraphics[clip,width=0.45\textwidth]{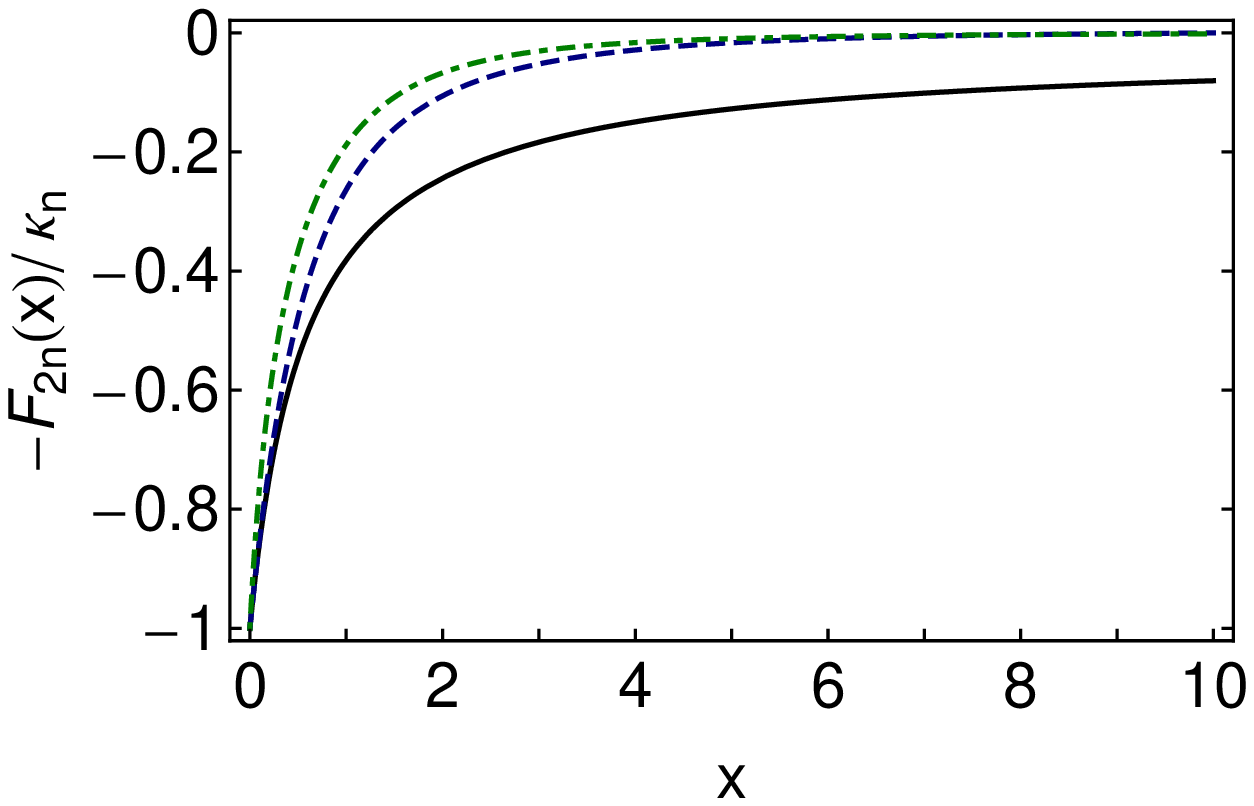}
\end{centering}
\caption{\label{fig:NeutronF1F2} Neutron Dirac (upper panel) and Pauli (lower panel) form factors, as a function of $x=Q^2/m_N^2$.  \emph{Solid curve} -- result obtained \protect\cite{Wilson:2011rj} using the contact-interaction, Eq.\,\protect\eqref{njlgluon}, and hence a dressed-quark mass-function and diquark Bethe-Salpeter amplitudes that are momentum-independent; \emph{dashed curve} -- result obtained in Ref.\,\protect\cite{Cloet:2008re}, which employed QCD-like momentum-dependence for the dressed-quark propagators and diquark Bethe-Salpeter amplitudes in solving the Faddeev equation; \emph{dot-dashed curve} -- a parametrisation of experimental data \protect\cite{Kelly:2004hm}.}
\end{figure}

\subsection{Nucleon form factors}
As we have seen in connection with the properties of mesons including at least one heavy quark, Sec.\,\ref{charmbeauty}, the computation of form factors can often be simplified by employing algebraic parametrisations of one or more elements (propagators and/or amplitudes).  Such parametrisations, based upon numerical solutions of the gap- and Bethe-Salpeter equations, have long been employed efficaciously \cite{Roberts:1994hh,Burden:1995ve,Roberts:1993ks}.  Following that path, Ref.\,\cite{Cloet:2008re} produced a comprehensive survey of nucleon electromagnetic form factors, the quality of which is characterised by the results depicted in Fig.\,\ref{fig:ProtonF1F2}.  With these results, Ref.\,\cite{Cloet:2008re} unified the computation of meson and nucleon form factors, and also their valence-quark distribution functions (see Sec.\,\ref{sec:VPDF}).

Figure~\ref{fig:ProtonF1F2} also depicts nucleon form factors obtained \cite{Wilson:2011rj} with a symmetry-preserving DSE-treatment of the contact interaction in Eq.\,\eqref{njlgluon} and hence the diquark form factors described in Sec.\,\ref{sec:diquarkFFs}.  These form factors characterise a nucleon that is constructed from diquarks whose Bethe-Salpeter amplitudes are momentum-independent and dressed-quarks with a momentum-independent mass-function, which inputs to the Faddeev equation yield a bound-state described by a momentum-independent Faddeev amplitude.  This last is the hallmark of a pointlike composite particle and explains the hardness of the computed form factors, which is evident in Figs.\,\ref{fig:ProtonF1F2}.  The hardness contrasts starkly with results obtained from a momentum-dependent Faddeev amplitude produced by dressed-quark propagators and diquark Bethe-Salpeter amplitudes with QCD-like momentum-dependence; and with experiment.  As we have remarked above, evidence for a connection between the momentum-dependence of each of these elements and the behaviour of QCD's $\beta$-function is accumulating; e.g., Refs.\,\cite{Roberts:2010rn,GutierrezGuerrero:2010md,Roberts:2011wy,Maris:2000sk,%
Eichmann:2008ef,Eichmann:2011vu,Bhagwat:2006pu}.
The comparisons in Figs.\,\ref{fig:ProtonF1F2} add to this evidence, in connection here with readily accessible observables, and support the view, presented herein, that experiment is a sensitive probe of the running of the $\beta$-function to infrared momenta.

\begin{figure}[t]
\includegraphics[clip,width=0.45\textwidth]{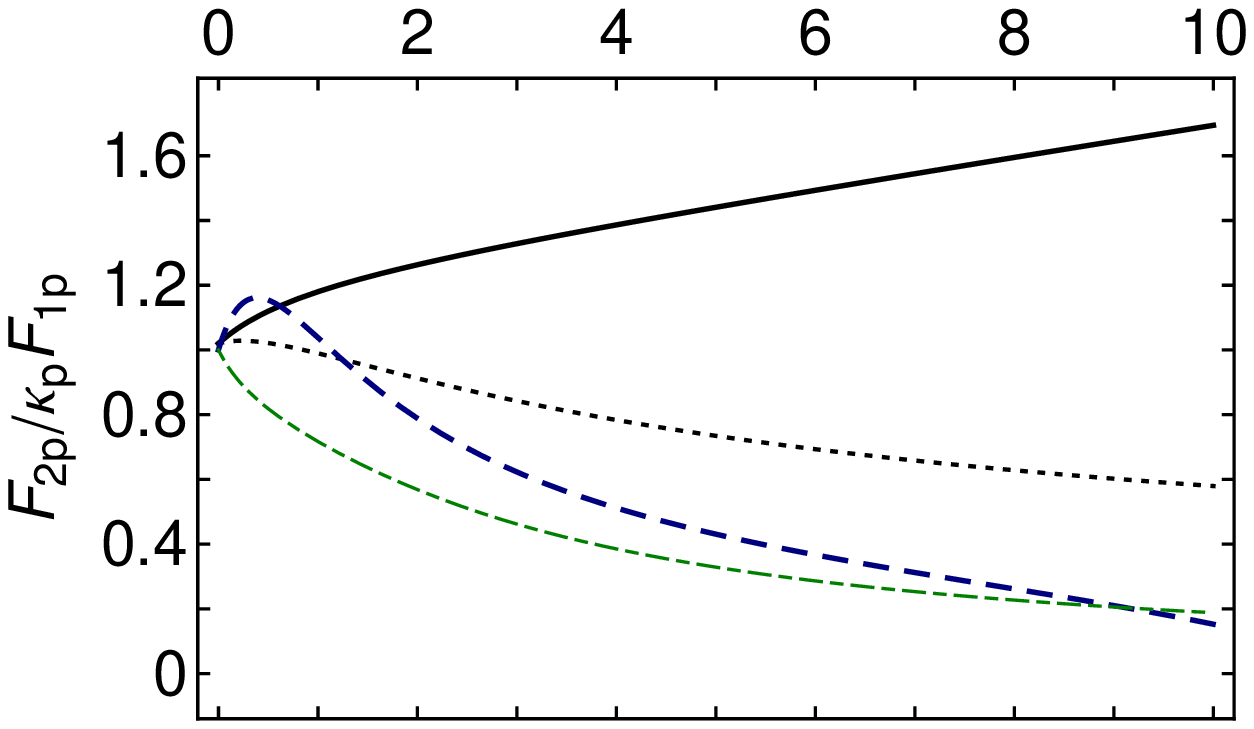}
\\[-12.4ex]

\includegraphics[clip,width=0.45\textwidth]{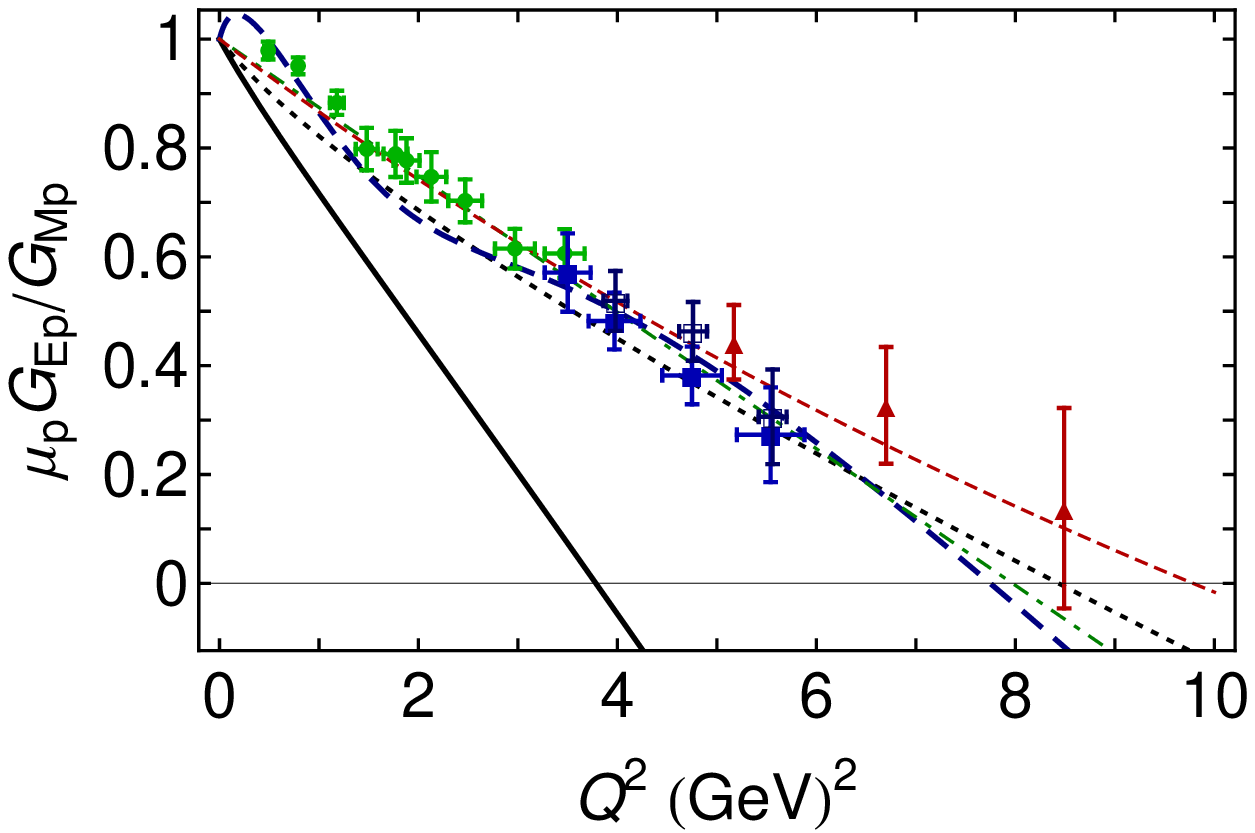}
\caption{\label{fig:GEGMp}
\emph{Upper panel}: Normalised ratio of proton Pauli and Dirac form factors.  \emph{Solid curve} -- contact interaction \protect\cite{Wilson:2011rj}; \emph{long-dashed curve} -- result from Ref.\,\protect\cite{Chang:2011tx}, which employed QCD-like momentum-dependence for the dressed-quark propagators and diquark Bethe-Salpeter amplitudes; \emph{long-dash-dotted curve} -- drawn from parametrisation of experimental data in Ref.\,\protect\cite{Kelly:2004hm}; and \emph{dotted curve} -- softened contact-interaction result, described in connection with Eq.\,\eqref{SoftF2p}.
\emph{Lower panel}: Normalised ratio of proton Sachs electric and magnetic form factors.  \emph{Solid curve} and \emph{long-dashed curve}, as above; \emph{dot-dashed curve} -- linear fit to data in Refs.\,\protect\cite{Jones:1999rz,Gayou:2001qd,Qattan:2004ht,Punjabi:2005wq,Puckett:2010ac}, constrained to one at $Q^2=0$; \emph{short-dashed curve} -- $[1,1]$-Pad\'e fit to that data; and \emph{dotted curve} -- softened contact-interaction result, described in connection with Eq.\,\eqref{SoftF2p}.
In addition, we have represented a selection of data explicitly: filled-squares \protect\cite{Gayou:2001qd}; circles \protect\cite{Punjabi:2005wq}; up-triangles \protect\cite{Puckett:2010ac}; and open-squares \protect\cite{Puckett:2011xg}.
}
\end{figure}

This returns us to the ratio $G_E^p(Q^2)/G_M^p(Q^2)$ described in the introduction to this section and depicted in Fig.\,\ref{fig:GEGMp}.  One observes that, independent of the nature of the interaction, computations of this ratio exhibit a zero.  Its location, however, is very sensitive to the interaction, although not to the electromagnetic size of the diquark correlations \cite{Cloet:2008re}.  (We note that the DSE prediction of the same ratio for the neutron has been confirmed by recent experiments \cite{Riordan:2010id}, as also is the trend of the nucleon's flavour separated Dirac and Pauli form factors \cite{Wilson:2011rj,Riordan:2010id,Cates:2011pz}.)

In order to assist in explaining the origin and location of a zero in the Sachs form factor ratio, in the top panel of Fig.\,\ref{fig:GEGMp} we depict the ratio of Pauli and Dirac form factors: both the actual contact-interaction result and that obtained when the Pauli form factor is artificially ``softened;'' viz.,
\begin{equation}
\label{SoftF2p}
F_{2p}(Q^2) \to \frac{F_{2p}(Q^2)}{1+Q^2/(4 m_N^2)}\,.
\end{equation}
As observed in Ref.\,\cite{Bloch:2003vn}, a softening of the proton's Pauli form factor has the effect of shifting the zero to larger values of $Q^2$.  In fact, if $F_{2p}$ becomes soft quickly enough, then the zero disappears completely.

The Pauli form factor is a gauge of the distribution of magnetisation within the proton. Ultimately, this magnetisation is carried by the dressed-quarks and influenced by correlations amongst them, which are expressed in the Faddeev wave-function.
If the dressed-quarks are described by a momentum-independent mass-function, then they behave as Dirac particles with constant Dirac values for their magnetic moments and produce a hard Pauli form factor.
%
Alternatively, suppose that the dressed-quarks possess a momentum-dependent mass-function, which is large at infrared momenta but vanishes as their momentum increases.  At small momenta they will then behave as constituent-like particles with a large magnetic moment, but their mass and magnetic moment will drop toward zero as the probe momentum grows.  (N.B.\ As described in Sec.\,\ref{spectrum2}, massless fermions do not possess a measurable magnetic moment.)  Such dressed-quarks will produce a proton Pauli form factor that is large for $Q^2 \sim 0$ but drops rapidly on the domain of transition between nonperturbative and perturbative QCD, to give a very small result at large-$Q^2$.  The precise form of the $Q^2$-dependence will depend on the evolving nature of the angular momentum correlations between the dressed-quarks.
From this perspective, existence, and location if so, of the zero in $\mu_p G_{Ep}(Q^2)/G_{Mp}(Q^2)$ are a fairly direct measure of the location and width of the transition region between the nonperturbative and perturbative domains of QCD as expressed in the momentum-dependence of the dressed-quark mass-function.

One may expect that a mass-function which rapidly becomes partonic -- namely, is very soft -- will not produce a zero; has seen that a constant mass-function produces a zero at a small value of $Q^2$, and knows that a mass-function which resembles that obtained in the best available DSE studies \cite{Qin:2011dd,Bhagwat:2006tu} and via lattice-QCD simulations \cite{Bowman:2005vx}, produces a zero at a location that is consistent with extant data.  There is plainly an opportunity here for very constructive feedback between future experiments and theory.  It is anticipated that experiments at JLab in the 12\,GeV era will establish conclusively whether or not this ratio possesses a zero.  The result will assist greatly in refining understanding of the dressed-quark mass function and therefrom QCD's $\beta$-function.

\subsection{Valence-quark distributions at $x=1$}
\label{sec:VPDF}
Before closing this section we would like to exploit a connection between the $Q^2=0$ values of elastic form factors and the Bjorken-$x=1$ values of the dimensionless structure functions of deep inelastic scattering, $F_2^{n,p}(x)$.  First recall that the $x=1$ value of a structure function is invariant under the evolution equations, Sec.\,\ref{FF2}.  Hence the value of
\begin{equation}
\label{dvuv1}
\left. \frac{d_v(x)}{u_v(x)}\right|_{x\to 1}\rule{-0.5em}{0ex}, \;\mbox{where} \rule{1em}{0ex}
\frac{d_v(x)}{u_v(x)} =
\frac{4 \frac{F_2^n(x)}{F_2^p(x)} - 1}{4- \frac{F_2^n(x)}{F_2^p(x)}},
\end{equation}
is a scale-invariant feature of QCD and a discriminator between models.  Next, when Bjorken-$x$ is unity, then $Q^2+2P\cdot Q=0$; i.e., one is dealing with elastic scattering.  Therefore, in the neighbourhood of $x=1$ the structure functions are determined by the target's elastic form factors.
The ratio in Eq.\,\eqref{dvuv1} expresses the relative probability of finding a $d$-quark carrying all the proton's light-front momentum compared with that of a $u$-quark doing the same or, equally, owing to invariance under evolution, the relative probability that a $Q^2=0$ probe either scatters from a $d$-quark or a $u$-quark; viz.,
\begin{equation}
\label{dvuvF1}
\left. \frac{d_v(x)}{u_v(x)}\right|_{x\to 1} = \frac{P_{1}^{p,d}}{P_{1}^{p,u}}.
\end{equation}
%

In constituent-quark models with $SU(6)$-symmetric spin-flavour wave-functions the right-hand-side of Eq.\,\eqref{dvuvF1} is $1/2$ because there is nothing to distinguish between the wave-functions of $u$- and $d$-quarks, and the proton is constituted from $u$-quarks and one $d$-quark.  On the other hand, when a Poincar\'e-covariant Faddeev equation is employed to describe the nucleon,
\begin{equation}
\label{dvuvF1result}
\frac{P_{1}^{p,d}}{P_{1}^{p,u}} =
\frac{\frac{2}{3} P_1^{p,a} + \frac{1}{3} P_1^{p,m}}
{P_1^{p,s}+\frac{1}{3} P_1^{p,a} + \frac{2}{3} P_1^{p,m}},
\end{equation}
where we have used the notation of Ref.\,\cite{Cloet:2008re}.  Namely,
$P_1^{p,s}=F_{1p}^s(Q^2=0)$ is the contribution to the proton's charge arising from diagrams with a scalar diquark component in both the initial and final state: $u[ud]\otimes \gamma \otimes u[ud]$.  The diquark-photon interaction is far softer than the quark-photon interaction and hence this diagram contributes solely to $u_v$ at $x=1$.
$P_1^{p,a}=F_{1p}^a(Q^2=0)$, is the kindred axial-vector diquark contribution; viz., $2 d\{uu\}\otimes \gamma\otimes d\{uu\}+u\{ud\} \otimes\gamma \otimes u\{ud\}$.  At $x=1$ this contributes twice as much to $d_v$ as it does to $u_v$.
$P_1^{p,m}=F_{1p}^m(Q^2=0)$, is the contribution to the proton's charge arising from diagrams with a different diquark component in the initial and final state.  The existence of this contribution relies on the exchange of a quark between the diquark correlations and hence it contributes twice as much to $u_v$ as it does to $d_v$.  If one uses the ``static approximation'' to the nucleon form factor, Eq.\,\eqref{staticexchange}, as with the treatment of the contact-interaction in Ref.\,\cite{Wilson:2011rj}, then $P_1^{p,m}\equiv 0$.  It is plain from Eq.\,\eqref{dvuvF1result} that $d_v/u_v=0$ in the absence of axial-vector diquark correlations; i.e., in scalar-diquark-only models of the nucleon, which were once common and, despite their weaknesses, still too often employed.

Using the probabilities presented in Refs.\,\cite{Cloet:2008re,Wilson:2011rj}, respectively, one obtains:
\begin{equation}
\label{compdvonuv}
\begin{array}{l|ccccc}
 & P_1^{p,s} & P_1^{p,a} & P_1^{p,m} & \frac{d_v}{u_v} & \frac{F_2^n}{F_2^p} \\\hline
M(p^2) & 0.60 & 0.25 & 0.15 & 0.28 & 0.49\\
%
%
\mbox{M=constant} &  0.78 & 0.22 & 0\rule{1.2em}{0ex} & 0.18 & 0.41\\
\end{array}\;,
\end{equation}
Both rows in Eq.\,\eqref{compdvonuv} are consistent with $d_v/u_v= 0.23\pm 0.09$ (90\% confidence level, $F_2^n/F_2^p = 0.45 \pm 0.08$) inferred recently via consideration of electron-nucleus scattering at $x>1$ \cite{Hen:2011rt}.  On the other hand, this is also true of the result obtained through a naive consideration of the isospin and helicity structure of a proton's light-front quark wave function at $x\sim 1$, which leads one to expect that $d$-quarks are five-times less likely than $u$-quarks to possess the same helicity as the proton they comprise; viz., $d_v/u_v=0.2$ \cite{Farrar:1975yb}.  Plainly, contemporary experiment-based analyses do not provide a particularly discriminating constraint.  Future experiments with a tritium target should help \cite{Holt:2010zz}, emphasising again the critical interplay between experiment and theory in elucidating the nature of the strong interaction.



\section{Strongly-Coupled Quarks In-Medium}
In this section we review a small selection of the recent progress made using the DSEs on QCD's phase transitions at nonzero temperature $(T)$ and chemical potential $(\mu)$, transitions which are the focus of much experimental and theoretical attention \cite{Muller:2006ee}.  In this regard, owing to the absence of a probability measure at $\mu\neq 0$ in lattice-QCD; i.e., the fermion sign problem \cite{Philipsen:2007rj}, the DSEs are a valuable tool because of their ability to connect confinement and DCSB in the continuum \cite{Roberts:2000aa,McLerran:2007qj}.

\subsection{Critical endpoint and phase coexistence}
Concerning the phase diagram in terms of temperature ($T$) and chemical potential ($\mu$), besides the general phase separation line, there are some other interesting possibilities,
such as: a domain of color superconductivity; and the existence of a critical endpoint (CEP) and an associated coexistence region.  In the chiral-limit theory, a CEP marks the end of a line of second-order chiral-symmetry-restoring (and possibly deconfining) transitions, originating on the temperature axis in the $(\mu,T)$-plane, and the beginning of a line of first-order transitions.  Such a critical endpoint would have observable consequences \cite{Stephanov:1998dy,Mohanty:2009vb}, so it is imperative to demonstrate its existence, determine its location and demarcate the subsequent domain of phase coexistence.

Despite the fermion sign problem, attempts have been made with lattice-QCD, using mathematical devices to extrapolate from $\mu=0$.  They yield \cite{Fodor:2004nz,deForcrand:2006ec}: $\mu^{\rm E}/T_c = 0.73\,$--$\,1.06$ and $T^{\rm E}/T_c \approx 0.95$,
and a signal for a material phase coexistence region \cite{deForcrand:2006ec}.  However, it is not yet certain whether the existence of a CEP survives in simulations with lattice parameters that more closely resemble the physical world \cite{deForcrand:2006pv}.

A wide variety of models have also been used to search for a CEP.
The Nambu--Jona-Lasinio (NJL) type yield \cite{Sasaki:2007qh,Costa:2008yh}: $\mu^{\rm E}/T_c \approx 1.7$, $T^{\rm E}/T_c \approx 0.4$; their Polyakov-loop extensions produce \cite{Fu:2007xc,Abuki:2008nm,Schaefer:2008hk,Costa:2009ae}: $\mu^{\rm E}/T_c = 1.5\,$--$\,1.8$, $T^{\rm E}/T_c = 0.3\,$--$\,0.8$;
and a chiral quark model gives \cite{Kovacs:2007sy} $(\mu^{\rm E},T^{\rm E})/T_c = (2.0,0.4)$.
On the other hand, by adding six- and eight-fermion interactions to the NJL model, one can obtain $(\mu^{\rm E},T^{\rm E})/T_c = (1.1,0.8)$, which is similar to the result produced by a Polyakov-loop-augmented chiral quark model \cite{Schaefer:2007pw} $(\mu^{\rm E},T^{\rm E})/T_c = (0.9,0.8)$.
The former, mutually consistent results for the CEP's location conflict markedly with those obtained from lattice-QCD: $\mu^{\rm E}/T_c$ is significantly larger and $T^{\rm E}/T_c$, much smaller.  If they are nevertheless correct, then finding the CEP in experiment will be difficult because modern colliders are restricted to exploration of the small-$\mu$ domain.
Given this observation, it is unsurprising that an analysis of flow data from the relativistic heavy ion collider leads to the estimate \cite{Lacey:2007na}: $\mu^{\rm E}/T_c \gtrsim 1.0$ and $T^{\rm E}/T_c \lesssim 1.0$.

Simple DSE truncations have also been applied to the CEP problem.  A confining zero-width momentum-space interaction, the antithesis of the NJL-model, produces \cite{Blaschke:1997bj} $\mu^{\rm E}/T_c =0$, $T^{\rm E}/T_c = 1$; and a separable-interaction \cite{He:2008yr}: $\mu^{\rm E}/T_c =1.09$, $T^{\rm E}/T_c = 0.78$.  However, neither study described a region of coexisting phases.  Notwithstanding that, in this chain of remarks about the model results, there is a hint that the length-scale characterising confinement in the quark-antiquark interaction markedly influences the location of the CEP.  Another notable feature of the studies is that the size of the coexistence regions associated with the CEP is very model dependent.


A novel principle for locating the CEP and demarcating the coexistence region was introduced in Ref.\,\cite{Qin:2010nq}.  The basic tools are the chiral susceptibility and the gap equation.  As we have seen, the gap equation's kernel is defined by a contraction of the dressed-gluon-propagator and -quark-gluon-vertex.  In order to exemplify the method, Ref.\,\cite{Qin:2010nq} used a propagator that can interpolate between models of the non-confining NJL-type and the confining interactions used in efficacious DSE studies of hadron observables \cite{Maris:2003vk,Roberts:2007jh,Holt:2010vj,Chang:2011vu}, whilst always providing a super-renormalisable interaction.
For the vertex, the studied used either the rainbow-truncation; i.e., the leading-order term in a symmetry-preserving scheme \cite{Bender:1996bb}, or a dressed-vertex \emph{Ansatz}.  The capacity to draw the phase diagram derived from an arbitrary dressed-vertex is an essentially new feature of the method.

At $T\neq 0 \neq \mu$, the gap equation is ($\tilde\omega_n= \omega_n + i \mu$)
\begin{eqnarray}
\label{eq:gap1}
S(\vec{p},\tilde\omega_n)^{-1} &=&  i\vec{\gamma}\cdot\vec{p}
+ i\gamma_4 \tilde \omega_{n} + m
  + \Sigma(\vec{p}, \tilde\omega_{n}) \, ,\\
\nonumber
\Sigma(\vec{p},\tilde\omega_n) &=& T\sum_{l=-\infty}^\infty \! \int\frac{d^3{q}}{(2\pi)^3}\; {g^{2}} D_{\mu\nu} (\vec{p}-\vec{q}, \Omega_{nl}; T, \mu)\\
& & \times \frac{\lambda^a}{2} {\gamma_{\mu}} S(\vec{q},
\tilde\omega_{l}) \frac{\lambda^a}{2}
\Gamma_{\nu} (\vec{q}, \tilde\omega_{l},\vec{p},\tilde\omega_{n})\, ,
\label{eq:gap2}
\end{eqnarray}
where: $\omega_n=(2n+1)\pi T$ is the fermion Matsubara frequency; $\Omega_{nl} = \omega_{n} - \omega_{l}$; $D_{\mu\nu}$ is the dressed-gluon propagator; and $\Gamma_{\nu}$ is the dressed-quark-gluon vertex.  (As an ultraviolet-finite model is employed, renormalisation is unnecessary and $m=0$ in Eq.\,(\ref{eq:gap1}) defines the chiral limit.)

In-medium, the gap equation's solution can be expressed as
\begin{eqnarray}
\nonumber
\lefteqn{S(\vec{p},\tilde\omega_n)^{-1} = i\vec{\gamma} \cdot \vec{p}\, A(\vec{p}\,^2, \tilde\omega_{n}^2) }\\
&  & \rule{4em}{0ex} +\, i\gamma_{4} \tilde\omega_{n} C(\vec{p}\,^2, \tilde\omega_{n}^2) + B(\vec{p}\,^2, \tilde\omega_{n}^2) \, ,
\label{eq:qdirac}
\end{eqnarray}
with, e.g., $B(\vec{p}\,^2, \tilde\omega_{n}^2)^\ast = B(\vec{p}\,^2, \tilde\omega_{-n-1}^2)$.
The dressed-gluon propagator has the form
\begin{equation}
g^2 D_{\mu\nu}(\vec{k}, \Omega_{nl}) = P_{\mu\nu}^{T} D_{T}(\vec{k}\,^2, \Omega_{nl}^2) + P_{\mu\nu}^{L} D_{L}(\vec{k}\,^2, \Omega_{nl}^2)\,,
\end{equation}
where $P_{\mu\nu}^{T,L}$ are, respectively, transverse and longitudinal projection operators, Eq.\,(\ref{projectionT}).  Whilst for $T\neq 0 \neq \mu$ it is generally true that $D_T \neq D_L$, there are indications \cite{Cucchieri:2007ta} that for $T<0.2\,$GeV, the domain relevant here, it is a good approximation to treat $D_T = D_L=:D_0$.  For the in-vacuum interaction, a simplified form of the interaction in Ref.\,\cite{Maris:1997tm} was used; viz., with $\kappa = \vec{k}\,^2 + \Omega_{nl}^2$,
\begin{equation}
\label{IRGs}
D_0(\kappa) = D \frac{4\pi^2}{\sigma^6} \kappa \, {\rm e}^{-\kappa/\sigma^2}.
\end{equation}
The parameters in Eq.\,(\ref{IRGs}) are $D$ and $\sigma$ but it has long been known that they are not independent: a change in $D$ can be compensated by an alteration of $\sigma$ \cite{Maris:2002mt}.  For $\sigma\in[0.3,0.5]\,$GeV, using Eq.\,(\ref{rainbowV}) below, ground-state pseudoscalar and vector-meson observables are roughly constant if $\sigma D  = (0.8 \, {\rm GeV})^3$. 
A value of $\sigma=0.5\,$GeV was usually employed in Ref.\,\cite{Qin:2010nq}.  N.B.\ Eq.\,(\ref{IRGs}) was used for illustrative simplicity, not out of necessity.  The most up-to-date interaction is described in Ref.\,\cite{Qin:2011dd} and the status of propagator and vertex studies can be tracked from Ref.\,\cite{Roberts:2007ji}.

As explained above, the gap equation is complete once the vertex is specified.
Here results obtained using the rainbow-truncation:
\begin{equation}
\label{rainbowV}
\Gamma_\nu(\vec{q}, \tilde\omega_{l},\vec{p},\tilde\omega_{n}) = \gamma_\nu\,,
\end{equation}
the first term in a symmetry-preserving scheme \cite{Bender:1996bb}, are compared with those produced by the Ball-Chiu (BC) \emph{Ansatz} \cite{Ball:1980ay,Maris:2000ig,Ayala:2001mb}:
\begin{eqnarray}
\nonumber
\lefteqn{
i\Gamma_{\mu}(\vec{q}, \tilde\omega_{l},\vec{p},\tilde\omega_{n}) = {i\gamma_{\mu}^{T}}{\Sigma_{A}} + {i\gamma_{\mu}^{L}} {\Sigma_{C}} }  \\
\nonumber
& & \rule{2em}{0ex}
+(\tilde{p}_n+\tilde{q}_l)_{\mu} \bigg[
\frac{i}{2}\gamma^{T}_\alpha (\tilde{p}_n + \tilde{q}_l)_\alpha {\Delta_{A}}\\
&& \rule{2em}{0ex} + \frac{i}{2}\gamma^{L}_\alpha (\tilde{p}_n+\tilde{q}_l)_\alpha {\Delta_{C}}
+ {\Delta_{B}} \bigg], \label{bcvtxM}
\end{eqnarray}
$\tilde{p}_n=(\vec{p},\omega_n+i\mu)$, $\tilde{q}_l=(\vec{q},\omega_l+i\mu)$, with ($F=A,B,C$)
\begin{eqnarray}
\Sigma_F(\vec{q}\,^2,\omega_l^2,\vec{p}\,^2,\omega_n^2) &= &\frac{1}{2}
[F(\vec{q}\,^2,\omega_l^2)+F(\vec{p}\,^2,\omega_n^2)], \rule{2em}{0ex}\\
\Delta_F(\vec{q}\,^2,\omega_l,\vec{p}\,^2,\omega_n) &=&
\frac{F(\vec{q}\,^2,\omega_l^2)-F(\vec{p}\,^2,\omega_n^2)}
{\tilde{q}_l^2-\tilde{p}_n^2}, \rule{2em}{0ex}\label{eq:BC_delta}
\end{eqnarray}
where, defining $u=(0,0,0,1)$, $\gamma_\mu^T=\gamma_\mu-u_\mu\gamma_\alpha u_\alpha$, $\gamma_\mu^L=u_\mu\gamma_\alpha u_\alpha$.
The comparison is natural because vertices of the type in Eq.\,(\ref{rainbowV}) have widely been used in studies of hadron observables \cite{Maris:2003vk,Roberts:2007jh,Holt:2010vj,Chang:2011vu}, and the BC \emph{Ansatz} provides a semi-quantitatively accurate representation of lattice-QCD results for important terms in $\Gamma_\mu$ at $T=0=\mu$ (see Sec.\,\ref{spectrum2}).
%
(It is plain from Eq.\,(\ref{eq:gap2}) that with the BC vertex the effective interaction strength is $\hat D=D A(0,0)$.  In contrast, owing to Eq.\,(\ref{rainbowV}), $\hat D=D$ in rainbow-ladder truncation.)

\begin{figure}[t]
\centering
\includegraphics[width=8.0cm]{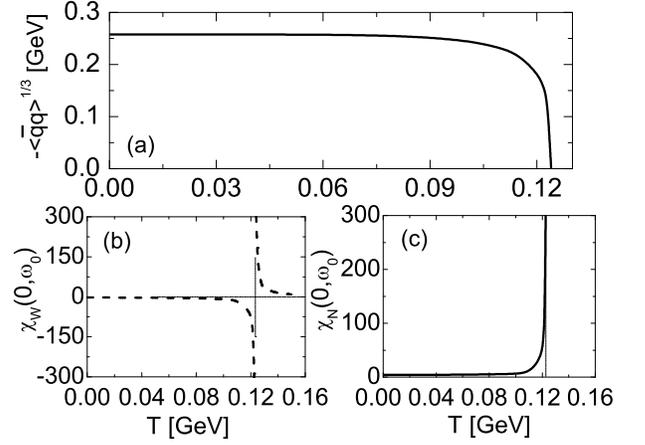}
\caption{Panel~(a) -- Temperature dependence of the chiral-symmetry order parameter in Eq.\,(\protect\ref{qbq}).  Chiral susceptibility computed in the Wigner phase -- Panel~(b), and in the Nambu phase -- Panel~(c).  In all panels: the Ball-Chiu vertex was used, Eq.\,(\protect\ref{bcvtxM}); and $\mu=0$, $D=0.5\,$GeV$^2$, $m=0$.
\label{fig:Teffect-BC}}
\end{figure}

The gap equation formulated above is readily solved and in Fig.\,\ref{fig:Teffect-BC} we depict the $T$-dependence of a chiral susceptibility \cite{Holl:1998qs} and a chiral-symmetry order parameter \cite{Brodsky:2010xf}
\begin{eqnarray}
\chi(0,\omega_0) & = &\frac{\partial}{\partial m} B(\vec{0},\omega_0^2)\,,\\
\label{qbq}
-\langle \bar q q \rangle^0 &= &N_c T \sum^{\infty}_{n= -\infty}\!\!\! {\rm tr}_{\rm D} \int \!
\frac{d^3 p}{(2\pi)^3} \, S_{m=0}(\vec{p},\omega_n) \, .\rule{2.5em}{0ex}
\end{eqnarray}
For $T<T^{\rm E}$ the behaviour of the order parameter is typical of models without long-range correlations in the gap equation's kernel \cite{Holl:1998qs}.  Namely, initially slow evolution from its $T=0$ value: $\langle \bar q q \rangle^0=(-0.258\,{\rm GeV})^3$, which signals chiral symmetry realised in the Nambu mode; i.e., dynamically broken chiral symmetry, and this followed by a mean-field transition to a phase with chiral symmetry restored; i.e., realised in the Wigner mode.  In the Wigner phase the dressed-quarks are not confined.

The lower panels in Fig.\,\ref{fig:Teffect-BC} show the chiral susceptibility of the Wigner and Nambu phases, which correspond to gap equation solutions that are, respectively, within the domain of attraction of the $B=0$ or $B\neq0$ solution \cite{Chang:2006bm}.  A phase is unstable in response to fluctuations if the susceptibility is negative, but stable and realisable otherwise.  With $\mu=0$, one sees the Nambu phase completely replaced by the Wigner phase at $T=124\,$MeV.

\begin{figure}[t]
\centering
\includegraphics[width=8.0cm]{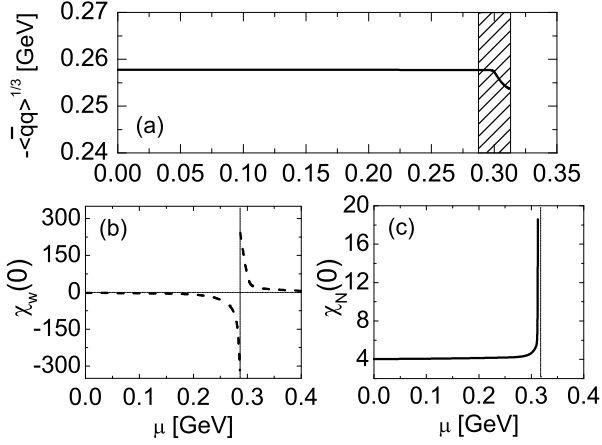}
\caption{Analogue of Fig.\,\ref{fig:Teffect-BC}, displaying evolution with chemical potential at $T=0$.
\label{fig:mueffect-BC}}
\end{figure}

This should be contrasted with the behaviour in Fig.\,\ref{fig:mueffect-BC}.  At $T=0$ the order parameter remains constant with increasing $\mu$ until $\mu_a = 0.30\,$GeV, which is the upper bound on the domain of analyticity for this gap equation's kernel \cite{Chen:2008zr}.  On a small domain beyond this; viz., $\mu\in (\mu_a,\mu_c^N)$, with $\mu_c^N=0.314\,$GeV, the order parameter diminishes smoothly, an effect that may be denominated a partial restoration of chiral symmetry.  For $\mu>\mu_c^N$ the order parameter vanishes so that chiral symmetry is completely restored via a first-order transition.

The lower panels of Fig.\,\ref{fig:mueffect-BC} provide extra information.  For $\mu<\mu_c^W = 0.286\,$GeV, the Wigner phase is unstable.  That changes at $\mu_c^W$, when $\chi_W$ switches sign and thereafter, on the domain $\mu_c^W < \mu < \mu_c^N$, both the Wigner- and Nambu-phase susceptibilities are positive.  This is the domain of phase coexistence, with a metastable Wigner phase for $\mu_c^W<\mu<\mu_a$ and a metastable Nambu phase for $\mu_a<\mu<\mu_c^N$.  The pressure of the phases is equal at $\mu=\mu_a$; and the Nambu phase is completely displaced by the Wigner phase for $\mu>\mu_c^N$.
Notably, with an \emph{Ansatz} for the dressed-quark gluon vertex, the diagrammatic content of the gap equation's kernel is generally unknown.  However, owing to the insights provided in Ref.\,\cite{Zhao:2008zzi}, one can draw these conclusions despite being unable to calculate an explicit expression for the thermodynamic pressure.

At the onset of the coexistence domain one can expect pockets of deconfined, chirally symmetric quark matter to appear in the confining Nambu medium.  Their number and average volume will increase with $\mu$.  The opposite situation occurs at the termination of the domain; i.e., it is the Nambu phase which exists only in pockets.  For $\mu\in (\mu_a,\mu_c^N)$, which is the domain of Nambu-phase metastability, the properties of observed hadrons will be affected by the partial restoration of chiral symmetry.  The coexistence domain is not of a quarkyonic nature \cite{McLerran:2007qj}.

\begin{figure}[t]
\centering
\includegraphics[width=7.0cm]{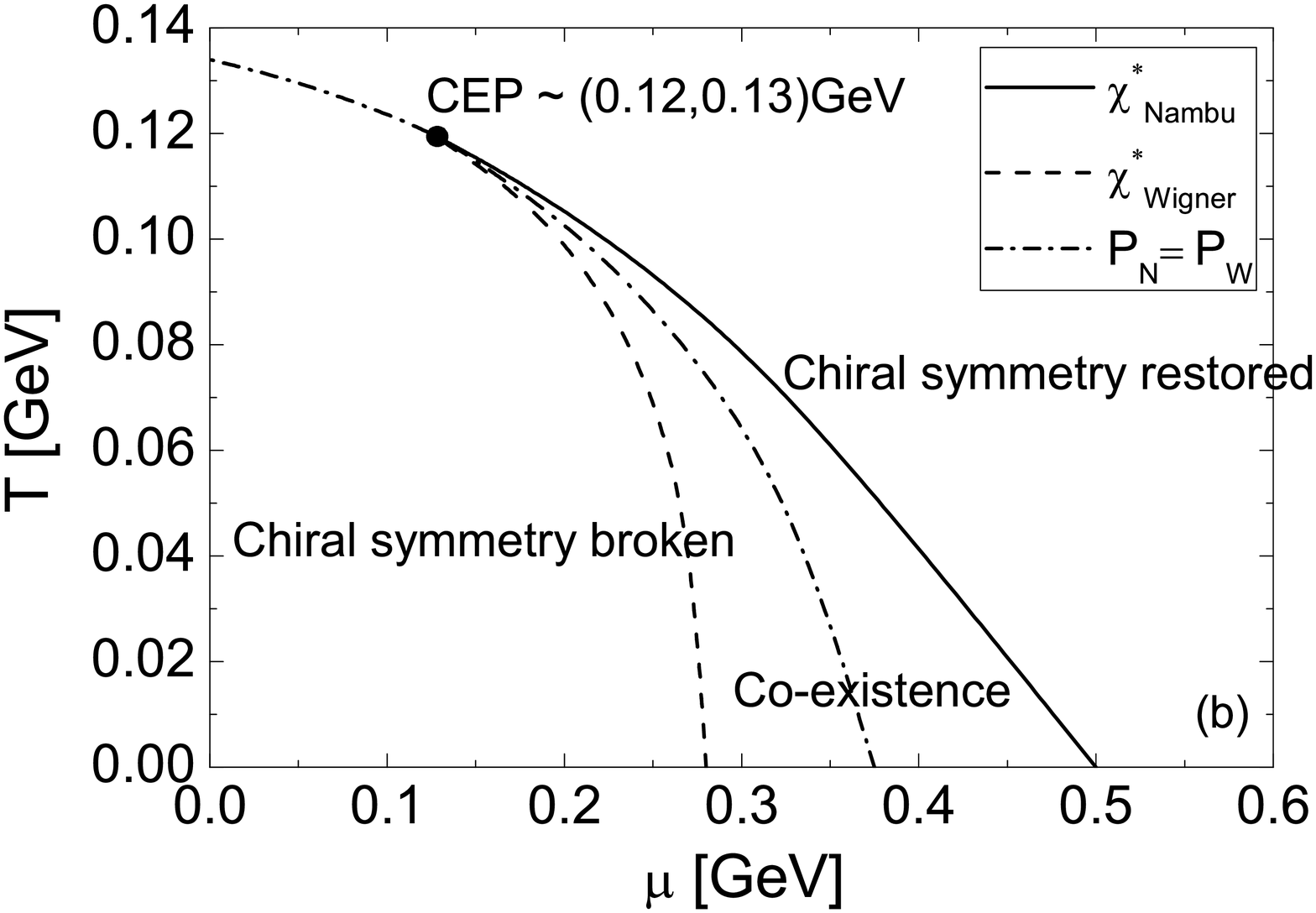}
\includegraphics[width=7.0cm]{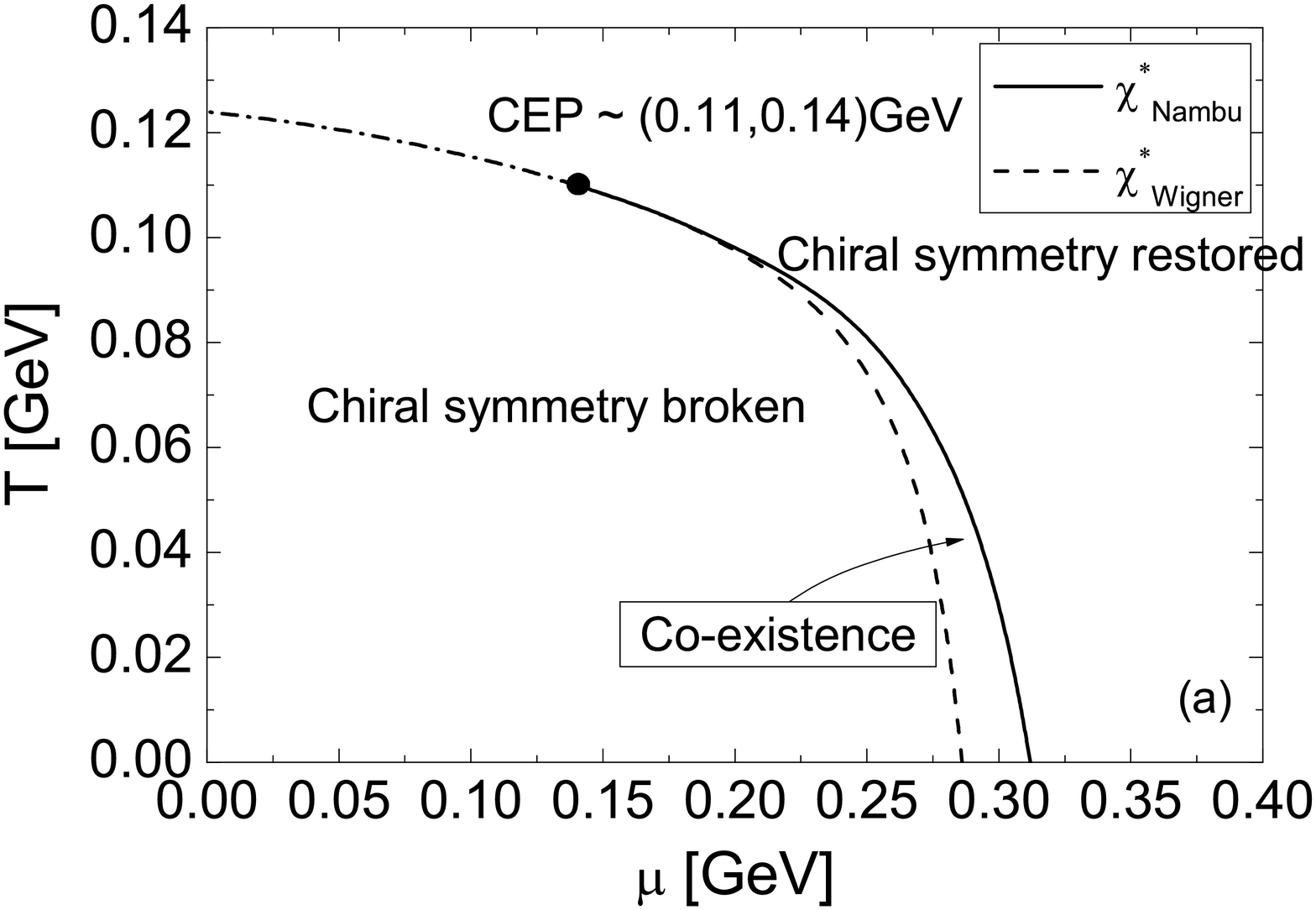}
\caption{Chiral-limit phase diagram in the tem\-pe\-ra\-ture/che\-mi\-cal-po\-ten\-tial plane for strongly-interacting quarks.  The critical endpoint (CEP) is marked explicitly.
\emph{Upper panel} -- rainbow vertex, Eq.\,(\ref{rainbowV}), with $D=1.0\,$GeV$^2$, $\sigma=0.5\,$GeV.  The $\mu=0$ critical temperature is $T_c^0=0.133\,$GeV.
\emph{Lower panel} -- Ball-Chiu vertex, Eq.\,(\ref{bcvtxM}), with $D=0.5\,$GeV$^2$, $\sigma=0.5\,$GeV.  The $\mu=0$ critical temperature is $T_c^0=0.124\,$GeV.
\label{fig:CEP}}
\end{figure}

The computed phase diagrams \cite{Qin:2010nq} are drawn in Fig.\,\ref{fig:CEP}.  The upper panel was obtained with the rainbow truncation, Eq.\,(\ref{rainbowV}), in which case the diagrammatic content of the gap equation's kernel is known and one can thus compute the dressed-quark component of the pressure.  It is given by the auxiliary-field effective action evaluated at its extrema \cite{Roberts:1994dr,Roberts:2000aa}.  Within the domain $(\mu<\mu^{\rm E},T>T^{\rm E})$ the system exhibits a mean-field transition, which is signalled both by: equality of the Wigner- and Nambu-phase pressures; and coincident singularities in the Wigner- and Nambu-phase chiral susceptibilities.  The curve tracking the singularity location in the Wigner and Nambu susceptibilities bifurcates at the CEP, with a domain of phase coexistence opening.  Naturally, the curve of equal thermodynamic pressure lies within this domain.

The lower panel in Fig.\,\ref{fig:CEP} was obtained using the dressed vertex, Eq.\,(\ref{bcvtxM}).  Its features are similar to those displayed in the upper panel but the domain of phase coexistence is smaller.  Here, one cannot derive a form for the dressed-quark pressure.  Thus it is only an appreciation of the information contained in the chiral susceptibilities that enables
a phase diagram to be drawn.

As explained in Sec.\,\ref{Sect:Conf}, confinement is expressed in the analytic structure of the dressed-quark propagator.  Measured in this way, it is significant that the models considered in this subsection are members of a class in which chiral symmetry restoration is accompanied by a coincident dressed-quark deconfinement transition in the chiral limit.

Table~\ref{tab:CEP} illustrates the response of the CEP's location to changing the vertex or the parameters.  Defining a confinement length-scale $r_\sigma=1/\sigma$, it is apparent that the CEP rotates toward the temperature axis as $r_\sigma$ is increased.  The extreme case is $r_\sigma = \infty$, which was computed in Ref.\,\cite{Blaschke:1997bj} and reported above: \mbox{($\mu_{\text{E}}^{},T_{\text{E}}^{})/T_c = (0,1)$}.  Models of the NJL-type, as they have usually been used in the current context, represent the opposite limiting case: they are expressed via a gap equation in which the confinement length-scale vanishes.  From this perspective, it is unsurprising that they produce a CEP whose angular separation from the $\mu$-axis is significantly smaller.

\begin{table}[t]
\begin{center}
\caption{Parameter- and vertex-dependence of the critical endpoint and coexistence region.  $\Delta_{\rm C}$ is the width of the coexistence region on the $\mu$-axis.  Each parameter set produces similar results for the in-vacuum values of the so-called vacuum quark condensate $\langle \bar q q\rangle^0$ and the pion's leptonic decay constant $f_\pi$ \protect\cite{Chang:2009zb}.  Lattice-QCD suggests ($\mu_{\text{E}}^{},T_{\text{E}}^{}$)$/T_c = (0.73\,$--$\,1.06,0.95)$.  (Dimensioned quantities in GeV.)
\label{tab:CEP}}
\begin{tabular}{|c|c|c|c|c|c|c|}
\hline
\multicolumn{3}{|c|}{model} & \multicolumn{4}{|c|}{result} \\
\hline
%
%
vertex & $\hat D^{1/2}$ & $\sigma$ & $T_c$ & $\Delta_{\rm C}$ &($\mu_{\text{E}}^{}$,$T_{\text{E}}^{}$)$/T_c$ & ${\mu_{\text{E}}^{}}/{T_{\text{E}}^{}}$
\\
\hline
BC   & 0.7 & 0.50 & 0.124 & 0.026 &($1.13,0.89$) & $1.27$                    \\
BC   & '' & 0.45 & 0.128 & 0.048 & ($0.69,0.92$) & $0.75$                    \\
BC   & '' & 0.40 & 0.139 & 0.076 & ($0.16,0.96$) & $0.17$                    \\
Bare & 1.0 & 0.50 & 0.133 & 0.220 & ($0.98,0.90$) & $1.08$                    \\
Bare & '' & 0.45 & 0.136 & 0.280 & ($0.81,0.89$) & $0.91$                    \\
Bare & '' & 0.40 & 0.148 & 0.360 & ($0.17,0.95$) & $0.18$                    \\
\hline
\end{tabular}
\end{center}
\end{table}

This subsection reviews a method, based on the chiral susceptibility, which enables one to draw a phase diagram in the chemical-potential/temperature plane for quarks whose interactions are described by any sensibly-constructed gap equation.  Thus, in attempting to chart the phase structure of QCD using the methods of continuum quantum field theory, one is no longer restricted to the simplest class of mean-field kernels: sophisticated quark-gluon vertices can be used.  The method is general and potentially useful in all branches of physics that explore the properties of dense fermionic systems.

A class of models that successfully describes in-vacuum properties of $\pi$- and $\rho$-mesons, exhibits a critical endpoint (CEP) in the neighbourhood $(\mu^{\rm E},T^{\rm E})\sim (1.0,0.9)T_c$.  The CEP's angular separation from the temperature axis is a measure of the confinement length-scale: the separation decreases as the confinement length-scale increases.  Furthermore, a domain of phase coexistence opens at the CEP.  It's size depends on the structure of the gap equation's kernel but, other aspects being equal, it increases in area as the confinement length-scale increases.  We are hopeful that illumination of the CEP and its consequences is within the reach of modern colliders.

\subsection{Quark spectral density and a sQGP}
It is believed that a primordial state of matter has been recreated by the relativistic heavy-ion collider (RHIC) \cite{sQGP}.  This substance appears to behave as a nearly-perfect fluid on some domain of temperature, $T$, above that required for its creation, $T_c$ \cite{Song:2008hj}.  An ideal fluid has zero shear-viscosity: $\eta=0$, and hence no resistance to the appearance and growth of transverse velocity gradients.  A perfect fluid with near-zero viscosity is the best achievable approximation to that ideal.  Graphene might provide a room temperature example \cite{graphene}.  From Newton's law for viscous fluid flow; viz., $dv/dz = (1/\eta) (F/A)$, it is apparent that in near-perfect fluids a macroscopic velocity gradient is achieved from a microscopically small pressure.  Strong interactions between particles constituting the fluid are necessary to achieve this outcome.  Hence the primordial state of matter is described as a strongly-coupled quark gluon plasma (sQGP).

We have explained that at $T=0$, QCD is characterised by confinement and dynamical chiral symmetry breaking (DCSB).  These phenomena are represented by a range of order parameters which all vanish in the sQGP.  Understanding the sQGP therefore requires elucidation of the behaviour and properties of quarks and gluons within this state.  Perturbative techniques have been developed for use far above $T_c$; viz., the hard thermal loop (HTL) expansion \cite{Pisarski:1988vd,Braaten:1990wp}, which has enabled the computation of gluon and quark thermal masses $m_T\sim g T$ and damping rates $\gamma_T\sim g^2 T$, with $g=g(T)$ being the strong running coupling.  It also suggests the existence of a collective plasmino or ``abnormal'' branch to the dressed-quark dispersion relation, which is characterised by antiparticle-like evolution at small momenta \cite{Blaizot:2001nr}.

Owing to asymptotic freedom, the running coupling in QCD increases as $T\to T_c^+$.  Therefore, a simple interpretation of the HTL results suggests the plasmino should disappear before $T_c$ is reached because $\gamma_T$ increases more rapidly than $m_T$ and $\gamma_T/m_T\sim 1$ invalidates a quasiparticle picture.  On the other hand, lattice-regularised quenched-QCD suggests that the plasmino branch persists in the vicinity of $T_c$ \cite{Karsch:2009tp}.  It is necessary to resolve the active degrees of freedom in the neighbourhood of $T_c$ because the spectral properties of the dressed-quark propagator are intimately linked with light-quark confinement (Sec.\,\ref{Sect:Conf}) and it is the long-range modes which might produce strong correlations.

We saw in the preceding section that below $T_c$ dressed-gluons and -quarks are confined and chiral symmetry is dynamically broken; and beyond $T_c$, deconfinement and chiral symmetry restoration occur via coincident second-order phase transitions.  The sQGP seems to occur in the neighbourhood of $T_c$, so in this section we review how the gap equation may be used to elucidate the active fermion quasiparticles for $T\gtrsim T_c$ \cite{Qin:2010pc}.

On the domain  $T>T_c$ the chiral-limit dressed-quark propagator can be written
\begin{equation}
\label{eq:qdiracA}
S(i\omega_n,\vec{p})= -i\vec{\gamma} \cdot \vec{p}\, \sigma_A(\omega_n,
\vec{p}\,^2) - i\gamma_4\omega_n\sigma_C(\omega_n, \vec{p}\,^2) \,.
\end{equation}
There is no Dirac-scalar part because chiral symmetry is realised in the Wigner mode but this does not mean that nonperturbative phenomena are excluded, as is apparent, for example, in the discussion of novel Wigner-mode solutions to the gap equation in Refs.\,\cite{Chang:2006bm,Williams:2006vva}.  The retarded real-time propagator is found by analytic continuation
\begin{equation}
\label{eq:qreal}
S^R(\omega,\vec{p})= \left. S(i{\omega_{n}},\vec{p})\right|_{i{\omega_{n}}
\rightarrow\omega+i\eta^+}
\end{equation}
and from this one obtains the spectral density
\begin{equation}
\label{eq:spec}
\rho(\omega, \vec{p})=-2 \Im\,S^R(\omega,\vec{p}) \, .
\end{equation}
Equations~(\ref{eq:qreal}) and (\ref{eq:spec}) are equivalent to the statement:
\begin{equation}
S(i\omega_n,\vec{p}) = \frac{1}{2\pi}\int_{-\infty}^{+\infty}\!\!\!\!\!\!
d\omega^\prime\,\frac{\rho(\omega^\prime,\vec{p})}{\omega^\prime-i{\omega_{n}}
} \, . \label{eq:mat_spec}
\end{equation}
NB.\ If one requires a nonnegative spectral density, then Eq.\,(\ref{eq:mat_spec}) is only valid for $T>T_c$; i.e., on the deconfined domain (Sec.\,\ref{Sect:Conf}).  Furthermore, including a light current-quark mass has no material impact on the primary conclusions of Ref.\,\cite{Qin:2010pc}.

For an unconfined dressed-quark propagator of the form in Eq.\,(\ref{eq:qdiracA}), the spectral density can be expressed
\begin{eqnarray}
\rho(\omega,\vec{p}) = {\rho_{+}} (\omega, \vec{p}\,^2) {P_{+}} +
{\rho_{-}} (\omega, \vec{p}\,^2) {P_{-}}  \,,
\end{eqnarray}
where $P_\pm=(\gamma_{4} \pm i\vec{\gamma}\cdot \vec{u}_p)/2$, $\vec{u}_p \cdot \vec{p} = |\vec{p}|$, are operators which project onto spinors with a positive or negative value for the ratio ${\cal H}:=\,$helicity/chirality: ${\cal H} = 1$ for a free positive-energy fermion.  The spectral density is interesting and expressive because it reveals the manner by which interactions distribute the single-particle spectral strength over momentum modes; and the behaviour at $T\neq 0$ shows how that is altered by a heat bath. As with many useful quantities, however, it is nontrivial to evaluate $\rho(\omega,|\vec{p}|)$.  Nonetheless, if one has at hand a precise numerical determination of the dressed-quark propagator in Eq.\,(\ref{eq:qdirac}), then it is possible to obtain an accurate approximation to the spectral density via the maximum entropy method (MEM) \cite{Nickel:2006mm,Mueller:2010ah}.

\begin{figure}
\includegraphics[width=0.41\textwidth]{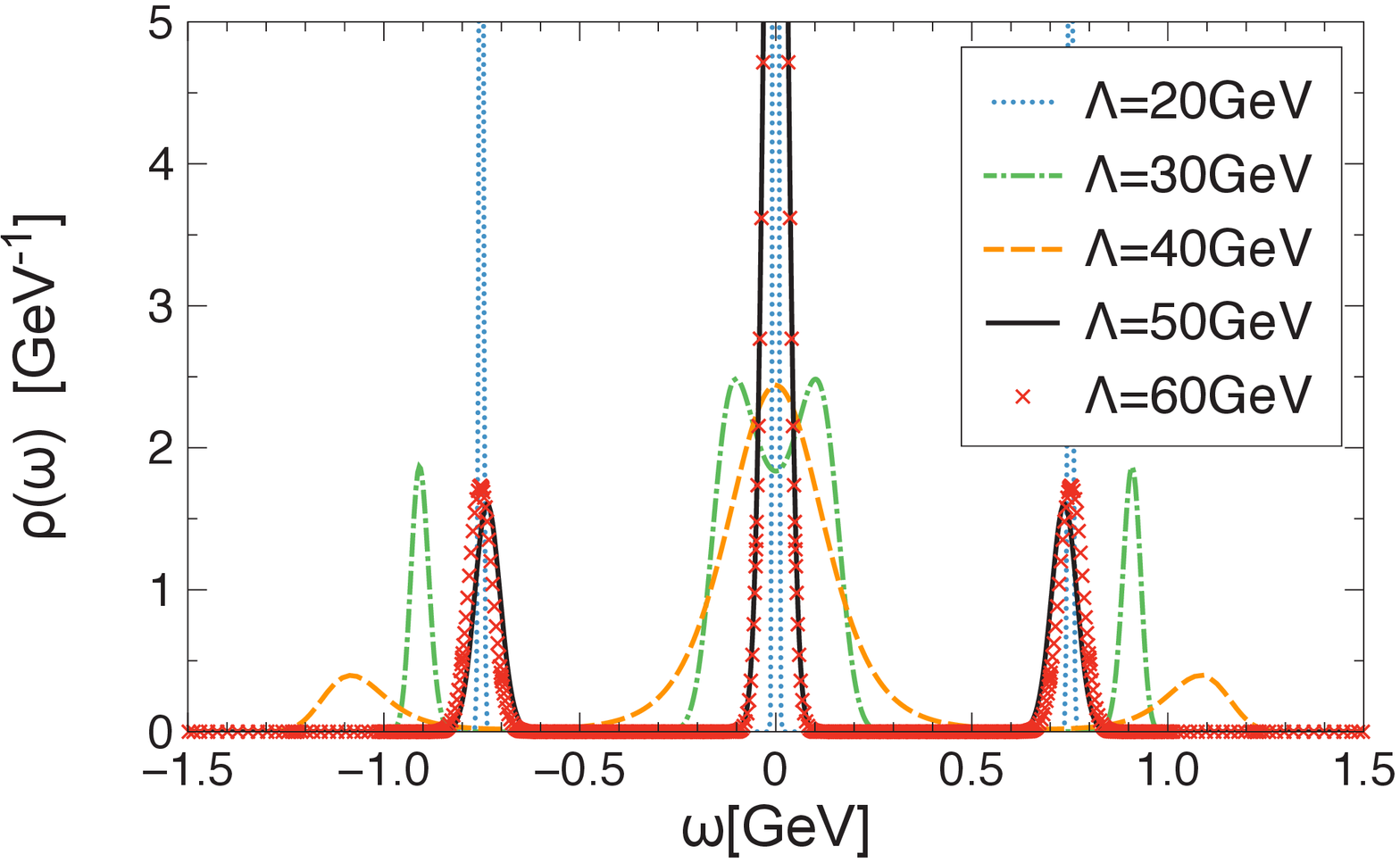}

\includegraphics[width=0.41\textwidth]{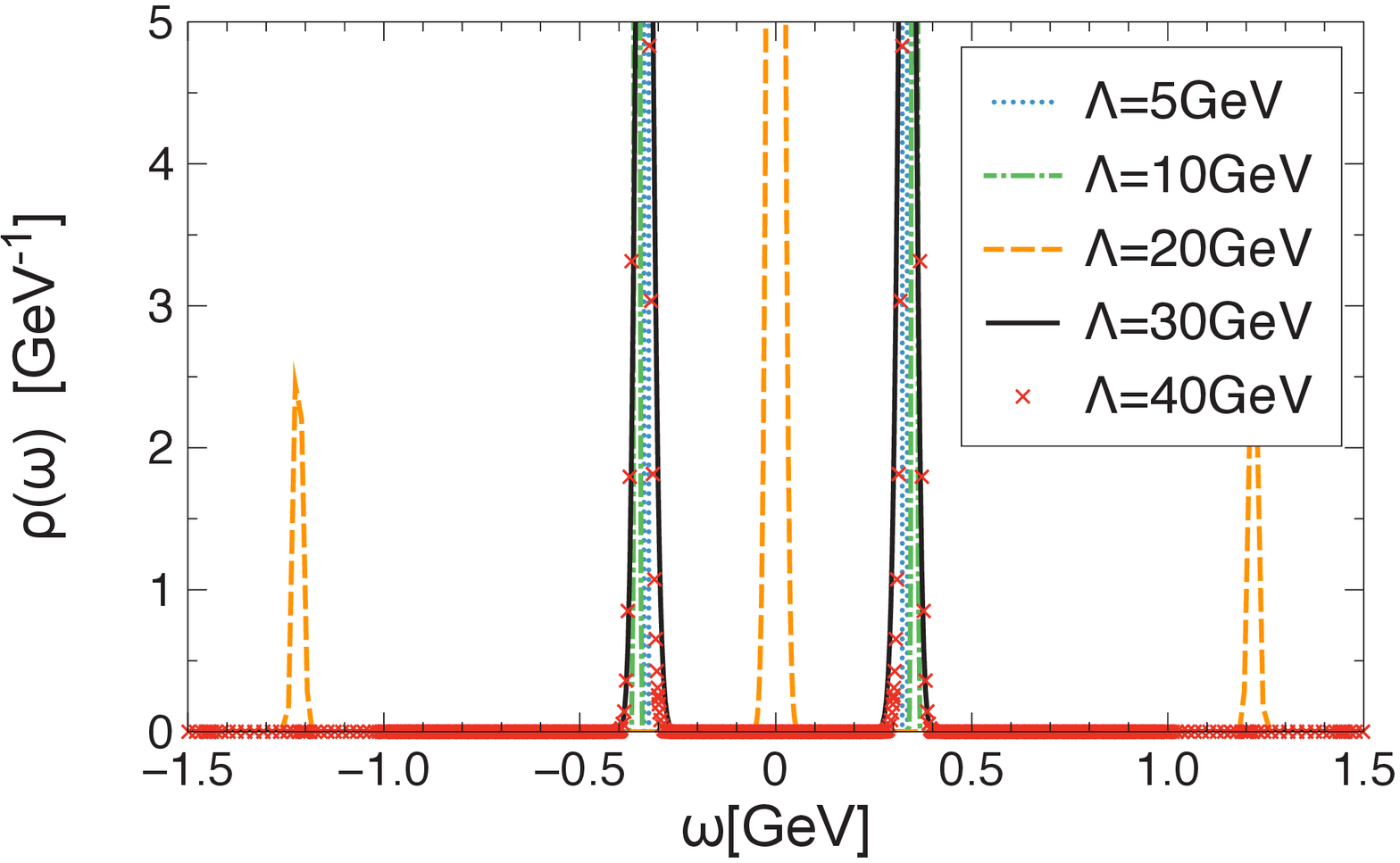}

\caption{\label{fig:zero} Dependence of the spectral density on the cutoff, $\Lambda$, in the default MEM model, Eq.\,(\ref{defaultm0}), at $T=1.1 \,T_{c}$ (\emph{upper panel}) and  $T=3.0 \, T_{c}$ (\emph{lower panel}), where $T_c$ is discussed in Eq.\,(\protect\ref{Tcvalue}).}
\end{figure}

In the MEM the entropy functional of the spectral function is defined as
\begin{equation}
S[\rho]=\int_{-\infty}^{+\infty}d\omega\left[\rho(\omega)-m(\omega)
-\rho(\omega)~\textrm{log}\frac{\rho(\omega)}{m(\omega)}\right] \, ,
\end{equation}
where $m(\omega)$ is the ``default model'' of the spectral function.  One typically adopts a simple form; viz.,
\begin{eqnarray}
\label{defaultm0}
m(\omega)=m_0 \, \theta(\Lambda^2-\omega^2) \, ,
\end{eqnarray}
which is a uniform distribution with no \emph{a priori} structure assumption.  It should be borne in mind that one can only claim a reliable result from the MEM if the spectral function produced is insensitive to $m_0$ and $\Lambda$.   The sensitivity to $m_0$ is uniformly weak in Ref.\,\cite{Qin:2010pc}.  However, on a material domain, the sensitivity to $\Lambda$ is strong. Nevertheless, for any given temperature, it was found that there is always a value $\Lambda$ above which a stable spectral function is obtained.  This is illustrated in Fig.\,\ref{fig:zero}.

In this instance one is interested in the chiral-limit dressed-quark propagator, which is obtained from ($F=A,C$)
\begin{eqnarray}
\nonumber
S(i\omega_n,\vec{p})^{-1} &=& Z_{2}^{A} i\vec{\gamma} \cdot \vec{p}
+Z_{2}i\gamma_4\omega_n+\Sigma^\prime(i \omega_n,\vec{p}) \,, \\
&&\\
\nonumber
\Sigma^\prime(i \omega_n,\vec{p}) &= & i\vec{\gamma} \cdot \vec{p} \, \Sigma_A^\prime(i \omega_n,\vec{p}) + i\gamma_4\omega_n \,\Sigma_C^\prime(i \omega_n,\vec{p})\,,\\
&&\\
\Sigma_F^\prime(i \omega_n,\vec{p}) &=&
 T\! \sum_l \int \frac{d^3 q}{(2\pi)^3}
g^2 D_{\mu\nu}({\omega_{n}}\! - \! {\omega_{l}},
\vec{p} \! - \! \vec{q})             \notag    \\
\nonumber
&  \times & \frac{1}{3} {\rm tr}_{\rm D}{\cal P}_F\gamma_\mu
S(i\omega_l,\vec{q})\Gamma_\nu(\omega_n,
\omega_l,\vec{p},\vec{q}),\\
&& \label{eq:gapeq}
\end{eqnarray}
where: ${\cal P}_A = -Z_1^A i\vec\gamma\cdot \vec{p}/\vec{p}\,^2$, ${\cal P}_C = -Z_1 i\gamma_4/\omega_n$; and $Z_{1,2}$, $Z_{1,2}^A$ are, respectively, the vertex and quark wave function renormalisation constants.  The regularisation and renormalisation procedures of  Refs.\,\cite{Bender:1996bm,Maris:2000ig} are followed.

Reference~\cite{Qin:2010pc} used the rainbow-ladder truncation and the phenomenologically-efficacious one-loop renormalisation-group-improved interaction of Ref.\,\cite{Maris:2000ig}.  Namely:
\begin{eqnarray}
\nonumber
& & g^{2} D_{\mu\nu}(\omega_{n} - \omega_{l},\vec{p}-\vec{q})
\Gamma_{\nu}({\omega_{n}}, {\omega_{l}},\vec{p},\vec{q}) \\
& = & [{P_{T}^{\mu\nu}}({k_{\Omega}}){D_{T}}({k_{\Omega}})
+{P_{L}^{\mu\nu}}({k_{\Omega}}){D_{L}}({k_{\Omega}})]{\gamma_{\nu}}, \rule{1em}{0ex} \label{eq:model}
\end{eqnarray}
where ${k_{\Omega}}:=(\Omega,\vec{k})=({\omega_{n}} -{\omega_{l}},\vec{p}-\vec{q})$;
\begin{eqnarray}
P_T^{\mu\nu}(k_{\Omega})=\left\{
\begin{aligned}
&0, \qquad \qquad \quad {\mu}\text{ and/or } {\nu} = 4 \, ,  \\
&\delta_{ij}-\frac{{k_{i}} {k_{j}}}{k^2}, \quad {\mu}, {\nu} =1,2,3
\, ,
\end{aligned}
\right.
\label{projectionT}
\end{eqnarray}
with $P_{L}^{\mu\nu} + P_{T}^{\mu\nu} = \delta_{\mu\nu} - k_\Omega^\mu k_\Omega^\nu /k_\Omega^2$; and
\begin{eqnarray}
{D_{T}(k_{\Omega})} &=&\mathcal{D}({k^{2}_{\Omega}},0), \;
{D_{L}(k_{\Omega})} =\mathcal{D}({k^{2}_{\Omega}},{m_{g}^{2}}) \, ,\rule{2.5em}{0ex}\\
\nonumber
\mathcal{D}({k^{2}_{\Omega}}, {m_{g}^{2}}) & = & 4{\pi^{2}} D
\frac{s_\Omega}{\omega^6}e^{-s_\Omega/\omega^2}\\
& & + \frac{8{\pi^{2}} {\gamma_{m}}}{{\ln}[ \tau \! + \! (1 \! + \!
s_\Omega/{\Lambda_{\text{QCD}}^{2}} ) ^{2} ] } \,
{\cal F}(s_\Omega)\,,
\end{eqnarray}
with ${\cal F}(s_\Omega) = (1-\exp(-s_\Omega/4 m_t^2)/s_\Omega$, $s_\Omega = \Omega^2 + \vec{k}\,^2 + m^2_g$, $\tau=e^2-1$, $m_t=0.5\,$GeV, $\gamma_m=12/25$, and $\Lambda^{N_f=4}_{\text{QCD}}=0.234$~GeV.  For pseudoscalar and vector mesons with masses$\,\lesssim 1\,$GeV, this interaction provides a uniformly good description of their $T=0$ properties \cite{Maris:2003vk} when $\omega = 0.4\,$GeV, $D=(0.96\,{\rm GeV})^2$.  In generalising to $T\neq 0$, Ref.\,\cite{Qin:2010pc} have followed perturbation theory and included a Debye mass in the longitudinal part of the gluon propagator: $m_g^2= (16/5) T^2$.
This kernel produces coincident second-order deconfinement and chiral symmetry restoring transitions for two massless flavors at 
\begin{eqnarray}
\label{Tcvalue}
T_c = 0.14\,{\rm GeV} \, ,
\end{eqnarray}
which is $10$\% smaller than that obtained in Ref.\,\cite{Aoki:2009sc}.  (As mentioned above, this interaction has only recently been superseded \cite{Qin:2011dd}.)

One insufficiency of the interaction defined above is that $D$, the parameter expressing its infrared strength, is assumed to be $T$-independent.  Since the nonperturbative part of the interaction should be screened for $T\gtrsim T_c$, one may remedy that by writing $D\to D(T)$ with
\begin{eqnarray}
D(T)=\left\{
\begin{array}{ll}
\displaystyle
D \,, &   T<T_{\text{p}} \, ,  \\
\displaystyle
\frac{a}{b+ \ln[T/\Lambda_{QCD}]}\,, &  T \ge
T_{\text{p}}
\end{array}
\right.\,, \label{DTfunction}
\end{eqnarray}
where $T_{\rm p}$ is a ``persistence'' temperature; i.e., a scale below which nonperturbative effects associated with confinement and dynamical chiral symmetry breaking are not materially influenced by thermal screening.  Logarithmic screening is typical of QCD and with $a=0.028$, $b=0.56$ the modified interaction yields $m_T = 0.8 \, T$ for $T \gtrsim 2\, T_c$; viz., a thermal quark mass consistent with lattice-QCD \cite{Karsch:2009tp}.  Unless stated otherwise, $T_{\rm p}=T_c$ herein.

The spectral density can be obtained by employing the MEM in connection with the solution of our gap equation.  Notably, the behaviour changes qualitatively at $T_c$.  Indeed, employing a straightforward generalisation of the inflexion point criterion introduced in Refs.\,\cite{Roberts:2007ji,Bashir:2008fk}, one can readily determine that reflection positivity is violated for $T<T_c$.  This signals confinement.  On the other hand, the spectral function is nonnegative for $T>T_c$.

\begin{figure}[t]
\centerline{\includegraphics[width=0.47\textwidth]{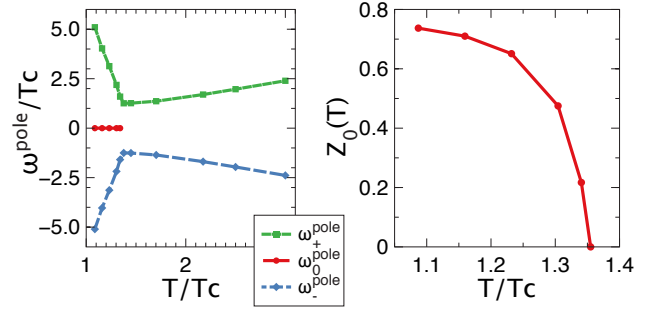}}
\caption{\label{fig:poles}
\emph{Left panel} -- Temperature-dependence of the dressed-quark thermal masses.  Notably, spectral strength is associated with a massless-mode.
\emph{Right panel} -- $T$-dependence of the residue associated with that zero mode.}
\end{figure}

Figure~\ref{fig:poles} depicts the $T>T_c$-dependence of the locations of the poles in $\rho(\omega,\vec{p}=0)$; i.e., the thermal masses.  As conventionally anticipated, spectral strength is located at $\omega_+(\vec{p}=0)$ and $\omega_-(\vec{p}=0) = - \omega_+(\vec{p}=0)$, corresponding to the fermion's normal and plasmino modes at nonzero temperature.  However, it is striking that on a measurable $T$-domain, spectral strength is also associated with a quasiparticle excitation described by $\omega_0(\vec{p}=0) = 0$.  The appearance of this zero mode is an essentially nonperturbative effect.  It is an outgrowth of the evolution in-medium of the gap equation's $T=0$ Wigner-mode solution and analogous to this solution's persistence at nonzero current-quark mass in vacuum \cite{Chang:2006bm}.  This feature is stable for sufficiently-large $\Lambda$ in Eq.\,(\ref{defaultm0}), as evident in Fig.\,\ref{fig:zero}.

\begin{figure}[t]
\centerline{\includegraphics[width=0.47\textwidth]{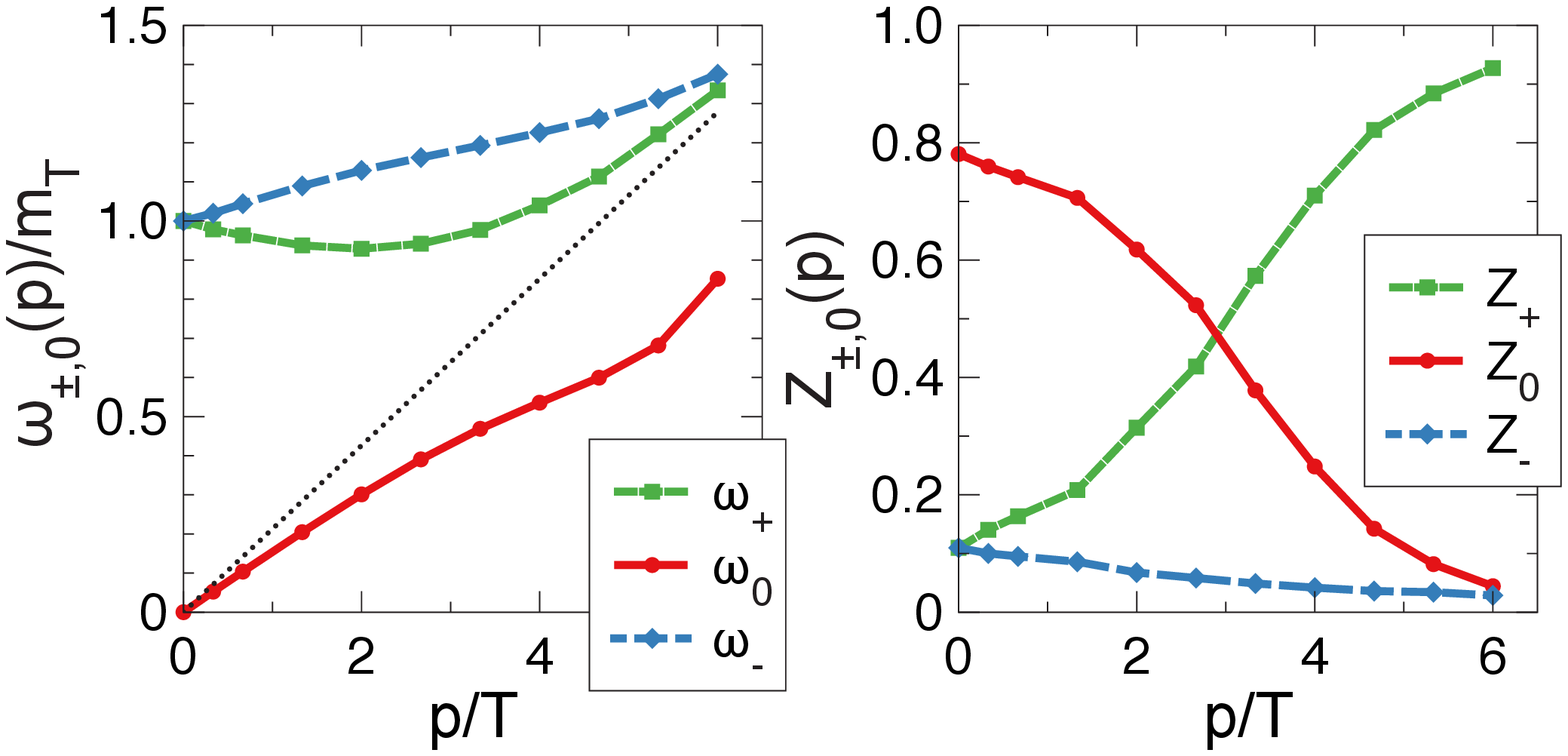}}
\centerline{\includegraphics[width=0.47\textwidth]{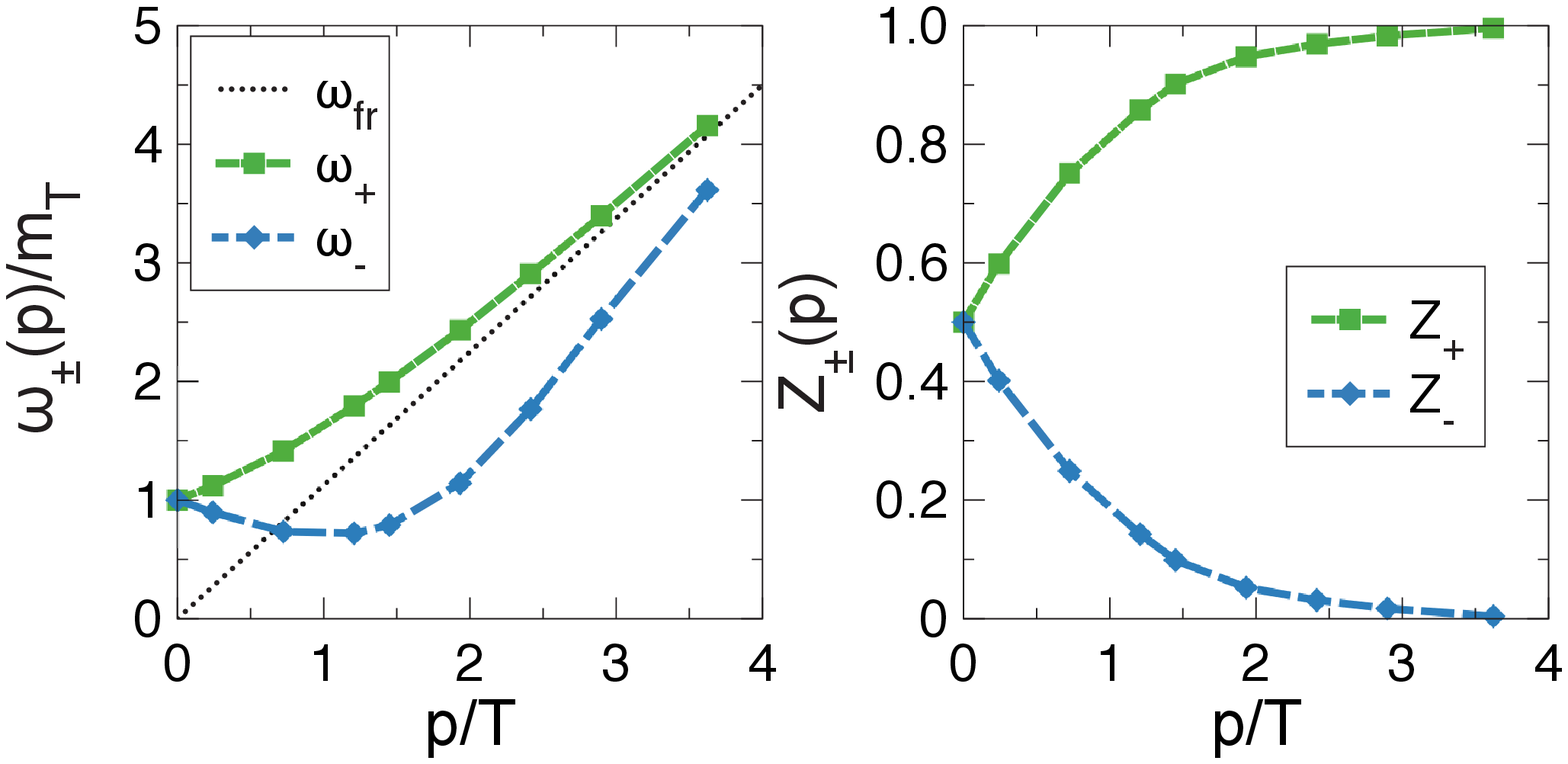}}
\caption{\label{fig:1.1Tc_rel}
\emph{Upper panel, left} -- Quasiparticle dispersion relations, $\omega_{\pm,0}(p)$, at temperature $T = 1.1 T_{c}$.  The diagonal dotted line is the free-fermion dispersion relation at this $T$.  \emph{Upper panel, right} -- The residues associated with these quasiparticle poles.
\emph{Lower panel} -- Same information for $T=3 T_c$, whereat the zero mode has vanished.}
\end{figure}

The spectral density possesses support associated with this zero mode on $T\in [0,T_s]$.  In fact: all the Wigner-phase spectral strength is located within this mode at $T=0$; it is the dominant contribution for $T\gtrsim T_c$; and, while it is dominant, it is the system's longest wavelength collective mode.
On the other hand, as evident in the right panel of Fig.\,\ref{fig:poles}, the mode's spectral strength diminishes uniformly with increasing $T$ and finally vanishes at $T_s \approx 1.35\,T_c$.  Then, for $T>T_s$ the quark's normal and plasmino modes exhibit behavior that is broadly consistent with HTL calculations.  This is apparent in Fig.\,\ref{fig:poles} and in a comparison between the upper and lower panels of Fig.\,\ref{fig:1.1Tc_rel}.
Given these observations, one may reasonably argue that the system should be considered a sQGP for $T\in [T_c,T_s]$, whereupon it contains a long-range collective mode.

It should be noted that the HTL approach is perturbative and only applicable for $T\gg T_c$.  Hence it could not have predicted the zero mode's existence.  Numerical simulations of lattice-QCD, on the other hand, are nonperturbative.  However, it is practically impossible in contemporary computations to exactly preserve chiral symmetry.  This can plausibly explain the absence of the zero mode in lattice simulations because any source of explicit chiral symmetry breaking heavily suppresses the mode \cite{Chang:2006bm}.

The upper-left panel of Fig.\,\ref{fig:1.1Tc_rel} depicts the dispersion relations for all dressed-quark modes that exist for $T<T_s$ and the behavior of their associated residues.  On this sQGP domain the dispersion relations are atypical, with
\begin{subequations}
\begin{eqnarray}
\omega_{\pm}(|\vec{p}|) &\stackrel{p\sim 0}{=}& m_T
\begin{array}{l}
-0.2\,|\vec{p}|\,,\\
+0.3\,|\vec{p}|\,,
\end{array}\\
%
%
%
\omega_{0}(|\vec{p}|) &\stackrel{p\sim 0}{=}& 0.80 \, |\vec{p}|\,.
\end{eqnarray}
\end{subequations}
Notwithstanding this, all realise free-particle behaviour for $|\vec{p}|\gg T$.  Moreover, the usual spectral sum rules are satisfied.  Indeed, the identity
\begin{equation}
\label{momrule}
\langle \omega \rangle := \frac{Z_2^2}{Z_2^A} \int_{-\infty}^\infty \frac{d\omega^\prime}{2\pi} \; \omega^\prime \rho_\pm(\omega^\prime,|\vec{p}|) = |\vec{p}|
\end{equation}
assists in understanding the momentum-dependence in the upper-left panel of Fig.\,\ref{fig:1.1Tc_rel}.  The upper-right panel displays the momentum-dependence of the pole residues: spectral support is located completely in the normal mode for $|\vec{p}|\gg T$; i.e., on the perturbative domain.

The lower panel of Fig.\,\ref{fig:1.1Tc_rel} characterises the behaviour of $\rho(\omega,|\vec{p}|)$ on the high-$T$ domain.  In agreement with HTL analysis \cite{Braaten:1990wp}, expected to be valid thereupon, one finds only normal and plasmino modes, with
\begin{equation}
\omega_{\pm}(|\vec{p}|) \stackrel{p\sim 0}{=} m_T \pm 0.33 |\vec{p}|\,.
\end{equation}
The plasmino dispersion law exhibits the expected minimum, in this case at $|\vec{p}|/T\simeq 1$; and both $\omega_{\pm}(|\vec{p}|)$ approach free-particle behaviour at $|\vec{p}| \gg T$, with that of the plasmino approaching this limit from below.  The lower-right panel shows that the contribution to the spectral density from the plasmino is strongly damped and contributes little for $p>2 T$.  These results are in-line with those obtained via simulations of lattice-QCD \cite{Karsch:2009tp}.

It is anticipated from HTL analyses that $T\neq 0$ propagators exhibit branch cuts whose appearance can be attributed to the opening of scattering channels that are absent at $T=0$ \cite{Weldon:2001vt}.  However, such branch cuts do not materially contribute to the nonperturbatively-determined spectral density described herein.  This is plausible because a branch point is a lower-order nonanalyticity than a pole; i.e., in numerical studies, poles are features with large height, small width and significant spectral strength, whilst branch points are low, broad features with lesser spectral strength.  Thus, compared with poles, branch points can be invisible to a numerical procedure.  Uncovering them requires fine tuning within the MEM, or any other method.

\begin{figure}[t]
\centerline{\includegraphics[width=0.45\textwidth]{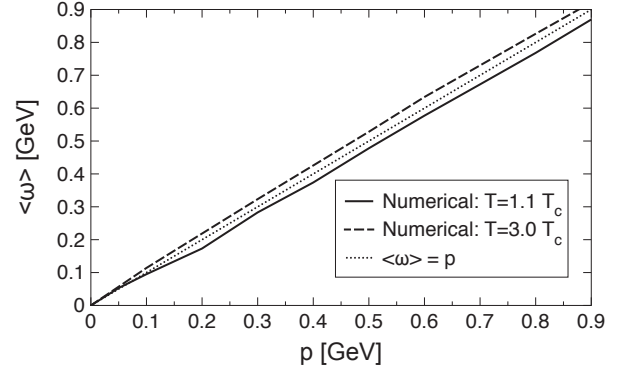}}
\caption{\label{fig:4}
Numerical check of the momentum sum rule in Eq.\,(\ref{momrule}): the curves would be indistinguishable with a perfect reconstruction of the spectral density.}
\end{figure}

One should nevertheless check whether the numerical result omits significant spectral strength, and this was done in Ref.\,\cite{Qin:2010pc}; e.g., it was verified that, to a high degree of accuracy, the spectral density satisfies the textbook relation:
\begin{equation}
Z_2\int_{-\infty}^{+\infty}\frac{d\omega'}{2\pi}~\rho_\pm(\omega',\text{p})\approx
Z_2\sum_{Q=0,+,-}Z_Q\approx 1
\end{equation}
where $Z_Q$ are the quasiparticle residues.  Other, more demanding sum rules, such as that in Eq.\,(\ref{momrule}), were also checked
with the result depicted in Fig.\,\ref{fig:4}.  These tests indicate that no significant contribution to the dressed-quark's spectral strength was overlooked in applying the MEM.

Equation~(\ref{DTfunction}) is a model and it is natural to enquire after its influence.  None of the results in this subsection are qualitatively altered by varying $T_{\rm p}$ but, as one would expect, the width of the sQGP domain expands slowly with increasing $T_{\rm p}$; e.g., a 50\% increase in $T_{\rm p}$ produces a 30\% increase in $T_s$.

The appearance of a third and long-wavelength mode in the dressed-fermion spectral density on a material temperature domain above $T_c$ has also been observed in one-loop computations of the fermion self-energy, irrespective of the nature of the boson which dresses the fermion \cite{Kitazawa:2005mp,Kitazawa:2006zi}.  This mode appears for $T>m_G$, where $m_G$ is an infrared mass-scale associated with the boson, and persists for fermion bare masses $\lesssim 0.2 m_G$ \cite{Kitazawa:2007ep}.  (NB.\ In the present case, $m_G \approx 0.17\,$GeV, hence $0.2\,m_G \approx 34\,$MeV, appreciably greater than the light-quark current-masses.)  Where a comparison is possible, the dependence of the spectral density described in this subsection on $(\omega,|\vec{p}|,T)$ is similar to that seen in the one-loop analyses of model gap equations.  In analogy with a similar effect in high-temperature superconductivity \cite{janko:1997}, that behaviour has been attributed to Landau damping, an interference phenomenon known from plasma physics.  Indeed, Landau damping is typical of in-medium self-energy corrections when the thermal energy of the fermion is commensurate with the mass-scale which characterises the dispersion law of the dressing boson.

A feature that unifies Ref.\,\cite{Qin:2010pc} and Refs.\,\cite{Kitazawa:2005mp,Kitazawa:2007ep} is the strictly nonperturbative phenomenon of DCSB.  Whilst Refs.\,\cite{Kitazawa:2005mp,Kitazawa:2007ep} extract spectral densities via one-loop estimates in Nambu--Jona-Lasinio- or Yukawa-like models, the couplings are tuned to mimic a world in which chiral symmetry is dynamically broken.  In this connection we re-emphasise that the zero mode appears only when the strength of the interaction is capable of producing DCSB in-vacuum; viz., when the gap equation's $T=0$ kernel has sufficient support at infrared momenta.  Hence, it is an essentially nonperturbative phenomenon.
Furthermore, given the qualitative differences between Refs.\,\cite{Kitazawa:2005mp,Kitazawa:2007ep} and Ref.\,\cite{Qin:2010pc}, the similarity between results suggests the possibility that the appearance of zero modes is model-independent.

It is notable that a coupling to meson-like correlations in the gap equation is not a precondition for the appearance of the zero mode because such correlations are absent in the rainbow truncation \cite{Holl:1998qs}.  On the other hand, the kernel of the gap equation used in this subsection is characterised by an interaction that features an infrared mass-scale $m_G \gtrsim T_c$ and supports dynamical chiral symmetry breaking at $T=0$.  We anticipate that the zero mode will markedly affect colour-singlet vacuum polarisations on $T\in [T_c,T_s]$.  This could be explicated using the methods of Refs.\,\cite{Chang:2008ec}.

Experiment has presented us with the fascinating possibility that a near-perfect fluid might have been a key platform in the universe's evolution.  This subsection reviews the possibility that a zero mode exists just above the critical temperature associated with the strong phase transition, $T_c$.  The mode contains the bulk of the spectral strength for $T\gtrsim T_c$ and so long this mode persists, the system may reasonably be described as a strongly-interacting state of matter.  If the existence of this long-wavelength mode is model-independent, then it is natural to anticipate that a strongly-interacting state of matter should precede the QCD phase transition.

\section{Epilogue}
\label{sec:Epilogue}
QCD is the most interesting part of the Standard Model and Nature's only example of an essentially nonperturbative fundamental theory.  Whilst confinement remains a puzzle, it is recognised that dynamical chiral symmetry breaking (DCSB) is a fact in QCD.  It is manifest in dressed-propagators and vertices, and, amongst other things, it is responsible for:
the transformation of the light current-quarks in QCD's Lagrangian into heavy constituent-like quarks, in terms of which order was first brought to the hadron spectrum;
the unnaturally small values of the masses of light-quark pseudoscalar mesons and the $\eta$-$\eta^\prime$ splitting;
the unnaturally strong coupling of pseudoscalar mesons to light-quarks -- $g_{\pi \bar q q} \approx 4.3$;
and the unnaturally strong coupling of pseudoscalar mesons to the lightest baryons -- $g_{\pi \bar N N} \approx 12.8 \approx 3 g_{\pi \bar q q}$.

Herein we have illustrated the dramatic impact that DCSB has upon observables in hadron physics.  A ``smoking gun'' for DCSB is the behaviour of the dressed-quark mass function.  The momentum dependence manifest in Fig.\,\ref{gluoncloud} is an essentially quantum field theoretical effect.  Exposing and elucidating its consequences therefore requires a nonperturbative and symmetry-preserving approach, where the latter means preserving Poincar\'e covariance, chiral and electromagnetic current-conservation, etc.  The Dyson-Schwinger equations (DSEs) provide such a framework.  We have explained the nature of some of the experimental and theoretical studies that are underway which can potentially identify observable signals of $M(p^2)$ and thereby confirm and explain the mechanism responsible for the vast bulk of visible mass in the Universe.

Along the way we have described a number of exact results proved in QCD using the DSEs, amongst them:
\begin{itemize}
\item Light-quark confinement is a dynamical phenomenon, which cannot in principle be expressed via a potential;
\item Goldstone's theorem is fundamentally an expression of equivalence between the one-body problem and the two-body problem in the pseudoscalar channel;
\item quarks are not Dirac particles -- they possess anomalous chromo- and electro-magnetic moments which are large at infrared momenta;
\item and gluons are nonperturbatively massive, being described by a mass-function which is large in the infrared but diminishes with power-law behaviour in the ultraviolet.
\end{itemize}
Numerous items could be added to this list, some of which are described above.

There are many reasons why this is an exciting time in hadron physics.  These lecture notes emphasise one.  Namely, through the DSEs, one is unifying phenomena as apparently diverse as: the hadron spectrum; hadron elastic and transition form factors, from small- to large-$Q^2$; parton distribution functions; the physics of hadrons containing one or more heavy quarks; and properties of the quark gluon plasma.  The key is an understanding of both the fundamental origin of visible mass and the far-reaching consequences of the mechanism responsible; namely, DCSB.  Through continuing feedback between experiment and theory, these studies should lead us to an explanation of confinement, the phenomenon that makes nonperturbative QCD the most interesting piece of the Standard Model.  They might also provide an understanding of nonperturbative physics that enables the formulation of a realistic extension of that model.


\begin{acknowledgments}
We are grateful to the organisers for the opportunity to be involved in this well conceived and executed conference on \emph{AdS/CFT and Novel Approaches to Hadron and Heavy Ion Physics}, for financial support from the Kavli Institute for Theoretical Physics China (KITPC), at the Chinese Academy of Sciences, which enabled our participation and, above all, for the kindness and hospitality of the KITPC staff.
We acknowledge valuable input from L.~X.~Guti\'errez-Guerrero, V.~Mokeev, M.~A.~Ivanov, A.~K{\i}z{\i}lers\"u, S.-x.~Qin, J.~Rodriguez-Quintero, S.\,M.~Schmidt and D.~J.~Wilson.
This work was supported by:
the Project of Knowledge Innovation Program of the Chinese Academy of Sciences, grant no.\ KJCX2.YW.W10;
Sistema Nacional de Investigadores;
CONACyT grant 46614-F;
the University of Adelaide and the Australian Research Council through grant no.~FL0992247;
Coordinaci\'on de la Investigaci\'on Cient\'ifica (UMSNH) grant 4.10;
the U.\,S.\ Department of Energy, Office of Nuclear Physics, contract no.~DE-AC02-06CH11357;
Funda\c{c}\~ao de Amparo \`a Pesquisa do Estado de S\~ao Paulo, grant nos.~2009/51296-1 and 2010/05772-3;
the National Natural Science Foundation of China, under contract nos.\ 10425521, 10675002, 10705002, 10935001 and 11075052;
the Major State Basic Research Development Program, under contract no.~G2007CB815000;
Forschungszentrum J\"ulich GmbH;
and the U.\,S.\ National Science Foundation, under grant no.\ PHY-0903991, in conjunction with a CONACyT Mexico-USA collaboration grant.
\end{acknowledgments}





\end{document}